\documentclass[12pt]{article}
\usepackage{amsmath}
\usepackage{graphicx}
\usepackage{enumerate}
\usepackage{natbib}
\usepackage{url} 
\usepackage{hyperref,color}
\usepackage{booktabs}
\usepackage{subcaption}  
\usepackage[linesnumbered,ruled,vlined]{algorithm2e}
\usepackage{setspace}

\begin{document}

\def\spacingset#1{\renewcommand{\baselinestretch}%
{#1}\small\normalsize} \spacingset{1}

%%%%%%%%%%%%%%%%%%%%%%%%%%%%%%%%%%%%%%%%%%%%%%%%%%%%%%%%%%%%%%%%%%%%%%%%%%%%%%

  \title{\bf A flexible Bayesian framework for detecting cross-sample spatial expression variability in heterogeneous tissues}
    \author{
        Meng Zhou \\
        School of Statistics and Data Science, \\
        Shanghai University of Finance and Economics
        \and
        Shuangge Ma \\
        Department of Biostatistics, Yale University
        \and
        Mengyun Wu \thanks{
  		The corresponding author, wu.mengyun@mail.shufe.edu.cn}\\
        School of Statistics and Data Science, \\
        Shanghai University of Finance and Economics
    }
     \date{}
  \maketitle

\bigskip
\begin{abstract}

Spatial transcriptomics measures gene expression alongside the spatial coordinates of each capture spot or cell across tissue samples. The detection of spatially variable (SV) genes, whose expression exhibits systematic spatial variation, enables the delineation of functional tissue domains and the identification of region‑specific transcriptional changes that underlie disease heterogeneity. However, empirical evidence from spatial transcriptomics data indicates that existing methods are hindered by two practical issues: reliance on predefined spatial patterns that often miss tissue complexity, and a lack of standardized approaches for multi‑sample integration, which undermines reproducibility and biological interpretability. To address these issues, we propose a novel integrated Bayesian hierarchical model that combines flexible nonparametric spatial modeling with information sharing across samples. The model uses an adaptive spatial process that can capture a wide range of spatial patterns while remaining interpretable. We also introduce a new prior that borrows strength across samples, enabling robust detection of SV genes from multiple tissue sections. An efficient variational approximation is developed for scalable posterior computation. Analyzing spatial transcriptomics data from human brain and skin cancer tissues, our framework identifies spatially structured SV genes, enabling the delineation of tissue domains and the discovery of functionally coherent gene clusters through pathway enrichment analysis.

\end{abstract}

\noindent%
\vfill

\newpage
\spacingset{1.9} 

\section{Introduction}
\label{sec:intro}

Spatial transcriptomics measures gene expression and its spatial coordinates for each capture spot or cell within a tissue. A key challenge is identifying spatially variable (SV) genes, whose expression exhibits systematic spatial variation \citep{yan2025categorization}, as they help delineate functional tissue domains and reveal region-specific alterations associated with pathological conditions. In this study, we focus on two representative datasets: the human dorsolateral prefrontal cortex (DLPFC) and squamous cell carcinoma (SCC). In particular, the DLPFC is one of the most studied brain regions due to its role in cognition and executive function, and characterizing its spatial expression is key to mapping layered cortical architecture and identifying neuropsychiatric disorder signatures. Likewise, SCC exhibits pronounced spatial heterogeneity in its tumor microenvironment, where SV genes can help uncover region‑specific immune evasion and stromal interactions. However, current investigations predominantly rely on RNA sequencing or single‑nucleus multiome profiling \citep{Ma2022, Scott2023,liu2023single}, inherently discarding spatial context. These approaches thus miss the opportunity to directly model spatial dependencies, leading to loss of architectural information and potential oversight of disease‑relevant genes. This highlights the need for statistical methods that explicitly leverage spatial coordinates to identify SV genes.

A variety of statistical methods have been developed for SV gene detection. These approaches typically integrate joint modeling of gene expression patterns and spatial coordinates to detect genes exhibiting distinctive spatial expression profiles. They broadly fall into parametric, semiparametric, and nonparametric frameworks. Parametric methods, including Gaussian process-based models that assess spatial covariance (e.g., SpatialDE \citep{svensson2018spatialde}, SPARK \citep{sun2020statistical}, BOOST-GP \citep{li2021bayesian}, nnSVG \citep{weber2023nnsvg}), and mean-variation detectors that test for mean shifts (e.g., CTSV \citep{yu2022identification}, NABM \citep{wu2024joint}), rely on a common set of prespecified candidate spatial patterns (e.g., linear, focal, or periodic) and quantify the statistical evidence for the presence of such patterns. While statistically principled, these preset assumptions may not fully capture the intricate spatial organizations observed in real tissues. Specifically, the DLPFC dataset exhibits spatial expression trends that are not well represented by simple linear or focal patterns, with variation occurring across multiple anatomical scales {\citep[][Fig.~2 and Discussion]{HuukiMyers2024_PFC_spatial}}. Similarly, the SCC dataset comprises spatially distinct tumor compartments with heterogeneous expression profiles {\citep[][Fig.~3]{ji2020multimodal}}. Such complex spatial structures often exceed the flexibility of standard spatial patterns.

Semiparametric methods, such as spVC \citep{yu2024spvc}, typically incorporate nonparametric spatial smoothing (e.g., via bivariate penalized splines) within a parametric distributional framework. Nonparametric approaches like SPARKX \citep{zhu2021sparkx} and HEARTSVG \citep{yuan2024heartsvg} forgo explicit distributional assumptions to enhance sensitivity to complex spatial signals. Other methods, including GASTON \citep{Chitra2025Mapping} and StarTrail \citep{chen2026startrail}, model one-dimensional tissue structure using deep learning and nearest-neighbor Gaussian processes, respectively. While these methods alleviate the reliance on predefined spatial patterns to some extent, semiparametric and nonparametric approaches may suffer from limited interpretability and often lack a principled way to adjust for spot-level covariates (e.g., cell-type mixture proportions), which can lead to inflated false discoveries or confounding effects, whereas deep learning-based or graph-based methods face computational scaling challenges and dependence on accurately pre‑identified spatial domains. These challenges, together with the complex spatial structures observed in real tissues, motivate the central question of our study: how to develop a flexible, data-adaptive framework that does not presuppose specific spatial forms, while maintaining interpretability and systematically accommodating covariate adjustment.

Most existing methods are designed for single-sample analysis, yet multi-section data are now routine in spatial transcriptomics. For instance, the DLPFC dataset comprises multiple sections per individual, and the SCC dataset includes comparable regions across specimens, with both shared structural features and distinct spatial expression patterns across sections, reflecting the fact that these sections are either as closely spaced slices from a single tissue (with defined spacing intervals) or as comparable regions from different individuals, retaining biological congruence while exhibiting inter‑section variability {\citetext{\citealp[][Figs.~2, 4]{Arora2023_ST_tumor_architecture};
\citealp[][Extended Data Fig.~2]{Maynard2021}}}. Analyzing a single arbitrarily selected section can yield highly variable results: our evaluation of seven established methods on the DLPFC and SCC datasets reveals low consensus (less than 50\% overlap) in detected SV genes across sections {(Figures S1--S3)}. Although some multi-sample approaches have been developed, they have important limitations: PASTE \citep{Zeira2022} integrates multiple layers via low-rank matrix factorization but performs integration and detection separately, which may propagate biases or information loss; DESpace \citep{cai2024despace} adopts a linear mean assumption and does not explicitly model spatial coordinates, which may restrict its ability to capture complex spatial signals. These observations collectively motivate our second objective: to develop a unified framework for multi-sample integrative analysis that enables coherent information sharing across sections while preserving section-specific spatial signals, thereby supporting reproducible and biologically interpretable identification of SV genes.

\begin{figure}[!ht]
    \centering
    \includegraphics[width=0.9\linewidth]{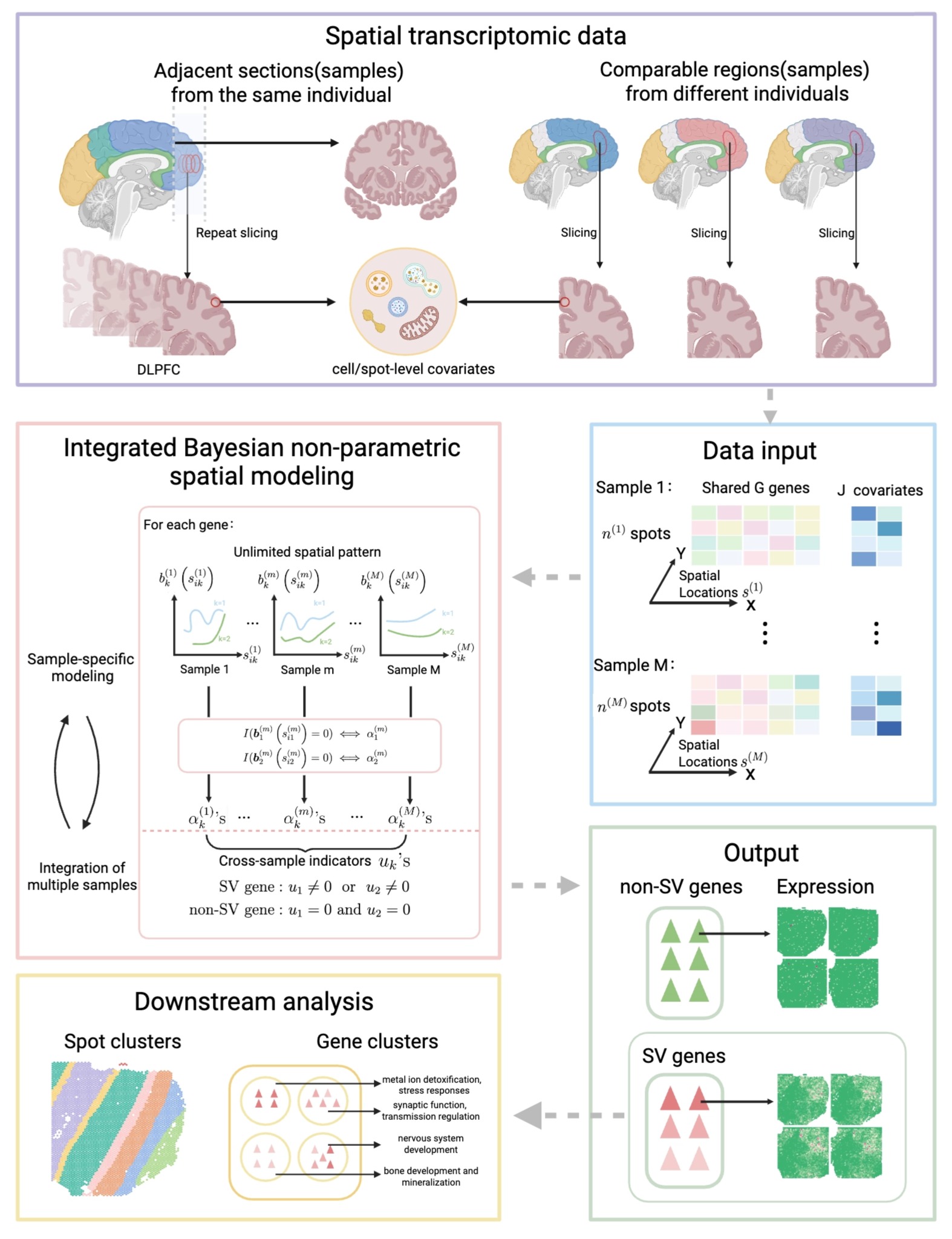}
    \caption{Workflow of the proposed method.}
    \label{fig:workflow}
\end{figure}

In this study, we propose an integrated Bayesian nonparametric spatial model for cross-sample SV gene identification (Figure~\ref{fig:workflow}). Built upon a zero-inflated negative binomial likelihood, our approach eliminates reliance on predefined spatial patterns and incorporates spot-level covariates to adjust for confounding. The core innovation is a hierarchical architecture with a bi-level shrinkage prior, which enables principled information sharing across tissue sections, preserving biological coherence while suppressing technical noise. To ensure scalability, we develop a variational inference algorithm that markedly accelerates computation relative to MCMC. Extensive simulations demonstrate that our method consistently achieves higher F1 scores than state-of-the-art alternatives. Analyzing the DLPFC and SCC datasets, our framework identifies biologically interpretable SV genes that enable tissue domain delineation and form functionally coherent clusters via pathway enrichment analysis. Collectively, this work offers a statistically principled and computationally efficient framework for integrative spatial transcriptomics analysis.

\section{Data description and scientific motivation}

We analyze three spatial transcriptomic scenarios derived from two biologically and technically distinct human datasets. These scenarios differ in tissue organization, spatial resolution, sample composition, and biological context. Specifically, the datasets include:

\begin{itemize}
\item The human dorsolateral prefrontal cortex (DLPFC) dataset \citep{Maynard2021} exhibits well-defined cortical laminar structures. Identifying genes associated with these layers helps characterize the molecular organization of the human cortex and may provide insight into neurological disorders in which layer-specific expression programs are disrupted. We examine the DLPFC data under two scenarios: multiple tissue sections from the same donor and tissue sections from different donors. 
\item The human squamous cell carcinoma (SCC) dataset \citep{ji2020multimodal} is a representative spatial transcriptomic dataset of tumor tissue. Identifying SV genes in SCC helps characterize the spatial organization of the tumor microenvironment and reveals molecular features associated with tumor progression, such as immune exclusion, hypoxic regions, and invasion fronts, providing insights into potential therapeutic targets and disease intervention strategies. Unlike the layered structure of the DLPFC, the SCC dataset provides a complementary setting for studying spatial expression patterns in tumor tissues.
\end{itemize}

Together, these datasets represent spatial transcriptomic studies with markedly different tissue architectures and spatial resolutions, with spot counts ranging from approximately 1,000 in SCC to an average of 4,000 in DLPFC. They also exhibit substantial zero inflation, reflecting the sparsity commonly observed in spatial transcriptomic measurements. Although we focus on these two specific datasets, the data characteristics they exhibit, namely complex spatial structures, zero inflation, and multi‑section heterogeneity, are representative of those encountered across a wide range of spatial transcriptomic applications. Additional details are provided in Supplementary Sections S2 and S3.

These data motivate two substantive scientific questions:

\begin{itemize}
    \item [(Q1)] Which genes characterize biologically important spatial structures, such as cortical layers in the DLPFC and distinct regions of the SCC tumor microenvironment?
    \item [(Q2)] Which spatially variable genes can be reproducibly identified across multiple tissue sections or donors, rather than being driven by noise or variation in individual samples?
\end{itemize}

In current spatial transcriptomic data analyses, answering these questions remains challenging with existing methods. First, many methods rely on predefined spatial patterns and may therefore miss genes associated with irregular tissue structures, limiting biological insight into complex tissues (Q1). In addition, commonly used approaches either analyze tissue sections separately or pool them without accounting for sample-to-sample heterogeneity, resulting in unstable gene discoveries and making it difficult to distinguish reproducible spatial patterns from sample-specific noise (Q2).

To address these challenges, we develop IBaySVG, a multi-sample Bayesian framework for identifying SV genes. IBaySVG combines a nonparametric representation to flexibly model diverse spatial expression patterns with a hierarchical bi-level technique that integrates information across tissue sections while accommodating sample-specific variation. Together, these components enable more reliable identification of biologically meaningful and reproducible spatial expression patterns.

\section{Methods} 
\label{sec:meth}
For a specific tissue, consider $M$ samples (e.g., either adjacent sections from the same individual or comparable regions from different individuals), each containing $n^{(m)}$ spots with $m=1,\cdots,M$. For each spot, measurements from the $G$ genes are observed. Specifically, for the spot $i$ in the $m$th sample, let $\boldsymbol{s}_i^{(m)}=\left(s_{i1}^{(m)},s_{i2}^{(m)}\right)$ represent its two-dimensional coordinate, and $y_{ig}^{(m)}$ denote the raw count of the gene $g~(g=1,\cdots, G)$ which frequently exhibits high zero inflation due to technical artifacts such as dropout events. When available, spot‑specific covariates, such as cell‑type proportions, cell‑state indicators, or regulatory‑factor activities, are included as a $J$ -dimensional vector $\boldsymbol{x}_i^{(m)} =\left(x_{i1}^{(m)},\cdots, x_{iJ}^{(m)}\right)^{\top}$.

Our exploratory analysis of the DLPFC and SCC datasets, along with similar spatial transcriptomics data, highlights three key characteristics (Section S3): sparse and overdispersed expression, complex and diverse spatial architectures, and shared yet heterogeneous cross-sample patterns. To simultaneously address (Q1) and (Q2) while robustly accounting for these features, we propose a zero-inflated negative binomial (ZINB) framework with nonparametric spatial modeling (we omit the gene index $g$ for notational clarity):
\begin{eqnarray}
\nonumber &&y_{i}^{(m)}\Big|\boldsymbol{x}_i^{(m)},\boldsymbol{s}_i^{(m)},{\zeta}^{(m)}\sim
	\pi^{(m)}\delta_0+\left(1-\pi^{(m)}\right)NB\left( \lambda_{i}^{(m)},\phi^{(m)}\right),\\
&&\text{with}~ \log\lambda_{i}^{(m)}=\eta^{(m)}+b_{1}^{(m)}\left(s_{i1}^{(m)}\right)+b_{2}^{(m)}\left(s_{i2}^{(m)}\right)+\left(\boldsymbol{x}_i^{(m)}\right)^{\top}\boldsymbol{\psi}^{(m)}.
\label{model1}
\end{eqnarray}
Here, $\delta_0$ denotes the Dirac delta function with point mass at zero, and $NB\left(\lambda_{i}^{(m)},\phi^{(m)}\right)$ represents a negative binomial (NB) distribution with mean parameter $\lambda_{i}^{(m)}$ and dispersion parameter $\phi^{(m)}$. $\pi^{(m)}$ quantifies the dropout-zero probability. For the analyzed gene in sample $m$, $\eta^{(m)}$ captures baseline expression levels accounting for batch effects and inter-sample heterogeneity, and $b_{1}^{(m)}\left(s_{i1}^{(m)}\right)$ and $b_{2}^{(m)}\left(s_{i2}^{(m)}\right)$ are nonparametric spatial effect functions. $\boldsymbol{\psi}^{(m)}=\left(\psi_{1}^{(m)},\cdots,\psi_{J}^{(m)}\right)^{\top}$ captures the covariate effects when covariate information is available. If such covariates are absent or not required for a given analysis, this term and its associated prior specification can be omitted without affecting the core inference procedure. The full parameter set for the $m$th sample is denoted by ${\zeta}^{(m)}$, which includes $\pi^{(m)},\phi^{(m)}$, $\eta^{(m)}$ and $\boldsymbol{\psi}^{(m)}$ (when applicable).

In model (\ref{model1}), we adopt the ZINB model. While it has been noted that zero-inflation modeling may not be universally necessary for spatial transcriptomic data \citep{zhao2022modeling}, the ZINB framework effectively accommodates two key characteristics of such data: its inherent count nature with over-dispersion and the prevalence of excess zeros often linked to technical dropout. Moreover, likelihood-ratio tests on the real datasets provide empirical support for including the zero-inflation component (Figure S34).
The parametric component $\boldsymbol{\psi}^{(m)}$ systematically adjusts for technical and biological confounders encoded in the covariate vector $x_{ij}^{(m)}$. Unlike prior methodologies that confine themselves to a limited set of pre-defined spatial structures (such as linear, focal, and periodic patterns), we introduce functions $b_{1}^{(m)}\left(s_{i1}^{(m)}\right)$ and $b_{2}^{(m)}\left(s_{i2}^{(m)}\right)$ without the constraint of formulaic assumptions. This functional representation enables data-adaptive modeling of complex spatial expression topographies while maintaining computational tractability through mean parameterization. Genes exhibiting null spatial effects with both $b_{1}^{(m)}\left(\cdot\right)=0$ and $b_{2}^{(m)}\left(\cdot\right)=0$ are probabilistically identified as non-spatially varying within sample $m$. %Zhao2025SPARROW,

\subsection{Integrated Bayesian model for cross-sample identification of SV genes}
\label{sec:meth:prior}

Based on (\ref{model1}), we first perform a basis expansion as:
$b_{1}^{(m)}\left(s_{i1}^{(m)}\right)=\left(\boldsymbol{\beta}_1^{(m)}\right)^{\top}\boldsymbol{\xi}\left(s_{i1}^{(m)}\right)$ and $b_{2}^{(m)}\left(s_{i2}^{(m)}\right)=\left(\boldsymbol{\beta}_2^{(m)}\right)^{\top}\boldsymbol{\xi}\left(s_{i2}^{(m)}\right)$, where $\boldsymbol{\xi}(\cdot)=\left(\xi_1(\cdot),\cdots,\xi_L(\cdot)\right)^{\top}$ is an $L$-dimensional basis function vector, and $\boldsymbol{\beta}_1^{(m)}=\left(\beta_{11}^{(m)},\cdots,\beta_{L1}^{(m)}\right)^{\top}$ and $\boldsymbol{\beta}_2^{(m)}=\left(\beta_{12}^{(m)},\cdots,\beta_{L2}^{(m)}\right)^{\top}$ are the corresponding coefficient vectors. As such, for $k=1, 2$, the spatial effect $b_{k}^{(m)}\left(\cdot\right)=0$ implies that the coefficient vector $\boldsymbol{\beta}_k^{(m)}=0$.
%This strategy is perhaps the most common technique for function approximation.

Denote $\Theta^{(m)}$ as the parameter set consisting of $\eta^{(m)}$, $\boldsymbol{\beta}_1^{(m)}$, $\boldsymbol{\beta}_2^{(m)}$, $\boldsymbol{\psi}^{(m)}$, and $\phi^{(m)}$. Then, to enable cross-sample integration and robust detection of SV genes that are shared across sections, we propose the integrated Bayesian model for cross-sample identification of SV genes (IBaySVG) as: for $m=1,\cdots,M$ and $k=1, 2$,
\begin{eqnarray}
 &&\nonumber y_{i}^{(m)}\Big|\boldsymbol{x}_i^{(m)},\boldsymbol{s}_i^{(m)},r_{i}^{(m)},\Theta^{(m)}\sim\delta_0^{r_{i}^{(m)}}NB\left(\lambda_i^{(m)},\phi^{(m)}\right)^{1-r_{i}^{(m)}},\\ &&\text{with~}\log\lambda_i^{(m)}=\eta^{(m)}+\sum_{k=1}^2\left(\boldsymbol{\beta}_k^{(m)}\right)^{\top}\boldsymbol{\xi}\left(s_{ik}^{(m)}\right)+\left(\boldsymbol{x}_i^{(m)}\right)^{\top}\boldsymbol{\psi}^{(m)},\label{model2}\\
\nonumber     &&r_i^{(m)}\sim Bern\left(\pi^{(m)}\right),~\pi^{(m)} \sim Beta\left(a_{\pi}^{(m)},b_{\pi}^{(m)}\right),\\
\nonumber &&\eta^{(m)} \sim N\left(0,\sigma_{\eta }^{2 (m)}\right),~\boldsymbol{\psi^{(m)}} \sim N\left(0,\sigma^{2(m)}_{\psi}\boldsymbol{I}\right),~\phi^{(m)} \sim Ga\left(a_{\phi}^{(m)},b_{\phi}^{(m)}\right), \\
&&\boldsymbol{\beta}_{k}^{(m)}|\sigma^{2{(m)}}_{k},\alpha_{k}^{(m)} \sim N\left(0,\sigma^{2 {(m)}}_{k}\boldsymbol{I}\right)^{\alpha_{k}^{(m)}}N(0,\Gamma_1^{2}\boldsymbol{I})^{1-\alpha_{k}^{(m)}}, \label{model3} \\
&& \sigma^{2(m)}_{k}|a_{k}^{(m) }\sim IG\left(\frac{1}{2},\frac{1}{a_{k}^{(m)}}\right),a_{k}^{(m)} \sim IG\left(\frac{1}{2},\frac{1}{A_k^2}\right), \label{model4}
\end{eqnarray}
\begin{eqnarray}
  &&\alpha_{k}^{(m)}|u_{k},q_{k} \sim Bern(q_{k})^{u_{k}}Bern(\Gamma_2)^{1-u_{k}}, \hspace{5cm} \label{model5}\\
 \nonumber       &&u_{k}|p_{k} \sim Bern(p_{k}), p_{k} \sim Beta(c_p,d_p),q_{k} \sim Beta(c_q,d_q),
\end{eqnarray}
with $\boldsymbol{I}$ being an identity matrix with potential different dimensions. 

To address zero-inflation, we introduce an indicator variable $r_{i}^{(m)}$, where $r_{i}^{(m)}=1$ indicates that $y_{i}^{(m)}$ is from the Dirac probability measure, and otherwise $y_{i}^{(m)}$ is from the NB distribution. A Bernoulli prior is assigned for $r_{i}^{(m)}$ with the hyperparameter $\pi^{(m)} \sim Beta\left(a_{\pi}^{(m)},b_{\pi}^{(m)}\right)$. Gaussian priors are assumed for $\eta^{(m)}$ and $\psi_j^{(m)}$ and a Gamma distribution is assumed for the dispersion parameter $\phi^{(m)}$.

To integrate multiple samples for SV gene identification, for each $\boldsymbol{\beta}_{k}^{(m)}$, we first innovatively introduce a sample-specific indicator variable $\alpha_{k}^{(m)}$ and also a sample-common indicator $u_{k}$, which indicates the spatial variability of the gene within the $m$th sample and the overall spatial variability across all the samples. Then, a bi-level shrinkage prior is introduced, which includes a group spike and slab Gaussian prior (\ref{model3}) for $\boldsymbol{\beta}_{k}^{(m)}$ based on $\alpha_{k}^{(m)}$ and a mixture of Bernoulli prior (\ref{model5}) for $\alpha_{k}^{(m)}$ based on $u_{k}$. Specifically, in (\ref{model3}) and (\ref{model4}), the standard deviation $\sigma^{{(m)}}_{k}$ in the slab part is further assigned a $\mathrm{Half\text{-}Cauchy}(A_k)$ prior and the variance $\Gamma_1^2$ in the spike part is a small constant close to zero. Here, $\mathrm{Half\text{-}Cauchy}(A_k)$ prior is the marginal distribution of the 
defined double Inverse-Gamma prior $\sigma^{2(m)}_{k}|a_{k}^{(m) }\sim IG\left(\frac{1}{2},\frac{1}{a_{k}^{(m)}}\right)$ and $a_{k}^{(m)} \sim IG\left(\frac{1}{2},\frac{1}{A_k^2}\right)$ in (\ref{model4}), which has been advanced by its robustness in parameter estimation and weak informativeness. With (\ref{model3}) and (\ref{model4}), when $\alpha_{k}^{(m)}=0$, the whole vector $\boldsymbol{\beta}_{k}^{(m)}$ will be shrunk towards zero with a high probability, and otherwise $\boldsymbol{\beta}_{k}^{(m)}\neq 0$. This strategy can effectively explore the sample-specific spatial effects of genes. Further, with (\ref{model5}) where $\Gamma_2$ is close to zero, all $\alpha_{k}^{(m)}$'s will be zero with a high probability when $u_{k}=0$, promoting consistent sparsity patterns across samples and enabling efficient information sharing. Benefiting from the Bayesian framework with uncertainty, the proposed model can effectively accommodate both the variability within the $m$th sample and that across samples. 

\subsection{Bayesian posterior inference}
\label{sec:meth:inference}
We first reorganize (\ref{model2}) as: $y_{i}^{(m)}|g_{i}^{(m)},r_{i}^{(m)} \sim  \delta_0^{r_{i}^{(m)}}Pois\left(g_{i}^{(m)}\right)^{1-r_{i}^{(m)}}$ and $g_{i}^{(m)}|\boldsymbol{x}_i^{(m)},\boldsymbol{s}_i^{(m)},\Theta^{(m)} \sim$ $Ga\left(\phi^{(m)},\phi^{(m)}\exp\left(-\eta^{(m)}-\sum_{k=1}^2\left(\boldsymbol{\beta}_k^{(m)}\right)^{\top}\boldsymbol{\xi}\left(s_{ik}^{(m)}\right)-\left(\boldsymbol{x}_i^{(m)}\right)^{\top}\boldsymbol{\psi}^{(m)}\right)\right)$, where the NB distribution is written as a Poisson distribution with a Gamma prior.

Denote the model parameter space as $\Omega=\left\{\Theta^{(m)}\text{'s}, r^{(m)}_i\text{'s}, g^{(m)}_i\text{'s}, \sigma^{(m)}_k\text{'s}, \alpha_k^{(m)}\text{'s}, a_k^{(m)}\text{'s}, u_k\text{'s},\right.$ $\left.q_k\text{'s}, p_k\text{'s}\right\}$, and $\boldsymbol{Y}$, $\boldsymbol{X}$ and $\boldsymbol{S}$ as the sets consisting of all $y_i^{(m)}$'s, $x_i^{(m)}$'s and $s_i^{(m)}$'s, respectively. Posterior inference is conducted via variational approximation, whereby the posterior distribution $p\left(\Omega \mid \boldsymbol{Y}, \boldsymbol{X}, \boldsymbol{S}\right)$ is approximated by a variational distribution $q(\Omega)$. Unlike MCMC, variational inference is computationally efficient for high-dimensional parameters. We minimize the Kullback-Leibler divergence between $q(\Omega)$ and the true posterior, equivalent to maximizing the Evidence Lower Bound (ELBO): $ELBO(q)=E_{q\left(\Omega\right)}\left(\log\left(p\left(\Omega,\boldsymbol{Y}\mid \boldsymbol{X}, \boldsymbol{S}\right)\right)\right)-E_{q\left(\Omega\right)}\left(\log\left(q(\Omega\right)\right)$, where $E_{q(\Omega)}$ denotes expectation under $q(\Omega)$.

We utilize the mean-field method to estimate $q(\Omega)$ as:
$q(\Omega)=\left\{\prod_{m=1}^{M}\left\{\prod_{i=1}^{n^{(m)}}q\left(g_i^{(m)}\right) \right.\right.$ $\left.q\left(r_{i}^{(m)}\right)\right\}
	q\left({\boldsymbol{\theta}^{(m)}}\right)q\left(\phi^{(m)}\right)\prod_{k=1}^{2}q\left(\sigma_{k}^{{(m)}}\right)
q\left(a_{k}^{(m)}\right)\left.q\left(\alpha_k^{(m)}\right)\right\}\prod_{k=1}^{2}q\left(u_k\right)q\left(p_k\right)q\left(q_k\right)$, where $\boldsymbol{\theta}^{(m)}=\left(\eta^{(m)},\left(\boldsymbol{\beta}_1^{(m)}\right)^{\top},\left(\boldsymbol{\beta}_2^{(m)}\right)^{\top},\left(\boldsymbol{\psi}^{(m)}\right)^{\top}\right)^{\top}$.
Then, for any $\tau \in \Omega$, maximizing ELBO gives the optimal variational distribution: $q^{*}(\tau) \propto \exp\left\{ E_{q(-\tau)} \log p\left(\tau \mid \Omega_{-\tau},\boldsymbol{Y}, \boldsymbol{X}, \boldsymbol{S}\right)\right\}$, where $\Omega_{-\tau}$ means $\Omega$ with $\tau$ excluded and $E_{q(-\tau)}$ denotes expectation with respect to the variational distribution $q(\cdot)$ of all parameters except $\tau$. For most parameters, the chosen priors are conjugate to the likelihood. Consequently, their optimal variational updates $q^{*}(\cdot)$
admit standard closed-form parametric distributions (e.g., Gamma, Beta, Bernoulli, and inverse-Gamma). In contrast, the variational density 
 $q^{*}(\phi^{(m)})$ for the dispersion parameter is not a common distribution and needs to be solved using numerical integration, such as the Gaussian quadrature technique. In addition, computation of $q^{*}\left(\boldsymbol{\theta}^{(m)}\right)$ is more challenging since it entails multivariate integrals that cannot be articulated in closed form due to the non-conjugate priors. Here, following \cite{knowles2011nonconjugate}, we apply the non-conjugate variational message passing approach to approximate the optimal variational density $q^{*}\left(\boldsymbol{\theta}^{(m)}\right)$ as a multivariate Gaussian distribution $N\left(\boldsymbol{\mu}_{\theta^{(m)}},\boldsymbol{\Sigma}_{\theta^{(m)}}\right)$ with $\boldsymbol{\mu}_{\theta^{(m)}}$ being a $(2L+1+J)$-dimension mean vector and $\boldsymbol{\Sigma}_{\theta^{(m)}}$ being a $(2L+1+J)\times(2L+1+J)$ covariance matrix. Details can be seen in {Section S4 of the Supplementary Materials.}

Using the variational distributions, we implement coordinate ascent variational inference (CAVI) to approximate the posterior. The algorithm iteratively updates each variational factor while holding others fixed, cycling through all parameters until convergence (ELBO change $<10^{-2}$). For functional approximation, we use B-spline basis functions for $\boldsymbol{\xi}(\cdot)$, a well-established approach. Full derivations and hyperparameter selection strategies are provided in {Supplementary Section S4}. %(\ref{variational density1}) and $q^{*}\left(\boldsymbol{\theta}^{(m)}\right)$

To identify SV genes, we compute the posterior expectations $E\left(u_k\right)$'s from the converged variational distributions $q^{*}\left(u_k\right)$'s. For each gene $g$, we define $\tilde{u}_g=\max\left(E\left(u_1\right),E\left(u_2\right)\right)$ and select SV genes as $\{g \in \{1,\dots,G\} : \tilde{u}_g \geq u_0\}$, where $u_0$ is determined by a Bayesian FDR procedure (see {Supplementary Section S4} for more details). The computational complexity of our variational inference algorithm is
$\mathcal{O}\Big(\text{max\_iter} \times M \times \bar{n} \times (1 + 2L + J)^2 \times G \Big)$ where \(\bar{n}\) is the average number of spatial locations per sample. This complexity is linear in the number of spatial locations and genes, quadratic in the combined spline basis and covariate dimension, and linear in the number of samples and iterations. Table S4 reports runtime on simulated data with varying spot counts and dropout rates using a 12-core Intel Xeon Gold CPU and 128 GB memory. IBaySVG integrates four spatial slices ($\sim$40,000 spots each) in $\sim$0.4 minutes per gene, demonstrating scalability for large-scale analyses. Parallel computing can further enhance multi-gene efficiency.

\subsection{Simulation}
\label{sec:simu}

We establish a systematic simulation framework to benchmark methods using spatially resolved transcriptomic data. The framework considers 4 tissue samples, each comprising 1,024 spots and 5,000 genes, including 500 SV genes exhibiting linear, focal, or periodic patterns. Scenarios vary the signal strength, degree of cross-sample heterogeneity, and dropout rate, giving 36 core settings.

We compare the IBaySVG method with nine existing approaches: single‑sample methods SPARK \citep{sun2020statistical}, SPARKX \citep{zhu2021sparkx}, HEARTSVG \citep{yuan2024heartsvg}, nnSVG \citep{weber2023nnsvg}, spVC \citep{yu2024spvc}, StarTrail \citep{chen2026startrail} and GASTON \citep{Chitra2025Mapping}; and multi‑sample integration methods PASTE \citep{Zeira2022} and DESpace \citep{cai2024despace}. These encompass parametric (SPARK, nnSVG, DESpace), semiparametric (spVC), and nonparametric (SPARKX, HEARTSVG) frameworks, with SPARKX, nnSVG and spVC additionally supporting covariate adjustment. For multi-slice analysis, each single-sample method is first applied independently to multiple tissue slices. The resulting per-slice outputs are then integrated using four late-integration strategies: ``Union'' (aggregating SV genes across slices), ``Inter'' (identifying consensus SV genes), ``Cauchy'' (via the Cauchy combination test \citep{liu2019acat}), and ``HMP'' (via the harmonic mean p-value method \citep{Wilson2019}). Because StarTrail and GASTON do not produce p-values, they are evaluated only under the Union and Inter strategies. Additionally, the single‑sample methods are run separately on the composite sample generated by PASTE. Detailed descriptions and implementations of all compared methods are provided in {Supplementary Section S5}.

Evaluation shows that our proposed method consistently achieves higher F1 scores and greater robustness compared to existing single-sample and integrative strategies. Furthermore, we extend the evaluation across a wider range of spatial configurations and under conditions without covariate information, which continues to demonstrate the method’s superior efficiency in identifying spatially variable structures. Complete simulation specifications, sensitivity analysis, all corresponding results, and detailed discussions are provided in {Supplementary Section S6.}

\section{Analysis of DLPFC and SCC data}
\label{sec:conc}

We analyze the DLPFC (same‑donor and cross‑donor sections) and SCC datasets described in Section 2 using the proposed method to address questions (Q1)-(Q2). In these analyses, we incorporate estimated cell‑type proportions as spot‑specific covariates, generated using the deconvolution tool Redeconve \citep{zhou2023redeconve}, a method validated for spatial transcriptomics. 

For the DLPFC data, we follow the annotation in \citet{Maynard2021} to derive cell‑type proportions for seven neuronal and glial classes (Astro, EndoMural, MicroOligo, Oligo, OPC, Excit, Inhib) from the associated snRNA‑seq reference. {As shown in Figure S30 (Supplementary Materials)}, the estimated proportions reflect expected spatial patterns: sections A \& B and C \& D, which are only 10 $\mu m$ apart, display more similar cellular distributions than the more distant pair B \& C (300 $\mu m$ apart). When integrating slices from different donors {(Figure S31)}, pronounced inter‑donor heterogeneity in cellular composition is evident, highlighting the challenge of cross‑donor integration. In the SCC dataset, we use a study‑matched scRNA‑seq reference (GEO GSE144236) comprising three major cell types: Keratinocyte, TSK, and Melanocyte, and similarly apply Redeconve to estimate spot‑wise proportions. The resulting composition {(Figure S32)} shows greater similarity between slices A and B than with slice C, consistent with their biological context. By treating these estimated cell‑type composition vectors as covariates, we explicitly account for cellular heterogeneity during the identification of SV genes.

We first perform outlier detection, model checks, stability analyses, and sensitivity analyses to evaluate model adequacy. Following the quality control procedures described in Section S2, only a few outliers are identified using randomized quantile residuals (Table S7 and Figure S33), with negligible impact on downstream results {(Table S8).} To preserve the integrity of the tissue sections, we retain the complete dataset for subsequent analyses. Likelihood ratio tests confirm the necessity of the zero-inflation component, with the ZINB model providing a significantly better fit than the standard NB model for most genes (Figure S34). Vuong tests indicate that the proposed nonparametric spatial structure offers a fit that is comparable to, or better than, common parametric forms (Tables S9--S11). Boundary analyses further demonstrate that the B-spline representation adequately captures spatial transitions without systematic loss of fit at tissue interfaces (Figures S35--S39). Algorithm stability, assessed via sample combination and spatial subsampling, is satisfactory and comparable to established parametric methods. Sensitivity analysis indicates that SV gene identification is robust across different hyperparameter ranges. Detailed methods, results, and discussions are provided in Supplementary Sections {S7.2--S7.5}.

Building on these diagnostic validations, we next turn to the biological interpretation of the identified SV genes. We benchmark our method against several state‑of‑the‑art approaches, highlighting SV genes uniquely detected by IBaySVG and examining their biological relevance through literature‑based evidence. We further perform gene clustering to reveal functionally coherent modules shared across tissue sections. In addition, we assess the utility of SV genes identified by competing methods for spatial domain identification, and separately investigate whether the uniquely selected SV genes from IBaySVG provide improved domain delineation.

\subsection{Benchmarking SV gene identification: the robust performance of IBaySVG}

As shown in {Figures S1--S3}, single-sample methods produce highly inconsistent results across slices, limiting interpretability and failing to address the need for cross-sample consistency. The identification results of different integration methods and their distinctions are presented in {Figures S43--S45}. Our method identifies 1,092, 1,590, and 1,051 SV genes across the three datasets, respectively, with counts consistently falling within the central range of the methods using union-based integration and substantially exceeding those from intersection-based integration. In contrast, single-sample methods show high inconsistency. Variations between intersection and union strategies are evident in the DLPFC data from the same donor (e.g., SPARKX-Inter: 509 vs. SPARKX-Union: 2,405; nnSVG-Inter: 96 vs. nnSVG-Union: 357). The gap widens markedly when different donors are included, for instance, SPARKX detects 650 (Inter) versus 4,030 (Union) genes. The Cauchy combination and HMP approaches consistently yield SV gene sets that lie between the intersection and union extremes, with Cauchy being more conservative than HMP. Integrative results using PASTE vary notably across datasets: in both same- and across-donor DLPFC datasets, nonparametric/semiparametric methods (SPARKX, HEARTSVG, spVC) identify $>$3,800 SV genes, whereas parametric ones (SPARK, nnSVG) find only (504, 86) and (770, 111); in SCC dataset, SPARKX again detects large numbers ($>$3,900), but spVC and HEARTSVG return only 104 and 256 SV genes, revealing that PASTE’s compatibility with single-slice methods depends strongly on the dataset. Furthermore, single-sample methods designed for the 1D structure of DLPFC tissue diverge sharply: StarTrail averages $<$100 SV genes per section, while GASTON exceeds 2,600. Across all datasets, DESpace identifies markedly higher SV gene counts (4,727, 4,094, and 3,269), far surpassing our method and most single-sample approaches.

For the DLPFC data, we further evaluate the performance of different methods in identifying layer-specific marker genes using established datasets, assessing whether cross-sample integration yields biologically interpretable and replicable SV genes. Following \citet{ma2022belayer}, we compile a reference set of marker genes by selecting the top 150 layer-enriched genes from \citet{Maynard2021} and incorporating known cortical markers from \citet{Zeng2012LargeScale} (see {Table S18} for details). Based on this set, the True Positive Rate (TPR), False Positive Rate (FPR), and F1 score for each method are calculated and presented in {Tables S19--S20}. We observe that while the F1 scores across all methods remain modest, the proposed IBaySVG method and SPARK demonstrate the strongest overall performance. The SV genes identified by these two methods show satisfactory biological relevance, consistent with prior knowledge. Notably, under both within- and across-donor scenarios, our approach successfully recovers canonical layer-specific genes highlighted by \citet{Maynard2021}, such as PCP4, MOBP, and SNAP25, as well as non-layer-specific genes involved in immune processes (e.g., HBB and IGKC) in the DLPFC, genes that are missed by some alternatives, such as nnSVG, StarTrail, and GASTON.

\subsection{IBaySVG leads to biologically relevant SV genes}

\begin{figure}[!ht]
    \centering
    \includegraphics[width=\linewidth]{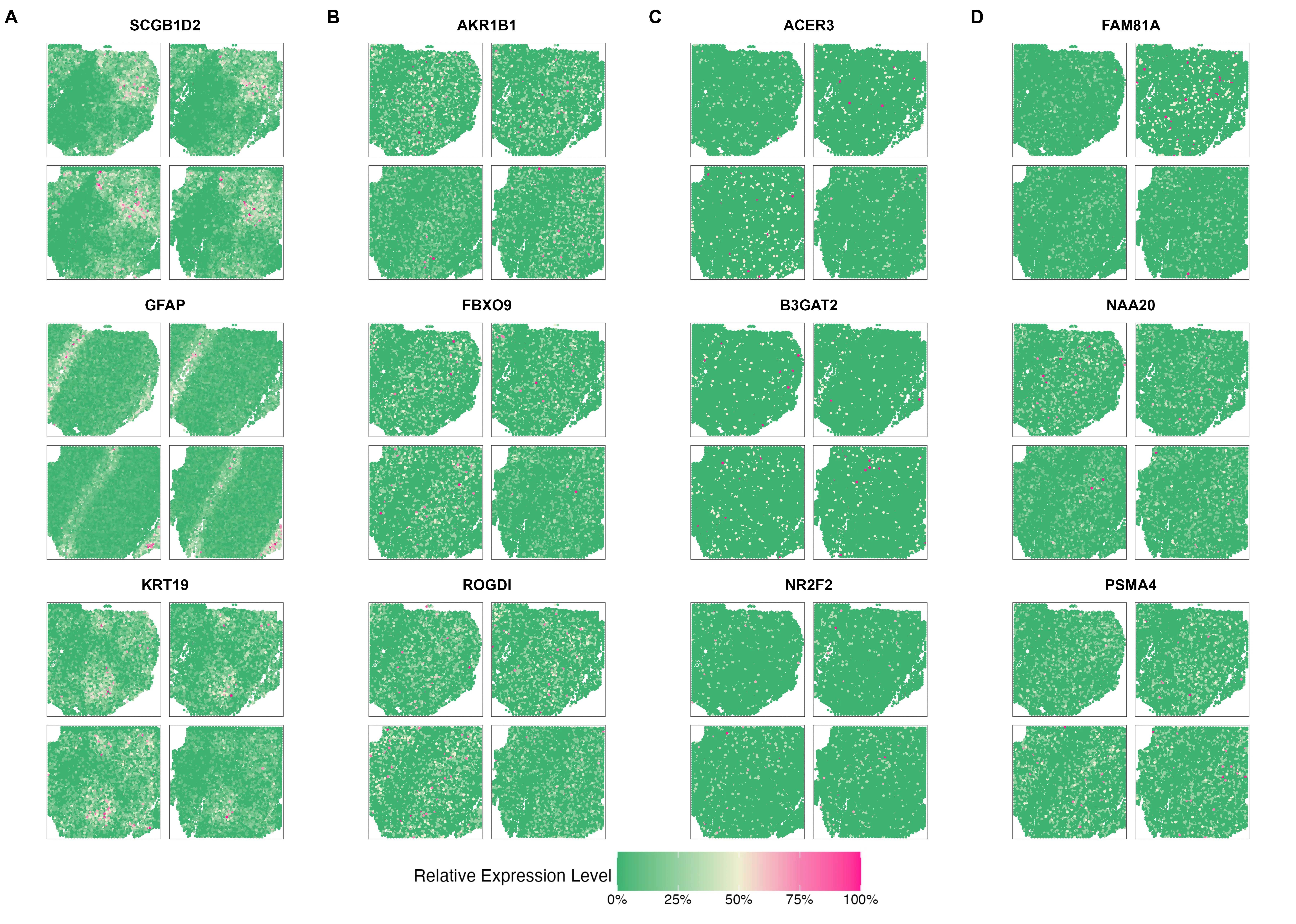}
    \caption{Spatial expression patterns of representative genes identified through comparative analysis for the within-donor DLPFC dataset.
(A) Three genes identified by the proposed method and most single-sample approaches using intersection-based integration, as well as the Cauchy and HMP integration. (B) Three genes detected by our method but overlooked by all single-sample approaches with intersection-based integration. (C) Three genes recognized by three or more single-sample methods using union-based integration but undetected by the proposed method. (D) Three genes uniquely identified by the proposed method but absent in all single-sample analyses (GASTON, which selects almost all genes, is excluded from this comparison) across individual slice examinations.}
    \label{fig:spatial-expression}
\end{figure}

We focus on the within-donor DLPFC dataset, while the across-donor DLPFC analysis and the SCC analysis are presented in {Supplementary Sections S7.9 and S7.10}. For the within-donor DLPFC dataset, only SCGB1D2 and GFAP are identified consistently by all seven single-sample methods. Notably, IBaySVG shares 26 SV genes with the five conventional single-sample methods (SPARK, SPARKX, HEARTSVG, nnSVG, spVC) under the ``Inter'', ``Cauchy'', and ``HMP'' integration strategies. Of these overlapping genes, GASTON-Inter and StarTrail-Inter respectively recover 11 and 15. The spatial expression patterns of three representative genes SCGB1D2, GFAP, and KRT19 are provided in Figure \ref{fig:spatial-expression} (A). It can be seen that these genes exhibit strong spatial differential expression across all four slices, resulting in their high detectability. Here, SCGB1D2 inhibits the growth of Borrelia burgdorferi and modulates susceptibility to Lyme disease, which can lead to neurological impairments \citep{Strausz2024}. GFAP has been extensively demonstrated in numerous studies to play a crucial role in structural support, neuroprotection, and injury repair within the central nervous system \citep{Abdelhak2022}. In addition, KRT19, as one of the intermediate filament proteins, plays a critical role in maintaining cellular structure and in the development of neurons \citep{Coulombe2004}.

Furthermore, among the 1,544 genes detected by at least one single-sample method under intersection integration, 561 are also identified by the proposed IBaySVG method. IBaySVG also identifies an additional 531 genes that are overlooked by all single-sample methods with intersection-based integration. Using the Cauchy method, 237 out of 531 genes are detected by at least one single-sample approach, leaving 294 genes unidentified. 
A thorough examination of these genes reveals that they typically exhibit strong spatial differential expression in a subset of the four slices, while exhibiting minimal spatial effects in the remaining slices. The spatial expression patterns of three representative genes, AKR1B1, FBXO9, and ROGDI, are illustrated in Figure \ref{fig:spatial-expression} (B). Specifically, AKR1B1 and FBXO9 demonstrate spatial differential expression in only two slices, but their effects diminish in the remaining slice, while ROGDI indicates strong spatial differential expression in three slices. Hence, they cannot be identified by the single-sample method with intersection integration. However, AKR1B1 is markedly upregulated in astrocytes following spinal cord injury and interacts with Akt protein to activate downstream signaling pathways \citep{Chen2018AKR1B1Upregulation}, thereby promoting astrocyte proliferation, which may influence neural repair and functional recovery after injury. FBXO9 promotes ubiquitin-mediated degradation of Neurog2, thereby shifting neural crest progenitors toward glial fate at the expense of neuronal differentiation \citep{Liu2020Fbxo9Functions}. Additionally, ROGDI is highly expressed in the brain, and loss-of-function mutations in ROGDI cause Kohlschütter–Tönz syndrome, a severe neurodevelopmental disorder characterized by epilepsy and psychomotor regression \citep{Schossig2012ROGDI}. These findings suggest that the intersection and Cauchy strategies may lead to the exclusion of numerous crucial genes. 

With the union-based integration, 1,795 genes are identified by three or more of the single-sample methods. Among them, 1,010 are undetected by IBaySVG. Under the HMP strategy, 427 out of the 1,010 genes are identified by at least three single-sample methods. Even under the relatively conservative Cauchy strategy, 274 genes are still inevitably detected by at least three single-sample methods. These genes are observed to involve high levels of sparsity and display some potential outlier expressions in some of the four slices. The spatial expression patterns of three representative genes are provided in Figure \ref{fig:spatial-expression} (C). This indicates that the union-based and p-value integration is susceptible to incorporating noise when analyzing the multi-sample data with single-sample techniques.

We further observe that IBaySVG identifies 76 genes that are missed by all single-sample methods in individual slice (GASTON, which selects almost all genes, is excluded from this comparison). Figure \ref{fig:spatial-expression} (D) demonstrates the spatial expression patterns of three representative genes. It can be seen that these genes exhibit weak spatial effects across all four slices, resulting in systematic omission by single-sample analysis methods. The proposed integration strategy can facilitate information borrowing across four slices and amplify biological signals with empirical Bayes estimation. Among these genes, 56 are supported by independent evidence from prior studies (Table S21), with reported roles in neuronal development and synaptic function and associations with Alzheimer's disease and schizophrenia, providing evidence for their biological relevance. Examining these three representative genes in Figure \ref{fig:spatial-expression} (D) more closely, FAM81A protein has been identified as a novel component of the postsynaptic density in the adult brain and plays an important role in synaptic function \citep{Kaizuka2024FAM81APostsynaptic}. Additionally, impaired NAA20 function disrupts the proteostatic balance of neurodevelopment-related proteins, leading to abnormal brain development and neurological dysfunction \citep{DOnofrio2023NovelBiallelic}. Furthermore, PSMA4 participates in protein degradation to maintain cellular protein homeostasis and has been identified as a potential key target in brain aneurysm-related diseases \citep{Wu2025PRCPPromising}.

Finally, DESpace, HEARTSVG-PASTE, spVC-PASTE, and SPARKX-PASTE select nearly all genes; consequently, the genes identified by our proposed method are largely covered by these approaches. A near-universal selection of genes as SV reduces the interpretability of SV gene identification as a feature selection procedure \citep{chen2024evaluating}.  On the other hand, the 86 genes identified by nnSVG-PASTE and 286 genes identified by StarTrail-PASTE are mostly included in the 1,544 genes detected by at least one single-sample method with intersection integration. SPARK-PASTE selects 194 genes from these 1,544 genes, while the remaining 310 genes sometimes involve high levels of sparsity and outlier expressions, such as ACER3, B3GAT2, and NR2F2 (as discussed in Figure \ref{fig:spatial-expression} (C)). 

\subsection{IBaySVG delineates coherent functional gene clusters across slices}

We perform two types of downstream clustering analyses using the identified SV genes to explore their biological implications. First, at the gene level, we apply hierarchical clustering to each slice based on spatial expression profiles; results are shown in Figures S48--S53. Coherent gene clusters can reveal functionally related gene groups and their association with tissue architecture, providing insight into the spatial organization captured by our method.

In the DLPFC dataset from the same donor, four dominant gene clusters, each showing distinct expression patterns, are consistently identified across all four slices, despite minor variations in the total number of clusters per slice (five, four, five, and four clusters for slices A, B, C, and D, respectively). For the DLPFC dataset from different donors, three gene clusters are obtained: clusters 1 and 2 retain largely consistent gene membership across slices, but their spatial expression profiles display clear inter‑donor heterogeneity. Similarly, in the SCC dataset, three gene clusters are formed with high consistency in gene grouping (average overlap$>$0.8); however, their expression patterns vary across slices, for example, cluster 2 genes show stronger expression in slice B, while cluster 3 genes are weaker in slices B and C. These results indicate that although certain co‑clustering relationships are conserved, the specific spatial expression architectures are shaped by tissue context and inter‑sample heterogeneity.

To elucidate the functional relevance of the identified gene clusters, we perform Gene Ontology (GO) enrichment analysis for biological processes in each dataset. {Tables S24--S26} present the top five significant GO terms shared by all slices for each cluster, highlighting their biological relevance. Notably, despite minor variations in gene clustering patterns, the GO results show high consistency, supporting our assumption that the spatial expression patterns of SV genes are preserved across slices. Specifically, in the DLPFC dataset from the same donor, Cluster 1 is enriched for synaptic signaling and neuronal excitability, Cluster 2 for nervous system development and glial function, Cluster 3 for metal ion detoxification and oxidative stress response, and Cluster 4 for immune signaling and extracellular matrix organization. In the DLPFC dataset from different donors, the three clusters maintain core neurobiological themes: Clusters 1 and 2 are associated with synaptic processes and glial development, respectively, while Cluster 3 includes broader terms such as organ development and immune regulation. In contrast, the three clusters in the SCC dataset reflect tumor microenvironment activities: Cluster 1 is enriched in ATP metabolism and mitochondrial function, Cluster 2 in extracellular matrix organization and epithelial morphogenesis, and Cluster 3 in cytoplasmic translation and oxidative phosphorylation. Together, these results indicate that core biological functions (neural processes in DLPFC and metabolic/structural programs in SCC) are conserved across slices and donors within a tissue type, while specific enriched terms closely reflect the distinct physiology of each tissue context.

\subsection{IBaySVG enables identification of anatomically meaningful spatial domains}

\begin{figure}[!ht]
    \centering
    \includegraphics[width=\linewidth]{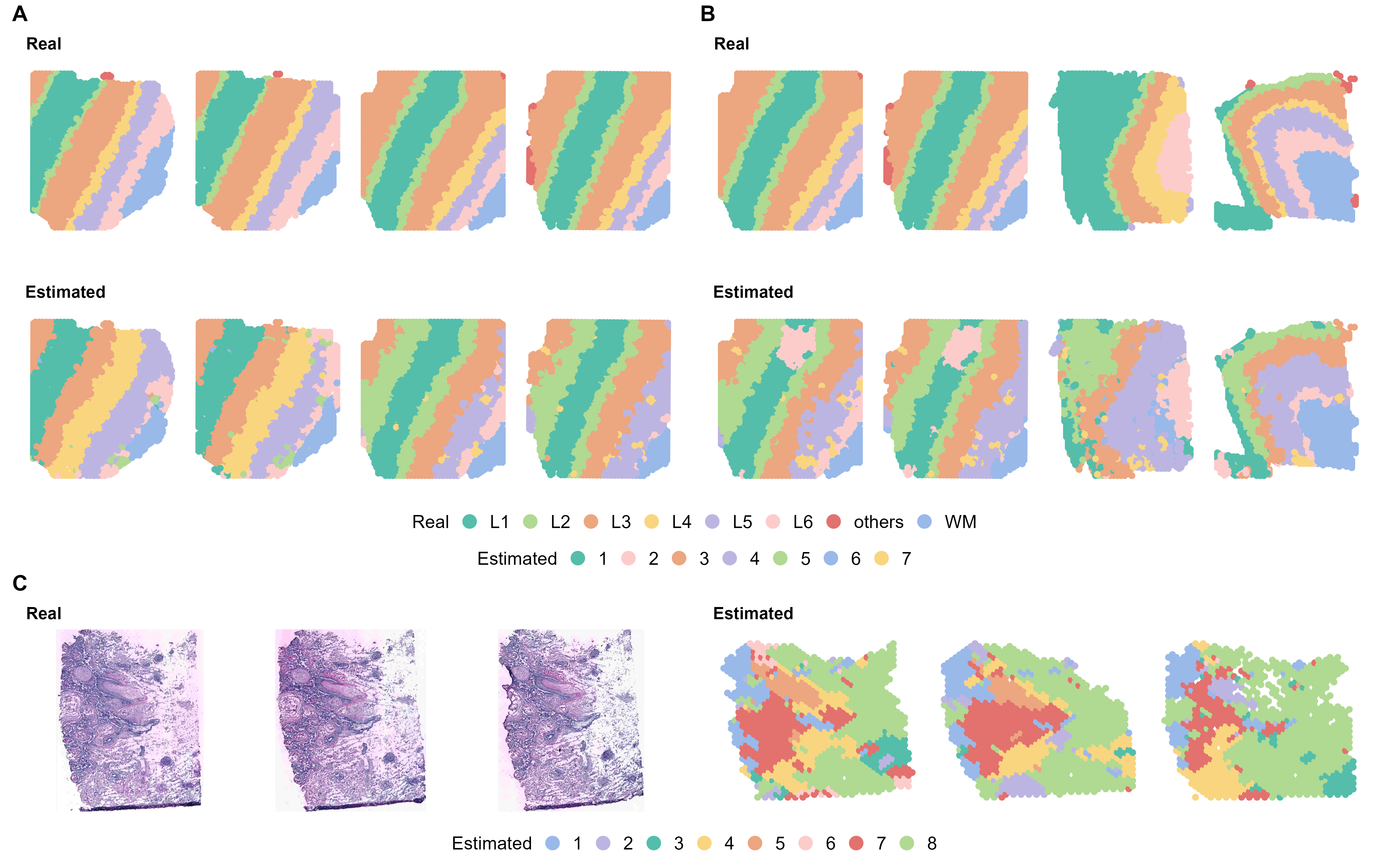}
    \caption{Spatial domains derived from SV genes identified by the proposed method. (A) DLPFC dataset (same donor), (B) DLPFC dataset (different donors), (C) SCC dataset. Each panel represents an individual tissue section within the respective dataset.}
    \label{spotcluster_our}
\end{figure}

At the spot level, we apply the BASS method \citep{Li2022BASS} to identify spatial domains based on the detected SV genes. In the DLPFC dataset from the same donor, seven distinct domains are resolved across the four slices (Figure \ref{spotcluster_our} A). While variability exists, several cortical layers, each with distinct cellular and functional profiles, are consistently delineated: the outermost layer L1 (dark green), primarily composed of glial cells and inhibitory interneurons for processing external inputs; layer L5 (purple), whose neurons project to subcortical and other cortical regions to support executive control; layer L6 (pink), which provides feedback to the thalamus; and the white matter WM (blue), containing myelinated axons that relay signals outside the cortex. For the intermediate layers (L2-L4), which are generally involved in intracortical communication, the method shows partial but meaningful discriminative capability: in sections 1 and 2, layers L3 and L4 are clearly resolved, whereas in sections 3 and 4, the method predominantly distinguishes layers L2 and L3, reflecting the subtle transcriptional gradients among these regions.

In the DLPFC dataset from different donors, seven domains are also identified (Figure \ref{spotcluster_our} B). Interestingly, for slices C and D derived from additional donors, the cortical spatial architecture displays a curved laminar pattern that differs from that of the first donor, posing additional technical challenges for accurate spatial domain identification. Although the detected domains show noticeable variability across the four slices, several cortical regions are consistently and clearly resolved. Specifically, in slices A and B from the first donor and slice D from the third donor, IBaySVG reliably distinguishes layers L1 (purple), L2 (yellow), L3 (dark green), L6 (blue), and WM (light green) in the human dorsolateral prefrontal cortex. In contrast, layers L4 (pink) and L5 (orange) cannot be clearly separated, likely due to their higher structural similarity. Turning to the SCC dataset, eight distinct spatial domains are obtained (Figure \ref{spotcluster_our} C). A clear boundary and a triangular-like structure are observed, suggesting that the identified SV genes drive functional heterogeneity across tumor regions. Collectively, these results demonstrate that our method effectively captures conserved broad spatial architecture, such as laminar organization in the cortex and regional compartmentalization in the tumor, while finer-resolution discrimination is influenced by tissue-specific transcriptional gradients and morphological features.

To comprehensively evaluate clustering performance, we assess spot clustering based on SV genes identified by different methods using six established metrics: Adjusted Rand Index (ARI), Normalized Mutual Information (NMI), Davies–Bouldin Index (DBI), Calinski–Harabasz (CH) Index, Silhouette Coefficient, and an ANOVA-based F‑statistic. Notably, the DLPFC dataset contains manually annotated cortical layers, which serve as a gold standard for evaluating SV gene detection performance, allowing the calculation of external metrics such as ARI and NMI. For reference, we also compute the ARI and NMI based on all genes. In the DLPFC dataset from the same donor, our method and SPARKX‑Inter achieve the highest mean ARI (0.535 and 0.548, respectively) and NMI (0.626 and 0.645), significantly outperforming clustering based on all genes (ARI=0.432, NMI=0.537). Notably, both methods also demonstrate superior performance in terms of within‑cluster compactness and between‑cluster separation across other internal metrics. In the DLPFC dataset from different donors, our method, along with nnSVG‑Union, SPARK‑Cauchy, and spVC‑Union, attains the highest ARI and NMI scores (ARI range: 0.459-0.498; NMI: 0.548-0.589). In contrast, SPARKX performs notably worse in this setting (ARI $<$0.4; NMI $<$0.54). Clustering using all genes again yields unsatisfactory results, with a mean ARI of 0.412 and NMI of 0.427. For the SCC dataset, which lacks such gold‑standard annotations, evaluation relies solely on internal metrics; here, although our method does not attain the best DBI, it consistently excels on the CH index, Silhouette coefficient, and F-statistic, demonstrating robust clustering performance. Across all datasets, our approach maintains strong and stable results, confirming the functional relevance of the identified SV genes for revealing spatial domains under varying tissue contexts and donor heterogeneity. Detailed metric definitions and results are provided in {Supplementary Section S7.12}.

Additional domain identification analyses are conducted to evaluate the contribution of SV genes uniquely identified by IBaySVG. Specifically, we compare two clustering configurations: one using the full set of detected SV genes, and the other excluding those uniquely selected by IBaySVG but not recovered by other competing approaches (after removing overly permissive methods that select nearly all genes). The removal of these unique SV genes leads to a consistent and noticeable reduction in domain identification performance across multiple evaluation metrics {(Tables S42--S44),} further supporting the functional and structural relevance of the SV genes uniquely captured by IBaySVG. Detailed results are provided in Supplementary Section S7.13.

\section{Discussion}

In this article, we developed an integrated Bayesian nonparametric spatial model for identifying SV genes. The proposed framework addressed critical challenges in spatial transcriptomic data analysis by incorporating nonparametric spatial modeling within a zero-inflated negative binomial distribution, effectively handling count-based measurements, sparsity, and overdispersion without relying on predefined kernel functions. Its bi-level shrinkage prior enabled robust cross-sample information sharing while filtering technical noise, surpassing conventional integration methods through unified model-based fusion. Computational efficiency was achieved via an optimized variational inference algorithm. Practical strategies for larger datasets include pre-filtering genes with low expression, high dropout, or low spatial variance, removing low-quality spots, and focusing on genes of interest (e.g., markers or pathway genes). The framework applies directly to emerging platforms such as Visium HD and Xenium, underscoring its algorithmic scalability and the growing need for adaptive preprocessing as data resolution increases. Validation through comprehensive simulations demonstrated superior SV gene detection performance compared to existing methods. 

We applied our method to within- and across-donor DLPFC slices and to SCC tissues, identifying SV genes that overlap with existing methods while also uncovering novel findings. Notably, the commonly identified genes across methods generally exhibit strong spatial variability across all slices. Compared with intersection-based and Cauchy combination post-integration methods, our method successfully detects SV genes with weak signals in certain slices, all of which have been validated as possessing important biological functions. In contrast to union-based and HMP-based post-integration analysis methods, our method effectively filtered out non-authentic SV genes caused by noise in specific slices, while simultaneously integrating weak signals across multiple slices to reveal novel findings. Furthermore, we observed substantial discrepancies in results obtained from single-slice analyses across the multiple slices, underscoring the necessity of integrative analysis. Through downstream analyses, we obtained functionally enriched gene clusters and identified spatially coherent domains aligned with the underlying tissue architecture, providing novel insights for further investigation of biological processes in the dorsolateral prefrontal cortex and squamous cell carcinoma. 

Our empirical findings further reveal secondary data features that point to several directions for future improvement. In this work, the metric $\max\big(E(u_1), E(u_2)\big)$ achieves our detection goals while aligning with the BFDR framework. Future studies may explore more comprehensive criteria for quantifying combined spatial effects. Although the B-spline basis provides a reasonable approximation for the datasets considered here, it may underfit datasets exhibiting true stepwise discontinuities. Future work will explore discontinuity-aware basis functions and adaptive-knot spline constructions to relax the global smoothness assumption. Our study focused on SV gene identification and conducted a post-clustering analysis. It is of interest to develop an integrated model that simultaneously detects SV genes and performs spatial clustering, which would address observed discrepancies between gene expression patterns and cluster boundaries. Incorporating spatial regularization techniques like the Potts model could optimize cluster coherence and biological interpretability. Expanding beyond our current same-tissue analysis, the proliferation of multi-tissue and multi-technology spatial transcriptomic datasets (spanning species and experimental platforms) creates unprecedented opportunities to systematically characterize biological conservation and variation through cross-tissue integration. In addition, integrating multimodal data sources, such as histopathological features from H\&E-stained images that encode critical tissue architecture information, could substantially enhance SV gene detection by complementing transcriptional profiles with structural microenvironment context, as demonstrated by emerging multimodal spatial analysis frameworks. While existing literature and downstream analyses provide supporting evidence for the identified SV genes, more rigorous independent validation would strengthen the biological interpretation of these findings.

\section*{Supplementary Materials}

In supplementary materials, the pdf file contains supplementary sections, figures, and tables referred in the paper. The zipped file includes the code and datasets used in the simulation and real application and provides detailed instructions
for reproducing results. The code is also available at \url{https://github.com/zhoumeng123456/IBaySVG}.

\section*{Acknowledgments}
The authors thank the editors and reviewers for their invaluable feedback and insightful
suggestions, which have significantly improved this paper.

\section*{Data Availability Statement}
The data that support the findings of this study are openly available in \url{https://github.com/zhoumeng123456/IBaySVG}. % https://github.com/***.

\section*{Funding}

This research was supported by the National Natural Science Foundation of China (12071273); MOE Project of Humanities and Social Sciences (25YJCZH291); Shanghai Science and Technology Development Funds (23JC1402100); National Institutes of Health (CA204120); Fundamental Research Funds for the Central Universities (CXJJ-2024-460); China Scholarship Council; and National Science Foundation (2209685).
%the National Natural Science Foundation of China (12071273); Shanghai Rising-Star Program (22QA1403500); 
\section*{Disclosure Statement}
The authors report there are no competing interests to declare.

\bibliographystyle{agsm}

\bibliography{Bibliography-MM-MC}

@article{li2021bayesian,
  title={{Bayesian modeling of spatial molecular profiling data via Gaussian process}},
  author={Li, Qiwei and Zhang, Minzhe and Xie, Yang and Xiao, Guanghua},
  journal={Bioinformatics},
  volume={37},
  number={22},
  pages={4129--4136},
  year={2021},
  publisher={Oxford University Press},
  doi={10.1093/bioinformatics/btab455}
}

@article{zhu2021sparkx,
  title={{SPARK-X: non-parametric modeling enables scalable and robust detection of spatial expression patterns for large spatial transcriptomic studies}},
  author={Zhu, J. and Sun, S. and Zhou, X.},
  journal={Genome Biology},
  volume={22},
  pages={184},
  year={2021},
  publisher={BioMed Central},
  doi={10.1186/s13059-021-02404-0}
}

@article{svensson2018spatialde,
  title={{SpatialDE: identification of spatially variable genes}},
  author={Svensson, V. and Teichmann, S. and Stegle, O.},
  journal={Nature Methods},
  volume={15},
  pages={343--346},
  year={2018},
  publisher={Nature Publishing Group},
  doi={10.1038/nmeth.4636}
}

@article{cai2024despace,
  title={{DESpace: spatially variable gene detection via differential expression testing of spatial clusters}},
  author={Cai, Peiying and Robinson, Mark D. and Tiberi, Simone},
  journal={Bioinformatics},
  volume={40},
  number={2},
  year={2024},
  pages={btae027},
  publisher={Oxford University Press},
  doi={10.1093/bioinformatics/btae027}
}

@article{weber2023nnsvg,
  title={{nnSVG for the scalable identification of spatially variable genes using nearest-neighbor Gaussian processes}},
  author={Weber, L.M. and Saha, A. and others},
  journal={Nature Communications},
  volume={14},
  pages={4059},
  year={2023},
  publisher={Nature Publishing Group},
  doi={10.1038/s41467-023-39748-z}
}

@article{sun2020statistical,
  title={Statistical analysis of spatial expression patterns for spatially resolved transcriptomic studies},
  author={Sun, S. and Zhu, J. and Zhou, X.},
  journal={Nature Methods},
  volume={17},
  pages={193--200},
  year={2020},
  publisher={Nature Publishing Group},
  doi={10.1038/s41592-019-0701-7}
}

@article{yuan2024heartsvg,
  title={{HEARTSVG}: a fast and accurate method for identifying spatially variable genes in large-scale spatial transcriptomics},
  author={Yuan, X. and Ma, Y. and Gao, R. and others},
  journal={Nature Communications},
  volume={15},
  pages={5700},
  year={2024},
  publisher={Nature Publishing Group},
  doi={10.1038/s41467-024-49846-1}
}

@article{yu2022identification,
  title={Identification of cell-type-specific spatially variable genes accounting for excess zeros},
  author={Yu, Jinge and Luo, Xiangyu},
  journal={Bioinformatics},
  volume={38},
  number={17},
  pages={4135--4144},
  year={2022},
  publisher={Oxford University Press},
  doi={10.1093/bioinformatics/btac457}
}

@article{yu2024spvc,
  title={{spVC} for the detection and interpretation of spatial gene expression variation},
  author={Yu, S. and Li, W.},
  journal={Genome Biology},
  volume={25},
  pages={103},
  year={2024},
  publisher={Springer Nature},
  doi={10.1186/s13059-024-03245-3}
}

@article{zhou2023redeconve,
  title={Spatial transcriptomics deconvolution at single-cell resolution using {Redeconve}},
  author={Zhou, Z. and Zhong, Y. and Zhang, Z. and others},
  journal={Nature Communications},
  volume={14},
  pages={7930},
  year={2023},
  publisher={Nature Publishing Group}
}

@article{Wilson2019,
  author  = {Wilson, Daniel J.},
  title   = {The harmonic mean p-value for combining dependent tests},
  journal = {Proceedings of the National Academy of Sciences},
  year    = {2019},
  volume  = {116},
  number  = {4},
  pages   = {1195--1200},
  doi     = {10.1073/pnas.1814092116}
}

@article{tadesse2005wavelet,
  title={Wavelet thresholding with Bayesian false discovery rate control},
  author={Tadesse, MG and Ibrahim, JG and Vannucci, M and Gentleman, R},
  journal={Biometrics},
  volume={61},
  number={1},
  pages={25--35},
  year={2005},
  publisher={Wiley},
  doi={10.1111/j.0006-341X.2005.031102.x}
}

@inproceedings{knowles2011nonconjugate,
 author = {Knowles, David and Minka, Tom},
 booktitle = {Advances in Neural Information Processing Systems},
 publisher = {Curran Associates, Inc.},
 title = {Non-conjugate Variational Message Passing for Multinomial and Binary Regression},
 volume = {24},
 year = {2011}
}

@article{wu2024joint,
  title={Joint identification of spatially variable genes via a network-assisted {Bayesian} regularization approach},
  author={Wu, Mingcong and Li, Yang and Ma, Shuangge and Wu, Mengyun},
  journal={The Annals of Applied Statistics},
  volume={19},
  number={4},
  pages={2705--2723},
  year={2025b},
  publisher={Institute of Mathematical Statistics}
}

@article{Zeira2022,
  author = {Zeira, R. and Land, M. and Strzalkowski, A. and others},
  title = {Alignment and integration of spatial transcriptomics data},
  journal = {Nature Methods},
  volume = {19},
  pages = {567--575},
  year = {2022},
  doi = {10.1038/s41592-022-01459-6}
}

@article{Strausz2024,
  author  = {Strausz, S. and Abner, E. and Blacker, G. and others},
  title   = {{SCGB1D2} inhibits growth of Borrelia burgdorferi and affects susceptibility to Lyme disease},
  journal = {Nature Communications},
  year    = {2024},
  volume  = {15},
  pages   = {2041}
}

@article{Abdelhak2022,
  author  = {Abdelhak, A. and Foschi, M. and Abu-Rumeileh, S. and others},
  title   = {Blood {GFAP} as an emerging biomarker in brain and spinal cord disorders},
  journal = {Nature Reviews Neurology},
  year    = {2022},
  volume  = {18},
  pages   = {158--172}  
}

@article{Coulombe2004,
  author = {Coulombe, P. and Wong, P.},
  title = {Cytoplasmic intermediate filaments revealed as dynamic and multipurpose scaffolds},
  journal = {Nature Cell Biology},
  volume = {6},
  number = {8},
  pages = {699--706},
  year = {2004},
  doi = {10.1038/ncb0804-699}
}

@article{Yan2024,
  author = {Yinqiao Yan and Xiangyu Luo},
  title = {Bayesian Integrative Region Segmentation in Spatially Resolved Transcriptomic Studies},
  journal = {Journal of the American Statistical Association},
  year = {2024},
  volume = {119},
  number = {535},
  pages = {1709--1721}
}

@article{Maynard2021,
  author = {Maynard, K.R. and Collado-Torres, L. and others},
  title = {Transcriptome-scale spatial gene expression in the human dorsolateral prefrontal cortex},
  journal = {Nature Neuroscience},
  year = {2021},
  volume = {24},
  pages = {425--436}
}

@article{chen2024evaluating,
  author    = {Chen, Chao and Kim, Hae Jong and Yang, Pei},
  title     = {Evaluating spatially variable gene detection methods for spatial transcriptomics data},
  journal   = {Genome Biology},
  year      = {2024},
  volume    = {25},
  pages     = {18},
  doi       = {10.1186/s13059-023-03145-y}
}

@article{yan2025categorization,
  author    = {Yan, Gang and Hua, Shao-Hui and Li, Jun-Jie},
  title     = {Categorization of 34 computational methods to detect spatially variable genes from spatially resolved transcriptomics data},
  journal   = {Nature Communications},
  year      = {2025},
  volume    = {16},
  pages     = {1141},
  doi       = {10.1038/s41467-025-56080-w}
}

@article{Scott2023,
  author    = {Scott, M. R. and Zong, W. and Ketchesin, K. D. and Seney, M. L. and Tseng, G. C. and others},
  title     = {Twelve-hour rhythms in transcript expression within the human dorsolateral prefrontal cortex are altered in schizophrenia},
  journal   = {PLOS Biology},
  year      = {2023},
  volume    = {21},
  number    = {1},
  pages     = {e3001688},
  doi       = {10.1371/journal.pbio.3001688}
}

@article{Ma2022,
  author    = {Ma, Shaojie and Skarica, Mario and Li, Qian and Xu, Chuan and others},
  title     = {Molecular and cellular evolution of the primate dorsolateral prefrontal cortex},
  journal   = {Science},
  year      = {2022b},
  volume    = {377},
  pages     = {eabo7257},
  doi       = {10.1126/science.abo7257}
}

@article{saussez2006galectin7,
  title={Galectin-7},
  author={Saussez, Serge and Kiss, Robert},
  journal={Cellular and Molecular Life Sciences},
  volume={63},
  number={6},
  pages={686--697},
  year={2006},
  publisher={Springer},
  doi={10.1007/s00018-005-5458-8},  
}

@article{sasahira2021identification,
  title={Identification of oral squamous cell carcinoma markers MUC2 and SPRR1B downstream of TANGO},
  author={Sasahira, Tadaaki and Kurihara-Shimomura, Mai and Shimomura, Hiroki and others},
  journal={Journal of Cancer Research and Clinical Oncology},
  volume={147},
  number={6},
  pages={1659--1672},
  year={2021},
  publisher={Springer},
  doi={10.1007/s00432-021-03568-9},  
}

@article{nagy2001s100a2,
  title={S100A2, a Putative Tumor Suppressor Gene, Regulates In Vitro Squamous Cell Carcinoma Migration},
  author={Nagy, Nandor and Brenner, Charles and Markadieu, Nicolas and others},
  journal={Laboratory Investigation},
  volume={81},
  number={5},
  pages={599--612},
  year={2001},
  publisher={Nature Publishing Group},
  doi={10.1038/labinvest.3780269},  
}

@article{liu2023single,
  title = {Single-cell dissection of cellular and molecular features underlying human cervical squamous cell carcinoma initiation and progression},
  author = {Liu, Chao and Yang, Yuanyuan and Wang, Lixue and Wang, Qian and Shi, Ying and Wang, Xiaoyan and Zhu, Yanyan and Guo, Junyao and others},
  journal = {Science Advances},
  volume = {9},
  number = {15},
  pages = {eadd8977},
  year = {2023},
  publisher = {American Association for the Advancement of Science},
  doi = {10.1126/sciadv.add8977}
}

@article{ji2020multimodal,
  title = {Multimodal analysis of composition and spatial architecture in human squamous cell carcinoma},
  author = {Ji, Alexander L and Rubin, Adam J and Thrane, Kristine and Jiang, Song and Reynolds, Daniel L and Meyers, Robert M and Guo, Mo and George, Joseph and Mollbrink, Annelie and others},
  journal = {Cell},
  volume = {182},
  number = {2},
  pages = {497--514.e22},
  year = {2020},
  publisher = {Elsevier},
  doi = {10.1016/j.cell.2020.05.039},
  pmid = {32579974}
}

@article{liu2023precast,
  author    = {Liu, Weizhong and Liao, Xiaoyu and Luo, Ziqi and Li, Rongrong and Jin, Xing and Zhang, Qing and Wang, Yixin and Li, Yuling and Hu, Yelin and Zhang, Zemin and others},
  title     = {Probabilistic embedding, clustering, and alignment for integrating spatial transcriptomics data with PRECAST},
  journal   = {Nature Communications},
  volume    = {14},
  number    = {1},
  pages     = {296},
  year      = {2023},
  publisher = {Nature Publishing Group},
  doi       = {10.1038/s41467-022-35578-6}
}

@article{Li2022BASS,
  author       = {Li, Z. and Zhou, X.},
  title        = {{BASS}: multi-scale and multi-sample analysis enables accurate cell type clustering and spatial domain detection in spatial transcriptomic studies},
  journal      = {Genome Biology},
  year         = {2022},
  volume       = {23},
  number       = {1},
  pages        = {168},
  doi          = {10.1186/s13059-022-02734-7}
}

@article{liu2019acat,
  title        = {{ACAT}: A Fast and Powerful p Value Combination Method for Rare-Variant Analysis in Sequencing Studies},
  author       = {Liu, Yuwei and Chen, Shu and Li, Zhihua and Morrison, Alanna C. and Boerwinkle, Eric and Lin, Xihong},
  journal      = {American Journal of Human Genetics},
  volume       = {104},
  number       = {3},
  pages        = {410--421},
  year         = {2019},
  publisher    = {Elsevier},
  doi          = {10.1016/j.ajhg.2019.01.002},
  pmid         = {30849328}
}

@incollection{orso2007ap,
  title={The AP-2a transcription factor regulates tumor cell migration and apoptosis},
  author={Orso, Francesca and Fassetta, Michela and Penna, Elisa and Solero, Alessandra and Filippo, Katia De and Sismondi, Piero and Bortoli, Michele De and Taverna, Daniela},
  booktitle={Advances in Molecular Oncology: Edited under the auspices of the European Institute of Oncology (IEO) and The FIRC Institute of Molecular Oncology Foundation (IFOM)},
  pages={87--95},
  year={2007},
  publisher={Springer}
}

@article{
Hemant2014dynamin,
author = {Hemant P. Joshi  and Indira V. Subramanian  and Erica K. Schnettler  and Goutam Ghosh  and Rajesha Rupaimoole  and Colleen Evans  and Manju Saluja  and Yawu Jing  and Ivan Cristina  and Sabita Roy  and Yan Zeng  and Vijay H. Shah  and Anil K. Sood  and Sundaram Ramakrishnan },
title = {Dynamin 2 along with microRNA-199a reciprocally regulate hypoxia-inducible factors and ovarian cancer metastasis},
journal = {Proceedings of the National Academy of Sciences},
volume = {111},
number = {14},
pages = {5331-5336},
year = {2014},
doi = {10.1073/pnas.1317242111},
eprint = {https://www.pnas.org/doi/pdf/10.1073/pnas.1317242111}}

@article{uhlen2015tissue,
  title={Tissue-based map of the human proteome},
  author={Uhl{\'e}n, Mathias and Fagerberg, Linn and Hallstr{\"o}m, Bj{\"o}rn M and Lindskog, Cecilia and Oksvold, Per and Mardinoglu, Adil and Sivertsson, {\AA}sa and Kampf, Caroline and Sj{\"o}stedt, Evelina and Asplund, Anna and others},
  journal={Science},
  volume={347},
  number={6220},
  pages={1260419},
  year={2015},
  publisher={American Association for the Advancement of Science},
  doi={10.1126/science.1260419}
}

@article{Chen2018AKR1B1Upregulation,
  author    = {Xin Chen and Chun Chen and Jun Hao and Rui Qin and Baoxin Qian and others},
  title     = {{AKR1B1} Upregulation Contributes to Neuroinflammation and Astrocytes Proliferation by Regulating the Energy Metabolism in Rat Spinal Cord Injury},
  journal   = {Neurochemical Research},
  year      = {2018},
  volume    = {43},
  number    = {8},
  pages     = {1491--1499},
  month     = {Aug},
  issn      = {1573-6903},
  doi       = {10.1007/s11064-018-2563-2},
  publisher = {Springer}
}

@article{Liu2020Fbxo9Functions,
  author    = {Jia A. Liu and Alvin Tai and Jess Hong and Maggie P. L. Cheung and Michael H. Sham and Kathryn S. E. Cheah and Chung Wai Cheung and Martin Cheung},
  title     = {{Fbxo9} functions downstream of {Sox10} to determine neuron-glial fate choice in the dorsal root ganglia through {Neurog2} destabilization},
  journal   = {Proceedings of the National Academy of Sciences},
  year      = {2020},
  volume    = {117},
  number    = {8},
  pages     = {4199--4210},
  month     = {Feb},
  day       = {25},
  issn      = {1091-6490},
  doi       = {10.1073/pnas.1916164117}, 
  publisher = {National Academy of Sciences}
}

@article{Kaizuka2024FAM81APostsynaptic,
  author    = {Takashi Kaizuka and Takahiro Hirouchi and Takeo Saneyoshi and Takuya Shirafuji and Mary O. Collins and others},
  title     = {{FAM81A} is a postsynaptic protein that regulates the condensation of postsynaptic proteins via liquid--liquid phase separation},
  journal   = {{PLOS Biology}},
  year      = {2024},
  volume    = {22},
  number    = {3},
  pages     = {e3002006},
  month     = {Mar},
  issn      = {1545-7885},
  doi       = {10.1371/journal.pbio.3002006},  
  publisher = {Public Library of Science ({PLoS})}
}

@article{DOnofrio2023NovelBiallelic,
  author    = {G. D'Onofrio and C. Cuccurullo and S. K. Larsen and others},
  title     = {Novel biallelic variants expand the phenotype of {NAA20}-related syndrome},
  journal   = {Clinical Genetics},
  year      = {2023},
  volume    = {104},
  number    = {3},
  pages     = {371--376},
  month     = {Sep},
  issn      = {1399-0004},
  doi       = {10.1111/cge.14359}, 
  publisher = {Wiley}
}

@article{Wu2025PRCPPromising,
  author={Wu, Jinghao and Mei, Yunyun and Li, XinYu and others},
  title={{PRCP} is a promising drug target for intracranial aneurysm rupture supported via multi-omics analysis},  
  journal={Stroke and Vascular Neurology},
  year={2025a},
  volume={10},
  number={2},
  pages = {e003076},  
  publisher={BMJ Publishing Group Ltd}
}

@article{Zeng2012LargeScale,
  author    = {Hongkui Zeng and Ed S. Shen and John G. Hohmann and Seung Wook Oh and Amy Bernard and Joshua J. Royall and Karl J. Glattfelder and others},
  title     = {Large-scale cellular-resolution gene profiling in human neocortex reveals species-specific molecular signatures},
  journal   = {Cell},
  year      = {2012},
  volume    = {149},
  number    = {2},
  pages     = {483--496},
  month     = {Apr},
  issn      = {1097-4172},
  doi       = {10.1016/j.cell.2012.02.052},
  publisher = {Elsevier}
}

@article{ma2022belayer,
  title     = {Belayer: Modeling discrete and continuous spatial variation in gene expression from spatially resolved transcriptomics},
  author    = {Ma, Chao and Chitra, Uthsav and Zhang, Siyu and others},
  journal   = {Cell Systems},
  volume    = {13},
  number    = {10},
  pages     = {786--797.e13},
  year      = {2022a},
  doi       = {10.1016/j.cels.2022.09.002}
}

@article{Foran1992MBP,
  author  = {Foran, D. R. and Peterson, A. C.},
  title   = {Myelin acquisition in the central nervous system of the mouse revealed by an MBP-Lac Z transgene},
  journal = {Journal of Neuroscience},
  year    = {1992},
  volume  = {12},
  number  = {12},
  pages   = {4890--4897},
  doi     = {10.1523/JNEUROSCI.12-12-04890.1992}
}

@article{Hashimoto2016,
  author    = {Hashimoto, N. and Sato, T. and Yajima, T. and Fujita, M. and Sato, A. and Shimizu, Y. and Ichikawa, H.},
  title     = {SPARCL1-containing neurons in the human brainstem and sensory ganglion},
  journal   = {Somatosensory \& Motor Research},
  year      = {2016},
  volume    = {33},
  number    = {2},
  pages     = {112--117},
  doi       = {10.1080/08990220.2016.1197115}
}

@article{Wang2022COX6c,
  title={Novel role of COX6c in the regulation of oxidative phosphorylation and diseases},
  author={Wang, C. and Lv, J. and Xue, C. and others},
  journal={Cell Death Discovery},
  volume={8},
  pages={336},
  year={2022},
  doi={10.1038/s41420-022-01130-1}
}

@article{Berg2015JAKMIP1,
  title   = {JAKMIP1, a Novel Regulator of Neuronal Translation, Modulates Synaptic Function and Autistic-like Behaviors in Mouse},
  author  = {Berg, Jamee M. and others},
  journal = {Neuron},
  year    = {2015},
  volume  = {88},
  number  = {6},
  pages   = {1173--1191},
  doi     = {10.1016/j.neuron.2015.11.024}
}

@article{Khodadadi2025,
  author    = {Hossein Khodadadi and Kamila {\L}uczy{\'n}ska and Dawid Winiarczyk and Pawe{\l} Leszczy{\'n}ski and Hiroaki Taniguchi},
  title     = {NFE2L1 as a central regulator of proteostasis in neurodegenerative diseases: interplay with autophagy, ferroptosis, and the proteasome},
  journal   = {Frontiers in Molecular Neuroscience},
  volume    = {18},
  year      = {2025},
  pages     = {1551571},
  doi       = {10.3389/fnmol.2025.1551571},
  issn      = {1662-5099}
}

@article{Edfawy2019,
  author       = {Edfawy, M. and Guedes, J. R. and Pereira, M. I. and Oliveira, C. R. and Oliveira, J. M. and Cavadas, C.},
  title        = {Abnormal mGluR-mediated synaptic plasticity and autism-like behaviours in Gprasp2 mutant mice},
  journal      = {Nature Communications},
  volume       = {10},
  pages        = {1431},
  year         = {2019},
  doi          = {10.1038/s41467-019-09382-9}
}

@article{Poduri2012AKT3,
  title   = {Somatic Activation of AKT3 Causes Hemispheric Developmental Brain Malformations},
  author  = {Poduri, Annapurna and Evrony, Gilad D. and Cai, Xuyu and Elhosary, Paula C. and Beroukhim, Rameen and Lehtinen, Maria K. and Hills, Lawrence B. and Heinzen, Erin L. and Hill, Robert S. and Hill, Alexander and Barry, Brenna J. and Bourgeois, Brigitte F. D. and Riviello, James J. and Barkovich, A. James and Kuzniecky, Ruben and Partlow, Jennifer and Poduri, C. and Sahin, Mustafa and Mefford, Heather C. and Michaud, Jacques L. and Scheffer, Ingrid E. and Berkovic, Samuel F. and Henegariu, Octavian and Shanske, Alan and Poduri, S. and Walsh, Christopher A.},
  journal = {Neuron},
  volume  = {74},
  number  = {1},
  pages   = {41--48},
  year    = {2012},
  doi     = {10.1016/j.neuron.2012.01.017}
}

@article{Piton2025NOVA,
  title   = {NOVA1/2 genes and alternative splicing in neurodevelopment},
  author  = {Piton, Am{\'e}lie},
  journal = {Current Opinion in Genetics \& Development},
  year    = {2025},
  volume  = {93},
  pages   = {102373},
  issn    = {0959-437X},
  doi     = {10.1016/j.gde.2025.102373}
}

@article{Fan2025EIF2S2,
  author    = {Fan, B. and Pan, Q. and Yuan, X. and et al.},
  title     = {EIF2S2 as a prognostic marker and therapeutic target in glioblastoma: insights into its role and downstream mechanisms},
  journal   = {Cancer Cell International},
  volume    = {25},
  pages     = {126},
  year      = {2025},
  doi       = {10.1186/s12935-025-03762-6}
}

@article{chen2026startrail,
  title   = {Investigating Spatial Dynamics in Spatial Omics Data with {StarTrail}},
  author  = {Chen, Jiawen and Xiong, Caiwei and Sun, Quan and Song, Yuxuan and Wang, Geoffery W. and Gupta, Gaorav P. and Halder, Aritra and Li, Yun and Li, Didong},
  journal = {Journal of the American Statistical Association},
  year    = {2026},
  pages   = {1--12},
  doi     = {10.1080/01621459.2026.2654225},  
}

@article{Chitra2025Mapping,
  title     = {Mapping the topography of spatial gene expression with interpretable deep learning},
  author    = {Chitra, Utsav and Arnold, Benjamin J. and Sarkar, Himadri and others},
  journal   = {Nature Methods},
  volume    = {22},
  number    = {2},
  pages     = {298--309},
  year      = {2025},
  doi       = {10.1038/s41592-024-02503-3},
  publisher = {Nature Publishing Group}
}

@article{Uribe2011EGFR,
  author  = {Uribe, Pablo and Gonzalez, Sergio},
  title   = {Epidermal growth factor receptor (EGFR) and squamous cell carcinoma of the skin: molecular bases for EGFR-targeted therapy},
  journal = {Pathology - Research and Practice},
  year    = {2011},
  volume  = {207},
  number  = {6},
  pages   = {337--342}
}

@article{Yang2021KLF5,
  author    = {Yang, S. and Feng, T. and Li, H.},
  title     = {KLF5, a Novel Therapeutic Target in Squamous Cell Carcinoma},
  journal   = {DNA and Cell Biology},
  year      = {2021},
  volume    = {40},
  number    = {12},
  pages     = {1503--1512},
  doi       = {10.1089/dna.2021.5580} 
}

@article{ONeill2025HIF1a,
  author    = {O’Neill, Wendi Quinn and Yu, Heping and Storl-Desmond, Marta and Pan, Quintin},
  title     = {HIF-1{\(\alpha\)} inhibition boosts the activity of anti-PD-1 antibody treatment in immune-cold HNSCC},
  journal   = {Oral Surgery, Oral Medicine, Oral Pathology and Oral Radiology},
  year      = {2025},
  volume    = {140},
  number    = {3},
  pages     = {e98},
  doi       = {10.1016/j.oooo.2025.04.173}
}

@article{zhao2022modeling,
  title={Modeling zero inflation is not necessary for spatial transcriptomics},
  author={Zhao, Peiyao and Zhu, Jiaqiang and Ma, Ying and Zhou, Xiang},
  journal={Genome Biology},
  volume={23},
  number={1},
  pages={118},
  year={2022},
  publisher={Springer}
}

@article{Schossig2012ROGDI,
  author  = {Schossig, A. and Wolf, N. I. and Fischer, C. and Fischer, M. and Stocker, G. and Pabinger, S. and Dander, A. and Steiner, B. and T\"onz, O. and others},
  title = {Mutations in {ROGDI} cause {Kohlsch{\"u}tter--T{\"o}nz} syndrome},
  journal = {American Journal of Human Genetics},
  year    = {2012},
  volume  = {90},
  number  = {4},
  pages   = {701--707},
  doi     = {10.1016/j.ajhg.2012.02.012},
  pmid    = {22424600},
  pmcid   = {PMC3322220}
}

@article{Bai2021RQR,
  author  = {Bai, Wenhan and Dong, Mengyuan and Li, Lei and others},
  title   = {Randomized quantile residuals for diagnosing zero-inflated generalized linear mixed models with applications to microbiome count data},
  journal = {BMC Bioinformatics},
  year    = {2021},
  volume  = {22},
  number  = {1},
  pages   = {564},
  doi     = {10.1186/s12859-021-04371-6},
  url     = {https://doi.org/10.1186/s12859-021-04371-6}
}

@article{Adams2010,
  author = {Adams and others},
  title = {The alpha-syntrophin PH and PDZ domains scaffold acetylcholine receptors, utrophin, and neuronal nitric oxide synthase at the neuromuscular junction},
  journal = {The Journal of Neuroscience},
  year = {2010},
  volume = {30},
  number = {33},
  pages = {11004-10},
  doi = {10.1523/jneurosci.1930-10.2010}
}

@article{Ahmed2022,
  author  = {Ahmed, M. and Kojima, Y. and Masai, I.},
  title   = {Strip1 regulates retinal ganglion cell survival by suppressing Jun-mediated apoptosis to promote retinal neural circuit formation},
  journal = {eLife},
  year    = {2022},
  volume  = {11},
  pages   = {e74650},
  doi     = {10.7554/eLife.74650}
}

@article{Ai2020,
  author = {Ai and others},
  title = {LINC00941 promotes oral squamous cell carcinoma progression via activating CAPRIN2 and canonical WNT/beta-catenin signaling pathway},
  journal = {Journal of Cellular and Molecular Medicine},
  year = {2020},
  volume = {24},
  number = {18},
  pages = {10512-10524},
  doi = {10.1111/jcmm.15667}
}

@article{ak2012,
  author = {Żak and others},
  title = {Ergic2, a brain specific interacting partner of Otoferlin},
  journal = {Cellular Physiology and Biochemistry},
  year = {2012},
  volume = {29},
  number = {5-6},
  pages = {941-8},
  doi = {10.1159/000188338}
}

@article{Akinyele2024,
  author  = {Akinyele, O. and others},
  title   = {Impaired polyamine metabolism causes behavioral and neuroanatomical defects in a mouse model of Snyder-Robinson syndrome},
  journal = {Disease Models \& Mechanisms},
  year    = {2024},
  volume  = {17},
  number  = {6},
  pages   = {dmm050639},
  doi     = {10.1242/dmm.050639}
}

@article{Almousa2024,
  author = {Almousa and others},
  title = {TRAPPC6B biallelic variants cause a neurodevelopmental disorder with TRAPP II and trafficking disruptions},
  journal = {Brain},
  year = {2024},
  volume = {147},
  number = {1},
  pages = {311-324},
  doi = {10.1093/brain/awad301}
}

@article{Aloulou2025,
  author = {Aloulou and others},
  title = {Dihydrolipoamide dehydrogenase deficiency in two unrelated Tunisian children},
  journal = {BMC Pediatrics},
  year = {2025},
  volume = {25},
  pages = {127},
  doi = {10.1186/s12887-024-05375-w}
}

@article{Alwadei2016,
  author = {Alwadei and others},
  title = {Loss-of-function mutation in RUSC2 causes intellectual disability and secondary microcephaly},
  journal = {Developmental Medicine and Child Neurology},
  year = {2016},
  volume = {58},
  number = {12},
  pages = {1317-1322},
  doi = {10.1111/dmcn.13250}
}

@article{Anderson2021,
  author = {Anderson and others},
  title = {Infralimbic cortex pyramidal neuron GIRK signaling contributes to regulation of cognitive flexibility but not affect-related behavior in male mice},
  journal = {Physiology Behavior},
  year = {2021},
  volume = {242},
  pages = {113597},
  doi = {10.1016/j.physbeh.2021.113597}
}

@article{AndresAlonso2019,
  author = {Andres-Alonso and others},
  title = {SIPA1L2 controls trafficking and local signaling of TrkB-containing amphisomes at presynaptic terminals},
  journal = {Nature Communications},
  year = {2019},
  volume = {10},
  number = {1},
  pages = {5448},
  doi = {10.1038/s41467-019-13224-z}
}

@article{Baccon2002,
  author = {Baccon and others},
  title = {Identification and characterization of Gemin7, a novel component of the survival of motor neuron complex},
journal = {The Journal of Biological Chemistry},
  year = {2002},
  volume = {277},
  number = {35},
  pages = {31957-62},
  doi = {10.1074/jbc.m203478200}
}

@article{Ban2021,
  author = {Ban and others},
  title = {Prickle promotes the formation and maintenance of glutamatergic synapses by stabilizing the intercellular planar cell polarity complex},
  journal = {Science Advances},
  year = {2021},
  volume = {7},
  number = {41},
  pages = {eabh2974},
  doi = {10.1126/sciadv.abh2974}
}

@article{Banks2018,
  author = {Banks and others},
  title = {A missense mutation in Katnal1 underlies behavioural, neurological and ciliary anomalies},
  journal = {Molecular Psychiatry},
  year = {2018},
  volume = {23},
  pages = {713-722},
  doi = {10.1038/mp.2017.54}
}

@article{Baum2024,
  author = {Baum and others},
  title = {CSMD1 regulates brain complement activity and circuit development},
  journal = {Brain, Behavior, and Immunity},
  year = {2024},
  volume = {119},
  pages = {317-332},
  doi = {10.1016/j.bbi.2024.03.041}
}

@article{Becker2021,
  author = {Becker and others},
  title = {Mice deficient in the NAAG synthetase II gene Rimkla are impaired in a novel object recognition task},
  journal = {Journal of Neurochemistry},
  year = {2021},
  volume = {157},
  pages = {2008-2023},
  doi = {10.1111/jnc.15333}
}

@article{Bosio2012,
  author = {Bosio and others},
  title = {PPP4R2 regulates neuronal cell differentiation and survival, functionally cooperating with SMN},
  journal = {European Journal of Cell Biology},
  year = {2012},
  volume = {91},
  number = {8},
  pages = {662-74},
  doi = {10.1016/j.ejcb.2012.03.002}
}

@article{Bowie2020,
  author = {Bowie, E and Goetz, SC},
  title = {TTBK2 and primary cilia are essential for the connectivity and survival of cerebellar Purkinje neurons},
  journal = {eLife},
  year = {2020},
  volume = {9},
  pages = {e51166},
  doi = {10.7554/elife.51166}
}

@article{Brand2020,
  author = {Brand and others},
  title = {FOCAD loss impacts microtubule assembly, G2/M progression and patient survival in astrocytic gliomas},
  journal = {Acta Neuropathologica},
  year = {2020},
  volume = {139},
  number = {1},
  pages = {175-192},
  doi = {10.1007/s00401-019-02067-z}
}

@article{Braune2025,
  author = {Braune and others},
  title = {VOPP1::EGFR fusion is associated with NFkappaB pathway activation in a glioneural tumor with histological features of ganglioglioma},
  journal = {Acta Neuropathologica Communications},
  year = {2025},
  volume = {13},
  number = {1},
  pages = {76},
  doi = {10.1186/s40478-025-01994-1}
}

@article{BrazoSilva2015,
  author = {Brazão-Silva and others},
  title = {Metallothionein gene expression is altered in oral cancer and may predict metastasis and patient outcomes},
  journal = {Histopathology},
  year = {2015},
  volume = {67},
  number = {3},
  pages = {358-67},
  doi = {10.1111/his.12660}
}

@article{Brown2024,
  author  = {Brown, M. and others},
  title   = {Regulation of Drosophila brain development and organ growth by the Minibrain/Rala signaling network},
  journal = {G3: Genes, Genomes, Genetics},
  year    = {2024},
  volume  = {14},
  number  = {11},
  pages   = {jkae219},
  doi     = {10.1093/g3journal/jkae219}
}

@article{Caglar2024,
  author = {Caglar and others},
  title = {Bioinformatics approach combined with experimental verification reveals OAS3 gene implicated in paclitaxel resistance in head and neck cancer},
  journal = {Head \& Neck},
  year = {2024},
  volume = {46},
  number = {9},
  pages = {2178-2196},
  doi = {10.1002/hed.27803}
}

@article{Cannon2005,
  author = {Cannon and others},
  title = {Association of DISC1/TRAX Haplotypes With Schizophrenia, Reduced Prefrontal Gray Matter, and Impaired Short- and Long-term Memory},
  journal = {Archives of General Psychiatry},
  year = {2005},
  volume = {62},
  number = {11},
  pages = {1205-1213},
  doi = {10.1001/archpsyc.62.11.1205}
}

@article{Chan2024,
  author = {Chan and others},
  title = {Upregulation of ENAH by a PI3K/AKT/beta-catenin cascade promotes oral cancer cell migration and growth via an ITGB5/Src axis.},
  journal = {Cellular \& Molecular Biology Letters},
  year = {2024},
  volume = {29},
  number = {1},
  pages = {136},
  doi = {10.1186/s11658-024-00651-0}
}

@article{Chand2022,
  author = {Chand and others},
  title = {Differential Sphingosine-1-Phosphate Receptor-1 Protein Expression in the Dorsolateral Prefrontal Cortex Between Schizophrenia Type 1 and Type 2},
  journal = {Frontiers in Psychiatry},
  year = {2022},
  volume = {13},
  pages = {827981},
  doi = {10.3389/fpsyt.2022.827981}
}

@article{Chen2021a,
  author = {Chen and others},
  title = {Loss of PIGK function causes severe infantile encephalopathy and extensive neuronal apoptosis},
  journal = {Human Genetics},
  year = {2021},
  volume = {140},
  number = {5},
  pages = {791-803},
  doi = {10.1007/s00439-020-02243-2}
}

@article{Chen2021b,
  author = {Chen and others},
  title = {TPM3 mediates epithelial-mesenchymal transition in esophageal cancer via MMP2/MMP9},
  journal = {Annals of Translational Medicine},
  year = {2021},
  volume = {9},
  number = {16},
  pages = {1338},
  doi = {10.21037/atm-21-4043}
}

@article{Chen2023,
  author = {Chen and others},
  title = {Aberrant epithelial cell interaction promotes esophageal squamous-cell carcinoma development and progression},
  journal = {Signal Transduction and Targeted Therapy},
  year = {2023},
  volume = {8},
  number = {1},
  pages = {453},
  doi = {10.1038/s41392-023-01710-2}
}

@article{Chen2024,
  author = {Chen and others},
  title = {TP63 transcriptionally regulates SLC7A5 to suppress ferroptosis in head and neck squamous cell carcinoma},
  journal = {Frontiers in Immunology},
  year = {2024},
  volume = {15},
  pages = {1445472},
  doi = {10.3389/fimmu.2024.1445472}
}

@article{Chen2025,
  author = {Chen and others},
  title = {CERS6 promotes esophageal squamous cell carcinoma proliferation by increasing the stability of RPN1},
  journal = {Cell Death Discovery},
  year = {2025},
  volume = {11},
  number = {1},
  pages = {512},
  doi = {10.1038/s41420-025-02727-y}
}

@article{Chen2026,
  author = {Chen and others},
  title = {Rad23b exacerbates pathological aggregates through disrupting proteasome functions in Spinocerebellar ataxia type 3},
  journal = {Neurotherapeutics},
  year = {2026},
  volume = {23},
  number = {3},
  pages = {e00928},
  doi = {10.1016/j.neurot.2026.e00928}
}

@article{Cheng2023,
  author = {Cheng and others},
  title = {Age-Associated UBE2O Reduction Promotes Neuronal Death in Alzheimer's Disease},
  journal = {Journal of Alzheimer's Disease},
  year = {2023},
  volume = {93},
  number = {3},
  pages = {1083-1093},
  doi = {10.3233/jad-221143}
}

@article{Cheng2025,
  author = {Cheng and others},
  title = {Uncovering chromatin factor landscapes in head and neck squamous cell carcinoma},
  journal = {Oral Oncology},
  year = {2025},
  volume = {170},
  pages = {107687},
  doi = {10.1016/j.oraloncology.2025.107687}
}

@article{Chien2024,
  author = {Chien and others},
  title = {Cyclic increase in the ADAMTS1-L1CAM-EGFR axis promotes the EMT and cervical lymph node metastasis of oral squamous cell carcinoma},
  journal = {Cell Death \& Disease},
  year = {2024},
  volume = {15},
  number = {1},
  pages = {82},
  doi = {10.1038/s41419-024-06452-9}
}

@article{Chintala2009,
  author = {Chintala and others},
  title = {The Vps33a gene regulates behavior and cerebellar Purkinje cell number},
  journal = {Brain Research},
  year = {2009},
  volume = {1266},
  pages = {18-28},
  doi = {10.1016/j.brainres.2009.02.035}
}

@article{Choi2023a,
  author = {Choi and others},
  title = {Transcription factor SOX15 regulates stem cell pluripotency and promotes neural fate during differentiation by activating the neurogenic gene Hes5},
  journal = {Journal of Biological Chemistry},
  year = {2023},
  volume = {299},
  number = {3},
  pages = {102996},
  doi = {10.1016/j.jbc.2023.102996}
}

@article{Choi2023b,
  author = {Choi and others},
  title = {Protective effect of GK2 fused BLVRA protein against oxidative stress-induced dopaminergic neuronal cell damage},
  journal = {The FEBS Journal},
  year = {2023},
  volume = {290},
  number = {11},
  pages = {2923-2938},
  doi = {10.1111/febs.16721}
}

@article{Chua2022,
  author = {Chua and others},
  title = {Myotubularin-related phosphatase 5 is a critical determinant of autophagy in neurons},
  journal = {Current Biology},
  year = {2022},
  volume = {32},
  number = {12},
  pages = {2581-2595.e6},
  doi = {10.1016/j.cub.2022.04.053}
}

@article{Dai2021,
  author = {Dai and others},
  title = {Aagab acts as a novel regulator of NEDD4-1-mediated Pten nuclear translocation to promote neurological recovery following hypoxic-ischemic brain damage},
 journal = {Cell Death and Differentiation},
  year = {2021},
  volume = {28},
  number = {8},
  pages = {2367-2384},
  doi = {10.1038/s41418-021-00757-4}
}

@article{Datta2020,
  author = {Datta and others},
  title = {Classical complement cascade initiating C1q protein within neurons in the aged rhesus macaque dorsolateral prefrontal cortex},
  journal = {Journal of Neuroinflammation},
  year = {2020},
  volume = {17},
  pages = {8},
  doi = {10.1186/s12974-019-1683-1}
}

@article{Datta2024,
  author = {Datta and others},
  title = {Key Roles of CACNA1C/Cav1.2 and CALB1/Calbindin in Prefrontal Neurons Altered in Cognitive Disorders},
  journal = {JAMA Psychiatry},
  year = {2024},
  volume = {81},
  number = {9},
  pages = {870-881},
  doi = {10.1001/jamapsychiatry.2024.1112}
}

@article{Davidson2013,
  author = {Davidson and others},
  title = {Mutations in ARL2BP, encoding ADP-ribosylation-factor-like 2 binding protein, cause autosomal-recessive retinitis pigmentosa},
  journal = {American Journal of Human Genetics},
  year = {2013},
  volume = {93},
  number = {2},
  pages = {321-9},
  doi = {10.1016/j.ajhg.2013.06.003}
}

@article{de2014,
  author = {de Bock and others},
  title = {Sperm-associated antigen 16 is a novel target of the humoral autoimmune response in multiple sclerosis},
  journal = {Journal of Immunology},
  year = {2014},
  volume = {193},
  number = {5},
  pages = {2147-56},
  doi = {10.4049/jimmunol.1401166}
}

@article{Deb2024,
  author = {Deb and others},
  title = {PSMD11 loss-of-function variants correlate with a neurobehavioral phenotype, obesity, and increased interferon response},
  journal = {American Journal of Human Genetics},
  year = {2024},
  volume = {111},
  number = {7},
  pages = {1352-1369},
  doi = {10.1016/j.ajhg.2024.05.016}
}

@article{Del2022,
  author = {Del Rosario and others},
  title = {TMEM120A/TACAN inhibits mechanically activated PIEZO2 channels},
  journal = {Journal of General Physiology},
  year = {2022},
  volume = {154},
  number = {8},
  pages = {e202213164},
  doi = {10.1085/jgp.202213164}
}

@article{Dharmapal2021,
  author = {Dharmapal and others},
  title = {beta-Tubulin Isotype, TUBB4B, Regulates The Maintenance of Cancer Stem Cells},
  journal = {Frontiers in Oncology},
  year = {2021},
  volume = {11},
  pages = {788024},
  doi = {10.3389/fonc.2021.788024}
}

@article{Ding2022,
  author = {Ding and others},
  title = {YY1 accelerates oral squamous cell carcinoma progression through long non-coding RNA Kcnq1ot1/microRNA-506-3p/SYPL1 axis},
  journal = {Journal of Ovarian Research},
  year = {2022},
  volume = {15},
  number = {1},
  pages = {77},
  doi = {10.1186/s13048-022-01000-5}
}

@article{Dixon2006,
  author = {Dixon and others},
  title = {Tcof1/Treacle is required for neural crest cell formation and proliferation deficiencies that cause craniofacial abnormalities},
  journal = {Proceedings of the National Academy of Sciences of the United States of America},
  year = {2006},
  volume = {103},
  number = {36},
  pages = {13403-13408},
  doi = {10.1073/pnas.0603730103}
}

@article{Dogra2025,
  author = {Dogra, A and Hasija, Y},
  title = {Unraveling prognostic biomarkers in oral squamous cell carcinoma: An approach based on explainable artificial intelligence},
  journal = {Cancer Genetics},
  year = {2025},
  volume = {296-297},
  pages = {163-171},
  doi = {10.1016/j.cancergen.2025.07.010}
}

@article{Drgonova2010,
  author = {Drgonova and others},
  title = {Effect of KEPI (Ppp1r14c) deletion on morphine analgesia and tolerance in mice of different genetic backgrounds: when a knockout is near a relevant quantitative trait locus},
  journal = {Neuroscience},
  year = {2010},
  volume = {165},
  number = {3},
  pages = {882-95},
  doi = {10.1016/j.neuroscience.2009.10.007}
}

@article{Drummond2012,
  author = {Drummond and others},
  title = {Upregulation of cornichon transcripts in the dorsolateral prefrontal cortex in schizophrenia},
  journal = {Neuroreport},
  year = {2012},
  volume = {23},
  number = {17},
  pages = {1031-4},
  doi = {10.1097/wnr.0b013e32835ad229}
}

@article{Dubey2023,
  author = {Dubey and others},
  title = {A novel de novo FEM1C variant is linked to neurodevelopmental disorder with absent speech, pyramidal signs and limb ataxia},
  journal = {Human Molecular Genetics},
  year = {2023},
  volume = {32},
  number = {7},
  pages = {1152-1161},
  doi = {10.1093/hmg/ddac276}
}

@article{Duche2025,
  author = {Duche and others},
  title = {Predictive gene expression signatures for Alzheimer's disease using post-mortem brain tissue},
  journal = {Frontiers in Aging Neuroscience},
  year = {2025},
  volume = {17},
  pages = {1591946},
  doi = {10.3389/fnagi.2025.1591946}
}

@article{Dutta2007,
  author = {Dutta and others},
  title = {Activation of the ciliary neurotrophic factor (CNTF) signalling pathway in cortical neurons of multiple sclerosis patients},
  journal = {Brain},
  year = {2007},
  volume = {130},
  number = {10},
  pages = {2566-2576},
  doi = {10.1093/brain/awm206}
}

@article{Elkouby2012,
  author = {Elkouby and others},
  title = {A hindbrain-repressive Wnt3a/Meis3/Tsh1 circuit promotes neuronal differentiation and coordinates tissue maturation},
  journal = {Development},
  year = {2012},
  volume = {139},
  number = {8},
  pages = {1487-97},
  doi = {10.1242/dev.072934}
}

@article{Endris2002,
  author = {Endris and others},
  title = {The novel Rho-GTPase activating gene MEGAP/srGAP3 has a putative role in severe mental retardation},
  journal = {Proceedings of the National Academy of Sciences of the United States of America},
  year = {2002},
  volume = {99},
  number = {18},
  pages = {11754-11759},
  doi = {10.1073/pnas.162241099}
}

@article{Etelinen2023,
  author = {Eteläinen and others},
  title = {A prolyl oligopeptidase inhibitor reduces tau pathology in cellular models and in mice with tauopathy},
  journal = {Science Translational Medicine},
  year = {2023},
  volume = {15},
  number = {691},
  pages = {eabq2915},
  doi = {10.1126/scitranslmed.abq2915}
}

@article{Fang2026a,
  author = {Fang and others},
  title = {The AGPS regulatory variant rs113671272 confers esophageal squamous cell carcinoma susceptibility through allele-specific MEIS3 binding and NF-kappaB activation in Chinese populations},
  journal = {EBioMedicine},
  year = {2026},
  volume = {127},
  pages = {106249},
  doi = {10.1016/j.ebiom.2026.106249}
}

@article{Fang2026b,
  author = {Fang and others},
  title = {The impact of heterodimeric G protein G12 family on proliferation, migration, and invasion in human oral squamous cell carcinoma cells},
  journal = {Journal of Stomatology, Oral and Maxillofacial Surgery},
  year = {2026},
  volume = {127},
  number = {3},
  pages = {102679},
  doi = {10.1016/j.jormas.2025.102679}
}

@article{Feinstein2014,
  author = {Feinstein and others},
  title = {VPS53 mutations cause progressive cerebello-cerebral atrophy type 2 (PCCA2)},
  journal = {Journal of Medical Genetics},
  year = {2014},
  volume = {51},
  number = {5},
  pages = {303-8},
  doi = {10.1136/jmedgenet-2013-101823}
}

@article{Feldcamp2017,
  author = {Feldcamp and others},
  title = {Pdxdc1 modulates prepulse inhibition of acoustic startle in the mouse},
  journal = {Translational Psychiatry},
  year = {2017},
  volume = {7},
  number = {5},
  pages = {e1125},
  doi = {10.1038/tp.2017.85}
}

@article{Feng2017,
  author = {Feng and others},
  title = {High expression of heat shock protein 10 (Hsp10) is associated with poor prognosis in oral squamous cell carcinoma},
  journal = {International Journal of Clinical and Experimental Pathology},
  year = {2017},
  volume = {10},
  number = {7},
  pages = {7784-7791},
  pmid = {31966626}
}

@article{Frye2024,
  author = {Frye and others},
  title = {Mediterranean diet protects against a neuroinflammatory cortical transcriptome: Associations with brain volumetrics, peripheral inflammation, social isolation, and anxiety in nonhuman primates (Macaca fascicularis)},
  journal = {Brain, Behavior, and Immunity},
  year = {2024},
  volume = {119},
  pages = {681-692},
  doi = {10.1016/j.bbi.2024.04.016}
}

@article{Fullard2018,
  author = {Fullard and others},
  title = {An atlas of chromatin accessibility in the adult human brain},
  journal = {Genome Research},
  year = {2018},
  volume = {28},
  number = {8},
  pages = {1243-1252},
  doi = {10.1101/gr.232488.117}
}

@article{Gana2024,
  author = {Gana and others},
  title = {LZTFL1, a rare cause of Bardet-Biedl syndrome: A new patient with severe short stature and moderate intellectual disability, more than casual associations?},
  journal = {American Journal of Medical Genetics. Part a},
  year = {2024},
  volume = {194},
  number = {10},
  pages = {e63723},
  doi = {10.1002/ajmg.a.63723}
}

@article{Gao2020,
  author = {Gao and others},
  title = {The RBP1-CKAP4 axis activates oncogenic autophagy and promotes cancer progression in oral squamous cell carcinoma},
  journal = {Cell Death \& Disease},
  year = {2020},
  volume = {11},
  number = {6},
  pages = {488},
  doi = {10.1038/s41419-020-2693-8}
}

@article{Gao2024,
  author  = {Gao, M. and others},
  title   = {Ribosomal Dysregulation in Metastatic Laryngeal Squamous Cell Carcinoma: Proteomic Insights and CX-5461's Therapeutic Promise},
  journal = {Toxics},
  year    = {2024},
  volume  = {12},
  number  = {5},
  pages   = {363},
  doi     = {10.3390/toxics12050363}
}

@article{German2025,
  author = {German and others},
  title = {Comprehensive genotypic, phenotypic, and biochemical characterization of GOT2 deficiency: A progressive neurodevelopmental disorder with epilepsy and abnormal movements},
  journal = {Genetics in Medicine},
  year = {2025},
  volume = {27},
  number = {12},
  pages = {101587},
  doi = {10.1016/j.gim.2025.101587}
}

@article{Grnhj2018,
  author = {Grønhøj and others},
  title = {Deep sequencing of human papillomavirus positive loco-regionally advanced oropharyngeal squamous cell carcinomas reveals novel mutational signature},
  journal = {BMC Cancer},
  year = {2018},
  volume = {18},
  number = {1},
  pages = {640},
  doi = {10.1186/s12885-018-4567-3}
}

@article{Groffen2010,
  author = {Groffen and others},
  title = {Doc2b Is a High-Affinity Ca2+ Sensor for Spontaneous Neurotransmitter Release},
  journal = {Science},
  year = {2010},
  volume = {327},
  pages = {1614-1618},
  doi = {10.1126/science.1183765}
}

@article{Gueneau2018,
  author = {Gueneau and others},
  title = {KIAA1109 Variants Are Associated with a Severe Disorder of Brain Development and Arthrogryposis},
  journal = {American Journal of Human Genetics},
  year = {2018},
  volume = {102},
  number = {1},
  pages = {116-132},
  doi = {10.1016/j.ajhg.2017.12.002}
}

@article{Gui2025,
  author = {Gui and others},
  title = {Proteome-Wide and Immune Cell Phenotype Mendelian Randomization Highlights Immune Involvement in Genetic Generalized Epilepsy},
  journal = {Brain and Behavior},
  year = {2025},
  volume = {15},
  number = {6},
  pages = {e70625},
  doi = {10.1002/brb3.70625}
}

@article{Guiberson2024,
  author = {Guiberson and others},
  title = {Disease-linked mutations in Munc18-1 deplete synaptic Doc2},
  journal = {Brain},
  year = {2024},
  volume = {147},
  number = {6},
  pages = {2185-2202},
  doi = {10.1093/brain/awae019}
}

@article{Guo2025HIPK4,
  author = {Guo and others},
  title = {HIPK4 accelerates cutaneous squamous cell carcinoma progression by phosphorylating TAp63 and inhibiting EFEMP1 expression},
  journal = {The Journal of Biological Chemistry},
  year = {2025},
  volume = {301},
  number = {7},
  pages = {108564},
  doi = {10.1016/j.jbc.2025.108564}
}

@article{Guo2026,
  author  = {Guo, L. and others},
  title   = {Impact of NPAS2 on mPFC dopamine synthesis and nap behavior},
  journal = {Nature Communications},
  year    = {2026},
  volume  = {17},
  pages   = {4014},
  doi     = {10.1038/s41467-026-70424-0}
}

@article{Ha2014,
  author  = {Ha, J. Y. and others},
  title   = {Tnfaip8l1/Oxi-$\beta$ binds to FBXW5, increasing autophagy through activation of TSC2 in a Parkinson's disease model},
  journal = {Journal of Neurochemistry},
  year    = {2014},
  volume  = {129},
  number  = {3},
  pages   = {527--538},
  doi     = {10.1111/jnc.12643}
}

@article{Hait2014,
  author = {Hait and others},
  title = {Active, phosphorylated fingolimod inhibits histone deacetylases and facilitates fear extinction memory},
  journal = {Nature Neuroscience},
  year = {2014},
  volume = {17},
  number = {7},
  pages = {971-80},
  doi = {10.1038/nn.3728}
}

@article{Hammerschmidt2023,
  author = {Hammerschmidt and others},
  title = {CerS6-dependent ceramide synthesis in hypothalamic neurons promotes ER/mitochondrial stress and impairs glucose homeostasis in obese mice},
  journal = {Nature Communications},
  year = {2023},
  volume = {14},
  number = {1},
  pages = {7824},
  doi = {10.1038/s41467-023-42595-7}
}

@article{Harada2021,
  author = {Harada and others},
  title = {Cell cycle arrest determines adult neural stem cell ontogeny by an embryonic Notch-nonoscillatory Hey1 module},
  journal = {Nature Communications},
  year = {2021},
  volume = {12},
  pages = {6562},
  doi = {10.1038/s41467-021-26605-0}
}

@article{Harich2020,
  author = {Harich and others},
  title = {From man to fly - convergent evidence links FBXO25 to ADHD and comorbid psychiatric phenotypes},
 journal = {Journal of Child Psychology and Psychiatry},
  year = {2020},
  volume = {61},
  number = {5},
  pages = {545-555},
  doi = {10.1111/jcpp.13161}
}

@article{Hasegawa2012,
  author = {Hasegawa and others},
  title = {Acetoacetyl-CoA synthetase is essential for normal neuronal development},
  journal = {Biochemical and Biophysical Research Communications},
  year = {2012},
  volume = {427},
  number = {2},
  pages = {398-403},
  doi = {10.1016/j.bbrc.2012.09.076}
}

@article{Hassan2025,
  author = {Hassan and others},
  title = {Cellular and subcellular distribution of the K+-dependent Na+/Ca2+-exchanger subtype 4, NCKX4, in mouse brain},
  journal = {Neuroscience},
  year = {2025},
  volume = {569},
  pages = {210-230},
  doi = {10.1016/j.neuroscience.2025.02.013}
}

@article{He2021,
  author = {He and others},
  title = {Increased VLCFA-lipids and ELOVL4 underlie neurodegeneration in frontotemporal dementia},
  journal = {Scientific Reports},
  year = {2021},
  volume = {11},
  number = {1},
  pages = {21348},
  doi = {10.1038/s41598-021-00870-x}
}

@article{Hsieh2023,
  author = {Hsieh and others},
  title = {Altered synaptic protein expression, aberrant spine morphology, and impaired spatial memory in Dlgap2 mutant mice, a genetic model of autism spectrum disorder},
  journal = {Cerebral Cortex},
  year = {2023},
  volume = {33},
  number = {8},
  pages = {4779-4793},
  doi = {10.1093/cercor/bhac379}
}

@article{Hsu2019,
  author = {Hsu and others},
  title = {Proteomic Profiling of Paired Interstitial Fluids Reveals Dysregulated Pathways and Salivary NID1 as a Biomarker of Oral Cavity Squamous Cell Carcinoma},
  journal = {Molecular \& Cellular Proteomics},
  year = {2019},
  volume = {18},
  number = {10},
  pages = {1939-1949},
  doi = {10.1074/mcp.ra119.001654}
}

@article{Hu2016,
  author = {Hu and others},
  title = {Postsynaptic SDC2 induces transsynaptic signaling via FGF22 for bidirectional synaptic formation},
  journal = {Scientific Reports},
  year = {2016},
  volume = {6},
  pages = {33592},
  doi = {10.1038/srep33592}
}

@article{Hu2023,
  author = {Hu and others},
  title = {HMGCS2-Induced Autophagic Degradation of Tau Involves Ketone Body and ANKRD24},
  journal = {Journal of Alzheimer's Disease},
  year = {2023},
  volume = {91},
  number = {1},
  pages = {407-426},
  doi = {10.3233/jad-220640}
}

@article{Hu2025,
  author = {Hu and others},
  title = {Inhibition of DDR1 potentiates carbon ion radiotherapy by promoting ferroptosis and immunogenic death in head and neck squamous cell carcinoma},
  journal = {Journal of Translational Medicine},
  year = {2025},
  volume = {23},
  number = {1},
  pages = {1011},
  doi = {10.1186/s12967-025-07062-5}
}

@article{Hu2026,
  author = {Hu and others},
  title = {RetSat Knockout Mitigates Hypoxia-Induced Microglial Activation by Enhancing Lipid Droplets Degradation},
  journal = {Glia},
  year = {2026},
  volume = {74},
  number = {2},
  pages = {e70118},
  doi = {10.1002/glia.70118}
}

@article{Huang2020,
  author = {Huang and others},
  title = {Progesterone receptor membrane component 1 is involved in oral cancer cell metastasis},
  journal = {Journal of Cellular and Molecular Medicine},
  year = {2020},
  volume = {24},
  number = {17},
  pages = {9737-9751},
  doi = {10.1111/jcmm.15535}
}

@article{Huang2025,
  author = {Huang and others},
  title = {Neoadjuvant Chemotherapy With Cisplatin Up-Regulates GSDMD to Enhance Oral Squamous Cell Carcinoma Metastasis Through MMP14-Mediated EMT Activation},
  journal = {Advanced Science},
  year = {2025},
  volume = {12},
  number = {25},
  pages = {e2501149},
  doi = {10.1002/advs.202501149}
}

@article{Hughes2020,
  author = {Hughes and others},
  title = {Loss-of-Function Variants in PPP1R12A: From Isolated Sex Reversal to Holoprosencephaly Spectrum and Urogenital Malformations.},
  journal = {American Journal of Human Genetics},
  year = {2020},
  volume = {106},
  number = {1},
  pages = {121-128},
  doi = {10.1016/j.ajhg.2019.12.004}
}

@article{Iguchi2021,
  author = {Iguchi and others},
  title = {Mutually Repulsive EphA7-EfnA5 Organize Region-to-Region Corticopontine Projection by Inhibiting Collateral Extension.},
  journal = {The Journal of Neuroscience},
  year = {2021},
  volume = {41},
  number = {22},
  pages = {4795-4808},
  doi = {10.1523/jneurosci.0367-20.2021}
}

@article{Ishino2022,
  author = {Ishino and others},
  title = {Coactivator-associated arginine methyltransferase 1 controls oligodendrocyte differentiation in the corpus callosum during early brain development},
  journal = {Developmental Neurobiology},
  year = {2022},
  volume = {82},
  number = {3},
  pages = {245-260},
  doi = {10.1002/dneu.22871}
}

@article{Ito2018,
  author = {Ito and others},
  title = {Pallidin is a novel interacting protein for cytohesin-2 and regulates the early endosomal pathway and dendritic formation in neurons},
  journal = {Journal of Neurochemistry},
  year = {2018},
  volume = {147},
  number = {2},
  pages = {153-177},
  doi = {10.1111/jnc.14579}
}

@article{Iuliano2018,
  author = {Iuliano and others},
  title = {Myosin 1b promotes axon formation by regulating actin wave propagation and growth cone dynamics},
  journal = {Journal of Cell Biology},
  year = {2018},
  volume = {217},
  number = {6},
  pages = {2033-2046},
  doi = {10.1083/jcb.201703205}
}

@article{Jackson2014,
  author = {Jackson and others},
  title = {A neuron autonomous role for the familial dysautonomia gene ELP1 in sympathetic and sensory target tissue innervation},
  journal = {Development},
  year = {2014},
  volume = {141},
  number = {12},
  pages = {2452-61},
  doi = {10.1242/dev.107797}
}

@article{Jia2023,
  author = {Jia and others},
  title = {Brain proteome-wide association study linking-genes in multiple sclerosis pathogenesis},
  journal = {Annals of Clinical and Translational Neurology},
  year = {2023},
  volume = {10},
  number = {1},
  pages = {58-69},
  doi = {10.1002/acn3.51699}
}

@article{Jiang2010,
  author = {Jiang and others},
  title = {Quantitative proteomics analysis of inborn errors of cholesterol synthesis: identification of altered metabolic pathways in DHCR7 and SC5D deficiency},
  journal = {Molecular \& Cellular Proteomics},
  year = {2010},
  volume = {9},
  number = {7},
  pages = {1461-75},
  doi = {10.1074/mcp.m900548-mcp200}
}

@article{Jiang2022,
  author = {Jiang and others},
  title = {lncRNA lnc-POP1-1 upregulated by VN1R5 promotes cisplatin resistance in head and neck squamous cell carcinoma through interaction with MCM5},
  journal = {Molecular Therapy},
  year = {2022},
  volume = {30},
  number = {1},
  pages = {448-467},
  doi = {10.1016/j.ymthe.2021.06.006}
}

@article{Jiang2025a,
  author  = {Jiang, M. and others},
  title   = {Smg5 Enhances Oligodendrocyte Differentiation via Nonsense-Mediated mRNA Decay of Hnrnpl Variant Transcripts},
  journal = {Journal of Neuroscience},
  year    = {2025},
  volume  = {45},
  number  = {43},
  pages   = {e0371252025},
  doi     = {10.1523/JNEUROSCI.0371-25.2025}
}

@article{Jiang2025b,
  author = {Jiang and others},
  title = {Case Report: Biallelic variants in MRPS36, encoding a component of the 2-oxoglutarate dehydrogenase complex, cause leigh syndrome},
  journal = {Frontiers in Pediatrics},
  year = {2025},
  volume = {13},
  pages = {1608840},
  doi = {10.3389/fped.2025.1608840}
}

@article{Jishi2024,
  author = {Jishi and others},
  title = {BCKDK loss impairs mitochondrial Complex I activity and drives alpha-synuclein aggregation in models of Parkinson's disease},
  journal = {Acta Neuropathologica Communications},
  year = {2024},
  volume = {12},
  number = {1},
  pages = {198},
  doi = {10.1186/s40478-024-01915-8}
}

@article{Jung2020,
  author  = {Jung, J. E. and others},
  title   = {MicroRNA-31 Regulates Expression of Wntless in Both Drosophila melanogaster and Human Oral Cancer Cells},
  journal = {International Journal of Molecular Sciences},
  year    = {2020},
  volume  = {21},
  number  = {19},
  pages   = {7232},
  doi     = {10.3390/ijms21197232}
}

@article{Kaizuka2024,
  author = {Kaizuka and others},
  title = {FAM81A is a postsynaptic protein that regulates the condensation of postsynaptic proteins via liquid-liquid phase separation},
  journal = {PLoS Biology},
  year = {2024},
  volume = {22},
  number = {3},
  pages = {e3002006},
  doi = {10.1371/journal.pbio.3002006}
}

@article{Kang2022,
  author = {Kang and others},
  title = {FBXL17/spastin axis as a novel therapeutic target of hereditary spastic paraplegia},
  journal = {Cell \& Bioscience},
  year = {2022},
  volume = {12},
  number = {1},
  pages = {110},
  doi = {10.1186/s13578-022-00851-1}
}

@article{Katz2001,
  author = {Katz and others},
  title = {Regulation of caspases and XIAP in the brain after asphyxial cardiac arrest in rats},
  journal = {Neuroreport},
  year = {2001},
  volume = {12},
  number = {17},
  pages = {3751-4},
  doi = {10.1097/00001756-200112040-00029}
}

@article{Key2025,
  author  = {Key, J. and others},
  title   = {Conditional ATXN2L-Null in Adult Frontal Cortex CamK2a+ Neurons Does Not Cause Cell Death but Restricts Spontaneous Mobility and Affects the Alternative Splicing Pathway},
  journal = {Cells},
  year    = {2025},
  volume  = {14},
  number  = {19},
  pages   = {1532},
  doi     = {10.3390/cells14191532}
}

@article{Khoury2021,
  author  = {Khoury, S. and others},
  title   = {Multi-ethnic GWAS and meta-analysis of sleep quality identify MPP6 as a novel gene that functions in sleep center neurons},
  journal = {Sleep},
  year    = {2021},
  volume  = {44},
  number  = {3},
  pages   = {zsaa211},
  doi     = {10.1093/sleep/zsaa211}
}

@article{Kim2000,
  author = {Kim and others},
  title = {Dolichol phosphate mannose synthase (DPM1) mutations define congenital disorder of glycosylation Ie (CDG-Ie)},
  journal = {The Journal of Clinical Investigation},
  year = {2000},
  volume = {105},
  number = {2},
  pages = {191-8},
  doi = {10.1172/jci7302}
}

@article{King2013,
  author = {King and others},
  title = {Topoisomerases facilitate transcription of long genes linked to autism},
  journal = {Nature},
  year = {2013},
  volume = {501},
  number = {7465},
  pages = {58-62},
  doi = {10.1038/nature12504}
}

@article{Kohli2011,
  author = {Kohli and others},
  title = {The neuronal transporter gene SLC6A15 confers risk to major depression},
  journal = {Neuron},
  year = {2011},
  volume = {70},
  number = {2},
  pages = {252-65},
  doi = {10.1016/j.neuron.2011.04.005}
}

@article{Konopacka2015,
  author  = {Konopacka, A. and others},
  title   = {RNA binding protein Caprin-2 is a pivotal regulator of the central osmotic defense response},
  journal = {eLife},
  year    = {2015},
  volume  = {4},
  pages   = {e09656},
  doi     = {10.7554/eLife.09656}
}

@article{Kwon2000,
  author = {Kwon and others},
  title = {Altered activity, social behavior, and spatial memory in mice lacking the NTAN1p amidase and the asparagine branch of the N-end rule pathway},
  journal = {Molecular and Cellular Biology},
  year = {2000},
  volume = {20},
  number = {11},
  pages = {4135-48},
  doi = {10.1128/mcb.20.11.4135-4148.2000}
}

@article{Lacrampe2025,
  author = {Lacrampe and others},
  title = {Myristoylation of TMEM106B by NMT1/2 regulates TMEM106B trafficking and turnover},
  journal = {The Journal of Biological Chemistry},
  year = {2025},
  volume = {301},
  number = {7},
  pages = {110322},
  doi = {10.1016/j.jbc.2025.110322}
}

@article{Lancaster2024,
  author = {Lancaster and others},
  title = {The RNA-binding protein Nab2 regulates levels of the RhoGEF Trio to govern axon and dendrite morphology},
  journal = {Molecular Biology of the Cell},
  year = {2024},
  volume = {35},
  number = {8},
  pages = {ar109},
  doi = {10.1091/mbc.e24-04-0150}
}

@article{Lehr2025,
  author  = {Lehr, A. W. and others},
  title   = {Phosphorylation of NLGN4X Regulates Spinogenesis and Synaptic Function},
  journal = {eNeuro},
  year    = {2025},
  volume  = {12},
  number  = {3},
  pages   = {ENEURO.0278-23.2025},
  doi     = {10.1523/ENEURO.0278-23.2025}
}

@article{Lei2024,
  author = {Lei and others},
  title = {Human retinal organoids with an OPA1 mutation are defective in retinal ganglion cell differentiation and function},
  journal = {Stem Cell Reports},
  year = {2024},
  volume = {19},
  number = {1},
  pages = {68-83},
  doi = {10.1016/j.stemcr.2023.11.004}
}

@article{Li2018,
  author = {Li and others},
  title = {Picroside II Exerts a Neuroprotective Effect by Inhibiting mPTP Permeability and EndoG Release after Cerebral Ischemia/Reperfusion Injury in Rats},
  journal = {Journal of Molecular Neuroscience},
  year = {2018},
  volume = {64},
  number = {1},
  pages = {144-155},
  doi = {10.1007/s12031-017-1012-z}
}

@article{Li2019,
  author = {Li and others},
  title = {Rap1A promotes esophageal squamous cell carcinoma metastasis through the AKT signaling pathway},
  journal = {Oncology Reports},
  year = {2019},
  volume = {42},
  number = {5},
  pages = {1815-1824},
  doi = {10.3892/or.2019.7309}
}

@article{Li2022,
  author = {Li and others},
  title = {MicroRNA-486-3p promotes the proliferation and metastasis of cutaneous squamous cell carcinoma by suppressing flotillin-2},
  journal = {Journal of Dermatological Science},
  year = {2022},
  volume = {105},
  number = {1},
  pages = {18-26},
  doi = {10.1016/j.jdermsci.2021.11.005}
}

@article{Li2023a,
  author = {Li and others},
  title = {USP25 Inhibits Neuroinflammatory Responses After Cerebral Ischemic Stroke by Deubiquitinating TAB2},
  journal = {Advanced Science},
  year = {2023},
  volume = {10},
  number = {28},
  pages = {e2301641},
  doi = {10.1002/advs.202301641}
}

@article{Li2023b,
  author = {Li and others},
  title = {The RNA polymerase II subunit B (RPB2) functions as a growth regulator in human glioblastoma},
  journal = {Biochemical and Biophysical Research Communications},
  year = {2023},
  volume = {674},
  pages = {170-182},
  doi = {10.1016/j.bbrc.2023.06.088}
}

@article{Li2023c,
  author = {Li and others},
  title = {Targeting ANXA7/LAMP5-mTOR axis attenuates spinal cord injury by inhibiting neuronal apoptosis via enhancing autophagy in mice},
  journal = {Cell Death Discovery},
  year = {2023},
  volume = {9},
  number = {1},
  pages = {309},
  doi = {10.1038/s41420-023-01612-w}
}

@article{Li2023d,
  author = {Li and others},
  title = {MicroRNA-128 suppresses tau phosphorylation and reduces amyloid-beta accumulation by inhibiting the expression of GSK3beta, APPBP2, and mTOR in Alzheimer's disease},
  journal = {CNS Neuroscience \& Therapeutics},
  year = {2023},
  volume = {29},
  number = {7},
  pages = {1848-1864},
  doi = {10.1111/cns.14143}
}

@article{Li2023e,
  author = {Li and others},
  title = {Regulation of Schwann cell proliferation and migration via miR-195-5p-induced Crebl2 downregulation upon peripheral nerve damage},
  journal = {Frontiers in Cellular Neuroscience},
  year = {2023},
  volume = {17},
  pages = {1173086},
  doi = {10.3389/fncel.2023.1173086}
}

@article{Li2024a,
  author = {Li and others},
  title = {Modulating the RPS27A/PSMD12/NF-kappaB pathway to control immune response in mouse brain ischemia-reperfusion injury},
  journal = {Molecular Medicine},
  year = {2024},
  volume = {30},
  number = {1},
  pages = {106},
  doi = {10.1186/s10020-024-00870-3}
}

@article{Li2024b,
  author = {Li and others},
  title = {Wee1 inhibitor PD0166285 sensitized TP53 mutant lung squamous cell carcinoma to cisplatin via STAT1},
  journal = {Cancer Cell International},
  year = {2024},
  volume = {24},
  number = {1},
  pages = {315},
  doi = {10.1186/s12935-024-03489-w}
}

@article{Liang2023,
  author = {Liang and others},
  title = {Mutation-associated transcripts reconstruct the prognostic features of oral tongue squamous cell carcinoma},
  journal = {International Journal of Oral Science},
  year = {2023},
  volume = {15},
  number = {1},
  pages = {1},
  doi = {10.1038/s41368-022-00210-3}
}

@article{Liao2021,
  author = {Liao and others},
  title = {Silencing SHMT2 inhibits the progression of tongue squamous cell carcinoma through cell cycle regulation},
  journal = {Cancer Cell International},
  year = {2021},
  volume = {21},
  number = {1},
  pages = {220},
  doi = {10.1186/s12935-021-01880-5}
}

@article{LibPhilippot2024,
  author = {Libé-Philippot and others},
  title = {Synaptic neoteny of human cortical neurons requires species-specific balancing of SRGAP2-SYNGAP1 cross-inhibition},
  journal = {Neuron},
  year = {2024},
  volume = {112},
  number = {21},
  pages = {3602-3617.e9},
  doi = {10.1016/j.neuron.2024.08.021}
}

@article{Lin2009,
  author = {Lin and others},
  title = {Neuron-derived FGF9 is essential for scaffold formation of Bergmann radial fibers and migration of granule neurons in the cerebellum},
  journal = {Developmental Biology},
  year = {2009},
  volume = {329},
  number = {1},
  pages = {44-54},
  doi = {10.1016/j.ydbio.2009.02.011}
}

@article{Lintell2007,
  author = {Lintell and others},
  title = {Analysis of SDHD and MMP12 in an affected solar keratosis and control cohort},
  journal = {Advances in Experimental Medicine and Biology},
  year = {2007},
  volume = {599},
  pages = {79-85},
  doi = {10.1007/978-0-387-71764-7_11}
}

@article{Liu2019,
  author = {Liu, Y and Ye, F},
  title = {Construction and integrated analysis of crosstalking ceRNAs networks in laryngeal squamous cell carcinoma},
  journal = {PeerJ},
  year = {2019},
  volume = {7},
  pages = {e7380},
  doi = {10.7717/peerj.7380}
}

@article{Liu2020a,
  author = {Liu and others},
  title = {TRA2A-induced upregulation of LINC00662 regulates blood-brain barrier permeability by affecting ELK4 mRNA stability in Alzheimer's microenvironment},
  journal = {RNA Biology},
  year = {2020},
  volume = {17},
  number = {9},
  pages = {1293-1308},
  doi = {10.1080/15476286.2020.1756055}
}

@article{Liu2020b,
  author = {Liu and others},
  title = {Proteomic Characterization of Proliferation Inhibition of Well-Differentiated Laryngeal Squamous Cell Carcinoma Cells Under Below-Background Radiation in a Deep Underground Environment},
  journal = {Frontiers in Public Health},
  year = {2020},
  volume = {8},
  pages = {584964},
  doi = {10.3389/fpubh.2020.584964}
}

@article{Liu2020c,
  author = {Liu and others},
  title = {lncRNA RASSF8‑AS1 suppresses the progression of laryngeal squamous cell carcinoma via targeting the miR‑664b‑3p/TLE1 axis},
  journal = {Oncology Reports},
  year = {2020},
  volume = {44},
  number = {5},
  pages = {2031-2044},
  doi = {10.3892/or.2020.7771}
}

@article{Liu2021,
  author = {Liu and others},
  title = {Identification of hub ubiquitin ligase genes affecting Alzheimer's disease by analyzing transcriptome data from multiple brain regions},
  journal = {Science Progress},
  year = {2021},
  volume = {104},
  number = {1},
  pages = {368504211001146},
  doi = {10.1177/00368504211001146}
}

@article{Liu2022MAP3K11,
  author = {Liu and others},
  title = {MAP3K11 facilitates autophagy activity and is correlated with malignancy of oral squamous cell carcinoma},
  journal = {Journal of Cellular Physiology},
  year = {2022},
  volume = {237},
  number = {11},
  pages = {4275-4291},
  doi = {10.1002/jcp.30881}
}

@article{Liu2023a,
  author = {Liu and others},
  title = {Mutation of the attractin gene impairs working memory in rats},
  journal = {Brain and Behavior},
  year = {2023},
  volume = {13},
  number = {2},
  pages = {e2876},
  doi = {10.1002/brb3.2876}
}

@article{Liu2023b,
  author = {Liu and others},
  title = {PSMA1, a Poor Prognostic Factor, Promotes Tumor Growth in Lung Squamous Cell Carcinoma},
  journal = {Disease Markers},
  year = {2023},
  volume = {2023},
  pages = {5386635},
  doi = {10.1155/2023/5386635}
}

@article{Liu2025a,
  author = {Liu and others},
  title = {ATP8A2 expression is reduced in the mPFC of offspring mice exposed to maternal immune activation and its upregulation ameliorates synapse-associated protein loss and behavioral abnormalities},
  journal = {Brain, Behavior, and Immunity},
  year = {2025},
  volume = {124},
  pages = {409-430},
  doi = {10.1016/j.bbi.2024.12.018}
}

@article{Liu2025b,
  author = {Liu and others},
  title = {SQLE amplification accelerates esophageal squamous cell carcinoma tumorigenesis and metastasis through oncometabolite 2,3-oxidosqualene repressing Hippo pathway},
  journal = {Cancer Letters},
  year = {2025},
  volume = {621},
  pages = {217528},
  doi = {10.1016/j.canlet.2025.217528}
}

@article{Long2025,
  author  = {Long, W. and others},
  title   = {SCFFBXO21-mediated ubiquitination and degradation of NMNAT2 regulates axon survival in nerve injury},
  journal = {Journal of Cell Biology},
  year    = {2025},
  volume  = {224},
  number  = {11},
  pages   = {e202501072},
  doi     = {10.1083/jcb.202501072}
}

@article{LopezDomenech2021,
  author = {Lopez-Domenech and others},
  title = {Loss of neuronal Miro1 disrupts mitophagy and induces hyperactivation of the integrated stress response},
  journal = {EMBO Journal},
  year = {2021},
  volume = {40},
  number = {14},
  pages = {e100715},
  doi = {10.15252/embj.2018100715}
}

@article{Lu2011,
  author = {Lu, PP and Ramanan, N},
  title = {Serum response factor is required for cortical axon growth but is dispensable for neurogenesis and neocortical lamination},
  journal = {The Journal of Neuroscience},
  year = {2011},
  volume = {31},
  number = {46},
  pages = {16651-64},
  doi = {10.1523/jneurosci.3015-11.2011}
}

@article{Lu2024,
  author = {Lu and others},
  title = {Overexpression of TPD52L2 in HNSCC: prognostic significance and correlation with immune infiltrates},
  journal = {BMC Oral Health},
  year = {2024},
  volume = {24},
  number = {1},
  pages = {1191},
  doi = {10.1186/s12903-024-04977-1}
}

@article{Lyu2025,
  author = {Lyu and others},
  title = {Integrated bioinformatics analysis of differences between EAC and ESCC},
  journal = {BMC Cancer},
  year = {2025},
  volume = {25},
  number = {1},
  pages = {1668},
  doi = {10.1186/s12885-025-15090-z}
}

@article{Ma2018,
  author = {Ma and others},
  title = {The integrated landscape of causal genes and pathways in schizophrenia},
  journal = {Translational Psychiatry},
  year = {2018},
  volume = {8},
  number = {1},
  pages = {67},
  doi = {10.1038/s41398-018-0114-x}
}

@article{Magrinelli2025,
  author  = {Magrinelli, F. and others},
  title   = {Biallelic NDUFA9 variants cause a progressive neurodevelopmental disorder with prominent dystonia and mitochondrial complex I deficiency},
  journal = {Brain Communications},
  year    = {2025},
  volume  = {7},
  number  = {5},
  pages   = {fcaf369},
  doi     = {10.1093/braincomms/fcaf369}
}

@article{Malhotra2021,
  author = {Malhotra and others},
  title = {De novo missense variants in LMBRD2 are associated with developmental and motor delays, brain structure abnormalities and dysmorphic features},
  journal = {Journal of Medical Genetics},
  year = {2021},
  volume = {58},
  number = {10},
  pages = {712-716},
  doi = {10.1136/jmedgenet-2020-107137}
}

@article{Marquez2023,
  author = {Marquez and others},
  title = {Expanding EMC foldopathies: Topogenesis deficits alter the neural crest},
  journal = {Genesis},
  year = {2023},
  volume = {61},
  number = {5},
  pages = {e23520},
  doi = {10.1002/dvg.23520}
}

@article{Martin2021,
  author = {Martin and others},
  title = {DYRK1A is required for maintenance of cancer stemness, contributing to tumorigenic potential in oral/oropharyngeal squamous cell carcinoma},
  journal = {Experimental Cell Research},
  year = {2021},
  volume = {405},
  number = {1},
  pages = {112656},
  doi = {10.1016/j.yexcr.2021.112656}
}

@article{Martin2025,
  author = {Martin and others},
  title = {Age-related nigral downregulation of the Parkinson's risk factor FAM49B primes human microglia for inflammaging},
  journal = {NPJ Aging},
  year = {2025},
  volume = {12},
  number = {1},
  pages = {1},
  doi = {10.1038/s41514-025-00296-z}
}

@article{MartinezdelaTorre2018,
  author = {Martinez-de-la-Torre and others},
  title = {An exercise in brain genoarchitectonics: Analysis of AZIN2-Lacz expressing neuronal populations in the mouse hindbrain},
  journal = {Journal of Neuroscience Research},
  year = {2018},
  volume = {96},
  number = {9},
  pages = {1490-1517},
  doi = {10.1002/jnr.24053}
}

@article{McGowan2009,
  author = {McGowan and others},
  title = {Epigenetic regulation of the glucocorticoid receptor in human brain associates with childhood abuse},
  journal = {Nature Neuroscience},
  year = {2009},
  volume = {12},
  number = {3},
  pages = {342-8},
  doi = {10.1038/nn.2270}
}

@article{McGrath2022,
  author = {McGrath and others},
  title = {Plasma EFEMP1 Is Associated with Brain Aging and Dementia: The Framingham Heart Study},
  journal = {Journal of Alzheimer's Disease},
  year = {2022},
  volume = {85},
  number = {4},
  pages = {1657-1666},
  doi = {10.3233/jad-215053}
}

@article{McGuigan2023,
  author  = {McGuigan, B. N. and others},
  title   = {Gene Expressions Preferentially Influence Cortical Thickness of Human Connectome Project Atlas Parcellated Regions in First-Episode Antipsychotic-Naïve Psychoses},
  journal = {Schizophrenia Bulletin Open},
  year    = {2023},
  volume  = {4},
  number  = {1},
  pages   = {sgad019},
  doi     = {10.1093/schizbullopen/sgad019}
}

@article{McHugh2020,
  author = {McHugh and others},
  title = {The Identification of Potential Therapeutic Targets for Cutaneous Squamous Cell Carcinoma},
  journal = {The Journal of Investigative Dermatology},
  year = {2020},
  volume = {140},
  number = {6},
  pages = {1154-1165.e5},
  doi = {10.1016/j.jid.2019.09.024}
}

@article{Melchior2010,
  author = {Melchior and others},
  title = {Dual induction of TREM2 and tolerance-related transcript, Tmem176b, in amyloid transgenic mice: implications for vaccine-based therapies for Alzheimer's disease},
  journal = {ASN Neuro},
  year = {2010},
  volume = {2},
  number = {3},
  pages = {e00037},
  doi = {10.1042/an20100010}
}

@article{Miao2025,
  author = {Miao and others},
  title = {A novel brain-derived peptide inhibits microglial pyroptosis through MBTPS1 in neonatal hypoxic-ischemic brain damage},
  journal = {Brain Research Bulletin},
  year = {2025},
  volume = {233},
  pages = {111664},
  doi = {10.1016/j.brainresbull.2025.111664}
}

@article{Milev2017,
  author = {Milev and others},
  title = {Mutations in TRAPPC12 Manifest in Progressive Childhood Encephalopathy and Golgi Dysfunction},
  journal = {American Journal of Human Genetics},
  year = {2017},
  volume = {101},
  number = {2},
  pages = {291-299},
  doi = {10.1016/j.ajhg.2017.07.006}
}

@article{Miroci2012,
  author = {Miroci and others},
  title = {Makorin ring zinc finger protein 1 (MKRN1), a novel poly(A)-binding protein-interacting protein, stimulates translation in nerve cells},
  journal = {The Journal of Biological Chemistry},
  year = {2012},
  volume = {287},
  number = {2},
  pages = {1322-34},
  doi = {10.1074/jbc.m111.315291}
}

@article{Mori2012,
  author = {Mori and others},
  title = {Ccdc85c encoding a protein at apical junctions of radial glia is disrupted in hemorrhagic hydrocephalus (hhy) mice},
  journal = {American Journal of Pathology},
  year = {2012},
  volume = {180},
  number = {1},
  pages = {314-27},
  doi = {10.1016/j.ajpath.2011.09.014}
}

@article{Morrison2021,
  author = {Morrison and others},
  title = {Missense NAA20 variants impairing the NatB protein N-terminal acetyltransferase cause autosomal recessive developmental delay, intellectual disability, and microcephaly},
  journal = {Genetics in Medicine},
  year = {2021},
  volume = {23},
  pages = {2213-2218},
  doi = {10.1038/s41436-021-01264-0}
}

@article{Murillo2015,
  author = {Murillo and others},
  title = {Zic2 Controls the Migration of Specific Neuronal Populations in the Developing Forebrain},
  journal = {Journal of Neuroscience},
  year = {2015},
  volume = {35},
  number = {32},
  pages = {11266-11280},
  doi = {10.1523/jneurosci.0779-15.2015}
}

@article{Nadarajah2026,
  author = {Nadarajah and others},
  title = {REV-ERB-alpha and -beta coordinately regulate astrocyte reactivity and proteostatic function},
  journal = {Proceedings of the National Academy of Sciences of the United States of America},
  year = {2026},
  volume = {123},
  number = {5},
  pages = {e2511093123},
  doi = {10.1073/pnas.2511093123}
}

@article{Ning2022,
  author = {Ning and others},
  title = {PART1 destabilized by NOVA2 regulates blood-brain barrier permeability in endothelial cells via STAU1-mediated mRNA degradation},
  journal = {Gene},
  year = {2022},
  volume = {815},
  pages = {146164},
  doi = {10.1016/j.gene.2021.146164}
}

@article{Noch2021,
  author = {Noch and others},
  title = {Distribution and localization of phosphatidylinositol 5-phosphate, 4-kinase alpha and beta in the brain},
  journal = {The Journal of Comparative Neurology},
  year = {2021},
  volume = {529},
  number = {2},
  pages = {434-449},
  doi = {10.1002/cne.24956}
}

@article{Oh2025,
  author = {Oh and others},
  title = {Homozygous missense variant in C2orf69 causes early-onset neurodegeneration, leukoencephalopathy and autoinflammation},
  journal = {Journal of Medical Genetics},
  year = {2025},
  volume = {62},
  number = {3},
  pages = {206-209},
  doi = {10.1136/jmg-2024-110419}
}

@article{Ohno2009,
  author = {Ohno and others},
  title = {Nardilysin regulates axonal maturation and myelination in the central and peripheral nervous system},
  journal = {Nature Neuroscience},
  year = {2009},
  volume = {12},
  number = {12},
  pages = {1506-13},
  doi = {10.1038/nn.2438}
}

@article{Oien2008,
  author = {Oien and others},
  title = {MsrA knockout mouse exhibits abnormal behavior and brain dopamine levels},
  journal = {Free Radical Biology \& Medicine},
  year = {2008},
  volume = {45},
  number = {2},
  pages = {193-200},
  doi = {10.1016/j.freeradbiomed.2008.04.003}
}

@article{Okur2016,
  author = {Okur and others},
  title = {De novo mutations in CSNK2A1 are associated with neurodevelopmental abnormalities and dysmorphic features},
  journal = {Human Genetics},
  year = {2016},
  volume = {135},
  number = {7},
  pages = {699-705},
  doi = {10.1007/s00439-016-1661-y}
}

@article{Osathanon2016,
  author = {Osathanon and others},
  title = {Expression and influence of Notch signaling in oral squamous cell carcinoma},
  journal = {Journal of Oral Science},
  year = {2016},
  volume = {58},
  number = {2},
  pages = {283-94},
  doi = {10.2334/josnusd.15-0535}
}

@article{Palavicini2013,
  author = {Palavicini and others},
  title = {RanBP9 Plays a Critical Role in Neonatal Brain Development in Mice},
  journal = {PLOS ONE},
  year = {2013},
  volume = {8},
  number = {6},
  pages = {e66908},
  doi = {10.1371/journal.pone.0066908}
}

@article{Park2008,
  author = {Park, TJ and Curran, T},
  title = {Crk and Crk-like play essential overlapping roles downstream of disabled-1 in the Reelin pathway},
  journal = {The Journal of Neuroscience},
  year = {2008},
  volume = {28},
  number = {50},
  pages = {13551-62},
  doi = {10.1523/jneurosci.4323-08.2008}
}

@article{Peng2021,
  author  = {Peng, S. and others},
  title   = {CAG RNAs induce DNA damage and apoptosis by silencing NUDT16 expression in polyglutamine degeneration},
  journal = {Proceedings of the National Academy of Sciences of the United States of America},
  year    = {2021},
  volume  = {118},
  number  = {19},
  pages   = {e2022940118},
  doi     = {10.1073/pnas.2022940118}
}

@article{Peng2025,
  author = {Peng and others},
  title = {EPHB2 Promotes the Progression of Oral Squamous Cell Carcinoma Cells Through the Activation of VPS4A-Mediated Autophagy},
  journal = {Cancer Science},
  year = {2025},
  volume = {116},
  number = {5},
  pages = {1308-1323},
  doi = {10.1111/cas.70033}
}

@article{Perland2016,
  author = {Perland and others},
  title = {The Putative SLC Transporters Mfsd5 and Mfsd11 Are Abundantly Expressed in the Mouse Brain and Have a Potential Role in Energy Homeostasis},
  journal = {PLOS ONE},
  year = {2016},
  volume = {11},
  number = {6},
  pages = {e0156912},
  doi = {10.1371/journal.pone.0156912}
}

@article{Petersen2004,
  author = {Petersen and others},
  title = {Continuing role for mouse Numb and Numbl in maintaining progenitor cells during cortical neurogenesis},
  journal = {Nature Neuroscience},
  year = {2004},
  volume = {7},
  number = {8},
  pages = {803-11},
  doi = {10.1038/nn1289}
}

@article{Piard2018,
  author = {Piard and others},
  title = {FRMPD4 mutations cause X-linked intellectual disability and disrupt dendritic spine morphogenesis},
  journal = {Human Molecular Genetics},
  year = {2018},
  volume = {27},
  number = {4},
  pages = {589-600},
  doi = {10.1093/hmg/ddx426}
}

@article{Poppinga2024,
  author  = {Poppinga, J. and others},
  title   = {Endosomal sorting protein SNX4 limits synaptic vesicle docking and release},
  journal = {eLife},
  year    = {2024},
  volume  = {13},
  pages   = {RP97910},
  doi     = {10.7554/eLife.97910}
}

@article{Qian2011,
  author = {Qian and others},
  title = {Knockout of Zn transporters Zip-1 and Zip-3 attenuates seizure-induced CA1 neurodegeneration},
  journal = {The Journal of Neuroscience},
  year = {2011},
  volume = {31},
  number = {1},
  pages = {97-104},
  doi = {10.1523/jneurosci.5162-10.2011}
}

@article{Qu2020,
  author = {Qu and others},
  title = {A homozygous mutation in CMAS causes autosomal recessive intellectual disability in a Kazakh family},
  journal = {Annals of Human Genetics},
  year = {2020},
  volume = {84},
  number = {1},
  pages = {46-53},
  doi = {10.1111/ahg.12349}
}

@article{Ramasubramanian2023,
  author = {Ramasubramanian and others},
  title = {High expression of novel biomarker TBRG4 promotes the progression and invasion of oral squamous cell carcinoma},
  journal = {Journal of Oral Pathology \& Medicine},
  year = {2023},
  volume = {52},
  number = {8},
  pages = {738-745},
  doi = {10.1111/jop.13470}
}

@article{Ramesh2025,
  author = {Ramesh Kumar and others},
  title = {Dysregulation of a novel m6A regulator YWHAG is correlated with metastasis and poor prognosis in oral squamous cell carcinoma - A cross-sectional study},
  journal = {Archives of Oral Biology},
  year = {2025},
  volume = {169},
  pages = {106090},
  doi = {10.1016/j.archoralbio.2024.106090}
}

@article{Rao2023,
  author = {Rao and others},
  title = {Aldose reductase inhibition decelerates optic nerve degeneration by alleviating retinal microglia activation},
  journal = {Scientific Reports},
  year = {2023},
  volume = {13},
  number = {1},
  pages = {5592},
  doi = {10.1038/s41598-023-32702-5}
}

@article{Rao2024,
  author  = {Rao, S. and others},
  title   = {RhoGEF Tiam2 Regulates Glutamatergic Synaptic Transmission in Hippocampal CA1 Pyramidal Neurons},
  journal = {eNeuro},
  year    = {2024},
  volume  = {11},
  number  = {7},
  pages   = {ENEURO.0500-21.2024},
  doi     = {10.1523/ENEURO.0500-21.2024}
}

@article{Rapoport2015,
  author = {Rapoport and others},
  title = {Coordinated Expression of Phosphoinositide Metabolic Genes during Development and Aging of Human Dorsolateral Prefrontal Cortex},
  journal = {PLOS ONE},
  year = {2015},
  volume = {10},
  number = {7},
  pages = {e0132675},
  doi = {10.1371/journal.pone.0132675}
}

@article{Raymundo2023,
  author  = {Raymundo, J. R. and others},
  title   = {KCTD1/KCTD15 complexes control ectodermal and neural crest cell functions, and their impairment causes aplasia cutis},
  journal = {Journal of Clinical Investigation},
  year    = {2023},
  volume  = {134},
  number  = {4},
  pages   = {e174138},
  doi     = {10.1172/JCI174138}
}

@article{Rder2005,
  author = {Rüder and others},
  title = {EBAG9 adds a new layer of control on large dense-core vesicle exocytosis via interaction with Snapin},
 journal = {Molecular Biology of the Cell},
  year = {2005},
  volume = {16},
  number = {3},
  pages = {1245-57},
  doi = {10.1091/mbc.e04-09-0817}
}

@article{Traylor2021,
  author  = {Traylor, Matthew and Persyn, Emma and Tomppo, Liisa and Klasson, Simon and Abedi, Vahab and Bakker, Maarten K. and et al.},
  title   = {Genetic basis of lacunar stroke: A pooled analysis of individual patient data and genome-wide association studies},
  journal = {The Lancet Neurology},
  year    = {2021},
  volume  = {20},
  number  = {5},
  pages   = {351--361},
  doi     = {10.1016/S1474-4422(21)00031-4}
}

@article{Ianzano2003,
  author  = {Ianzano, Luigi and Zhao, X. C. and Minassian, Berge A. and Scherer, Stephen W.},
  title   = {Identification of a novel protein interacting with laforin, the EPM2a progressive myoclonus epilepsy gene product},
  journal = {Genomics},
  year    = {2003},
  volume  = {81},
  number  = {6},
  pages   = {579--587},
  doi     = {10.1016/S0888-7543(03)00094-6}
}

@article{Ren2020,
  author = {Ren and others},
  title = {LncRNA PITPNA-AS1 boosts the proliferation and migration of lung squamous cell carcinoma cells by recruiting TAF15 to stabilize HMGB3 mRNA},
  journal = {Cancer Medicine},
  year = {2020},
  volume = {9},
  number = {20},
  pages = {7706-7716},
  doi = {10.1002/cam4.3268}
}

@article{Rivire2012,
  author = {Rivière and others},
  title = {De novo germline and postzygotic mutations in AKT3, PIK3R2 and PIK3CA cause a spectrum of related megalencephaly syndromes},
  journal = {Nature Genetics},
  year = {2012},
  volume = {44},
  number = {8},
  pages = {934-40},
  doi = {10.1038/ng.2331}
}

@article{Roberts2009,
  author = {Roberts, RK and Appel, B},
  title = {Apical polarity protein PrkCi is necessary for maintenance of spinal cord precursors in zebrafish},
  journal = {Developmental Dynamics},
  year = {2009},
  volume = {238},
  pages = {1638-1648},
  doi = {10.1002/dvdy.21970}
}

@article{Robertson2009,
  author = {Robertson and others},
  title = {Regulation of Postsynaptic Structure and Function by an A-Kinase Anchoring Protein-Membrane-Associated Guanylate Kinase Scaffolding Complex},
  journal = {Journal of Neuroscience},
  year = {2009},
  volume = {29},
  number = {24},
  pages = {7929-7943},
  doi = {10.1523/jneurosci.6093-08.2009}
}

@article{Roy2017,
  author = {Roy and others},
  title = {Small RNA sequencing revealed dysregulated piRNAs in Alzheimer's disease and their probable role in pathogenesis},
  journal = {Molecular BioSystems},
  year = {2017},
  volume = {13},
  number = {3},
  pages = {565-576},
  doi = {10.1039/c6mb00699j}
}

@article{Sakurai2024,
  author = {Sakurai and others},
  title = {Importin alpha4 deficiency induces psychiatric disorder-related behavioral deficits and neuroinflammation in mice},
  journal = {Translational Psychiatry},
  year = {2024},
  volume = {14},
  number = {1},
  pages = {426},
  doi = {10.1038/s41398-024-03138-w}
}

@article{SantosCortez2018,
  author = {Santos-Cortez and others},
  title = {Novel candidate genes and variants underlying autosomal recessive neurodevelopmental disorders with intellectual disability},
  journal = {Human Genetics},
  year = {2018},
  volume = {137},
  number = {9},
  pages = {735-752},
  doi = {10.1007/s00439-018-1928-6}
}

@article{Sarkar2020,
  author = {Sarkar and others},
  title = {PLA2G4A/cPLA2-mediated lysosomal membrane damage leads to inhibition of autophagy and neurodegeneration after brain trauma},
  journal = {Autophagy},
  year = {2020},
  volume = {16},
  number = {3},
  pages = {466-485},
  doi = {10.1080/15548627.2019.1628538}
}

@article{Scheidecker2015,
  author = {Scheidecker and others},
  title = {Mutations in TUBGCP4 alter microtubule organization via the gamma-tubulin ring complex in autosomal-recessive microcephaly with chorioretinopathy},
  journal = {American Journal of Human Genetics},
  year = {2015},
  volume = {96},
  number = {4},
  pages = {666-74},
  doi = {10.1016/j.ajhg.2015.02.011}
}

@article{Schwarz2025,
  author  = {Schwarz, C. and others},
  title   = {RAB24 Missense Variant in Dogs with Cerebellar Ataxia},
  journal = {Genes},
  year    = {2025},
  volume  = {16},
  number  = {8},
  pages   = {934},
  doi     = {10.3390/genes16080934}
}

@article{Shao2023,
  author = {Shao and others},
  title = {DUSP2 deletion with CRISPR/Cas9 promotes Mauthner cell axonal regeneration at the early stage of zebrafish},
  journal = {Neural Regeneration Research},
  year = {2023},
  volume = {18},
  number = {3},
  pages = {577-581},
  doi = {10.4103/1673-5374.350208}
}

@article{Shi2012,
  author = {Shi and others},
  title = {Upregulated miR-29b promotes neuronal cell death by inhibiting Bcl2L2 after ischemic brain injury},
  journal = {Experimental Brain Research},
  year = {2012},
  volume = {216},
  number = {2},
  pages = {225-30},
  doi = {10.1007/s00221-011-2925-3}
}

@article{Shi2024,
  author = {Shi and others},
  title = {DDX5 promotes esophageal squamous cell carcinoma growth through sustaining VAV3 mRNA stability},
  journal = {Oncogene},
  year = {2024},
  volume = {43},
  number = {44},
  pages = {3240-3254},
  doi = {10.1038/s41388-024-03162-6}
}

@article{Shi2025,
  author = {Shi and others},
  title = {Discovery of a beta-arrestin-biased CCKBR agonist that blocks CCKBR-dependent long-term potentiation},
  journal = {Nature Communications},
  year = {2025},
  volume = {16},
  pages = {10938},
  doi = {10.1038/s41467-025-65962-y}
}

@article{Singh2020,
  author = {Singh and others},
  title = {Differentially expressed full-length, fusion and novel isoforms transcripts-based signature of well-differentiated keratinized oral squamous cell carcinoma},
  journal = {Oncotarget},
  year = {2020},
  volume = {11},
  number = {34},
  pages = {3227-3243},
  doi = {10.18632/oncotarget.27693}
}

@article{Skopkova2017,
  author = {Skopkova and others},
  title = {EIF2S3 Mutations Associated with Severe X-Linked Intellectual Disability Syndrome MEHMO},
  journal = {Human Mutation},
  year = {2017},
  volume = {38},
  number = {4},
  pages = {409-425},
  doi = {10.1002/humu.23170}
}

@article{Smits2011,
  author = {Smits and others},
  title = {Mutation in mitochondrial ribosomal protein MRPS22 leads to Cornelia de Lange-like phenotype, brain abnormalities and hypertrophic cardiomyopathy},
  journal = {European Journal of Human Genetics},
  year = {2011},
  volume = {19},
  number = {4},
  pages = {394-9},
  doi = {10.1038/ejhg.2010.214}
}

@article{Song2015,
  author = {Song and others},
  title = {Regulation of axon regeneration by the RNA repair and splicing pathway},
  journal = {Nature Neuroscience},
  year = {2015},
  volume = {18},
  number = {6},
  pages = {817-25},
  doi = {10.1038/nn.4019}
}

@article{Song2022,
  author = {Song and others},
  title = {Epithelial-Mesenchymal Transition Gene Signature Is Associated with Neoadjuvant Chemoradiotherapy Resistance and Prognosis of Esophageal Squamous Cell Carcinoma},
  journal = {Disease Markers},
  year = {2022},
  volume = {2022},
  pages = {3534433},
  doi = {10.1155/2022/3534433}
}

@article{Stegner2019,
  author = {Stegner and others},
  title = {Loss of Orai2-Mediated Capacitative Ca2+ Entry Is Neuroprotective in Acute Ischemic Stroke},
  journal = {Stroke},
  year = {2019},
  volume = {50},
  number = {11},
  pages = {3238-3245},
  doi = {10.1161/strokeaha.119.025357}
}

@article{Steinlein1995,
  author = {Steinlein and others},
  title = {A missense mutation in the neuronal nicotinic acetylcholine receptor alpha 4 subunit is associated with autosomal dominant nocturnal frontal lobe epilepsy},
  journal = {Nature Genetics},
  year = {1995},
  volume = {11},
  number = {2},
  pages = {201-3},
  doi = {10.1038/ng1095-201}
}

@article{Su2023,
  author = {Su and others},
  title = {The Evaluation of Prognostic Value and Immune Characteristics of Ferroptosis-Related Genes in Lung Squamous Cell Carcinoma},
  journal = {Global Medical Genetics},
  year = {2023},
  volume = {10},
  number = {4},
  pages = {285-300},
  doi = {10.1055/s-0043-1776386}
}

@article{Sugatha2023,
  author  = {Sugatha, J. and others},
  title   = {Insights into cargo sorting by SNX32 and its role in neurite outgrowth},
  journal = {eLife},
  year    = {2023},
  volume  = {12},
  pages   = {e84396},
  doi     = {10.7554/eLife.84396}
}

@article{Sugawara2025,
  author = {Sugawara and others},
  title = {A p.N92K variant of the GTPase RAC3 disrupts cortical neuron migration and axon elongation},
  journal = {Journal of Biological Chemistry},
  year = {2025},
  volume = {301},
  number = {4},
  pages = {108346},
  doi = {10.1016/j.jbc.2025.108346}
}

@article{Sun2021,
  author = {Sun and others},
  title = {Gpr125 Marks Distinct Cochlear Cell Types and Is Dispensable for Cochlear Development and Hearing},
  journal = {Frontiers in Cell and Developmental Biology},
  year = {2021},
  volume = {9},
  pages = {690955},
  doi = {10.3389/fcell.2021.690955}
}

@article{Szretter2012,
  author = {Szretter and others},
  title = {2'-O methylation of the viral mRNA cap by West Nile virus evades ifit1-dependent and -independent mechanisms of host restriction in vivo},
  journal = {PLoS Pathogens},
  year = {2012},
  volume = {8},
  number = {5},
  pages = {e1002698},
  doi = {10.1371/journal.ppat.1002698}
}

@article{Tajima2025,
  author = {Tajima and others},
  title = {A humanized NOVA1 splicing factor alters mouse vocal communications},
  journal = {Nature Communications},
  year = {2025},
  volume = {16},
  pages = {1542},
  doi = {10.1038/s41467-025-56579-2}
}

@article{Tan2025,
  author = {Tan and others},
  title = {EIF3B stabilizes MAP2K2 to activate the ERK pathway and promote the progression of laryngeal squamous cell carcinoma},
  journal = {Cell Death Discovery},
  year = {2025},
  volume = {11},
  number = {1},
  pages = {333},
  doi = {10.1038/s41420-025-02634-2}
}

@article{Taslimifar2022,
  author = {Taslimifar and others},
  title = {Analysis of L-leucine amino acid transporter species activity and gene expression by human blood brain barrier hCMEC/D3 model reveal potential LAT1, LAT4, B0AT2 and y+LAT1 functional cooperation},
  journal = {Journal of Cerebral Blood Flow and Metabolism},
  year = {2022},
  volume = {42},
  number = {1},
  pages = {90-103},
  doi = {10.1177/0271678x211039593}
}

@article{Teoh2020,
  author = {Teoh and others},
  title = {Arfgef1 haploinsufficiency in mice alters neuronal endosome composition and decreases membrane surface postsynaptic GABAA receptors},
  journal = {Neurobiology of Disease},
  year = {2020},
  volume = {134},
  pages = {104632},
  doi = {10.1016/j.nbd.2019.104632}
}

@article{TomasRoca2015,
  author = {Tomas-Roca and others},
  title = {De novo mutations in PLXND1 and REV3L cause Möbius syndrome},
  journal = {Nature Communications},
  year = {2015},
  volume = {6},
  pages = {7199},
  doi = {10.1038/ncomms8199}
}

@article{Ucar2017,
  author = {Ucar and others},
  title = {Previously Unreported Biallelic Mutation in DNAJC19: Are Sensorineural Hearing Loss and Basal Ganglia Lesions Additional Features of Dilated Cardiomyopathy and Ataxia (DCMA) Syndrome?},
  journal = {JIMD Reports},
  year = {2017},
  volume = {35},
  pages = {39-45},
  doi = {10.1007/8904_2016_23}
}

@article{van2017,
  author = {van der Werf and others},
  title = {Mutations in two large pedigrees highlight the role of ZNF711 in X-linked intellectual disability},
  journal = {Gene},
  year = {2017},
  volume = {605},
  pages = {92-98},
  doi = {10.1016/j.gene.2016.12.013}
}

@article{Vetro2023,
  author = {Vetro and others},
  title = {Stretch-activated ion channel TMEM63B associates with developmental and epileptic encephalopathies and progressive neurodegeneration},
  journal = {American Journal of Human Genetics},
  year = {2023},
  volume = {110},
  number = {8},
  pages = {1356-1376},
  doi = {10.1016/j.ajhg.2023.06.008}
}

@article{Viswanathan2017,
  author = {Viswanathan and others},
  title = {Molecularly Defined Subplate Neurons Project Both to Thalamocortical Recipient Layers and Thalamus},
  journal = {Cerebral Cortex},
  year = {2017},
  volume = {27},
  number = {10},
  pages = {4759-4768},
  doi = {10.1093/cercor/bhw271}
}

@article{Wang2006,
  author = {Wang and others},
  title = {Long-term outcome and neuroradiological findings of 31 patients with 6-pyruvoyltetrahydropterin synthase deficiency},
  journal = {Journal of Inherited Metabolic Disease},
  year = {2006},
  volume = {29},
  number = {1},
  pages = {127-34},
  doi = {10.1007/s10545-006-0080-y}
}

@article{Wang2015,
  author = {Wang and others},
  title = {Immunohistochemical localization of DPP10 in rat brain supports the existence of a Kv4/KChIP/DPPL ternary complex in neurons},
  journal = {The Journal of Comparative Neurology},
  year = {2015},
  volume = {523},
  number = {4},
  pages = {608-28},
  doi = {10.1002/cne.23698}
}

@article{Wang2018,
  author = {Wang and others},
  title = {The ubiquitin conjugating enzyme Ube2W regulates solubility of the Huntington's disease protein, huntingtin},
  journal = {Neurobiology of Disease},
  year = {2018},
  volume = {109},
  number = {Pt A},
  pages = {127-136},
  doi = {10.1016/j.nbd.2017.10.002}
}

@article{Wang2020coexist,
  author = {Wang and others},
  title = {Coexisting overexpression of STOML1 and STOML2 proteins may be associated with pathology of oral squamous cell carcinoma},
 journal = {Oral Surgery, Oral Medicine, Oral Pathology and Oral Radiology},
  year = {2020},
  volume = {129},
  number = {6},
  pages = {591-599.e3},
  doi = {10.1016/j.oooo.2020.01.011}
}

@article{Wang2021a,
  author = {Wang and others},
  title = {A novel 12-gene signature as independent prognostic model in stage IA and IB lung squamous cell carcinoma patients},
  journal = {Clinical \& Translational Oncology},
  year = {2021},
  volume = {23},
  number = {11},
  pages = {2368-2381},
  doi = {10.1007/s12094-021-02638-1}
}

@article{Wang2021b,
  author = {Wang and others},
  title = {RUNX1 and REXO2 are associated with the heterogeneity and prognosis of IDH wild type lower grade glioma},
  journal = {Scientific Reports},
  year = {2021},
  volume = {11},
  number = {1},
  pages = {11836},
  doi = {10.1038/s41598-021-91382-1}
}

@article{Wang2022,
  author = {Wang and others},
  title = {Ahi1 regulates serotonin production by the GR/ERbeta/TPH2 pathway involving sexual differences in depressive behaviors},
  journal = {Cell Communication and Signaling},
  year = {2022},
  volume = {20},
  number = {1},
  pages = {74},
  doi = {10.1186/s12964-022-00894-4}
}

@article{Wang2023a,
  author = {Wang and others},
  title = {Hypoxia promotes EV secretion by impairing lysosomal homeostasis in HNSCC through negative regulation of ATP6V1A by HIF-1alpha},
  journal = {Journal of Extracellular Vesicles},
  year = {2023},
  volume = {12},
  number = {2},
  pages = {e12310},
  doi = {10.1002/jev2.12310}
}

@article{Wang2023b,
  author = {Wang and others},
  title = {KLF5 promotes esophageal squamous cell cancer through the transcriptional activation of FGFBP1},
  journal = {Medical Oncology},
  year = {2023},
  volume = {41},
  number = {1},
  pages = {17},
  doi = {10.1007/s12032-023-02244-x}
}

@article{Wang2024a,
  author = {Wang and others},
  title = {NR3C2 activates LCN2 transcription to promote endoplasmic reticulum stress and cell apoptosis in ischemic cerebral infarction},
  journal = {Brain Research},
  year = {2024},
  volume = {1822},
  pages = {148632},
  doi = {10.1016/j.brainres.2023.148632}
}

@article{Wang2024b,
  author = {Wang and others},
  title = {Alternative splicing of latrophilin-3 controls synapse formation},
  journal = {Nature},
  year = {2024},
  volume = {626},
  number = {7997},
  pages = {128-135},
  doi = {10.1038/s41586-023-06913-9}
}

@article{Weckhuysen2016,
  author = {Weckhuysen and others},
  title = {Involvement of GATOR complex genes in familial focal epilepsies and focal cortical dysplasia},
  journal = {Epilepsia},
  year = {2016},
  volume = {57},
  number = {6},
  pages = {994-1003},
  doi = {10.1111/epi.13391}
}

@article{Wei2016,
  author = {Wei and others},
  title = {Loss of ZNF32 augments the regeneration of nervous lateral line system through negative regulation of SOX2 transcription},
  journal = {Oncotarget},
  year = {2016},
  volume = {7},
  number = {43},
  pages = {70420-70436},
  doi = {10.18632/oncotarget.11895}
}

\end{document}

\def\spacingset#1{\renewcommand{\baselinestretch}%
{#1}\small\normalsize} \spacingset{0.5}

%%%%%%%%%%%%%%%%%%%%%%%%%%%%%%%%%%%%%%%%%%%%%%%%%%%%%%%%%%%%%%%%%%%%%%%%%%%%%%

\if1\blind
{
  \title{\bf Supplementary Materials for
  	``A flexible Bayesian framework for detecting cross-sample spatial expression variability in heterogeneous tissues''}
  \author{
        Meng Zhou \\
        School of Statistics and Data Science, \\
        Shanghai University of Finance and Economics
        \and
        Shuangge Ma \\
        Department of Biostatistics, \\
        Yale University
        \and
        Mengyun Wu \\
        School of Statistics and Data Science, \\
        Shanghai University of Finance and Economics
    }
    \date{}
  \maketitle
} \fi

\if0\blind
{
  \bigskip
  \bigskip
  \bigskip
  \begin{center}
    {\LARGE\bf Supplementary Materials for
    	``A flexible Bayesian framework for detecting cross-sample spatial expression variability in heterogeneous tissues''}
\end{center}
  \medskip
} \fi

\bigskip
%\begin{abstract}
%The text of your abstract. 200 or fewer words.
%\end{abstract}
%
%\noindent%
%{\it Keywords:}  3 to 6 keywords, that do not appear in the title
%\vfill

\newpage
\spacingset{1.5}  % DON'T change the spacing!

\section{Motivating Examples}

\begin{figure}[!ht]
    \centering
    \includegraphics[width=\linewidth]{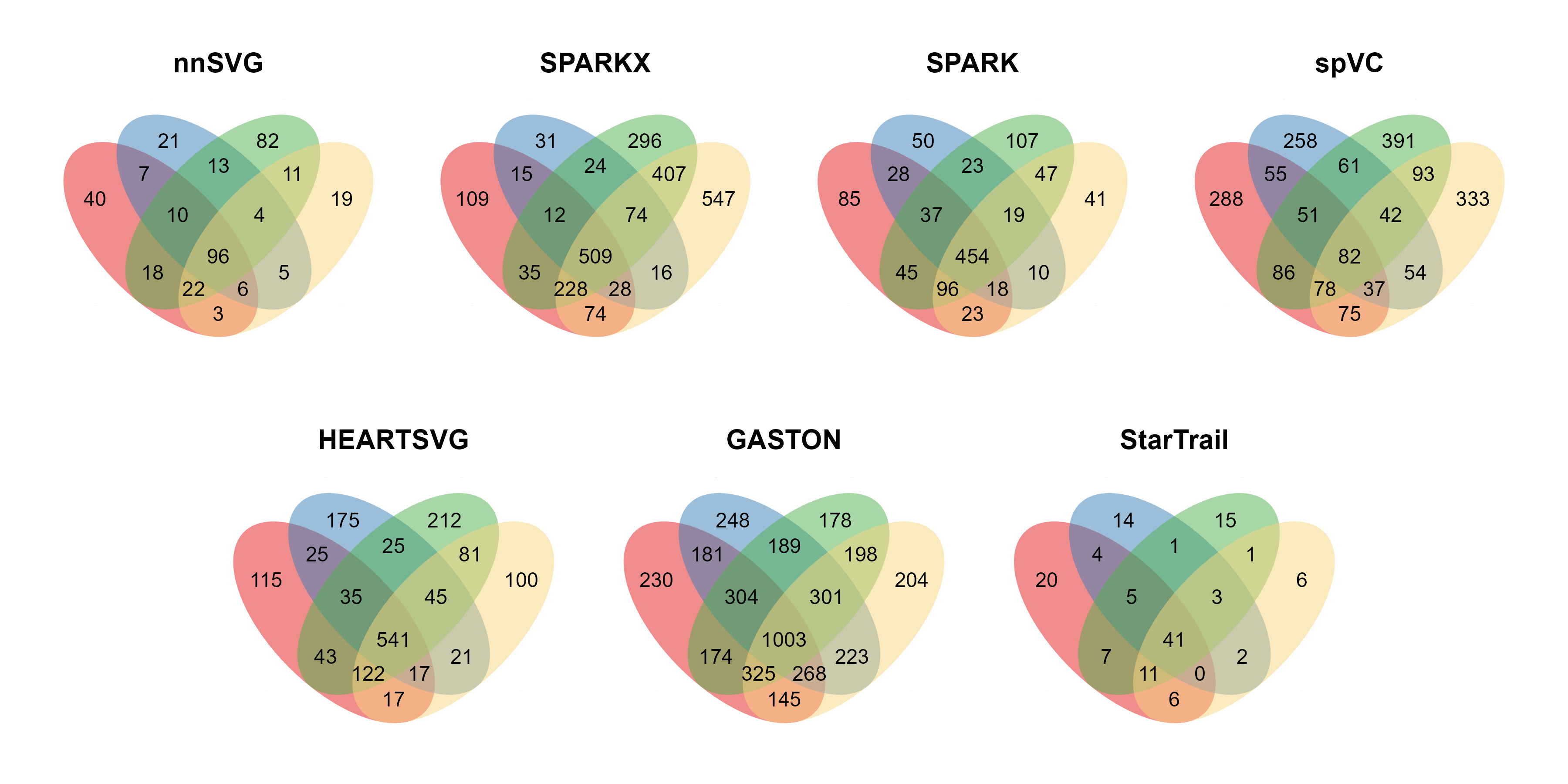}
    \caption{Venn plot of SV genes identified by seven single-sample methods in the DLPFC dataset using multiple sections from the same donor. Colors represent different sections: red, blue, green, and yellow.}
    \label{fig:venn_result_of_LIBD}
\end{figure}

\begin{figure}[!ht]
    \centering
    \includegraphics[width=\linewidth]{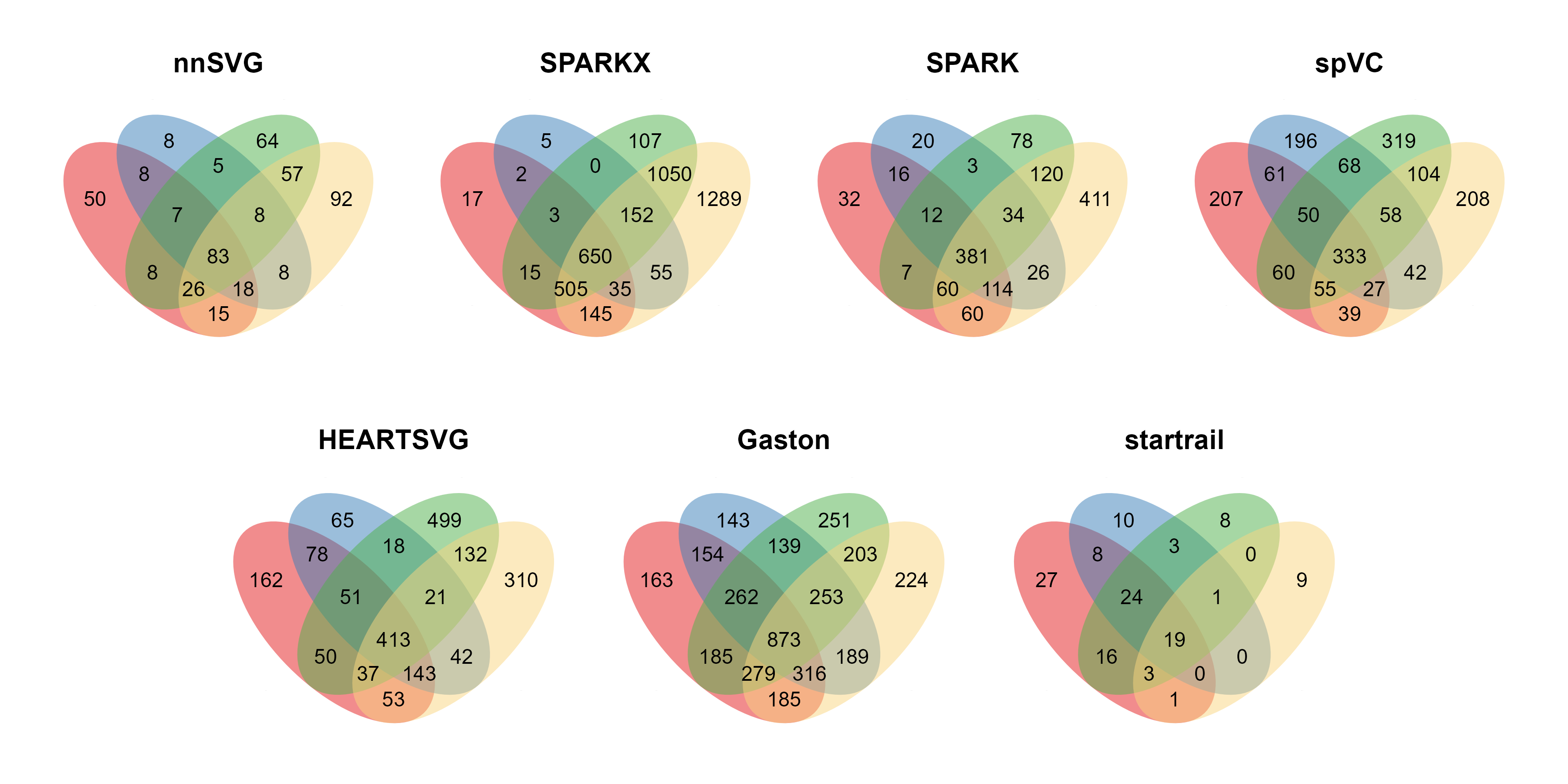}
    \caption{Venn plot of SV genes identified by seven single-sample methods in the DLPFC dataset using sections across different donors. Colors represent different sections: red, blue, green, and yellow.}
    \label{fig:venn_result_of_LIBD_across}
\end{figure}

\begin{figure}[!ht]
    \centering
    \includegraphics[width=\linewidth]{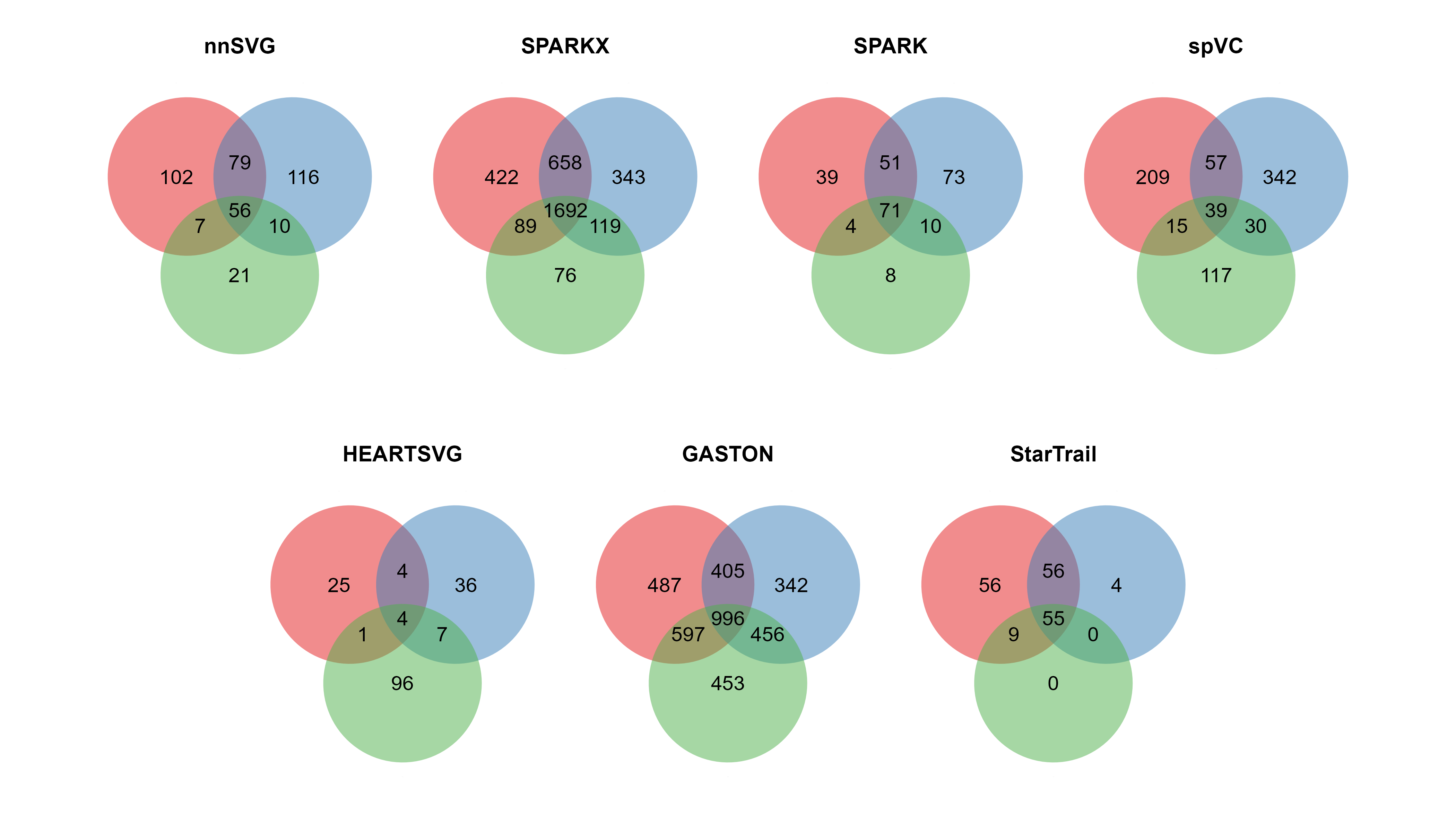}
    \caption{Venn plot of SV genes identified by seven single-sample methods in SCC dataset. Colors represent different sections: red, blue, and green.}
    \label{fig:venn_result_of_scc}
\end{figure}

\section{Details of the three distinct spatial transcriptomic scenarios}\label{real-details}

The DLPFC dataset (available at http://spatial.libd.org/spatialLIBD/) comprises 10X Genomics Visium spatial transcriptomics data from three individuals, each with four slices. For each individual, slices A and B, as well as slices C and D, are separated by 10 $\mu m$, while slices B and C are separated by 300 $\mu m$. Consequently, slice pairs $(A,B)$ and $(C,D)$ exhibit greater similarity than the $(B,C)$ pair. 

In the same‑donor scenario, we analyze the four slices from the first donor (Sample IDs: 151507, 151508, 151509, and 151510). These slices share 33,538 genes and contain 4,226, 4,383, 4,787, and 4,634 spots, respectively.

In the across‑donor scenario, we examine four tissue slices obtained from three donors (Sample IDs: 151509 and 151510 from the first donor; 151671 from the second donor; 151675 from the third donor). These slices also share 33,538 genes and contain 4,787, 4,634, 4,110, and 3,592 spots, respectively.

The human squamous cell carcinoma (SCC) dataset originates from traditional spatial transcriptomics (ST) and comprises 12 tissue slices across four patients (GEO accession GSE144240). We focus on three slices from Patient 9, which share 15,414 genes and contain 1,145, 1,071, and 1,182 spots, respectively.

To improve computational efficiency and stability for all three data scenarios, we follow published studies \citep{Yan2024} and apply a three‑step prescreening procedure: (1) removing genes expressed in fewer than 100 spots and spots with fewer than 500 expressed genes; (2) selecting the top 8,000 highly variable genes (HVGs) per slice; and (3) identifying the consensus HVGs across all slices.

After prescreening, we obtain the following filtered sets:
\begin{itemize}
\item Same‑donor DLPFC: 4,908 shared genes and 4,147, 4,148, 4,700, and 4,546 spatially resolved spots.
\item Across‑donor DLPFC: 4,406 shared genes and 4,700, 4,546, 3,988, and 3,565 spatially resolved spots.
\item SCC dataset: 4,437 shared genes and 1,143, 1,070, and 1,108 spatially resolved spots.
\end{itemize}

\newpage
\section{Exploratory analysis of the real spatial transcriptomic data}
\label{Exploratory analysis-real}
\subsection{Exploration of data sparsity and overdispersion}
\begin{figure}[!ht]
    \centering
    \includegraphics[width=0.6\linewidth]{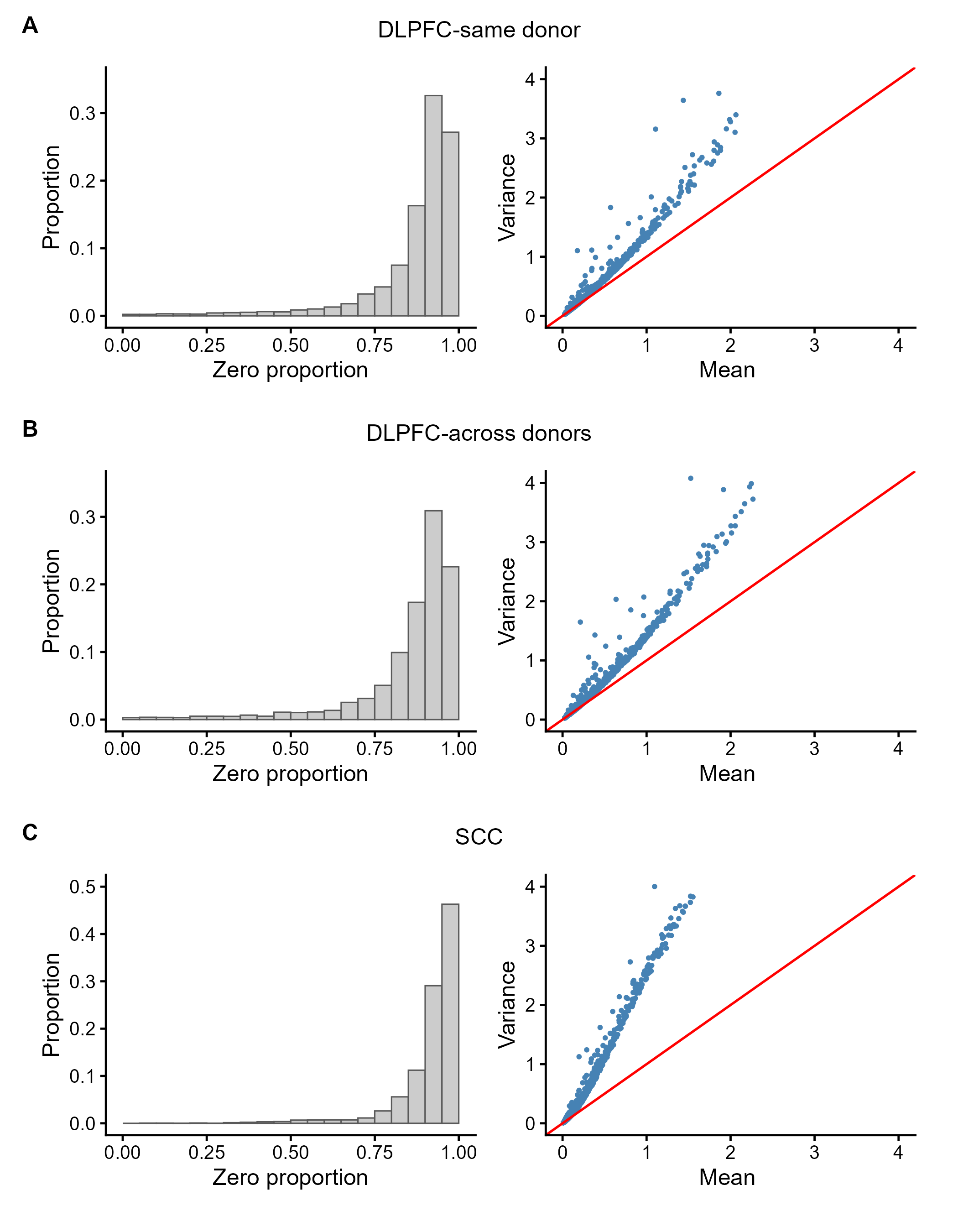}
    \caption{
    Empirical characteristics of the three real datasets: (A) DLPFC same-donor dataset; (B) DLPFC across-donor dataset; (C) SCC dataset. The left panels show the distributions of gene-wise zero proportions. The right panels compare the gene-wise mean expression and variance on the $\log(1+x)$ scale. The red line represents the Poisson relationship, $\mathrm{Var}(Y)=E(Y)$.
    }
    \label{fig:zeroinf_promotion}
\end{figure}

To characterize the distributional properties of the three real spatial transcriptomic datasets, we first perform an exploratory analysis of the gene-wise zero proportions and the relationship between the gene-wise mean and variance (Figure \ref{fig:zeroinf_promotion}).

In Figure \ref{fig:zeroinf_promotion}, the left panels show that a large proportion of genes have high zero proportions, revealing widespread sparsity across all three datasets. The right panels display the relationship between the gene-wise mean and variance, where the red reference line represents the Poisson relationship, $\mathrm{Var}(Y)=E(Y)$. Most genes are distributed well above the reference line, indicating that the variance substantially exceeds the mean and suggesting pronounced overdispersion. These exploratory observations show that the real spatial transcriptomic data are characterized by both excessive zeros and overdispersion, motivating the use of a zero-inflated negative binomial (ZINB) model in IBaySVG.

\subsection{Exploration of spatial expression heterogeneity and the need for flexible spatial effects}
We explore the spatial expression patterns of representative genes in DLPFC and SCC datasets to gain an initial understanding of the spatial effects present in biological tissues.

\begin{figure}[htbp]
\centering
\includegraphics[width=0.8\textwidth]{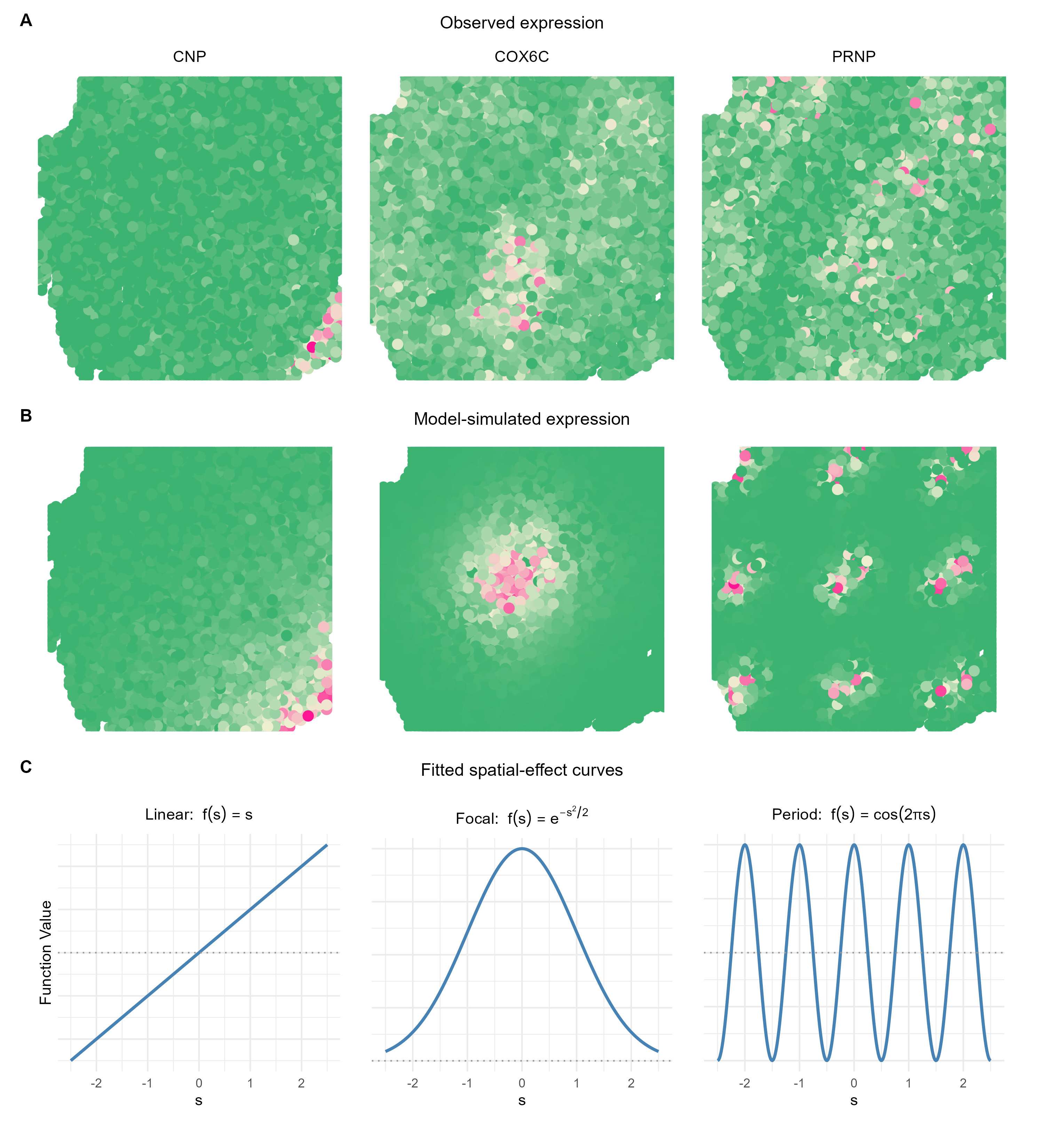}
\caption{
Illustration of three representative spatial patterns in the DLPFC dataset.
(A) Representative genes (CNP, COX6C, PRNP) whose spatial expression patterns are consistent with the three predefined spatial functions.
(B) Simulated spatial expressions generated from the corresponding linear, focal, and periodic functions shown in (C).
(C) Predefined spatial effect functions representing linear, focal, and periodic patterns.
}
\label{fig:basic_spatial_exam}
\end{figure}

For the DLPFC dataset, Figure \ref{fig:basic_spatial_exam} presents the expression of three representative genes (CNP, COX6C, and PRNP) in one of the tissue sections analyzed in our study (\citep{Peirce2006,Wang2022COX6c,Zhang2016PRNP}). By comparing Panels (A) and (B), we observe that the simulated spatial expression generated from the corresponding linear, focal, and periodic spatial functions in Panel (C) closely resembles the observed expression patterns. These examples suggest that the spatial variation of these genes can be adequately captured by commonly used parametric spatial structures.

\begin{figure}[htbp]
\centering
\includegraphics[width=0.8\textwidth]{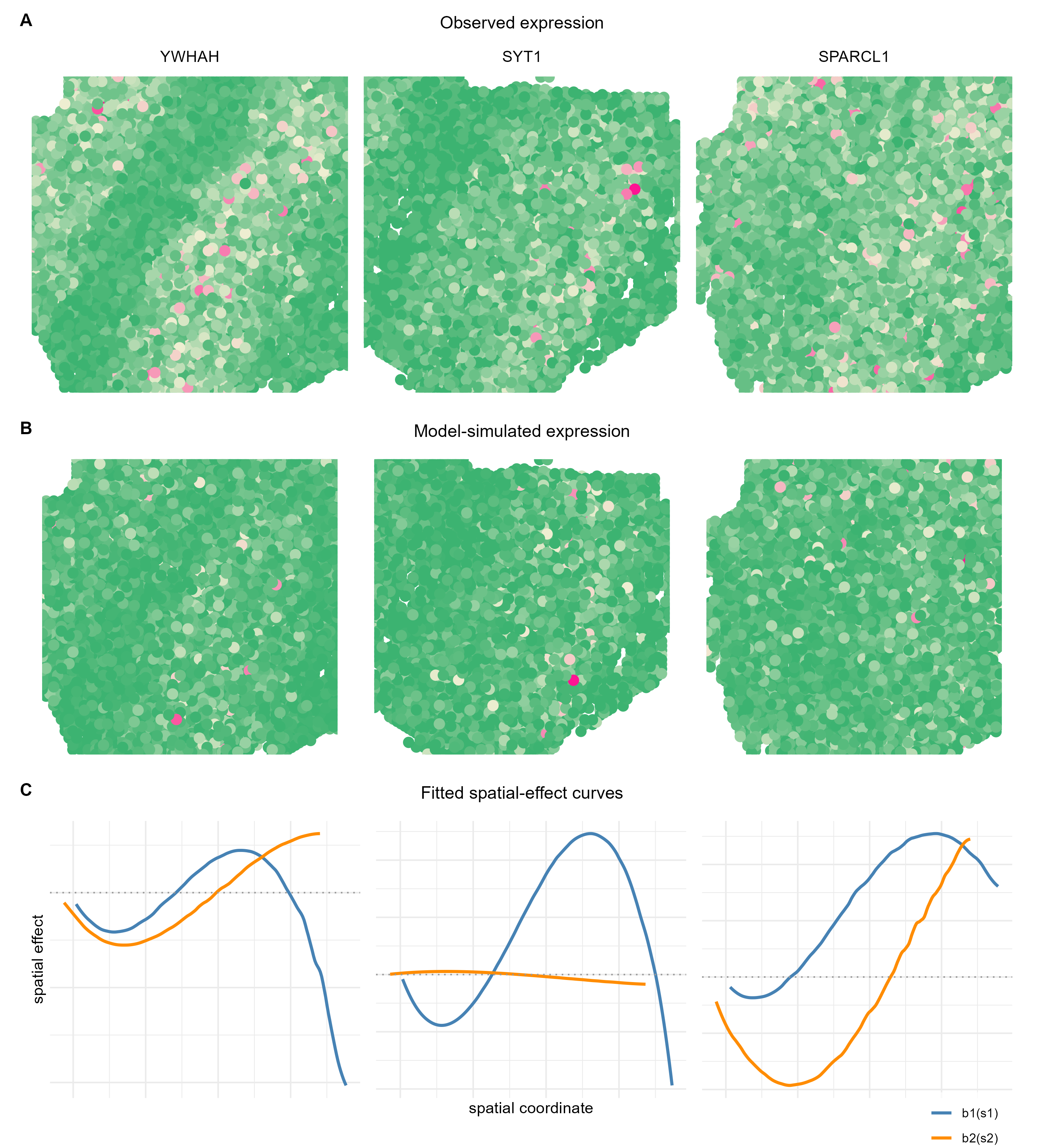}
\caption{
Illustration of complex nonparametric spatial effects and model‑fitted expressions for representative genes in the DLPFC same‑donor dataset.
(A) Observed spatial expression of representative genes YWHAH, SYT1, and SPARCL1.
(B) Model‑fitted spatial expressions generated from the estimated nonparametric effects.
(C) Estimated nonparametric spatial effects along the two spatial coordinates.
}
\label{fig:nonparametric_effect_same}
\end{figure}

\begin{figure}[htbp]
\centering
\includegraphics[width=0.8\textwidth]{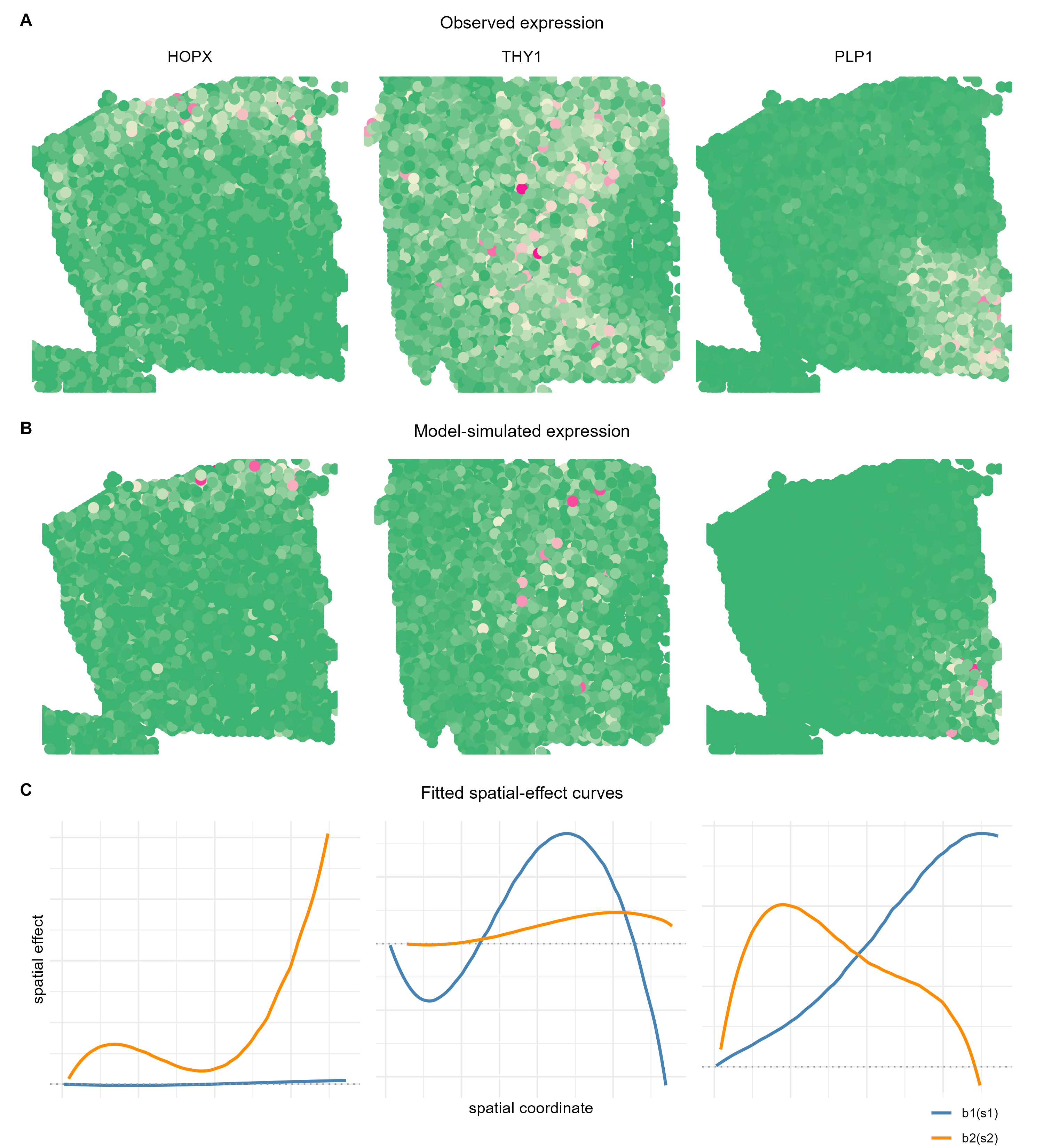}
\caption{
Illustration of complex nonparametric spatial effects and model‑fitted expressions for representative genes in the DLPFC across‑donor dataset.
(A) Observed spatial expression of representative genes HOPX, THY1, and PLP1.
(B) Model‑fitted spatial expressions generated from the estimated nonparametric effects.
(C) Estimated nonparametric spatial effects along the two spatial coordinates.
}
\label{fig:nonparametric_effect_across}
\end{figure}

\begin{figure}[htbp]
\centering
\includegraphics[width=0.8\textwidth]{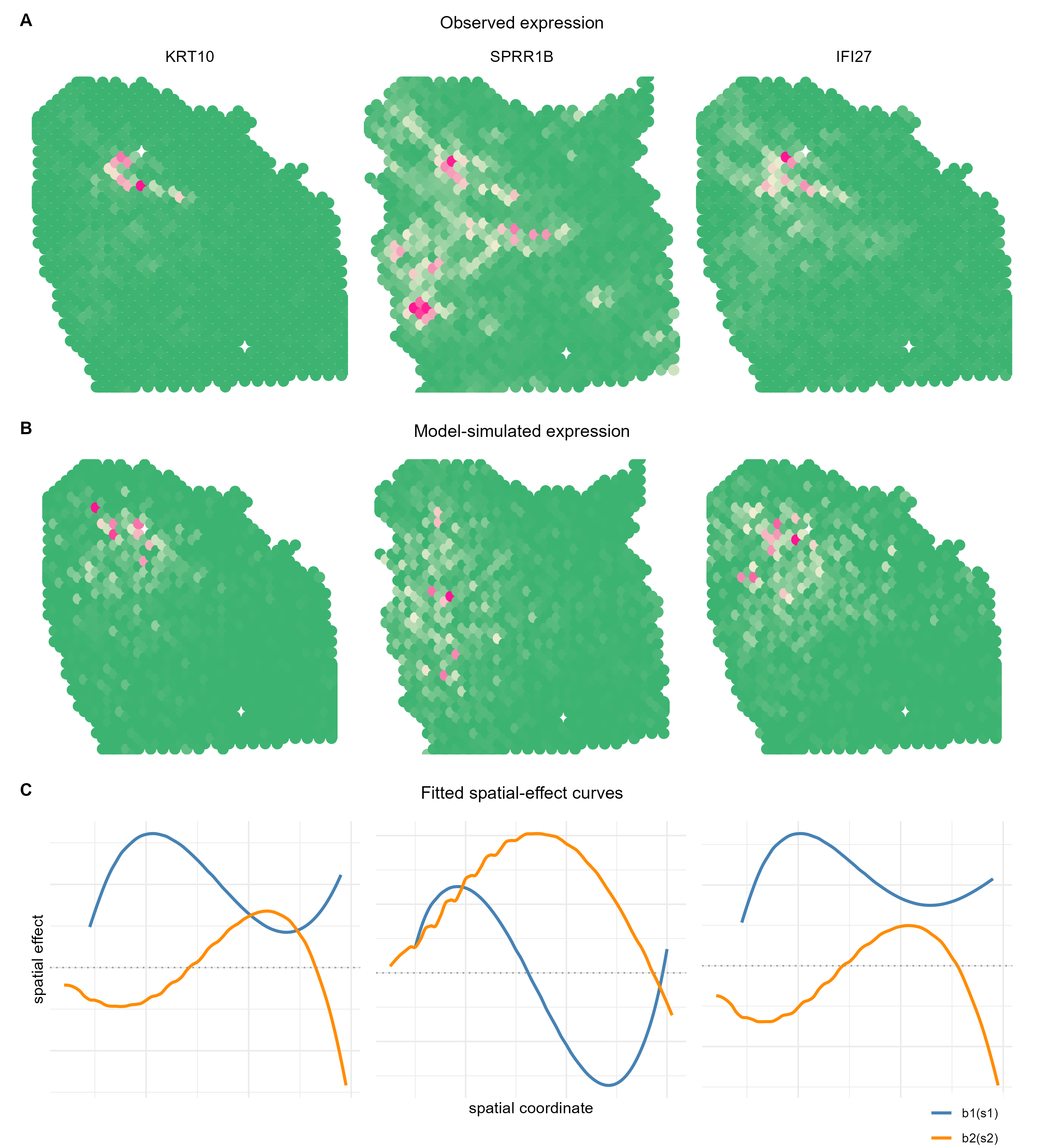}
\caption{
Illustration of complex nonparametric spatial effects and model‑fitted expressions for representative genes in the SCC dataset.
(A) Observed spatial expression of representative genes KRT10, SPRR1B, and IFI27.
(B) Model‑fitted spatial expressions generated from the estimated nonparametric effects.
(C) Estimated nonparametric spatial effects along the two spatial coordinates.
}
\label{fig:nonparametric_effect_scc}
\end{figure}

We next investigate additional representative genes from three real spatial transcriptomics datasets. For each gene, we select a representative slice from multiple available sections. These genes include YWHAH, SYT1, and SPARCL1 in the DLPFC same-donor dataset (\cite{Nguyen2023YWHAH,Riggs2022SYT1,Li2025SPARCL1}), HOPX, THY1, and PLP1 in the DLPFC across-donor dataset (\cite{Li2015HOPX,Webster1999Thy1L1,Santarelli2019}), and KRT10, SPRR1B, and IFI27 in the SCC dataset (\cite{Karantza2011Keratins,Michifuri2013SPRR1B,Wang2025IFI27}). As shown in Figures \ref{fig:nonparametric_effect_same}, \ref{fig:nonparametric_effect_across} and \ref{fig:nonparametric_effect_scc}, these genes exhibit substantially more complicated spatial organizations that are difficult to describe using conventional linear, focal, or periodic spatial patterns in Figure \ref{fig:basic_spatial_exam}. Their spatial effects frequently exhibit multiple local peaks, asymmetric spatial distributions, irregular and diffuse spatial boundaries, as well as localized high-intensity regions embedded within broader spatial gradients, indicating substantial diverse spatial heterogeneity.

These empirical observations suggest that real spatial transcriptomics data contain a much broader spectrum of spatial structures than those represented by a small collection of predefined parametric functions. Consequently, a more flexible representation of spatial effects is desirable for accommodating the complex and heterogeneous spatial organization observed in biological tissues.

\newpage
\subsection{Exploration of shared and sample‑specific spatial patterns}
\begin{figure}[htbp]
    \centering
    \includegraphics[width=\linewidth]{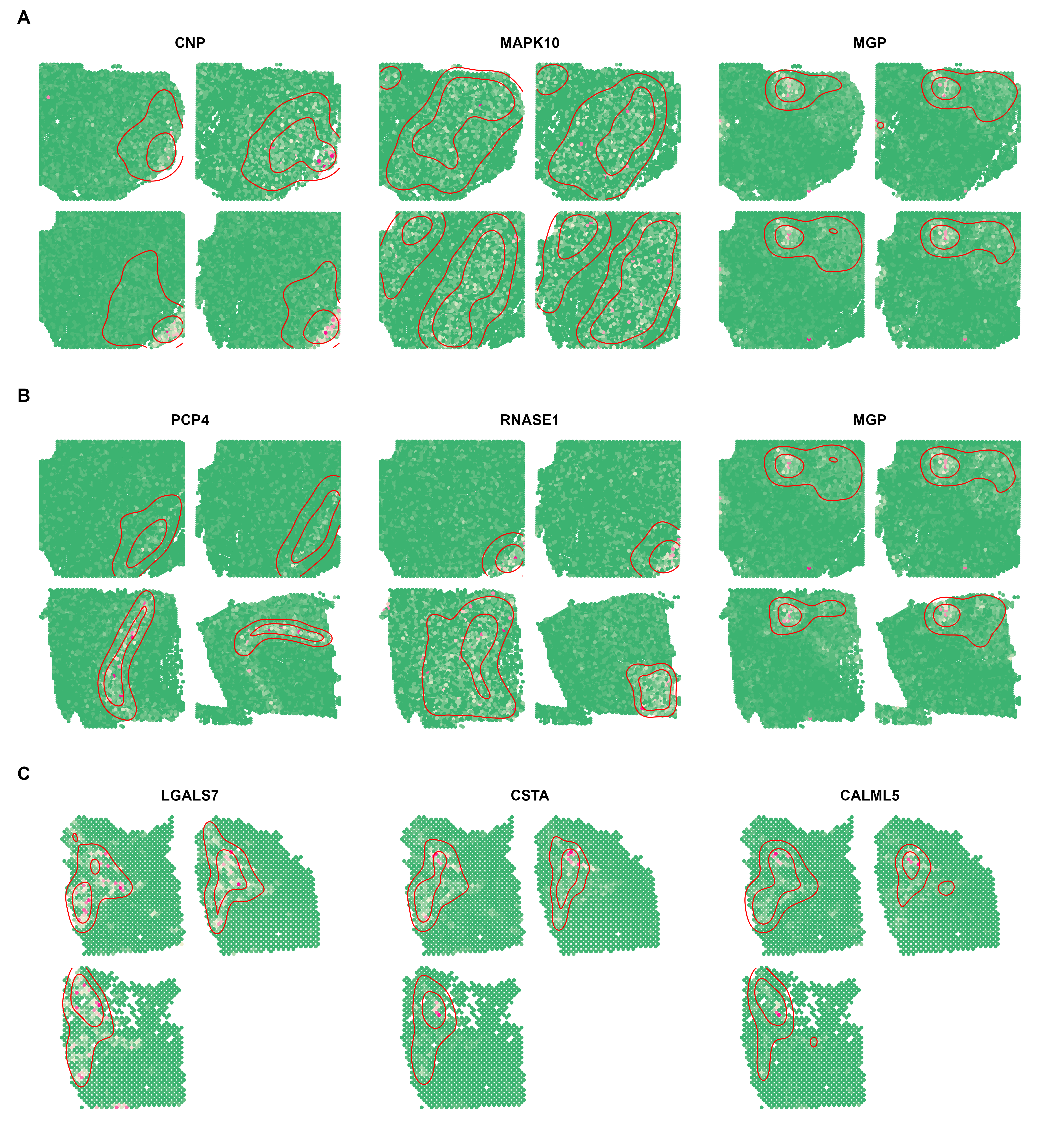}
    \caption{Spatial expression patterns of representative genes across multiple samples. (A) DLPFC same-donor dataset. (B) DLPFC across-donor dataset. (C) SCC dataset. Red curves denoting the kernel-density contours of the high expression spots (the top $15\%$ of non-zero expression).}
    \label{fig:bilevel_promotion}
\end{figure}

{We next visualize the spatial organization of representative genes across tissue sections and characterize the variation in their spatial expression patterns. To facilitate the comparison of spatial localization patterns, we summarize the dominant high-expression regions of representative genes using kernel density contours. Compared with the original expression maps, these contour representations provide a clearer view of the overall spatial localization and geometric variation of gene expression patterns across samples.

Figure \ref{fig:bilevel_promotion} presents several representative genes from the three real datasets, including CNP, MAPK10, and MGP for the DLPFC same-donor dataset \citep{Peirce2006,Musi2020JNK3,Mertsch2009MGP}, PCP4, RNASE1, and MGP for the DLPFC across-donor dataset \citep{Harashima2011PCP4, Kramer2022RNASE1}, and LGALS7, CSTA, and CALML5 for the SCC dataset \citep{An2022LGALS7,Butler2011CSTA,Taniwaki2025CALML5}. These genes are widely recognized as markers of neuronal, glial, vascular, or epithelial cell populations.

Across these examples, we observe two recurring spatial characteristics as shown in Figure \ref{fig:bilevel_promotion}. First, the dominant high-expression regions show consistent localization across samples, suggesting the presence of shared spatial organizations. Second, these regions also exhibit sample-to-sample variation in their spatial extent, shape, and local boundaries, reflecting heterogeneous spatial organization among tissue sections.

These empirical observations reveal that spatial expression patterns across samples contain both shared and sample-specific structures. A representation that captures only the common spatial organization may overlook biologically meaningful variations, whereas modeling only individual patterns may fail to preserve reproducible spatial signals. These findings motivate a cross-sample integration that simultaneously characterizes shared spatial structures and sample-specific spatial variations.}

\newpage
\section{Details of the proposed variational inference algorithm}

In this section, we provide the detailed derivations of variational updates, functional approximation, hyperparameter selection strategies, and BFDR controlling procedure for the proposed posterior inference algorithm. The computer time of the proposed algorithm is also examined.

\subsection{Variational posteriors of parameters}
The posterior distribution is as follows:
\begin{equation*}
\begin{aligned}
&p\left(\Omega \mid \boldsymbol{Y}, \boldsymbol{X}, \boldsymbol{S}\right)
\propto \int\limits  \prod_{m=1}^{M} \prod_{i=1}^{n^{(m)}} \prod_{k=1}^{K}p\left(y_i^{(m)} \mid g^{(m)}_i, r_{i}^{(m)}\right) p\left(g^{(m)}_i \mid \boldsymbol{x}_i^{(m)},\boldsymbol{s}_i^{(m)},\Theta^{(m)}\right) p\left(r_{i}^{(m)} \mid \pi^{(m)}\right) \\ 
& p\left(\pi^{(m)}\right) p\left(\eta^{(m)}\right) p\left(\boldsymbol{\psi}^{(m)}\right)p\left(\phi^{(m)} \mid a_{\phi}^{(m)}, b_{\phi}^{(m)}\right)p\left(\boldsymbol{\beta}_k^{(m)} \mid \sigma^{(m)}_{k}, \alpha_k^{(m)}\right) p\left(\sigma^{(m)}_{k} \mid a_{k}^{(m)}\right) p\left(a_{k}^{(m)}\right) \\
&p\left(\alpha_k^{(m)} \mid q_k,u_k\right)p\left(q_k\right) p\left(u_k \mid p_k\right) p\left(p_k\right)  \mathrm{d}\boldsymbol{\pi}.
\label{post distribution}
\end{aligned}
\end{equation*}

Denote $\boldsymbol{C}^{(m)}_i=\left(1,\boldsymbol{\xi}\left({s_{i1}^{(m)}}\right),\boldsymbol{\xi}\left({s_{i2}^{(m)}}\right),\left(\boldsymbol{x}^{(m)}_i\right)^{\top}\right)$ and $\boldsymbol{C}^{(m)}=\left(\boldsymbol{C}^{(m)}_1,\cdots,\boldsymbol{C}^{(m)}_{n^{(m)}}\right)^{\top}$ for $m = 1,...,M$. Then, we have 
$\log\left(\lambda_{i}^{(m)}\right)=\boldsymbol{C}^{(m)}_i\boldsymbol{\theta}^{(m)}$. The variational posteriors of the parameters are as follows.
 
(1) $q^*\left(g^{(m)}_i\right)$.

We first have:
\begin{equation*}
    \begin{aligned}
        &\log p\left(g^{(m)}_i \mid \Omega_{-\tau},\boldsymbol{Y}, \boldsymbol{X}, \boldsymbol{S}\right)\\
        \propto& \left(1-r_i^{(m)}\right)\log\left(\frac{\left(g_i^{(m)}\right)^{y_i^{(m)}}\exp\left\{-g_i^{(m)}\right\}}{\left[y_i^{(m)}\right]!} \frac{\left(\phi^{(m)} \right)^{\phi^{(m)}}\left(g_i^{(m)}\right)^{\phi^{(m)}-1}\exp\left\{-\boldsymbol{C}^{(m)}_i\boldsymbol{\theta}^{(m)} \cdot \phi^{(m)}\right\}}
        {\Gamma\left(\phi^{(m)}\right)\exp \left\{\phi^{(m)} g_i^{(m)} \exp\left\{-\boldsymbol{C}^{(m)}_i\boldsymbol{\theta}^{(m)}\right\} \right\} }\right)\\
        =&\left(1-r_{i}^{(m)}\right)\left(\left(y_i^{(m)}+\phi^{(m)}-1\right)\log\left(g_i^{(m)}\right)-g_i^{(m)}-\phi^{(m)} g_i^{(m)} \exp\left\{-\boldsymbol{C}^{(m)}_i\boldsymbol{\theta}^{(m)}\right\}\right)+\text{const},
    \end{aligned}
\end{equation*}
where ``$\text{const}$'' represents some constant.  
Then,
\begin{equation*}
    \begin{aligned}
        \log\left(q^*\left(g^{(m)}_i\right)\right)\propto &E_{-g^{(m)}_i}\left(\log p\left(g^{(m)}_i \mid \Omega_{-\tau},\boldsymbol{Y}, \boldsymbol{X}, \boldsymbol{S}\right)\right)\\        
        =&\left(1-u_{q\left(r_{i}^{(m)}\right)}\right)\left(\left(y_i^{(m)}+u_{q\left(\phi^{(m)}\right)}-1\right)\log\left(g_i^{(m)}\right)\right.\\
        -&\left.\left(1+u_{q(\phi^{(m)})}E_{q(\boldsymbol{\theta}^{(m)})}\left(\exp\left\{-\boldsymbol{C}^{(m)}_i\boldsymbol{\theta}^{(m)}\right\}
        \right)\right)g_i^{(m)}
        \right)+\text{const}.    
    \end{aligned}
\end{equation*}
Thus, we have
\begin{equation}\label{q_gi}
    q^*\left(g^{(m)}_i\right) \sim Ga(a_i,b_i),  
\end{equation}
where 
\begin{equation*}
    \begin{aligned}
        &a_i=\left(y^{(m)}_i+u_{q\left(\phi^{(m)}\right)}-1\right)\left(1-u_{q(r_{i}^{(m)})}\right)+1,\\
        &b_i=\left(1-u_{q(r_{i}^{(m)})}\right) \left(u_{q\left(\phi^{(m)}\right)}  \exp \left\{-\boldsymbol{C}^{(m)}_i u_{\boldsymbol{\theta}^{(m)}}+\frac{1}{2}  \boldsymbol{C}^{(m)}_i \Sigma_{\boldsymbol{\theta}^{(m)}} \boldsymbol{C}^{(m)\top}_i\right\}+1\right).
    \end{aligned}
\end{equation*}

Based on (\ref{q_gi}), we compute the following expectations, which can be used in further inference: 
\begin{eqnarray*}
&E_{q\left(g^{(m)}_i\right)}\left(g^{(m)}_i\right)=\frac{a_i}{b_i}\triangleq u_{q\left(g^{(m)}_i\right)}, \\
&E_{q\left(g^{(m)}_i\right)}\left(\log\left(g^{(m)}_i\right)\right)=\text{digamma}(a_i)-\log(b_i).
\end{eqnarray*}
Here, the digamma function is given by $\text{digamma}(x) \equiv \frac{d}{d x} \log \{\Gamma(x)\}$ with $\Gamma(x)$ being the Gamma function.

(2) $q^*\left(r_i^{(m)}\right)$.

Here, we only concern the situation $y_i^{(m)}=0$ since $q\left(r_i^{(m)}\right)=1$ when $y_i^{(m)}>0$. We begin by marginalizing over $\pi^{(m)}$ to obtain a tractable form:
\begin{equation*}
\begin{aligned}
    p\left(r_{i}^{(m)} \right) 
	&= \int p\left(r_{i}^{(m)}|\pi^{(m)}\right)\times p\left(\pi^{(m)}\right)\,d\pi^{(m)}\\
    &=\int \frac{\left(\pi^{(m)}\right)^{r_{i}^{(m)}}\left(1-\pi^{(m)}\right)^{1-r_{i}^{(m)}}\left(\pi^{(m)}\right)^{a_{\pi}^{(m)}-1}\left(1-\pi^{(m)}\right)^{b_{\pi}^{(m)}-1}}{B\left(a_{\pi}^{(m)},b_{\pi}^{(m)}\right)}d\pi^{(m)}\\
    &=\frac{B\left(a_{\pi}^{(m)}+r_{i}^{(m)},b_{\pi}^{(m)}-r_{i}^{(m)}+1\right)}{B\left(a_{\pi}^{(m)},b_{\pi}^{(m)}\right)},
\end{aligned}
\end{equation*}
where \( B(\cdot, \cdot) \) is the Beta function with $
B(\alpha, \beta) = \int_0^1 t^{\alpha-1} (1-t)^{\beta-1} \, dt \quad (\alpha > 0, \beta > 0)
$.

Then, it is explicit to compute the posterior of $r_i^{(m)}$ and determine the variational posterior: 

\begin{equation*}
\begin{aligned}
    &\log\left(q^{*}\left(r_{i}^{(m)} \mid y_i^{(m)}=0\right)\right)\\
    \propto&E_{-r_{i}^{(m)}}\left(\log p\left(r_{i}^{(m)} \mid \Omega_{-\tau},\boldsymbol{Y}, \boldsymbol{X}, \boldsymbol{S}\right)\right)\\
    \propto&E_{-r_{i}^{(m)}}\log \left(p\left(y_i^{(m)}=0 \mid r_i^{(m)},g^{(m)}_i\right)p\left(r_{i}^{(m)}  \right)\right)\\
    =&E_{-r_{i}^{(m)}}\log \left(\frac{B\left(a_{\pi}^{(m)}+r_i^{(m)},1+b_{\pi}^{(m)}-r_i^{(m)}\right)}{B\left(a_{\pi}^{(m)},b_{\pi}^{(m)}\right)}\cdot         \left[\exp\left\{-g_i^{(m)}\right\} \right]^{1-r_i^{(m)}}  
 \right)\\
    =&-\left(1-r_i^{(m)}\right)u_{q\left(g^{(m)}_i\right)}+\log \left(\frac{B\left(a_{\pi}^{(m)}+r_i^{(m)},1+b_{\pi}^{(m)}-r_i^{(m)}\right)}{B\left(a_{\pi}^{(m)},b_{\pi}^{(m)}\right)}\right).
    \end{aligned}
\end{equation*}

Thus, we have
\begin{equation}\label{r}
    q^{*}\left(r_{i}^{(m)}\right) \sim Bern\left(\frac{B\left(a_{\pi}^{(m)}+1,b_{\pi}^{(m)}\right)}{B\left(a_{\pi}^{(m)}+1,b_{\pi}^{(m)}\right)+\exp\left\{-u_{q\left(g^{(m)}_i\right)}\right\}\cdot B\left(a_{\pi}^{(m)},b_{\pi}^{(m)}+1\right)}\right).
\end{equation}

Based on (\ref{r}), we get the expectation $u_{q\left(r_{i}^{(m)}\right)}=E_{q\left(r^{(m)}_i\right)}\left(r^{(m)}_i\right)$ as 
\begin{equation}
    u_{q\left(r_{i}^{(m)}\right)}=\frac{B\left(a_{\pi}^{(m)}+1,b_{\pi}^{(m)}\right)}{B\left(a_{\pi}^{(m)}+1,b_{\pi}^{(m)}\right)+\exp\left\{-u_{q\left(g^{(m)}_i\right)}\right\}\cdot B\left(a_{\pi}^{(m)},b_{\pi}^{(m)}+1\right)}.
\end{equation}

(3) $q^*(\phi^{(m)})$.

Denote $\omega = \left(\phi^{(m)}\right)^{-\frac{1}{2}}$ for easier inference. We have:
\begin{equation*}
    \begin{aligned}
        &\log p\left(\omega^2 \mid \Omega_{-\tau},\boldsymbol{Y}, \boldsymbol{X}, \boldsymbol{S}\right)  \\
        \propto \ &\sum_{i=1}^{n^{(m)}}\log\left( p\left(g_{i}^{(m)}|\ \boldsymbol{x}_i^{(m)},\boldsymbol{s}_i^{(m)},\Theta^{(m)}\right) p\left(w^2 \mid a_\phi^{(m)},b_\phi^{(m)}\right)\right) \\
        \propto \ &\sum_{i=1}^{n^{(m)}}\log\left\{\left[ \frac{\left(\frac{1}{\omega^2} \right)^{\frac{1}{\omega^2}}\left(g_i^{(m)}\right)^{\frac{1}{\omega^2}-1}\exp\left\{-\boldsymbol{C}_i^{(m)}\boldsymbol{\theta}^{(m)} \cdot \frac{1}{\omega^2}\right\}}{\Gamma\left(\frac{1}{\omega^2}\right)\exp \left\{\frac{1}{\omega^2} g_i^{(m)} \exp\left(-\boldsymbol{C}_i^{(m)}\boldsymbol{\theta}^{(m)}\right) \right\} }\right]^{1-r_i^{(m)}}\right\} \\
        & \cdot\left(b_{\phi}^{(m)} \right)^{a_{\phi}^{(m)}}\left(w^2\right)^{-a_{\phi}^{(m)}-1} \exp \left\{-\frac{b_{\phi}^{(m)}}{\omega^2}\right\} \frac{1}{\Gamma\left(a_{\phi}^{(m)}\right)}\\
     \end{aligned}
\end{equation*}   
\begin{equation}
    \begin{aligned}
                 =& \sum_{i=1}^{n^{(m)}}\left(1-r_i^{(m)}\right)\left\{-\frac{1}{\omega^2} \log\left( \omega^2 \exp \left\{\boldsymbol{C}_i^{(m)}\boldsymbol{\theta}^{(m)}\right\}\right)-\log \Gamma \left(\frac{1}{\omega^2}\right)+\frac{1}{\omega^2} \log g_i^{(m)}\right. \\
        & \quad \left. -\frac{g_i^{(m)}}{\omega^2 \exp \left\{\boldsymbol{C}_i^{(m)}\boldsymbol{\theta}^{(m)}\right\}}\right\}-\left(a_{\phi}^{(m)}+1\right) \log \omega^2-\frac{b_{\phi}^{(m)}}{ \omega^2}+\text{const}\\
        =& \frac{1}{\omega^2}\left[\sum_{i=1}^{n^{(m)}}\left(1-r_i^{(m)}\right)\left(-\boldsymbol{C}_i^{(m)}\boldsymbol{\theta}^{(m)}+\log g_i^{(m)}-\frac{g_i^{(m)}}{\exp \left\{\boldsymbol{C}_i^{(m)}\boldsymbol{\theta}^{(m)}\right\}}\right)-b_{\phi}^{(m)}\right] \\
        & \quad -\left(\log \Gamma\left(\frac{1}{\omega^2}\right)+\frac{\log \omega^2}{\omega^2} \right)\sum_{i=1}^{n^{(m)}}\left(1-r_i^{(m)}\right) -\left(a_{\phi}^{(m)}+1\right)\log \omega^2 
        +\text{const.}
    \label{phi1}
    \end{aligned}
\end{equation}
Then, we compute the optimal density of $\omega^2$ by integrating over parameters other than $\omega^2$.
\begin{equation*}
    \begin{aligned}
    \log q^*\left(\omega^2\right) & \propto E_{-\omega^2}\left(\log p\left(\omega^2 \mid \Omega_{-\tau},\boldsymbol{Y}, \boldsymbol{X}, \boldsymbol{S}\right)\right) \\
            & =-\left(\log \Gamma \left(\frac{1}{\omega^2}\right)+\frac{\log \omega^2}{\omega^2} \right)\sum_{i=1}^{n^{(m)}}\left(1-u_{q\left(r_{i}^{(m)}\right)}\right) -(a_{\phi}^{(m)}+1) \log \omega^2 \\
            & +\frac{1}{\omega^2}\left[\sum_{i=1}^{n^{(m)}} \left(1-u_{q\left(r_{i}^{(m)}\right)}\right)\left(E_{q\left(g^{(m)}_i\right)}\left(\log\left(g^{(m)}_i\right)\right)-\boldsymbol{C}_i^{(m)} u_{\boldsymbol{\theta}^{(m)}}\right.\right.\\
            &\left.\left.-u_{q\left(g_i^{(m)}\right)} E\left(\exp \left\{-\boldsymbol{C}_i^{(m)}\boldsymbol{\theta}^{(m)}\right\}\right)\right)-b_{\phi}^{(m)}\right]+\text{const}
    \end{aligned}
\end{equation*}
\begin{equation*}
    \begin{aligned}   
        \ \  \ \ \   \ \ \ &=-\left(\log \Gamma \left(\frac{1}{\omega^2}\right)+\frac{\log \omega^2}{\omega^2} \right)\sum_{i=1}^{n^{(m)}}\left(1-u_{q\left(r_{i}^{(m)}\right)}\right) -(a_{\phi}^{(m)}+1) \log \omega^2 -\frac{c_1}{\omega^2}+\text{const}.
    \end{aligned}
\end{equation*}
where
\begin{equation*}
    \begin{aligned}
        & c_1=\sum_{i=1}^{n^{(m)}}\left[\left(1-u_{q\left(r_{i}^{(m)}\right)}\right)\left(\boldsymbol{C}^{(m)}_i u_{\boldsymbol{\theta}^{(m)}}-E_{q\left(g^{(m)}_i\right)}\left(\log\left(g^{(m)}_i\right)\right)\right)+b_{\phi}^{(m)} \right. \\
        &\left. +\left(1-u_{q\left(r_{i}^{(m)}\right)}\right)  u_{q\left(g^{(m)}_i\right)} \exp \left\{-\boldsymbol{C}^{(m)}_i u_{\boldsymbol{\theta}^{(m)}}+\frac{1}{2}  \boldsymbol{C}^{(m)}_i \Sigma_{\boldsymbol{\theta}^{(m)}} \boldsymbol{C}^{(m)\top}_i\right\}\right]. 
    \end{aligned}
\end{equation*}

Afterwards, we get the variational posterior of $\phi^{(m)}$ by exponentiation and transformation. Specifically,
\begin{equation*}
    q^*\left(w^2\right) \propto \left\{\frac{\left(\frac{1}{\omega^2}\right)^{\frac{1}{\omega^2}}}{\Gamma\left(\frac{1}{\omega^2}\right)}\right\}^{N_{\pi}^{(m)}}\left(\omega^2\right)^{-\left(a_{\phi}^{(m)}+1\right)} \exp \left\{-\frac{c_1 }{\omega^2}\right\}
\end{equation*}	
and
\begin{equation}
    q^*\left(\phi^{(m)}\right) \propto \left\{\frac{\left(\phi^{(m)}\right)^{\phi^{(m)}}}{\Gamma\left(\phi^{(m)}\right)}\right\}^{N_{\pi}^{(m)}}\left(\phi^{(m)}\right)^{a_{\phi}^{(m)}-1} \exp \left\{-\ c_1 \cdot \phi^{(m)}\right\},
    \label{phi3}
\end{equation}
with $N_{\pi}^{(m)}=\sum_{i=1}^{n^{(m)}}\left(1-u_{q\left(r_{i}^{(m)}\right)}\right)$.

Since Equation (\ref{phi3}) is not a regularized distribution, we normalize it with numerical integration:
\begin{equation}
    q^{*}\left(\phi^{(m)}\right) = \left\{\frac{\left[\phi^{(m)}\right]^{\phi^{(m)}}}{\Gamma\left(\phi^{(m)}\right)}\right\}^{N_{\pi}^{(m)}} \left(\phi^{(m)}\right)^{a_{\phi}^{(m)}-1} \frac{\exp \left\{-c_1 \cdot \phi^{(m)}\right\} }{\mathcal{H}\left(a_{\phi}^{(m)}-1, 0, 1, N_{\pi}^{(m)}, c_1\right)},
    \label{phi4}
\end{equation}
where 
$\mathcal{H}\left(p, q, r, s, t\right) \equiv \int_0^{\infty} x^p \log \left(1+r x\right)^q\left\{x^x / \Gamma(x)\right\}^s \exp \left(-t x\right) dx
$
for $p \in\left\{-\frac{1}{2}, \frac{1}{2}\right\}, q \in\{0,1\}$ and $r, s, t>0$. 

Finally, we compute the expectation of the variational posterior $q(\phi^{(m)})$ for iteration:
\begin{equation*}
    \begin{aligned}
        u_{q(\phi^{(m)})}&=\frac{\mathcal{H}\left(a_{\phi}^{(m)}, 0,1, N_{\pi}^{(m)}, c_1\right)}{\mathcal{H}\left(a_{\phi}^{(m)}-1, 0,1, N_{\pi}^{(m)}, c_1\right)}\\
        &=\exp \left\{\log \left(\mathcal{H}\left(a_{\phi}^{(m)}, 0,1, N_{\pi}^{(m)}, c_1\right)\right)-\log \left(\mathcal{H}\left(a_{\phi}^{(m)}-1, 0,1, N_{\pi}^{(m)}, c_1\right)\right)\right\}.
    \end{aligned}
\end{equation*}

(4) $q^*(\sigma_{k}^{2(m)})$.

We first have:
\begin{equation*}
    \begin{aligned}
        &\log \left(p\left(\sigma_{ k}^{2(m)} \mid \Omega_{-\tau},\boldsymbol{Y}, \boldsymbol{X}, \boldsymbol{S}\right)\right)
        \propto \log \left(p\left(\boldsymbol{\beta}_k^{(m)} \ | \ \sigma_{k}^{2(m)},\alpha_k^{(m)}\right) p\left(\sigma_{ k}^{2(m)} \ | \ a_{ k }^{(m)}\right)\right)\\
        =&\log\left(\left[\frac{1}{[\sigma_{k}^{(m)}]^G} \exp \left\{-\frac{(\boldsymbol{\beta}_k^{(m)})^{T} \boldsymbol{\beta}_k^{(m)}}{2 \sigma_{k}^{2(m)}}\right\}\right]^{\alpha_k^{(m)}}   \left[\frac{1}{\Gamma_1^G} \exp \left\{-\frac{(\boldsymbol{\beta}_k^{(m)})^{T} \boldsymbol{\beta}_k^{(m)}}{2 \Gamma_1^2}\right\}\right]^{1-\alpha_k^{(m)}} 
        \right.\\
        &\left.\cdot\frac{\left(\sigma_{k}^{2(m)}\right)^{-\frac{3}{2}} \exp \left\{-\frac{1}{a_{k}^{(m)} \sigma_{k}^{2(m)}}\right\}}{\left(a_{k}^{(m)}\right)^{\frac{1}{2}}\Gamma \left(\frac{1}{2}\right) }\right)\\
        =&- \alpha_k^{(m)} \frac{\boldsymbol{\beta}_k^{(m)T} \boldsymbol{\beta}_k^{(m)}}{2 \sigma_{k}^{2(m)}} - \frac{G \cdot \alpha_k^{(m)}}{2} \log \left(\sigma_{k}^{2(m)}\right)- \left(1 - \alpha_k^{(m)}\right) \frac{\boldsymbol{\beta}_k^{(m)T} \boldsymbol{\beta}_k^{(m)}}{2 \Gamma_1^2}- \frac{1}{2} \log \left(a_{k}^{(m)}\right)  \\
        & - \frac{3}{2} \log \left(\sigma_{k}^{2(m)}\right) - \frac{1}{a_{k}^{(m)} \sigma_{k}^{2(m)}}- \left(1 - \alpha_{k}^{(m)}\right) \cdot G \cdot \log \left(\Gamma_1\right)-\log\Gamma(\frac{1}{2}).   
    \end{aligned}
\end{equation*}
Then, we compute the optimal density of $\sigma_{ k}^{2(m)}$ by integrating over parameters other than $\sigma_{ k}^{2(m)}$:
\begin{equation}
    \begin{aligned}
        \log q^*\left(\sigma_{ k}^{2(m)}\right) 
        \propto& E_{-\sigma_{ k}^{2(m)}}\left(\log p\left(\sigma_{ k}^{2(m)} \mid \Omega_{-\tau},\boldsymbol{Y}, \boldsymbol{X}, \boldsymbol{S}\right)\right)\\
        =&-\left(\frac{u_{q\left(\alpha_k^{(m)}\right)}E_{q\left(\boldsymbol{\theta}^{(m)}\right)}\left(\boldsymbol{\beta}_k^{(m)T} \boldsymbol{\beta}_k^{(m)}\right)}{2} +u_{q\left(1/a_{k}^{(m)}\right)}\right)\cdot\frac{1}{\sigma_{k}^{2(m)}}\\
        &-\frac{G \cdot u_{q\left(\alpha_k^{(m)}\right)}+3}{2} \log \left(\sigma_{k}^{2(m)}\right)+\text{const}.
    \end{aligned}
\end{equation}
Thus, we have $q^{*}\left(1/\sigma_{ k}^{2(m)}\right) \sim Ga\left(\tilde{a},\tilde{b}\right)$ with the parameters
\begin{equation}\label{sigma_beta}
\left\{\begin{array}{ll}
&\tilde{a}  =\frac{1}{2}\left(G \cdot u_{q\left(\alpha_{k}^{(m)}\right)}+1\right), \\
&\tilde{b}  =\frac{1}{2}u_{q\left(\alpha_{k}^{(m)}\right)} E_{q\left(\boldsymbol{\theta}^{(m)}\right)}\left(\boldsymbol{\beta}_k^{(m)T} \boldsymbol{\beta}_k^{(m)}\right) + u_{q\left(1 / a_{k}^{(m)}\right)}.
\end{array}
\right.
\end{equation}
Then, we can compute the expectation $u_{q\left(1 / \sigma_{k}^{2(m)}\right)}$  based on (\ref{sigma_beta}) and (\ref{nvmp5}) for further inference:
\begin{equation}
\begin{aligned}
        u_{q\left(1/\sigma_{k}^{2(m)}\right)} &= \frac{\tilde{a}}{\tilde{b}},\\
E_{q\left(1/\sigma_{k}^{2(m)}\right)}\left(\log\left(1/\sigma_{k}^{2(m)}\right)\right)&=\text{digamma}(\tilde{a})-\log\left(\tilde{b}\right).
\end{aligned}
\end{equation}

(5) $q^*\left(a_{k}^{(m)}\right)$.

We first have:
\begin{equation*}
    \begin{aligned}
    & \log p\left(a_{k}^{(m)} \mid \Omega_{-\tau},\boldsymbol{Y}, \boldsymbol{X}, \boldsymbol{S}\right)\\
        \propto&\log\left(p\left( \sigma_{k}^{2(m)} \mid   a_{k}^{(m)}\right)p\left(a_{k}^{(m)}\right)\right)  \\
        =&\log\left(\frac{\left(\sigma_{k}^{2(m)}\right)^{-\frac{3}{2}} \exp \left\{-\frac{1}{a_{k}^{(m)} \sigma_{k}^{2(m)}}\right\}}{\left(a_{k}^{(m)}\right)^{\frac{1}{2}}\Gamma \left(\frac{1}{2}\right) } \cdot \frac{\left(a_{k}^{(m)}\right)^{-\frac{3}{2}} \exp \left\{-\frac{1}{A_k^2 a_{k}^{(m)}}\right\}}{\left(A_k^2\right)^{\frac{1}{2}} \Gamma \left(\frac{1}{2}\right)}\right)\\
        =& - \frac{3}{2} \log \left(\sigma_{k}^{2(m)}\right) - \frac{1}{a_{k}^{(m)} \sigma_{k}^{2(m)}}- \frac{1}{2} \log \left(a_{k}^{(m)}\right) - \frac{3}{2} \log \left(a_{k}^{(m)}\right) - \frac{1}{A_k^2 a_{k}^{(m)}} +\text{const.}
    \end{aligned}
\end{equation*}
Then, we compute $q^*\left(a_{k}^{(m)}\right)$ by integrating over parameters other than $a_{k}^{(m)}$.
\begin{equation}
    \begin{aligned}
        \log q^*\left(a_{k}^{(m)}\right) 
        \propto& E_{-a_{k}^{(m)}}\left(\log p\left(a_{k}^{(m)} \mid \Omega_{-\tau},\boldsymbol{Y}, \boldsymbol{X}, \boldsymbol{S}\right)\right)\\
        =&-\left(u_{q\left(1/\sigma_{k}^{2(m)}\right)}+\frac{1}{A_k^2}\right)\cdot \frac{1}{a_{k}^{(m)}}-2 \log \left(a_{k}^{(m)}\right) +\text{const.}
    \end{aligned}
\end{equation}
Thus, we have:
\begin{equation*}
        q^{*}\left(a_{k}^{(m)}\right) \sim IG\left(1,u_{q\left(1/ \sigma_{k}^{2(m)}\right)}+\frac{1}{A_k^{2}}\right).
\end{equation*} 
Then, we obtain the expectation $u_{q\left(1/a_{k}^{(m)}\right)}$ of the variational posterior for further inference:
\begin{equation*}
        u_{q\left(1/a_{k}^{(m)}\right)}=\frac{1}{u_{q\left(1/\sigma_{k}^{2(m)}\right)}+\frac{1}{A_k^{2}}}.
\end{equation*}

(6) $q^*\left(\alpha_k^{(m)}\right)$.

We first have:
\begin{equation*}
    \begin{aligned}
        \quad \quad& \log\left( p\left(\alpha_k^{(m)} \mid \Omega_{-\tau},\boldsymbol{Y},\boldsymbol{X}, \boldsymbol{S}\right)\right)\\
        \propto&\log\left(p\left(\boldsymbol{\beta}_{k}^{(m)}  \mid   \sigma_{k}^{(m)},\alpha_k^{(m)}\right)p\left(\alpha_k^{(m)} \mid q_k,u_k\right)\right)  \\
        =&\log\left(\left[\frac{1}{[\sigma_{k}^{(m)}]^G} \exp \left\{-\frac{(\boldsymbol{\beta}_k^{(m)})^{T} \boldsymbol{\beta}_k^{(m)}}{2 \sigma_{k}^{2(m)}}\right\}\right]^{\alpha_k^{(m)}}   \left[\frac{1}{\Gamma_1^G} \exp \left\{-\frac{(\boldsymbol{\beta}_k^{(m)})^{T} \boldsymbol{\beta}_k^{(m)}}{2 \Gamma_1^2}\right\}\right]^{1-\alpha_k^{(m)}} 
        \right.\\ 
        &\left. \left[q_k^{\alpha_k^{(m)}} \left(1-q_k\right)^{1-\alpha_k^{(m)}}\right]^{u_k}
        \left[\Gamma_2^{\alpha_k^{(m)}}\left(1-\Gamma_2\right)^{1-\alpha_k^{(m)}}\right]^{1-u_k}\right)\\
        =&- \alpha_k^{(m)} \frac{\boldsymbol{\beta}_k^{(m)T} \boldsymbol{\beta}_k^{(m)}}{2 \sigma_{k}^{2(m)}} - \frac{G \cdot \alpha_k^{(m)}}{2} \log \left(\sigma_{k}^{2(m)}\right)- \left(1 - \alpha_k^{(m)}\right) \frac{\boldsymbol{\beta}_k^{(m)T} \boldsymbol{\beta}_k^{(m)}}{2 \Gamma_1^2}        \\
        &- \left(1 - \alpha_{k}^{(m)}\right) \cdot G \cdot \log \left(\Gamma_1\right) + u_k \cdot \left(\alpha_k^{(m)} \log \left(q_k\right) + \left(1 - \alpha_k^{(m)}\right) \log \left(1 - q_k\right)\right)\\
         &+ (1 - u_k) \cdot \left(\alpha_k^{(m)} \log \left(\Gamma_2\right) + \left(1 - \alpha_k^{(m)}\right) \log \left(1 - \Gamma_2\right)\right).
    \end{aligned}
\end{equation*}
Then, we compute $q^*\left(\alpha_k^{(m)}\right)$ by integrating over parameters other than $\alpha_k^{(m)}$.
\begin{equation*}
    \begin{aligned}
        &\log q^*\left(\alpha_k^{(m)}\right) \\
        \propto& E_{-\alpha_k^{(m)}}\left(\log p\left(\alpha_k^{(m)} \mid \Omega_{-\tau},\boldsymbol{Y}, \boldsymbol{X}, \boldsymbol{S}\right)\right)\\
        =&-   \left[\frac{1}{2}E_{q\left(\boldsymbol{\theta}^{(m)}\right)}\left(\boldsymbol{\beta}_k^{(m)T} \boldsymbol{\beta}_k^{(m)}\right)u_{q\left(1/\sigma_{k}^{2(m)}\right)}+\frac{1}{2}G \cdot E_{q^{*}\left(1/\sigma_{ k}^{2(m)}\right)}\left(\log \left(\sigma_{k}^{2(m)}\right)\right)\right.
    \end{aligned}
\end{equation*}
\begin{equation*}
    \begin{aligned}
    & \quad\quad\left.-u_{q(u_k)}E_{q(q_k)}\left(\log\left(q_k\right) \right)-\left(1-u_{q(u_k)}\right)\log(\Gamma_2)\right]\cdot\alpha_k^{(m)}\\
        &-\left[\frac{E_{q\left(\boldsymbol{\theta}^{(m)}\right)}\left(\boldsymbol{\beta}_k^{(m)T} \boldsymbol{\beta}_k^{(m)}\right)}{2\Gamma_1^2}+G\log\left(\Gamma_1\right)-u_{q(u_k)}E_{q(q_k)}\left(\log\left(1-q_k\right) \right)\right.\\
        &\quad \quad\left.-\left(1-u_{q(u_k)}\right)\log\left(1-\Gamma_2\right)\right]\cdot\left(1-\alpha_k^{(m)}\right).
    \end{aligned}
\end{equation*}
Thus, we have:
\begin{equation*}
    q^{*}\left(\alpha_{k}^{(m)}\right) \sim Bern\left(\frac{\text{part1}}{\text{part1}+\text{part2}}\right).
\end{equation*}
where
\begin{equation}		
    \begin{aligned}
    \operatorname{part1} &=\exp\left\{-   \frac{1}{2}E_{q\left(\boldsymbol{\theta}^{(m)}\right)}\left(\boldsymbol{\beta}_k^{(m)T} \boldsymbol{\beta}_k^{(m)}\right)u_{q\left(1/\sigma_{k}^{2(m)}\right)}+u_{q(u_k)}E_{q(q_k)}\left(\log\left(q_k\right) \right)\right.\\
        &\quad \quad\quad\quad \left.-\frac{1}{2}G\cdot E_{q\left(1/\sigma_{ k}^{2(m)}\right)}\left(\log \left(\sigma_{k}^{2(m)}\right)\right)+\left(1-u_{q(u_k)}\right)\log(\Gamma_2)
    \right\},\\
    \operatorname{part2} &=\exp\left\{-\frac{E_{q\left(\boldsymbol{\theta}^{(m)}\right)}\left(\boldsymbol{\beta}_k^{(m)T} \boldsymbol{\beta}_k^{(m)}\right)}{2\Gamma_1^2}+u_{q(u_k)}E_{q(q_k)}\left(\log\left(1-q_k\right) \right)\right.\\
        &\quad \quad\quad\quad\left.+\left(1-u_{q(u_k)}\right)\log\left(1-\Gamma_2\right)-G\cdot\log\left(\Gamma_1\right)\right\}.
    \end{aligned}
    \label{alpha_post}
\end{equation}

Then, we obtain the expectation $u_{q\left(\alpha_{k}^{(m)}\right)}= \frac{\operatorname{part1}}{\operatorname{part1} + \operatorname{part2}}$ by the variational posterior for further inference.

(7) $q^*(u_k)$.

We first have:
\begin{equation*}
    \begin{aligned}
        & \log p\left(u_k \mid \Omega_{-\tau},\boldsymbol{Y}, \boldsymbol{X}, \boldsymbol{S}\right)\\
        \propto&\log\left(\prod_{m=1}^M p\left(\alpha_k^{(m)} \mid u_k,\ q_k\right)p\left(u_k \mid p_k\right)\right)   \\
        =&\log\left(\prod_{m=1}^M   
        \left[q_k^{\alpha_k^{(m)}} \left(1-q_k\right)^{1-\alpha_k^{(m)}}\right]^{u_k}
        \left[\Gamma_2^{\alpha_k^{(m)}}\left(1-\Gamma_2\right)^{1-\alpha_k^{(m)}}\right]^{1-u_k}\cdot (p_k)^{u_k} (1-p_k)^{(1-u_k)}       \right) \\
        =&\left(\sum_{m=1}^M \left(\alpha_k^{(m)}\log\left(q_k\right)+\left(1-\alpha_k^{(m)}\right)\log\left(1-q_k\right)\right)+\log\left(p_k\right)\right)\cdot u_k \\
        &+ \left(\sum_{m=1}^M \left(\alpha_k^{(m)}\log(\Gamma_2)+\left(1-\alpha_k^{(m)}\right)\log\left(1-\Gamma_2\right)\right)+\log\left(1-p_k\right)\right)\cdot\left(1-u_k\right).  \\
    \end{aligned}
\end{equation*}
Then, we compute $q^*(u_k)$ by integrating over parameters other than $u_k$.
\begin{equation*}
    \begin{aligned}
        &\log q^*(u_k) \\
        \propto& E_{-u_k}\left(\log p\left(u_k \mid \Omega_{-\tau},\boldsymbol{Y}, \boldsymbol{X}, \boldsymbol{S}\right)\right)\\
        =&\left(\sum_{m=1}^M \left(u_{q\left(\alpha_{k}^{(m)}\right)}E_{q(q_k)}\left(\log\left(q_k\right)\right)
        +\left(1-u_{q\left(\alpha_{k}^{(m)}\right)}\right)E_{q(q_k)}\left(\log\left(1-q_k\right)\right)\right)+E_{q(p_k)}\left(\log\left(p_k\right)\right)\right)\cdot u_k\\
        &+ \left(\sum_{m=1}^M \left(u_{q\left(\alpha_{k}^{(m)}\right)}\log(\Gamma_2)+\left(1-u_{q\left(\alpha_{k}^{(m)}\right)}\right)\log\left(1-\Gamma_2\right)\right)+E_{q(p_k)}\left(\log\left(1-p_k\right)\right)\right)\cdot\left(1-u_k\right).
    \end{aligned}
\end{equation*}
Thus, we have:
\begin{equation*}
    q^{*}(u_k)\sim Bern\left(\frac{\operatorname{part1}}{\operatorname{part1}+\operatorname{part2}}\right).
\end{equation*}
where
\begin{equation}
\begin{aligned}
        \operatorname{part1}&=\exp\left\{\sum_{m=1}^M \left(u_{q\left(\alpha_{k}^{(m)}\right)}E_{q(q_k)}\left(\log\left(q_k\right)\right)+\left(1-u_{q\left(\alpha_{k}^{(m)}\right)}\right)E_{q(q_k)}\left(\log\left(1-q_k\right)\right)\right)\right.\\
        &\quad \quad \quad \quad \left.+E_{q(p_k)}\left(\log\left(p_k\right)\right)\right\},\\
    \operatorname{part2}&=\exp\left\{\sum_{m=1}^M \left(u_{q\left(\alpha_{k}^{(m)}\right)}\log(\Gamma_2)+\left(1-u_{q\left(\alpha_{k}^{(m)}\right)}\right)\log\left(1-\Gamma_2\right)\right)+E_{q(p_k)}\left(\log\left(1-p_k\right)\right) \right\}.
\end{aligned}
\label{u_iteration}
\end{equation}

Then, we obtain the expectation $u_{q(u_k)}=\frac{part1}{part1+part2}$ by the variational posterior for further inference.

(8) $q^*(p_{k})$.

We first have:
\begin{equation*}
    \begin{aligned}
     \log p\left(p_{k} \mid \Omega_{-\tau},\boldsymbol{Y}, \boldsymbol{X}, \boldsymbol{S}\right)
        \propto&\log\left(p\left(u_{k}|p_{k}\right)p\left(p_{k}\right)\right)\\
       =&\log\left(p_k^{u_k} (1-p_k)^{(1-u_k)} \cdot 
        \frac{1}{B\left(c_p, d_p\right)} p_k^{c_p-1}\left(1-p_k\right)^{d_p-1} \right)\\
        =&\left( u_k \log \left(p_k\right) + (1 - u_k) \log \left(1 - p_k\right) + \left(c_p - 1\right) \log \left(p_k\right)\right.\\
        &\left.+\left(d_p - 1\right) \log \left(1 - p_k\right)\right)+\text{const.}\\
    \end{aligned}
\end{equation*}
Then, we compute $q^*(p_{k})$ by integrating over parameters other than $p_{k}$.
\begin{equation*}
    \begin{aligned}
    \log q^*(p_{k}) 
        \propto& E_{-p_{k}}\left(\log p\left(p_{k} \mid \Omega_{-\tau},\boldsymbol{Y}, \boldsymbol{X}, \boldsymbol{S}\right)\right)\\
        =&\left(u_{q(u_k)}+c_p-1\right)\log \left(p_k\right)+\left(d_p-u_{q(u_k)}\right)\log \left(1-p_k\right)+\text{const.}
    \end{aligned}
\end{equation*}
Thus, we have:
\begin{equation}
    q^{*}\left(p_{k}\right) \sim Beta\left(a_p,b_p\right),
    \label{q_pk}
\end{equation}
where
\begin{equation*}
    \begin{cases}
        a_p=u_{q(u_k)}+c_p,\\
        b_p=d_p-u_{q(u_k)}+1.
    \end{cases}
\end{equation*}
Based on (\ref{q_pk}), we compute the following expectations, which can be used in further inference: 
\begin{equation}
    \begin{aligned}
        u_{q(p_k)}&=\frac{a_p}{a_p+b_p},\\
        E_{q(p_k)}\left(\log\left(p_k\right)\right)&=\text{digamma}(a_p)-\text{digamma}(a_p+b_p),\\
        E_{q(p_k)}\left(\log\left(1-p_k\right)\right)&=\text{digamma}(b_p)-\text{digamma}(a_p+b_p).\\
    \end{aligned}
    \label{exp_pk}
\end{equation}

(9) $q^*(q_{k})$.
We first have:
\begin{equation*}
    \begin{aligned}
        \quad\quad\quad\quad\quad&\log \left(p\left(q_{k} \mid \Omega_{-\tau},\boldsymbol{Y}, \boldsymbol{X}, \boldsymbol{S}\right)\right)\\
        \propto&\log\left(\prod_{m=1}^{M}p\left(\alpha_{k}^{(m)}|u_{k},q_{k}\right) \times
	p(q_{k})\right)\\
    =&\log\left(\prod_{m=1}^{M}\left[q_k^{\alpha_k^{(m)}} \left(1-q_k\right)^{1-\alpha_k^{(m)}}\right]^{u_k}
        \left[\Gamma_2^{\alpha_k^{(m)}}\left(1-\Gamma_2\right)^{1-\alpha_k^{(m)}}\right]^{1-u_k}            \right)\\
    &+\log\left( \frac{1}{B\left(c_q, d_q\right)} q_k^{c_q-1}\left(1-q_k\right)^{d_q-1}      \right)
    \end{aligned}
\end{equation*}
\begin{equation*}
    \begin{aligned}
    =& \sum_{m=1}^{M} \left\{\left(\alpha_k^{(m)}\log\left(q_k\right)+\left(1-\alpha_k^{(m)}\right)\log\left(1-q_k\right)\right)\cdot u_k
         \right.\\
    & \left. + \left(\alpha_k^{(m)}\log(\Gamma_2)+\left(1-\alpha_k^{(m)}\right)\left(1-\Gamma_2\right)\right)\cdot\left(1-u_k\right)       \right\}\\
    &+\left(c_q - 1\right)\log \left(q_k\right) +  \left(d_q - 1\right) \log \left(1 - q_k\right)+\text{const.}
    \end{aligned}
\end{equation*}
Then, we compute $q^*(q_{k})$ by integrating over parameters other than $q_{k}$.
\begin{equation*}
    \begin{aligned}
    \log q^*(q_{k}) 
        \propto& E_{-q_{k}}\left(\log p\left(q_{k} \mid \Omega_{-\tau},\boldsymbol{Y}, \boldsymbol{X}, \boldsymbol{S}\right)\right)\\
        =&\left(\sum_{m=1}^{M} u_{q\left(\alpha_k^{(m)}\right)}u_{q(u_k)}+c_q - 1\right) \log\left(q_k\right)\\
    &+\left(\sum_{m=1}^{M}u_{q(u_k)}\left( 1-\alpha_k^{(m)}\right)+d_q-1 \right) \log\left(1-q_k\right)+\text{const.}
    \end{aligned}
\end{equation*}
Thus, we have:
\begin{equation}
    q^{*}\left(q_{k}\right) \sim Beta\left(a_q,b_q\right),
    \label{q_qk}
\end{equation}
where
\begin{equation*}
    \begin{cases}
        a_q=u_{q(u_k)}\sum_{m=1}^{M}u_{q\left(\alpha_{ k}^{(m)}\right)}+c_q,\\
        b_q=M \cdot u_{q(u_k)}+d_q-u_{q(u_k)}\sum_{m=1}^{M}u_{q\left(\alpha_{ k}^{(m)}\right)}.
    \end{cases}
\end{equation*}
Based on (\ref{q_qk}), we compute the following expectations, which can be used in further inference: 
\begin{equation}
    \begin{aligned}
        u_{q(q_k)}&=\frac{a_q}{a_q+b_q},\\
        E_{q(q_k)}\left(\log\left(q_k\right)\right)&=\text{digamma}(a_q)-\text{digamma}(a_q+b_q),\\
        E_{q(q_k)}\left(\log\left(1-q_k\right)\right)&=\text{digamma}(b_q)-\text{digamma}(a_q+b_q).
    \end{aligned}
    \label{exp_qk}
\end{equation}

(10) $q^*\left(\boldsymbol{\theta}^{(m)}\right)$.

As the posterior of $\boldsymbol{\theta}^{(m)}$ has no explicit form, we complement the Non-conjugate Variational Message passing, which approximates the posterior distribution as a multi-normal distribution $q\left(\boldsymbol{\theta}^{(m)}\right) \sim N\left(\boldsymbol{u_{\boldsymbol{\theta}^{(m)}}},\boldsymbol{\Sigma_{\boldsymbol{\theta}^{(m)}}}\right)$ with $\boldsymbol{\mu}_{\boldsymbol{\theta}^{(m)}}$ being a $(1+2L+J)$-dimension mean vector and $\boldsymbol{\Sigma}_{\boldsymbol{\theta}^{(m)}}$ being a $(1+2L+J)\times(1+2L+J)$ covariance matrix. The calculation of $\boldsymbol{\mu}_{\boldsymbol{\theta}^{(m)}}$ and $\boldsymbol{\Sigma}_{\boldsymbol{\theta}^{(m)}}$ can be computed via equation (\ref{compute of nvmp}).
\begin{equation}
    \begin{aligned}
    & \boldsymbol{\Sigma}_{\boldsymbol{\theta}^{(m)}} =\left\{-2 \operatorname{vec}^{-1}\left(\left[\mathrm{D}_{\operatorname{vec}\left(\boldsymbol{\boldsymbol{\Sigma_{\boldsymbol{\theta}^{(m)}}}}\right)} E_{\boldsymbol{\theta}^{(m)}}\left\{\log p\left(\Omega^{(m)}, \boldsymbol{Y}^{(m)}\mid\boldsymbol{X}^{(m)},\boldsymbol{S}^{(m)}\right)\right\}\right]^{\top}\right)\right\}^{-1} ,\\
    & \boldsymbol{\mu}_{\boldsymbol{\theta}^{(m)}} = \boldsymbol{\mu}_{\boldsymbol{\theta}^{(m)}}+\boldsymbol{\boldsymbol{\Sigma_{\boldsymbol{\theta}^{(m)}}}}\left[\mathrm{D}_{\boldsymbol{\mu_{\boldsymbol{\theta}^{(m)}}}} E_{\boldsymbol{\theta}^{(m)}}\left\{\log p\left(\Omega^{(m)}, \boldsymbol{Y}^{(m)}\mid \boldsymbol{X}^{(m)},\boldsymbol{S}^{(m)}\right)\right\}\right]^{\top},
    \end{aligned}
    \label{compute of nvmp}
\end{equation}
where $\operatorname{vec}(\boldsymbol{A})$ denotes a vector formed by stacking the columns of matrix $\boldsymbol{A}$ underneath each other in order from left to right and $\operatorname{vec}^{-1}(\boldsymbol{a})$ is a $d \times d$ matrix formed from listing the entries of the $d^2 \times 1$ vector $\boldsymbol{a}$ in a column-wise fashion in order from left to right. $\mathrm{D}_{\boldsymbol{x}} f$ is the derivative vector of $f$ with respect to $\boldsymbol{x}$.

In (\ref{compute of nvmp}), we first have
\begin{equation}
    \begin{aligned}
        &E_{\boldsymbol{\theta}^{(m)}}\left\{\log p\left(\Omega^{(m)}, \boldsymbol{Y}^{(m)}\mid \boldsymbol{X}^{(m)},\boldsymbol{S}^{(m)}\right)\right\}\\
        =&E_{\boldsymbol{\theta}^{(m)}}\left\{\log p\left(\boldsymbol{\theta}^{(m)} \mid \boldsymbol{Y}^{(m)},\boldsymbol{X}^{(m)},\boldsymbol{S}^{(m)},\Omega^{(m)}/\boldsymbol{\theta}^{(m)}\right)\right\}+\text{terms not involving }  \boldsymbol{\theta}^{(m)}.
    \end{aligned}
    \label{nvmp1}
\end{equation}
Then, we just need to compute the posterior distribution of $\boldsymbol{\theta}^{(m)}$:
\begin{equation}
    \begin{aligned}
        & p\left(\boldsymbol{\theta}^{(m)} \mid \boldsymbol{Y}^{(m)},\boldsymbol{X}^{(m)},\boldsymbol{S}^{(m)},\Omega^{(m)}/ \boldsymbol{\theta}^{(m)}\right)\\
        &\propto \prod_{i=1}^{n^{(m)}} \left\{
        \left[\frac{\left(g_i^{(m)}\right)^{y_i^{(m)}}\exp\left\{-g_i^{(m)}\right\}}{\left[y_i^{(m)}\right]!} \frac{\left(\phi^{(m)} \right)^{\phi^{(m)}}\left(g_i^{(m)}\right)^{\phi^{(m)}-1}\exp\left\{-\boldsymbol{C}^{(m)}_i\boldsymbol{\theta}^{(m)} \cdot \phi^{(m)}\right\}}
        {\Gamma\left(\phi^{(m)}\right)\exp \left\{\phi^{(m)} g_i^{(m)} \exp\left\{-\boldsymbol{C}^{(m)}_i\boldsymbol{\theta}^{(m)}\right\} \right\} }\right]^{1-r_i^{(m)}}\right\} \\
        &\quad \cdot \prod_{k=1}^K \left\{ \left[\frac{1}{\sigma_{k}^{(m)G}} \exp \left\{-\frac{(\boldsymbol{\beta}_k^{(m)})^{T} \boldsymbol{\beta}_k^{(m)}}{2 \sigma_{k}^{2(m)}}\right\}\right]^{\alpha_k^{(m)}}   \left[\frac{1}{\Gamma_1^G} \exp \left\{-\frac{(\boldsymbol{\beta}_k^{(m)})^{T} \boldsymbol{\beta}_k^{(m)}}{2 \Gamma_1^2}\right\}\right]^{1-\alpha_k^{(m)}} \right\} \\
                &\quad \cdot\frac{1}{\sigma_\eta^{(m)}} \exp \left\{-\frac{\eta^{2(m)}}{2 \sigma_\eta^{2(m)}}\right\} \cdot \frac{1}{\sigma_{\psi}^{(m)J}} \exp \left\{-\frac{\boldsymbol{\psi}^{(m)T}\boldsymbol{\psi}^{(m)}}{2 \sigma_\psi^{2(m)}}\right\}.
    \end{aligned}
    \label{nvmp2}
\end{equation}
Accordingly, we have:
\begin{equation}
    \begin{aligned}
        &\log p\left(\boldsymbol{\theta}^{(m)} \mid \boldsymbol{Y}^{(m)},\boldsymbol{X}^{(m)},\boldsymbol{S}^{(m)},\Omega^{(m)}/ \boldsymbol{\theta}^{(m)}\right) \\
        &=\sum _ { i = 1 } ^ { n ^ { (m) } } ( 1 - r _ { i } ^ { ( m ) } ) \left[y_i^{(m)} \log g_i^{(m)}-\log \left(y_i^{(m)} !\right)-g_i^{(m)}+\phi^{(m)} \log \phi^{(m)}+\left(\phi^{(m)}-1\right) \log g_i^{(m)}\right. \\
        &\quad \left. -\log \left(\Gamma\left(\phi^{(m)}\right)\right)-\phi^{(m)} \boldsymbol{C}^{(m)}_i\boldsymbol{\theta}^{(m)}-\frac{\phi^{(m)}g_i^{(m)}}{\exp\left\{ \boldsymbol{C}^{(m)}_i\boldsymbol{\theta}^{(m)}\right\}}\right]-\frac{\eta^{2(m)}}{2 \sigma_\eta^{2(m)}}-\alpha_1^{(m)} \frac{(\boldsymbol{\beta}_1^{(m)})^T \boldsymbol{\beta}_1^{(m)}}{2 \sigma^{2(m)}_{1}} \\
        &\quad -\left(1-\alpha_1^{(m)}\right) \frac{(\boldsymbol{\beta}_1^{(m)})^T \boldsymbol{\beta}_1^{(m)}}{2 \Gamma_1^2}-\alpha_2^{(m)} \frac{(\boldsymbol{\beta}_2^{(m)})^T \boldsymbol{\beta}_2^{(m)}}{2 \sigma^{2(m)}_{2}}-\left(1-\alpha_2^{(m)}\right) \frac{(\boldsymbol{\beta}_2^{(m)})^T \boldsymbol{\beta}_2^{(m)}}{2 \Gamma_1^2}-\frac{\boldsymbol{\psi}^{(m)T}\boldsymbol{\psi}^{(m)}}{2 \sigma_\psi^{2(m)}}\\
        &\quad+\text{terms not involving } \boldsymbol{\theta}^{(m)}.
        \label{nvmp3}
    \end{aligned}
\end{equation}
Subsequently, we compute the expectation with respect to $\boldsymbol{\theta}^{(m)}$:
\begin{equation}
    \begin{aligned}
    & E_{\boldsymbol{\theta}^{(m)}}\left(\log p\left(\Omega^{(m)}, \boldsymbol{Y}^{(m)}\mid\boldsymbol{X}^{(m)},\boldsymbol{S}^{(m)}\right)\right) \\
    = & \sum_{i=1}^{n^{(m)}}\left(1-u_{q\left(r_i^{(m)}\right)}\right)\left(-u_{q\left(\phi^{(m)}\right)}\boldsymbol{C}^{(m)}_i u_{\boldsymbol{\theta}^{(m)}}-u_{q\left(\phi^{(m)}\right)} u_{q\left(g_i^{(m)}\right)} E_{\boldsymbol{\theta}^{(m)}}\left(\exp \left\{-\boldsymbol{C}^{(m)}_i \boldsymbol{\theta}^{(m)}\right\}\right)\right) \\
    & -\frac{1}{2} E_{\boldsymbol{\theta}^{(m)}}\left(\boldsymbol{\theta}^{(m) T} M_{q\left(1/\sigma^{2{(m)}}\right)} \boldsymbol{\theta}^{(m)}\right)+\text{terms not involving }  \boldsymbol{\theta}^{(m)},
    \end{aligned}
    \label{nvmp4}
\end{equation}
where $
	M_{q\left(1/\sigma^{2(m)}\right)} = \operatorname{diag}\left[\frac{1}{\sigma_{\eta}^{2(m)}},\left(u_{q\left(\alpha_1^{(m)}\right)}u_{q\left(1/\sigma_{ 1}^{2{(m)}}\right)}+\frac{1-u_{q\left(\alpha_1^{(m)}\right)}}{ \Gamma_1^2}\right) \boldsymbol{1}_L
    ,\right.\left. \left(u_{q\left(\alpha_2^{(m)}\right)}u_{q\left(1/\sigma_{2}^{2{(m)}}\right)}\right.\right. $ $\left.\left.+\frac{1-u_{q\left(\alpha_2^{(m)}\right)}}{ \Gamma_1^2}\right)\boldsymbol{1}_L,\frac{1}{\sigma_{\psi}^{2(m)}}\boldsymbol{1}_J 
    \right]$, $\boldsymbol{1}_J$  represents a $J$-dimension column vector filled with ones, and $\operatorname{diag}[v]$ denotes a diagonal matrix whose diagonal elements are given by the elements of the vector $v$.

Based on the properties of the multivariate normal distribution, the following specific expression in expectation (\ref{nvmp4}) can be computed by:
\begin{equation}
    \begin{aligned}
E_{\boldsymbol{\theta}^{(m)}}\left(\boldsymbol{\theta}^{(m) T} M_{q\left(1/\sigma^{2{(m)}}\right)} \boldsymbol{\theta}^{(m)}\right)
	&=\boldsymbol{u_{\boldsymbol{\theta}^{(m)}}}^T M_{q\left(1/\sigma^{2{(m)}}\right)} \boldsymbol{u_{\boldsymbol{\theta}^{(m)}}}+tr\left(M_{q\left(1/\sigma^{2{(m)}}\right)}\boldsymbol{\Sigma_{\boldsymbol{\theta}^{(m)}}}\right)\\
    &=tr\left(M_{q\left(1/\sigma^{2{(m)}}\right)}\left(\boldsymbol{u_{\boldsymbol{\theta}^{(m)}}} \boldsymbol{u_{\boldsymbol{\theta}^{(m)}}}^T+\boldsymbol{\Sigma_{\boldsymbol{\theta}^{(m)}}}\right)\right),\\
    E_{\boldsymbol{\theta}^{(m)}}\left(\exp \left\{-\boldsymbol{C}^{(m)}_i \boldsymbol{\theta}^{(m)}\right\}\right)&= \exp \left\{-\boldsymbol{C}^{(m)}_i \boldsymbol{u_{\boldsymbol{\theta}^{(m)}}}+\frac{1}{2} \boldsymbol{C}^{(m)}_i \boldsymbol{\Sigma}_{\boldsymbol{\theta}^{(m)}}\boldsymbol{C}^{(m)\top}_i\right\},\\
    E_{\boldsymbol{\theta}^{(m)}}\left(\exp \left\{-\boldsymbol{C}^{(m)} \boldsymbol{\theta}^{(m)}\right\}\right)&=\exp \left\{-\boldsymbol{C}^{(m)} \boldsymbol{u_{\boldsymbol{\theta}^{(m)}}}+\frac{1}{2} \operatorname{diagonal}\left(\boldsymbol{C}^{(m)} \boldsymbol{\Sigma}_{\boldsymbol{\theta}^{(m)}}\boldsymbol{C}^{(m)\top}\right)\right\},
   \label{nvmp5} 
    \end{aligned}
\end{equation}
where $\operatorname{diagonal}(M)$ represents vector of diagonal entries of the matirx $M$.

Denote $u_{q\left(r^{(m)}\right)}=\left(u_{q\left(r_{1}^{(m)}\right)},...,u_{q\left(r_{n^{(m)}}^{(m)}\right)}\right)^{\top}$ for $m = 1,\cdots,M$, we further have:
\begin{equation}
    \begin{aligned}
        & E_{\boldsymbol{\theta}^{(m)}}\left(\log p\left(\Omega^{(m)}, \boldsymbol{Y}^{(m)}\mid \boldsymbol{X}^{(m)},\boldsymbol{S}^{(m)}\right)\right)\\
        &=-u_{q(\phi^{(m)})}(1-u_{q\left(r^{(m)}\right)})^{\top} \boldsymbol{C}^{(m)} \boldsymbol{u_{\boldsymbol{\theta}^{(m)}}}-\frac{1}{2} \operatorname{tr}\left(M_{q \left( 1/\sigma^{2(m)}\right)}\left(\boldsymbol{u_{\boldsymbol{\theta}^{(m)}}} \boldsymbol{u_{\boldsymbol{\theta}^{(m)}}}^{\top}+\boldsymbol{\Sigma}_{\boldsymbol{\theta}^{(m)}}\right)\right)\\
        &\quad -u_{q\left(\phi^{(m)}\right)}\left(\left(1-u_{q\left(r^{(m)}\right)}\right) \odot u_{q(g^{(m)})}\right)^{\top} E_{\boldsymbol{\theta}^{(m)}}\left(\exp \left\{-\boldsymbol{C}^{(m)} \boldsymbol{\theta}^{(m)}\right\}\right)\\
        &\quad+\text{terms not involving }  \boldsymbol{\theta}^{(m)}. 
        \label{nvmp6}
    \end{aligned}
\end{equation}

Then, we follow the ideas of \cite{knowles2011nonconjugate} and compute the first and second derivatives of the expectation:
\begin{equation}
    \begin{aligned}
    &\left[\mathrm{D}_{\boldsymbol{\mu_{\boldsymbol{\theta}^{(m)}}}} E_{\boldsymbol{\theta}^{(m)}}\left\{\log p\left(\Omega^{(m)}, \boldsymbol{Y}^{(m)}\mid\boldsymbol{X}^{(m)},\boldsymbol{S}^{(m)}\right)\right\}\right]^T \\
    &= u_{q\left(\phi^{(m)}\right)} \boldsymbol{C}^{(m)\top}\left[\left(1-u_{q\left(r^{(m)}\right)}\right)\odot u_{q(g^{(m)})} \odot E_{\boldsymbol{\theta}^{(m)}}\left(\exp \left\{-\boldsymbol{C}^{(m)} \boldsymbol{\theta}^{(m)}\right\}\right)\right.\\
    & \left.\quad-\left(1-u_{q\left(r^{(m)}\right)}\right)\right] -M_{q\left(1/\sigma^{2(m)}\right)} \boldsymbol{u_{\boldsymbol{\theta}^{(m)}}}
    \end{aligned}
    \label{nvmp7}
\end{equation}
and
\begin{equation}
    \begin{aligned}
&\quad\operatorname{vec}^{-1}\left(\left[\mathrm{D}_{\operatorname{vec}\left(\boldsymbol{\Sigma_{\boldsymbol{\theta}^{(m)}}}\right)} E_{\boldsymbol{\theta}^{(m)}}\left\{\log p\left(\Omega^{(m)}, \boldsymbol{Y}^{(m)}\mid \boldsymbol{X}^{(m)},\boldsymbol{S}^{(m)}\right)\right\}\right]^T\right)\\
&=\operatorname{vec}^{-1}\left(  -\frac{1}{2}\operatorname{vec}\left( u_{q\left(\phi^{(m)}\right)} \boldsymbol{C}^{(m)\top} \operatorname{diag}\left[\left(1-u_{q\left(r^{(m)}\right)}\right)\odot u_{q\left(g^{(m)}\right)} \odot E_{\boldsymbol{\theta}^{(m)}}\left(\exp \left\{-\boldsymbol{C}^{(m)} \boldsymbol{\theta}^{(m)}\right\}\right)\right] \right. \right.\\
    &\left.\left. \quad\boldsymbol{C}^{(m)} +  M_{q\left(1/\sigma^{2(m)}\right)}\right) \right) \\
    & =-\frac{1}{2}  u_{q\left(\phi^{(m)}\right)} \boldsymbol{C}^{(m)\top} \operatorname{diag}\left[\left(1-u_{q\left(r^{(m)}\right)}\right)\odot  u_{q\left(g^{(m)}\right)} \odot E_{\boldsymbol{\theta}^{(m)}}\left(\exp \left\{-\boldsymbol{C}^{(m)} \boldsymbol{\theta}^{(m)}\right\}\right)\right] \boldsymbol{C}^{(m)} \\
    & \quad -\frac{1}{2}  M_{q\left(1/\sigma^{2(m)}\right)}.
    \end{aligned}
    \label{nvmp9}
\end{equation}

Hence, we can iterate $\boldsymbol{u_{\boldsymbol{\theta}^{(m)}}}$ and $ \boldsymbol{\Sigma_{\boldsymbol{\theta}^{(m)}}}$ by equation (\ref{compute of nvmp}) , (\ref{nvmp7}), and (\ref{nvmp9}).

\subsection{Computation of ELBO}

Based on above results, we have the ELBO as:
\[ELBO(q)=E_{q\left(\Omega\right)}\left(\log\left(p\left(\Omega,\boldsymbol{Y}\mid \boldsymbol{X}, \boldsymbol{S}\right)\right)\right)-E_{q\left(\Omega\right)}\left(\log\left(q(\Omega\right)\right),\]
where the first part
\begin{equation*}
    \begin{aligned}
        &E_{q\left(\Omega\right)}\left(\log\left(p\left(\Omega,\boldsymbol{Y}\mid \boldsymbol{X}, \boldsymbol{S}\right)\right)\right)\\
        =&E_{q\left(\Omega\right)}\left[ \sum_{m=1}^M\left\{\sum_{i=1}^{N^{m}}\left\{\log p\left(y_{i}^{(m)}|g_{i}^{(m)},r_{i}^{(m)}\right)+\log p\left(g_{i}^{(m)}|\boldsymbol{x}_i^{(m)},\boldsymbol{s}_i^{(m)},\Theta^{(m)} \right)+\log p\left(r_{i}^{(m)} \right)\right\} \right.\right.\\
        &+\sum_{k=1}^2 \left\{\log p\left(\boldsymbol{\beta}_{k}^{(m)}|\sigma^{2{(m)}}_{k},\alpha_{k}^{(m)}  \right)+   \log p\left(\sigma^{2(m)}_{k}|a_{k}^{(m) }  \right)+ \log p\left(\alpha_{k}^{(m)}|u_{k},q_{k}  \right)\right\}+\log p\left(a_{k}^{(m)}\right) \\
        &\left.+\left.\log p\left(\eta^{(m)}\right)+\log p\left(\psi^{(m)}\right)+\log p\left(\phi^{(m)}\right)\right\}
        +\sum_{k=1}^2 \left\{\log p\left(u_{k}|p_{k}  \right)+\log p\left(p_k  \right)+\log p\left( q_k \right)  \right\}\right]
    \end{aligned}
\end{equation*}
and the second part
\begin{equation*}
    \begin{aligned}
        &E_q\left(\log q\left(\Omega\right)\right) \\
        =&E_{q\left(\Omega\right)}\left[ \sum_{m=1}^M\left\{\sum_{i=1}^{N^{m}}\left\{\log \text{Ga}\left(g_{i}^{(m)}|a_i,b_i \right)+\log \text{Bern}\left(r_{i}^{(m)} \mid u_{q(r_{i}^{(m)})} \right)\right\}+\log\mathcal{N}\left(\boldsymbol{\theta}^{(m)}\mid u_{\boldsymbol{\theta}^{(m)}},\boldsymbol{\Sigma}_{\boldsymbol{\theta}^{(m)}}\right)\right.\right.\\
        &+\sum_{k=1}^2 \left\{  \log \text{IG}\left(\sigma^{2(m)}_{k}|\tilde{a},\tilde{b}   \right)+ \log \text{Bern}\left(\alpha_{k}^{(m)}|u_{q(\alpha_{k}^{(m)})}  \right)\right\}+\log \text{IG}\left(a_{k}^{(m)}\mid 1,u_{q\left(1/ \sigma_{k}^{2(m)}\right)}+\frac{1}{A_k^{2}}\right) \\
        &\left.+\left.\log q^*\left(\phi^{(m)}\right) \right\}
        +\sum_{k=1}^2 \left\{\log \text{Bern}\left(u_{k}|u_{q(u_{k})}  \right)+\log \text{Beta}\left(p_k \mid a_p,b_p \right)+\log \text{Beta}\left( q_k \mid a_q,b_q\right)  \right\}\right].
    \end{aligned}
\end{equation*}

\clearpage
\subsection{Algorithm Flowchart}

\begin{algorithm}[ht!]
\setstretch{0.95}
\caption{Variational inference for obtaining the optimal densities $q^*(\Omega)$. }
\KwIn{Data $\boldsymbol{Y}$, Covariates $\boldsymbol{X}$, Location $\boldsymbol{S}$; Initial parameters $\Omega_0$ and set ELBO$=$0.}
\KwOut{Optimal posterior distribution of $q^{*}(u_k)$.}
Initialize parameters $\Omega \leftarrow \Omega_0$ \;
\Repeat{ELBO converges}{
   \For{$m$ in $\{1, 2, \dots, M\}$}{
    Assume that $\boldsymbol{\theta}^{(m)}=\left(\eta^{(m)},\left(\boldsymbol{\beta}_1^{(m)}\right)^{\top},\left(\boldsymbol{\beta}_2^{(m)}\right)^{\top},\left(\boldsymbol{\psi}^{(m)}\right)^{\top}\right)^{\top}$\;
    $\boldsymbol{\Sigma}_{\boldsymbol{\theta}^{(m)}} =\left\{-2 \operatorname{vec}^{-1}\left(\left[\mathrm{D}_{\operatorname{vec}\left(\boldsymbol{\boldsymbol{\Sigma_{\boldsymbol{\theta}^{(m)}}}}\right)} E_{\boldsymbol{\theta}^{(m)}}\left\{\log p\left(\Omega^{(m)}, \boldsymbol{Y}^{(m)}\mid\boldsymbol{X}^{(m)},\boldsymbol{S}^{(m)}\right)\right\}\right]^T\right)\right\}^{-1}$\;
    $\boldsymbol{\mu}_{\boldsymbol{\theta}^{(m)}} =\boldsymbol{\mu}_{\boldsymbol{\theta}^{(m)}}+\boldsymbol{\boldsymbol{\Sigma_{\boldsymbol{\theta}^{(m)}}}}\left[\mathrm{D}_{\boldsymbol{\mu_{\boldsymbol{\theta}^{(m)}}}} E_{\boldsymbol{\theta}^{(m)}}\left\{\log p\left(\Omega^{(m)}, \boldsymbol{Y}^{(m)}\mid \boldsymbol{X}^{(m)},\boldsymbol{S}^{(m)}\right)\right\}\right]^T$\;
    Update $ u_{q\left(\phi^{(m)}\right)}$ using numerical integration\;
    Update $ u_{q\left(a_\phi^{(m)}\right)}$ using current $q\left(a_\phi^{(m)}\right)$\;
    Update $u_{q\left(g^{(m)}_i\right)}$,$u_q\left(\log(g^{(m)}_i)\right)$ using current $q\left(g^{(m)}_i\right)$\;
        \For{$i$ in $\{1, 2,\dots, n^{(m)} \}$}{
            Update $u_{q\left(r_{i}^{(m)}\right)}=\frac{Beta\left(a_{\pi}^{(m)}+1,b_{\pi}^{(m)}\right)}{Beta\left(a_{\pi}^{(m)}+1,b_{\pi}^{(m)}\right)+\exp \left\{-u_{q}\left(g^{(m)}_i\right)\right\}\cdot Beta\left(a_{\pi}^{(m)},b_{\pi}^{(m)}+1\right)}$ 
        }
        \For{$k$ in $\{1, 2\}$}{
            Update $u_q\left(1/\sigma_{k}^{2(m)}\right)$ using current $q\left(\sigma_{k}^{2(m)}\right)$\;
            Update $u_q\left(a_{k}^{(m)}\right)$ using current $q\left(a_{k}^{(m)}\right)$\;
            Update $u_{q\left(\alpha_{k}^{(m)}\right)}$ using current $q\left(\alpha_{k}^{(m)}\right)$\;          
        }
    }
    \For{$k$ in $\{1, 2\}$}{
    Update $u_{q\left(q_{k}\right)}=\frac{u_{q\left(u_{k}\right)}\sum_{m=1}^{M}u_{q}\left(\alpha_{k}^{(m)}\right)+c_q}{u_{q(u_{k})}\sum_{m=1}^{M}u_{q}\left(\alpha_{k}^{(m)}\right)+c_q+u_{q(u_{k})}M+d_q-u_{q(u_{k})}\sum_{m=1}^{M}u_{q}\left(\alpha_{k}^{(m)}\right)}$ \;
    Update $u_{q(p_{k})}=\frac{u_{q(u_{k})}+c_p}{c_p+d_p+1}$\;
    }
    Update $u_{q(u_k)}$ using the current $q(u_k)$\;
}
Return $u_{q(u_k)}$ as the optimal expectation of $q(u_k)$.
\end{algorithm}

\clearpage

\subsection{Basis function selection}
For the basis expansion
$b_{1}^{(m)}\left(s_{i1}^{(m)}\right)=\left(\boldsymbol{\beta}_1^{(m)}\right)^{\top}\boldsymbol{\xi}\left(s_{i1}^{(m)}\right)$ and $b_{2}^{(m)}\left(s_{i2}^{(m)}\right)=\left(\boldsymbol{\beta}_2^{(m)}\right)^{\top}\boldsymbol{\xi}\left(s_{i2}^{(m)}\right)$, we employ B-spline basis functions in our numerical implementation, while noting that alternative basis systems (e.g., Fourier bases) remain viable options. The B-spline construction requires specification of two key parameters: the degree of freedom (df) and spline degree, which are set to be equal in order to reduce the computational cost, while maintaining the flexibility of the base function. To determine the optimal spline degree, we implement the following procedure. First, for each tissue slice, given the candidate degree set $\{1,2,3,4\}$, we fit Model (1) in main text to gene expression data, using the $\boldsymbol{zeroinfl}$ function from the R package $\boldsymbol{pscl}$ for slices with dropout rates $< 0.7$ and the $\boldsymbol{hurdle}$ function from the R package $\boldsymbol{pscl}$ for slices with dropout rates $\geq 0.7$. Then, select the optimal degree minimizing the AIC criterion based on the fitted model object from $\boldsymbol{zeroinfl}$ and $\boldsymbol{hurdle}$. Given our biological assumption that SV genes exhibit consistent expression patterns across adjacent slices with only intensity variations, we adopt the maximum optimal degree identified across all slices.

\subsection{Detailed settings for the hyperparameter set}\label{hyperpara-set}

Concerning the hyperparameter configuration, we initially establish $a_{\pi}^{(m)}=b_{\pi}^{(m)}=1$ to utilize the latent variable ${r_i^{(m)}}$ as a non-informative prior. For the dispersion parameter $\phi^{(m)}$, we set $a_{\phi}^{(m)}=0.001$ and $b_{\phi}^{(m)}=0.001$, resulting in a diffuse distribution with a mean and variance of 1 and 1000, respectively. Additionally, set the variances of the priors for $\eta^{(m)}$ and $\psi_j^{(m)}$ as $\sigma_{\eta}^{2(m)}=1$ and $\sigma_{\psi_j}^{2(m)}=1$. $\Gamma_1$ is assigned a small value of $\sqrt{0.01}$. We set $\Gamma_2$ to 0.01, 0.005, and 0.0001 to represent integration scenarios involving more than four, three, and two or fewer slices, respectively. This scaling balances the contribution of individual slices to $u_k$ under the iterative updating scheme in Equation \eqref{u_iteration}. For cross-donor datasets, additional shrinkage is recommended to account for biological heterogeneity. Specifically, we use 80\%, 60\%, and 50\% of the baseline $\Gamma_2$ value when integrating two, three, and four or more donors, respectively. This adjustment further stabilizes posterior probability estimation within the hierarchical model. In addition, following \cite{tadesse2005wavelet}, we restrict the beta-binomial prior's hyperparameters sum to 2 for a non-informative prior and establish $c_q = d_q = 1$. Moreover, set $c_p=0.2$ and $d_p=1.8$, which yields an expected prior proportion of SV genes being $\frac{c_p}{c_p+d_p}=10\%$. For $A_k$, we set its values to 0.08, 0.05, 0.04, and 0.03 for spline degrees 1, 2, 3, and 4, respectively.

\subsection{BFDR control process}
Given the composite statistic $\tilde{u}_g$, we derive the Bayesian false discovery rate (BFDR) through:
$BFDR(u_{0})=\frac{\Sigma_{g=1}^{G}\left(1-\tilde{u}_g\right)I\left(1-\tilde{u}_g<u_{0}\right)}{\Sigma_{g=1}^{G}I\left(1-\tilde{u}_g<u_{0}\right)}$,
where $u_{0}$ denotes the decision threshold. To enhance false positive control, we establish the significance criterion as $BFDR(u_{0})=\frac{0.05}{2G}$, introducing an adaptive threshold that scales inversely with both gene and location dimensionality.

\subsection{Assessing the utility of $\max\big(E(u_1), E(u_2)\big)$ in SV gene detection}

In our study, we employ $\max\big(E(u_1), E(u_2)\big)$ to jointly account for spatial variation along both dimensions, an approach that is intrinsically linked to the role of the latent indicators $u_k$ in our variational inference framework. Under the ELBO formulation and the iterative optimization procedure, the posterior expectations $E(u_k)$ converge to nearly binary values (close to 0 or 1). Thus, the primary goal of SV gene identification is to reliably distinguish non-SV genes—characterized by $E(u_1) = E(u_2) = 0$—from SV genes, where $E(u_1) = 1$ and/or $E(u_2) = 1$.

We acknowledge that, in addition to $\max\big(E(u_1), E(u_2)\big)$, alternative measures such as $\sqrt{E(u_1)^2 + E(u_2)^2}$ can also effectively achieve this separation. Both metrics return a value of 0 when $E(u_1) = E(u_2) = 0$, and yield a value $\geq 1$ whenever any $E(u_k) = 1$. Consequently, each serves the key objective of detecting whether spatial variation exists along one or both axes. In practice, the two metrics produce highly concordant gene rankings and therefore lead to nearly identical sets of selected SV genes.

To empirically verify this, using simulation data, we compare the SV identification results obtained under the two criteria using the same ranking cutoff (see Tables \ref{tab:criteria-inf1}–\ref{tab:criteria-inf5}). Details regarding the simulation settings and evaluation measures are provided in Section \ref{sec:simu}. The results show that the two criteria identify nearly identical sets of SV genes, achieving almost the same F1 score, TPR, and FPR. This confirms that the choice between the two summary measures has minimal impact on the final identification of SV genes.

Nevertheless, $\max\big(E(u_1), E(u_2)\big)$ is bounded within $[0, 1]$, which aligns more naturally with our BFDR-based cutoff selection procedure, compared with $\sqrt{E(u_1)^2 + E(u_2)^2}$ whose range is $[0, \sqrt{2}]$. The $[0, 1]$ scale of the maximum metric provides a more direct and interpretable basis for false discovery rate control in practice.

\begin{table}[htbp]
\footnotesize
\centering
\caption{Comparison of the performance between two statistics, $\max\big(E(u_1), E(u_2)\big)$ and $\sqrt{E(u_1)^2 + E(u_2)^2}$, under simulation scenarios with a 10\% dropout rate (values are shown as mean (SD)).}
\label{tab:criteria-inf1}
\renewcommand{\tabcolsep}{0.2pc}
\begin{tabular}{lccc ccc}
\toprule
Scenario  & TPR & FPR & F1 &
TPR & FPR& F1 \\
\midrule
&&$\max\big(E(u_1), E(u_2)\big)$ &&&$\sqrt{E(u_1)^2 + E(u_2)^2}$ &\\
\cline{2-4} \cline{5-7}
Linear\\
  1 & 1.000 (0.000) & 0.007 (0.001) & 0.969 (0.005) & 1.000 (0.000) & 0.007 (0.001) & 0.969 (0.005) \\
 2 & 1.000 (0.000) & 0.007 (0.001) & 0.969 (0.006) & 1.000 (0.001) & 0.007 (0.001) & 0.969 (0.006) \\
 3 & 1.000 (0.000) & 0.007 (0.001) & 0.968 (0.005) & 1.000 (0.001) & 0.007 (0.001) & 0.968 (0.005) \\
 4 & 1.000 (0.001) & 0.007 (0.002) & 0.970 (0.007) & 1.000 (0.001) & 0.007 (0.002) & 0.970 (0.007) \\
 \midrule
 Focal\\
  1 & 1.000 (0.000) & 0.007 (0.001) & 0.969 (0.006) & 1.000 (0.001) & 0.007 (0.001) & 0.969 (0.006) \\
  2 & 1.000 (0.001) &0.007 (0.001) & 0.970 (0.006) & 1.000 (0.002) & 0.007 (0.001) & 0.970 (0.006) \\
 3 & 1.000 (0.001) & 0.007 (0.002) & 0.969 (0.006) & 1.000 (0.001) & 0.007 (0.002) & 0.970 (0.006) \\
 4 & 0.856 (0.096) & 0.004 (0.002) & 0.900 (0.054) & 0.882 (0.101) & 0.001 (0.002) & 0.930 (0.057) \\
 \midrule
 Period\\
 1 & 1.000 (0.000) & 0.007 (0.002) & 0.968 (0.008) & 1.000 (0.000) & 0.007 (0.002) & 0.968 (0.008) \\
 2 & 0.999 (0.003) & 0.007 (0.002) & 0.968 (0.007) & 1.000 (0.002) & 0.007 (0.002) & 0.969 (0.007) \\
 3 & 0.998 (0.005) & 0.007 (0.002) & 0.967 (0.008) & 0.998 (0.004) & 0.007 (0.002) & 0.967 (0.008) \\
 4 & 0.968 (0.074) & 0.007 (0.002) & 0.954 (0.040) & 0.974 (0.066) & 0.006 (0.002) & 0.958 (0.032) \\
\bottomrule
\end{tabular}
\label{statis_compare1}
\end{table}

\begin{table}[htbp]
\footnotesize
\centering
\caption{Comparison of the performance between two statistics, $\max\big(E(u_1), E(u_2)\big)$ and $\sqrt{E(u_1)^2 + E(u_2)^2}$, under simulation scenarios with a 30\% dropout rate (values are shown as mean (SD)).}
\label{tab:criteria-inf3}
\renewcommand{\tabcolsep}{0.2pc}
\begin{tabular}{lccc ccc}
\toprule
Scenario  & TPR & FPR & F1 &
TPR & FPR& F1 \\
\midrule
&&$\max\big(E(u_1), E(u_2)\big)$ &&&$\sqrt{E(u_1)^2 + E(u_2)^2}$ &\\
\cline{2-4} \cline{5-7}
Linear\\
1 & 1.000 (0.000) & 0.003 (0.001) & 0.984 (0.004) & 1.000 (0.000) & 0.003 (0.001) & 0.985 (0.004) \\
2 & 1.000 (0.000) & 0.004 (0.001) & 0.984 (0.005) & 1.000 (0.002) & 0.004 (0.001)  & 0.984 (0.005) \\
3 & 1.000 (0.000) & 0.004 (0.001) & 0.983 (0.004) & 1.000 (0.000) & 0.004 (0.001)  & 0.983 (0.004) \\
4 & 0.985 (0.032) & 0.003 (0.001) & 0.979 (0.015) & 0.990 (0.028) & 0.002 (0.002)  & 0.984 (0.012) \\
 \midrule
 Focal\\
1 & 1.000 (0.000) & 0.004 (0.001) & 0.984 (0.004) & 1.000 (0.000) & 0.004 (0.001)  & 0.984 (0.004) \\
2 & 0.998 (0.004) & 0.003 (0.001) & 0.984 (0.004) & 0.999 (0.002) & 0.003 (0.001)   & 0.985 (0.005) \\
3 & 0.995 (0.008) & 0.003 (0.001) & 0.982 (0.005) & 0.997 (0.006) & 0.003 (0.001)   & 0.985 (0.005) \\
4 & 0.630 (0.141) & 0.002 (0.001) & 0.757 (0.105) & 0.644 (0.144) & 0.000 (0.000) & 0.768 (0.112) \\
 \midrule
 Period\\
1 & 0.999 (0.002) & 0.004 (0.002) & 0.983 (0.007) & 1.000 (0.002) & 0.004 (0.002)  & 0.983 (0.007) \\
2 & 0.993 (0.019) & 0.004 (0.001) & 0.980 (0.011) & 0.995 (0.016) & 0.003 (0.001)  & 0.982 (0.009) \\
3 & 0.987 (0.035) & 0.004 (0.002) & 0.976 (0.020) & 0.988 (0.033) & 0.004 (0.002)  & 0.978 (0.018) \\
4 & 0.914 (0.136) & 0.003 (0.002) & 0.931 (0.088) & 0.916 (0.137) & 0.003 (0.002)  & 0.939 (0.085) \\
\bottomrule
\end{tabular}
\label{statis_compare2}
\end{table}

\begin{table}[htbp]
\footnotesize
\centering
\caption{Comparison of the performance between two statistics, $\max\big(E(u_1), E(u_2)\big)$ and $\sqrt{E(u_1)^2 + E(u_2)^2}$, under simulation scenarios with a 50\% dropout rate (values are shown as mean (SD)).}
\label{tab:criteria-inf5}
\renewcommand{\tabcolsep}{0.2pc}
\begin{tabular}{lccc ccc}
\toprule
Scenario  & TPR & FPR & F1 &
TPR & FPR& F1 \\
\midrule
&&$\max\big(E(u_1), E(u_2)\big)$ &&&$\sqrt{E(u_1)^2 + E(u_2)^2}$ &\\
\cline{2-4} \cline{5-7}
Linear\\
1 & 1.000 (0.000) & 0.004 (0.001) & 0.982 (0.004) & 1.000 (0.000) & 0.004 (0.001) & 0.982 (0.004) \\
2 & 1.000 (0.000) & 0.004 (0.001) & 0.982 (0.004) & 1.000 (0.000) & 0.004 (0.001) & 0.982 (0.004) \\
3 & 1.000 (0.000) & 0.004 (0.001) & 0.982 (0.004) & 1.000 (0.000) & 0.004 (0.001) & 0.982 (0.004) \\
4 & 0.858 (0.093) & 0.002 (0.001) & 0.912 (0.055) & 0.874 (0.092) & 0.001 (0.001) & 0.926 (0.052) \\
 \midrule
 Focal\\
1 & 0.997 (0.009) & 0.004 (0.001) & 0.980 (0.006) & 0.998 (0.007) & 0.004 (0.001) & 0.982 (0.005) \\
2 & 0.935 (0.057) & 0.003 (0.001) & 0.951 (0.030) & 0.944 (0.057) & 0.002 (0.001)   & 0.962 (0.030) \\
3 & 0.835 (0.127) & 0.003 (0.001) & 0.891 (0.081) & 0.841 (0.128) & 0.002 (0.001)   & 0.899 (0.082) \\
4 & 0.307 (0.113) & 0.002 (0.001) & 0.451 (0.120) & 0.316 (0.112) & 0.001 (0.001)   & 0.466 (0.120) \\
 \midrule
 Period\\
1 & 0.972 (0.054) & 0.004 (0.001) & 0.969 (0.029) & 0.974 (0.054) & 0.004 (0.001)  & 0.970 (0.029) \\
2 & 0.932 (0.102) & 0.003 (0.001) & 0.946 (0.062) & 0.932 (0.102) & 0.003 (0.001)  & 0.948 (0.061) \\
3 & 0.885 (0.156) & 0.003 (0.001) & 0.917 (0.104) & 0.888 (0.159) & 0.003 (0.001)  & 0.919 (0.105) \\
4 & 0.679 (0.218) & 0.003 (0.001) & 0.766 (0.179) & 0.676 (0.218) & 0.002 (0.001)  & 0.774 (0.181) \\
\bottomrule
\end{tabular}
\label{statis_compare3}
\end{table}

\clearpage

\subsection{Computer time}
\begin{table}[!ht]
\centering
\caption{ Average execution time per gene (in minutes) under different spots numbers ($n$) and dropout rates ($\pi$) when integrating four samples.}
\begin{tabular}{lccccc}
\toprule
 & \multicolumn{5}{c}{\textbf{Dropout Rate ($\pi$)}} \\
\cmidrule(lr){2-6}
\textbf{Sample Size ($n$)} & 0.1 & 0.3 & 0.5 & 0.7 & 0.9 \\
\midrule
625(25 $\times$ 25) & 0.0044 & 0.0037 & 0.0035 & 0.0032 & 0.0029 \\
900(30 $\times$ 30) & 0.0049 & 0.0047 & 0.0041 & 0.0041 & 0.0039 \\
1024(32 $\times$ 32) & 0.0055 & 0.0053 & 0.0044 & 0.0040 & 0.0085 \\
1600(40 $\times$ 40) & 0.0136 & 0.0130 & 0.0123 & 0.0111 & 0.0098 \\
2500(50 $\times$ 50) & 0.0286 & 0.0241 & 0.0240 & 0.0226 & 0.0208 \\
3600(60 $\times$ 60) & 0.0450 & 0.0427 & 0.0396 & 0.0349 & 0.0300 \\
6400(80 $\times$ 80) & 0.0684 & 0.0662 & 0.0645 & 0.0555 & 0.0451 \\
10000(100 $\times$ 100) & 0.1032 &0.0970 &0.0924 & 0.0827 & 0.0695 \\
14400(120 $\times$ 120) & 0.1606 & 0.1315 & 0.1310 & 0.1237 & 0.1058 \\
22500(150 $\times$ 150) & 0.2545 & 0.2012 & 0.1975& 0.1753 & 0.1521 \\
40000(200 $\times$ 200) &0.5092 & 0.4983  & 0.4347 & 0.4171 & 0.3801 \\
\bottomrule
\end{tabular}
\label{tab:performance_comparison}
\end{table}

\section{Implementation details of the competing methods}

$~~$ (a) SPARK, a Poisson-Gaussian process hybrid model: The source code for SPARK is publicly available on \url{https://github.com/xzhoulab/SPARK}. Following the instructions, we filter genes that are expressed in less than 10\% spots and spots with total observations less than 10. The SPARK Object is first constructed by employing the function \textbf{CreateSPARKObject()} and \textbf{spark.vc()}, and the function \textbf{spark.test()} is subsequently employed for SV gene detection. Finally, we identify the SV genes with adjusted p-value less than 0.05.

(b) SPARKX, a scalable non-parametric test based on robust covariance analysis: The source code for SPARKX is publicly available on \url{https://github. com/xzhoulab/SPARK}. SV gene detection is directly implemented with the function \textbf{sparkx()} and choose the default value \textbf{option=``mixture"} for multiple kernel testing. To account for the impact of covariates, we assign the covariates of each spot to the argument $X$. Finally, we identify the SV genes with adjusted p-value from \textbf{res\_mtest} less than 0.05.

(c) HEARTSVG, a distribution-free approach that identifies SV genes through systematic exclusion of non-SV candidates: The source code for HEARTSVG is publicly available on \url{https://github.com/cz0316/HEARTSVG}. Detection of SV genes is conducted through the main function \textbf{heartsvg()} with the augment scale as recommended. Finally, we identify the SV genes with adjusted p-value less than 0.05.

(d) nnSVG, leveraging nearest-neighbor Gaussian processes for spatial covariance parameter estimation: The source code for nnSVG is publicly available on 
\url{https://bioconductor.org/packages/nnSVG}. We first filter the genes by function \textbf{filter\_genes()} and transform  \textbf{SpatialExperimentObject} with \textbf{computeLibraryFactors()} and \textbf{logNormCounts()}. To account for the impact of covariates, we assign the covariates of each spot to the argument $X$ and apply the main function \textbf{nnSVG()}. Finally, we identify the SV genes with adjusted p-value less than 0.05.

(e) spVC, a generalized Poisson model incorporating spatially varying coefficients for cell/spot-level covariates: The source code for spVC is publicly available on 
\url{https://github.com/shanyu-stat/spVC}. Following the instructions, we use the \textbf{identify()} function to identify the boundary of the spots and construct the triangle mesh. Then we filter genes that are expressed in less than 100 spots and apply the main function \textbf{test.spVC}. Finally, we identify the SV genes with adjusted p-value from \textbf{results.constant} less than 0.05.

(f) StarTrail: a spatial gradient–based method designed to identify critical biological boundaries and SV genes. The source code for StarTrail is  publicly available on \url{https://github.com/JiawenChenn/StarTrail}. Following the instructions, we first define boundaries based on the covariates using the elpigraph package and used \textbf{fit\_NNGP()} to estimate each gene’s spatial gradient. Then, we identify cliff genes that exhibit significant changes across at least one boundary, exceeding the predefined threshold in the study.

(g) GASTON: an unsupervised deep neural network method that identifies spatial domains and SV genes by learning the isodepth structure of tissue slices. The source code for GASTON is publicly available on 
\url{https://github.com/raphael-group/GASTON}. We first train the neural network until convergence and select the optimal model to obtain the tissue isodepth. We then perform a piecewise linear fit for each gene, identifying continuous and discontinuous genes, and take their union as the set of SV genes.

(h) DESpace, which directly detects spatially consistent SV patterns across multiple samples: The source code for DESpace is publicly available on 
\url{https://bioconductor.org/packages/DESpace}. As a two-step framework based on spatial clustering, we follow the simulation setting and the manually annotated or PRECAST processed \citep{liu2023precast} clusters in real datasets to retain the augment spatial cluster. Then we apply the main function \textbf{DESpace\_test()} and get the p-value in \textbf{gene\_results}. Finally, we identify the SV genes with p-value less than 0.05.

(i) PASTE, which constructs a composite sample from multiple samples through probabilistic point-to-point alignment for subsequent single-sample SV analysis: The source code for PASTE is publicly available on 
\url{https://github.com/raphael-group/paste}. Following the instructions, we filter the genes and spots by default and randomly select one slice as the \textbf{initial\_slice} which is the central layer of the mapping in the algorithm. We then compute $\lambda$ by default and integrate the four slices selected by function \textbf{pst.center\_align()}. Finally, we keep one decimal place for the results of the integrated dataset.

\clearpage

\section{Detailed simulation settings and results}\label{sec:simu}

\begin{figure}[htbp]
    \centering
    \includegraphics[width=0.5\linewidth]{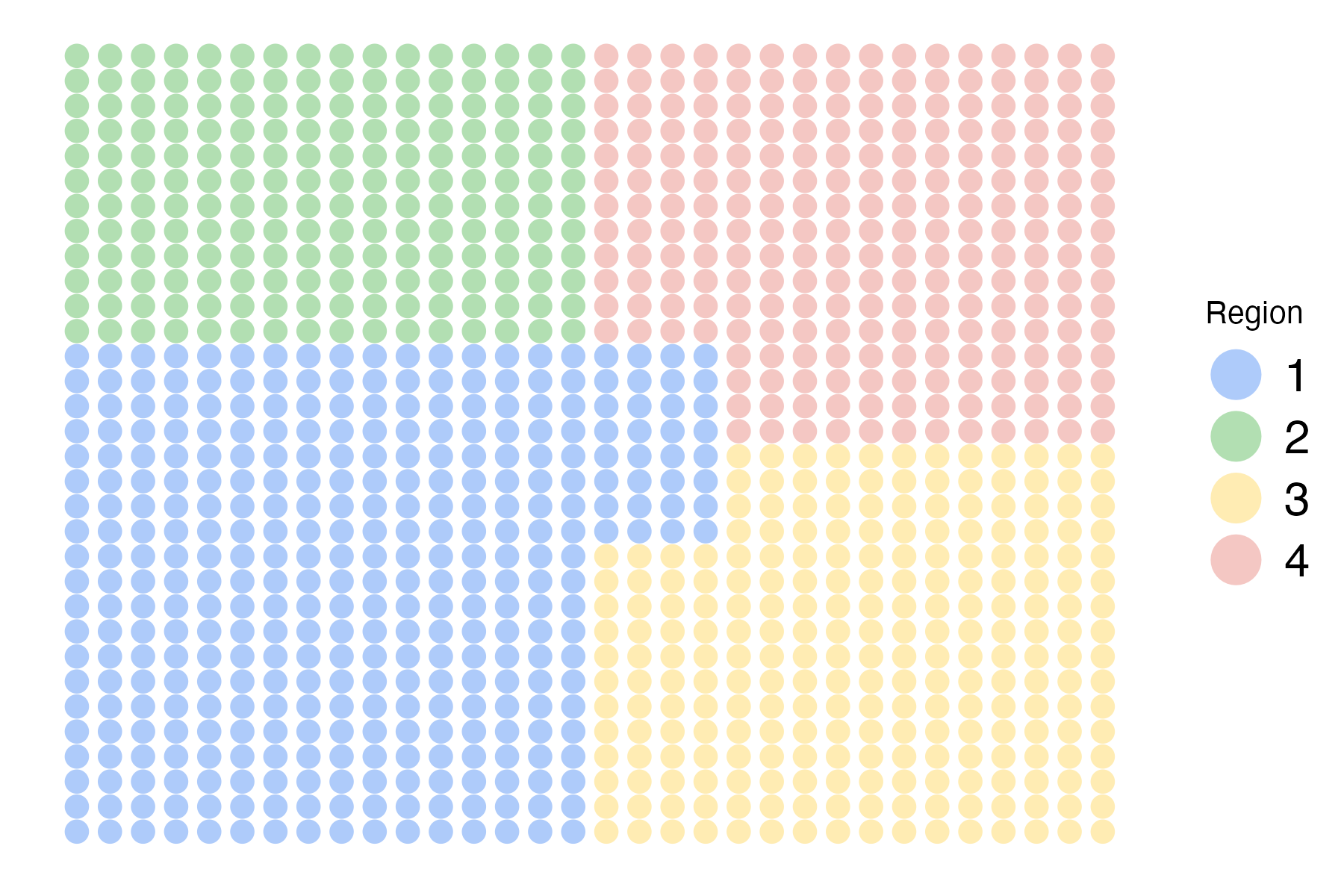}
    \caption{Partitioned spot regions considered in the basic simulations. Regions are denoted by different colors, where the cellular compositions $x_i^{(m)}$ are independently drawn from Dirichlet distributions \textbf{Dirc(1, 1, 1, 1, 1, 1)} (Region 1), \textbf{Dirc(1, 3, 5, 7, 9, 11)} (Region 2), \textbf{Dirc(14, 12, 10, 8, 6, 4)} (Region 3), and \textbf{Dirc(1, 4, 4, 4, 4, 1)} (Region 4).}
    \label{fig:spot_region}
\end{figure}

We establish the following simulation framework to emulate prevalent sequencing-based spatial molecular profiling technologies. First, we simulate $M=4$ slide samples, each containing $n = 1,024$ spots arranged in a $32$ by $32$ square lattice partitioned into four distinct regions. Each spot comprises a cellular ensemble of six cell types, with region-specific compositions sampled from Dirichlet distributions: $Dirc(1, 1, 1, 1, 1, 1)$ (Region 1), $Dirc(1, 3, 5, 7, 9, 11)$ (Region 2), $Dirc(14, 12, 10, 8, 6, 4)$ (Region 3), and $Dirc(1, 4, 4, 4, 4, 1)$ (Region 4). The spatial organization is visually summarized in Figure \ref{fig:spot_region}. Second, we model $G=5,000$ genes with 500 SV genes and 4,500 non-SV genes. Raw count data are generated through model (1). For the SV genes, three types of spatial patterns are considered, including linear with $b_{k}\left(s_{ik}\right)=\beta_0\times s_{ik}$, focal with $\beta_0\exp(-s_{ik}^2)/2$, and periodic with $\beta_0 \cos\left(2 \pi s_{ik}\right)$, which have also been usually examined in the literature. For each pattern, signal strength is parameterized with four levels (high/medium/low/extremely low), with $\beta_0$ being 0.8, 0.5, 0.2, and 0.05 for linear, 0.6, 0.4, 0.2, and 0.05 for focal, and 0.8, 0.6, 0.4, and 0.2 for periodic. To capture cross-sample heterogeneity, we implement four signal configurations across slides: Setting 1: (high, middle, middle, and middle), Setting 2: (high, middle, low, low), Setting 3: (middle, middle, low, low), and Setting 4 : (middle, low, extremely low, extremely low). Non-SV genes are modeled with $b_{k}\left(s_{ik}\right)=0$. Third, the 6-dimensional cell type ratio vector serves as spot-specific covariates $\boldsymbol{x}_i^{(m)}$. The covariate effects $\psi_j^{(m)}$'s and baseline parameter $\eta^{(m)}$ are simulated independently from $N(0, 1)$ and $N(2, 0.5^2)$, respectively. In addition, set $\phi^{(m)} = 15$. Fourth, three levels of dropout rate are examined, with $\pi^{(m)}=0.1$ (low), 0.3 (medium) and 0.5 (high). This comprehensive design spans 36 experimental scenarios, systematically probing diverse spatial patterns, effect magnitudes, cross-sample heterogeneity levels, and technical noise profiles.

\begin{figure}[!ht]
    \centering
    \includegraphics[width=1\linewidth]{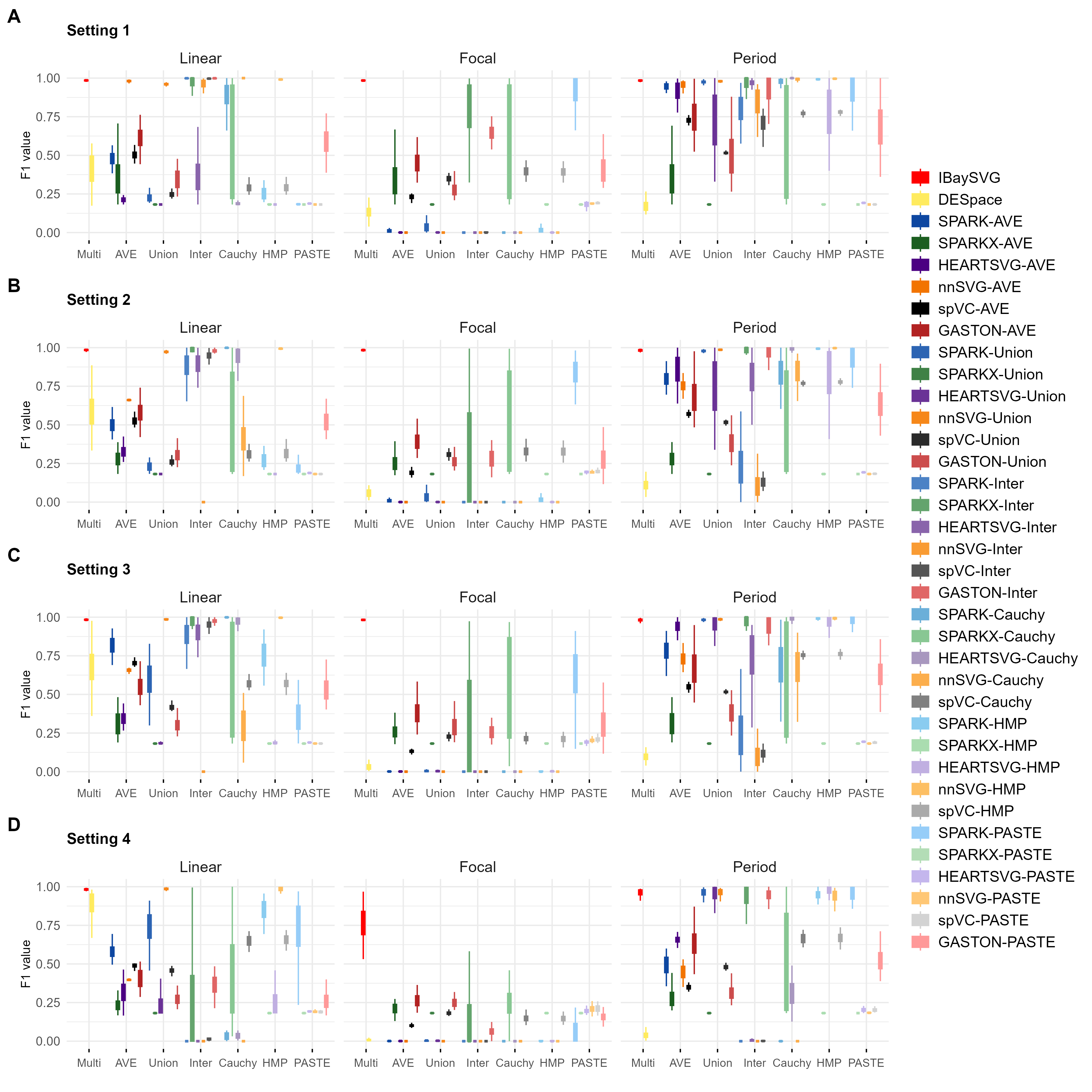}
    \caption{Boxplots of the F1 scores over 50 replicates with different methods under the scenarios with the medium dropout rate. (A) Setting 1, (B) Setting 2, (C) Setting 3, and (D) Setting 4. In each subfigure, three types of spatial patterns (linear, focal, and periodic) are examined.}
    \label{simu_F1_inf3}
\end{figure}

\begin{figure}[!ht]
    \centering
    \includegraphics[width=1\linewidth]{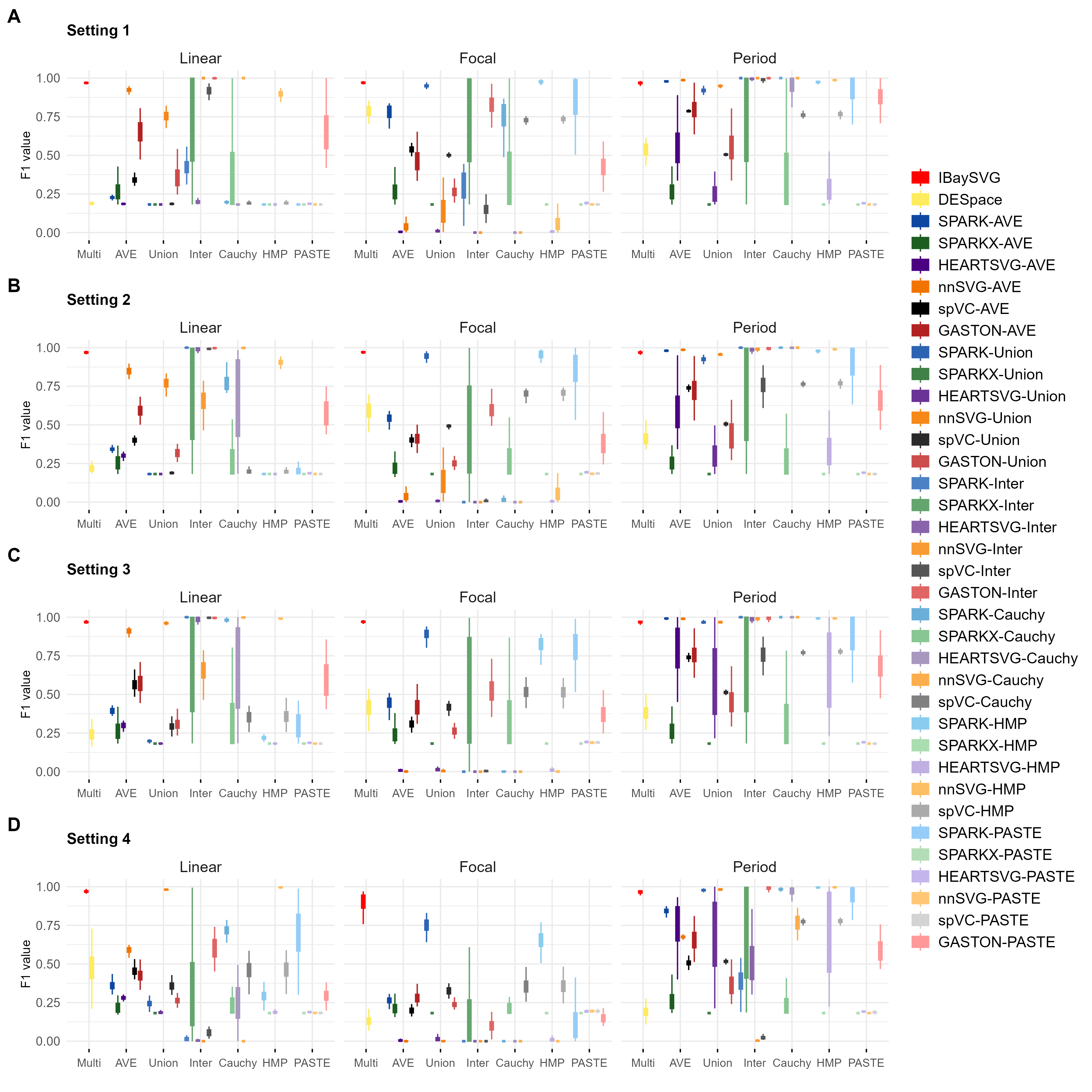}
    \caption{Boxplots of the F1 scores over 50 replicates with different methods under the scenarios with the low dropout rate. (A) Setting 1, (B) Setting 2, (C) Setting 3, and (D) Setting 4. In each subfigure, three types of spatial patterns (linear, focal, and periodic) are examined.}
    \label{simu_F1_inf1}
\end{figure}

\begin{figure}[!ht]
    \centering
    \includegraphics[width=1\linewidth]{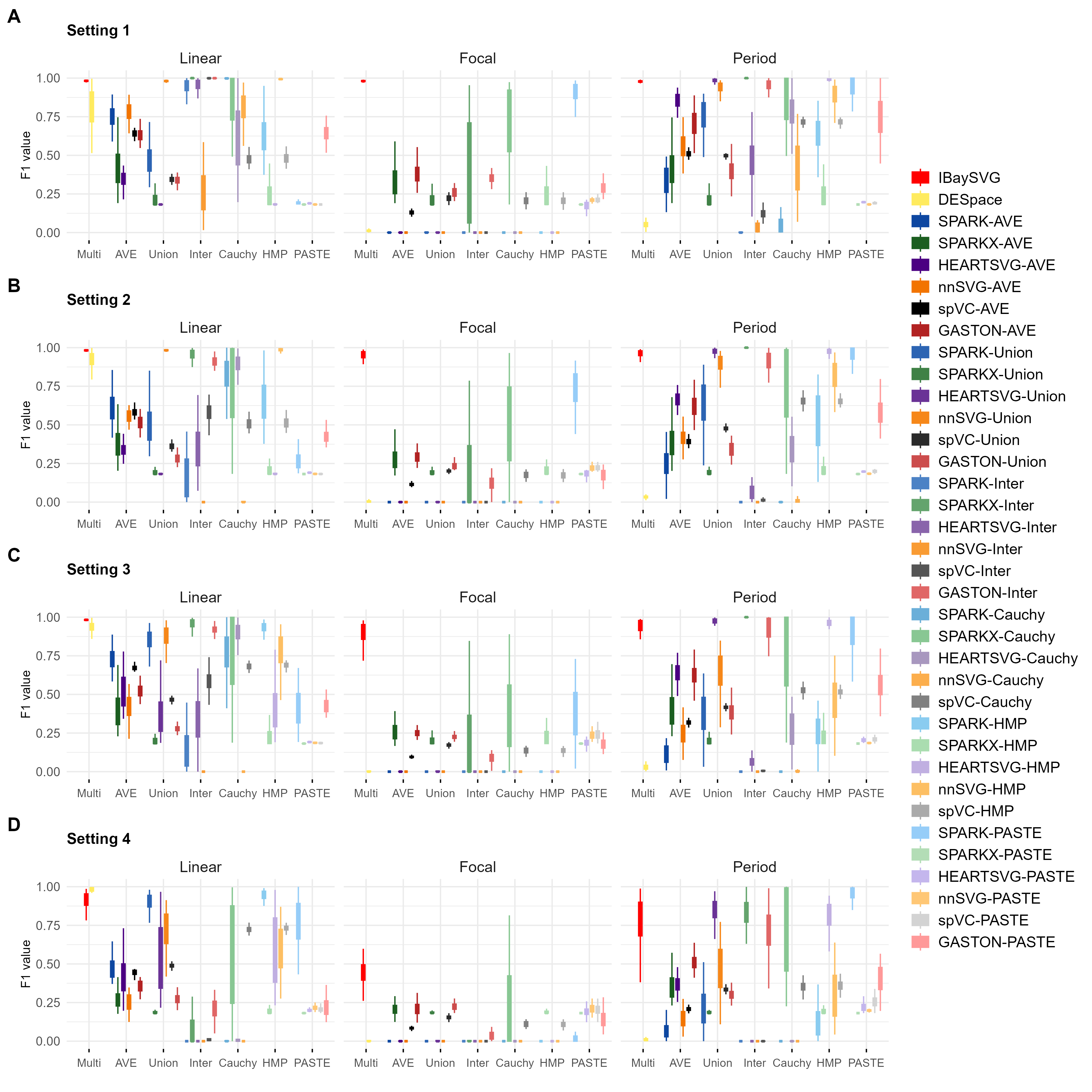}
    \caption{Boxplots of the F1 scores over 50 replicates with different methods under the scenarios with the high dropout rate. (A) Setting 1, (B) Setting 2, (C) Setting 3, and (D) Setting 4. In each subfigure, three types of spatial patterns (linear, focal, and periodic) are examined.}
    \label{simu_F1_inf5}
\end{figure}

\begin{figure}[!ht]
    \centering
    \includegraphics[width=1\linewidth]{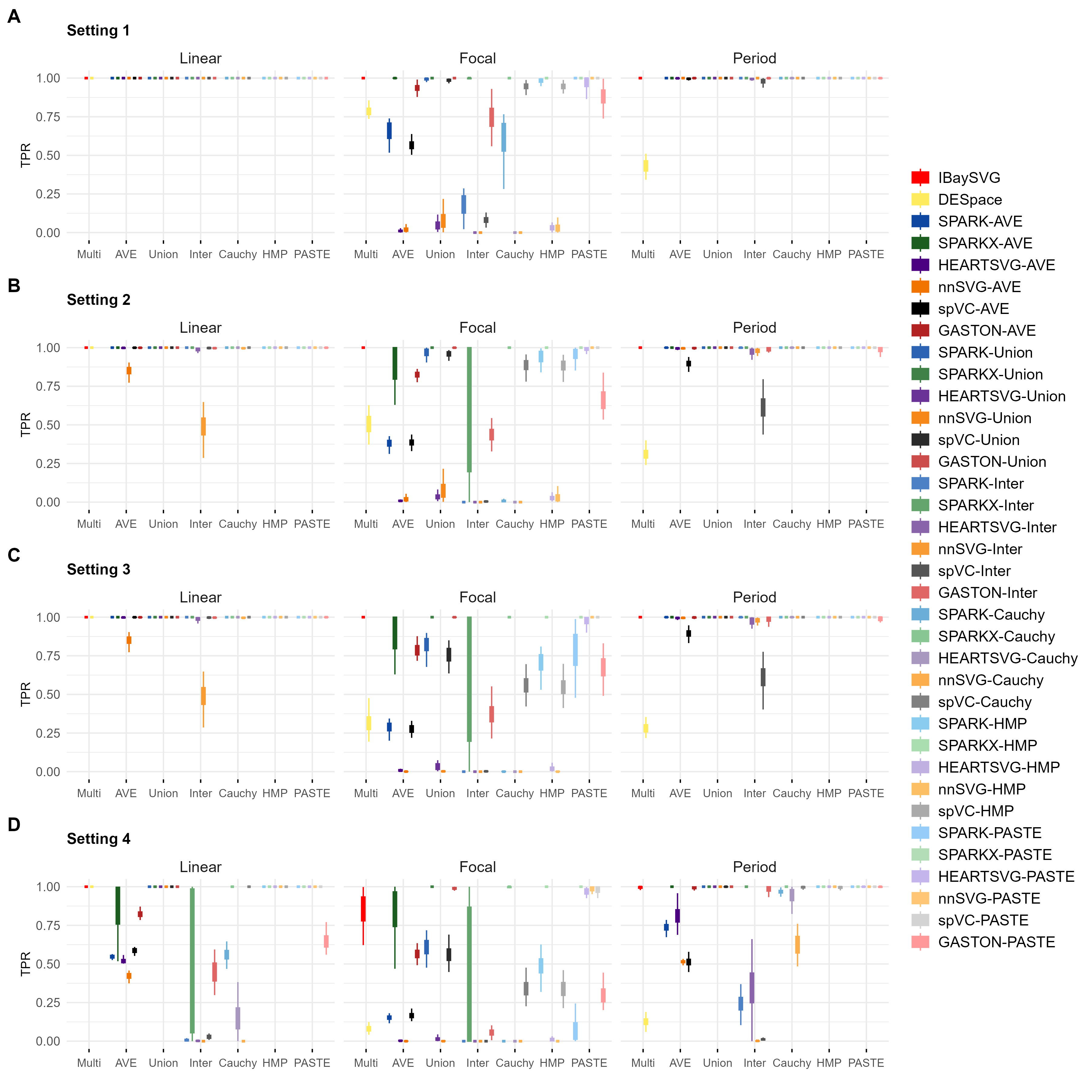}
    \caption{Boxplots of the TPR values over 50 replicates with different methods under the scenarios with the low dropout rate. (A) Setting 1, (B) Setting 2, (C) Setting 3, and (D) Setting 4. In each subfigure, three types of spatial patterns (linear, focal, and periodic) are examined.}
    \label{simu_TPR_inf1}
\end{figure}

\begin{figure}[!ht]
    \centering
    \includegraphics[width=1\linewidth]{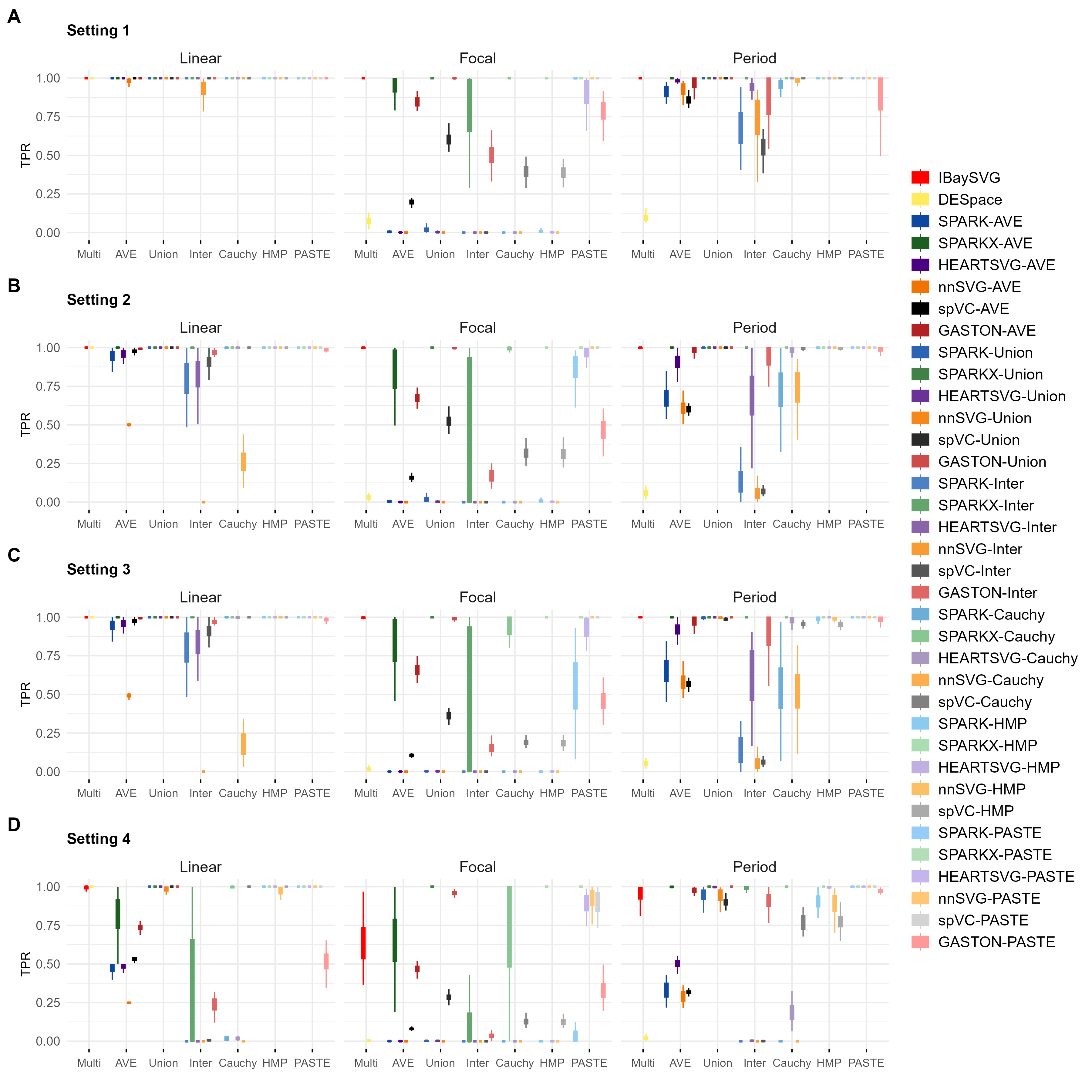}
     \caption{Boxplots of the TPR values over 50 replicates with different methods under the scenarios with the medium dropout rate. (A) Setting 1, (B) Setting 2, (C) Setting 3, and (D) Setting 4. In each subfigure, three types of spatial patterns (linear, focal, and periodic) are examined.}
    \label{simu_TPR_inf3}
\end{figure}

\begin{figure}[!ht]
    \centering
    \includegraphics[width=1\linewidth]{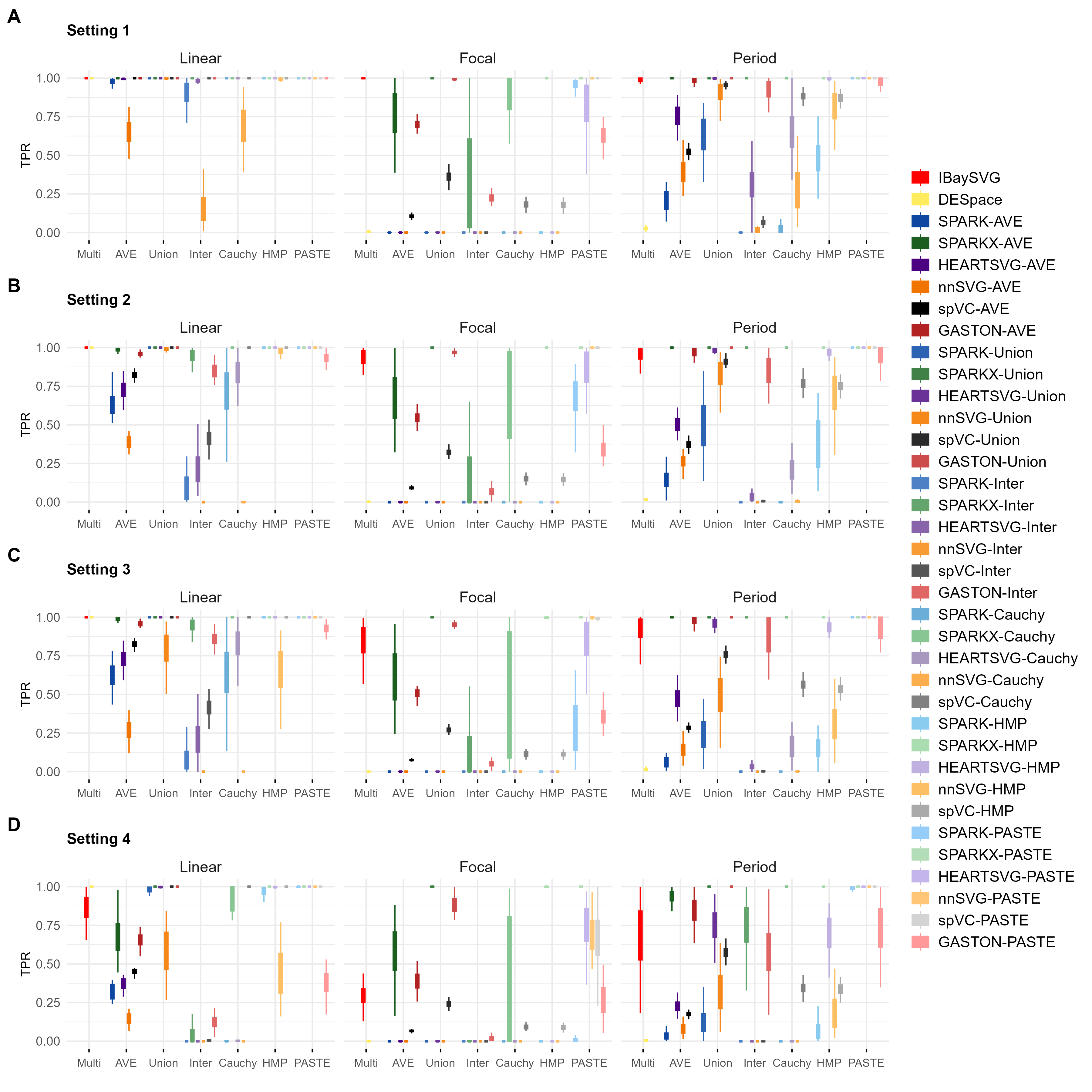}
    \caption{Boxplots of the TPR values over 50 replicates with different methods under the scenarios with the high dropout rate. (A) Setting 1, (B) Setting 2, (C) Setting 3, and (D) Setting 4. In each subfigure, three types of spatial patterns (linear, focal, and periodic) are examined.}
    \label{simu_TPR_inf5}
\end{figure}

\begin{figure}[!ht]
    \centering
    \includegraphics[width=1\linewidth]{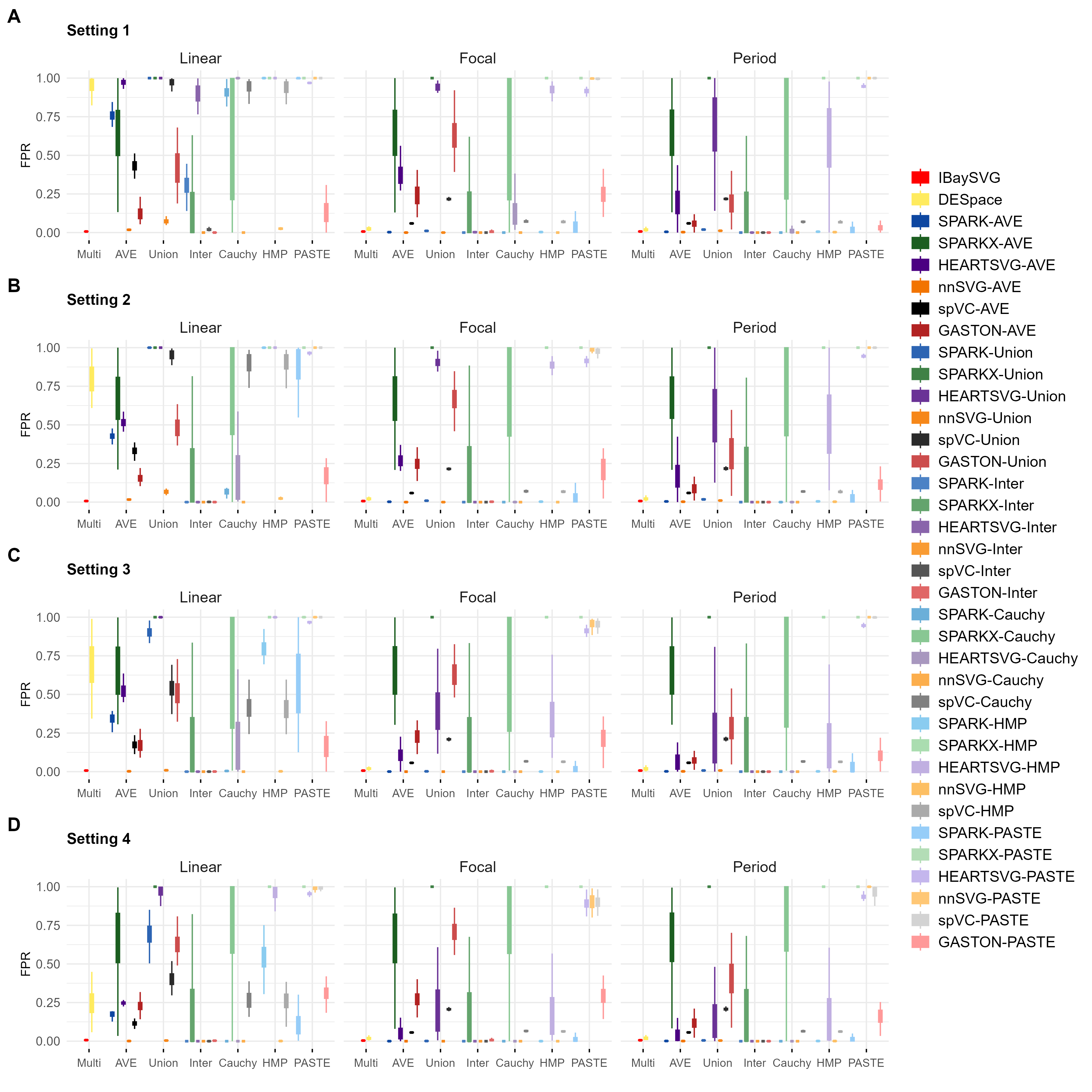}
    \caption{Boxplots of the FPR values over 50 replicates with different methods under the scenarios with the low dropout rate. (A) Setting 1, (B) Setting 2, (C) Setting 3, and (D) Setting 4. In each subfigure, three types of spatial patterns (linear, focal, and periodic) are examined.}
    \label{simu_FPR_inf1}
\end{figure}

\begin{figure}[!ht]
    \centering
    \includegraphics[width=1\linewidth]{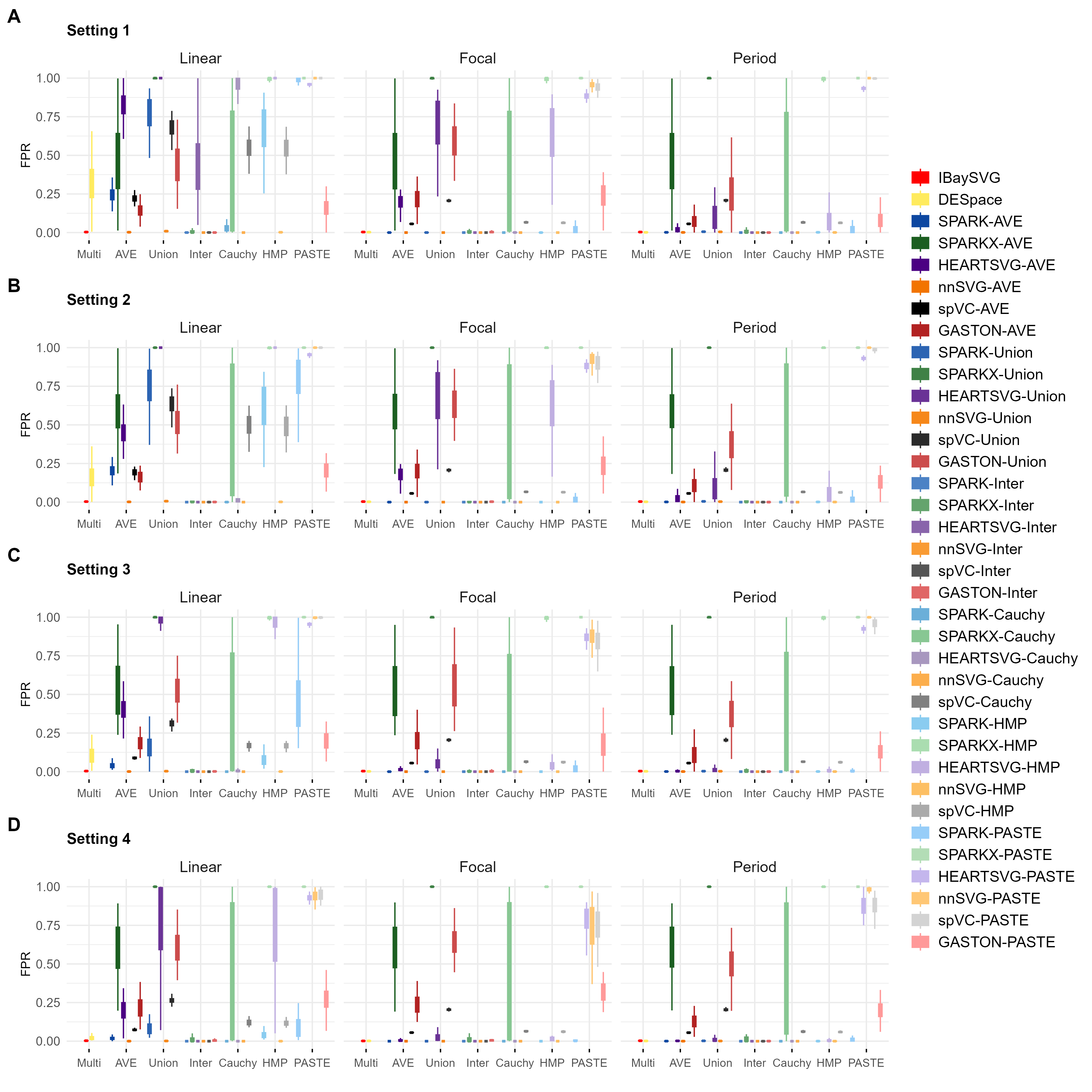}
     \caption{Boxplots of the FPR values over 50 replicates with different methods under the scenarios with the medium dropout rate. (A) Setting 1, (B) Setting 2, (C) Setting 3, and (D) Setting 4. In each subfigure, three types of spatial patterns (linear, focal, and periodic) are examined.}
    \label{simu_FPR_in3}
\end{figure}

\begin{figure}[!ht]
    \centering
    \includegraphics[width=1\linewidth]{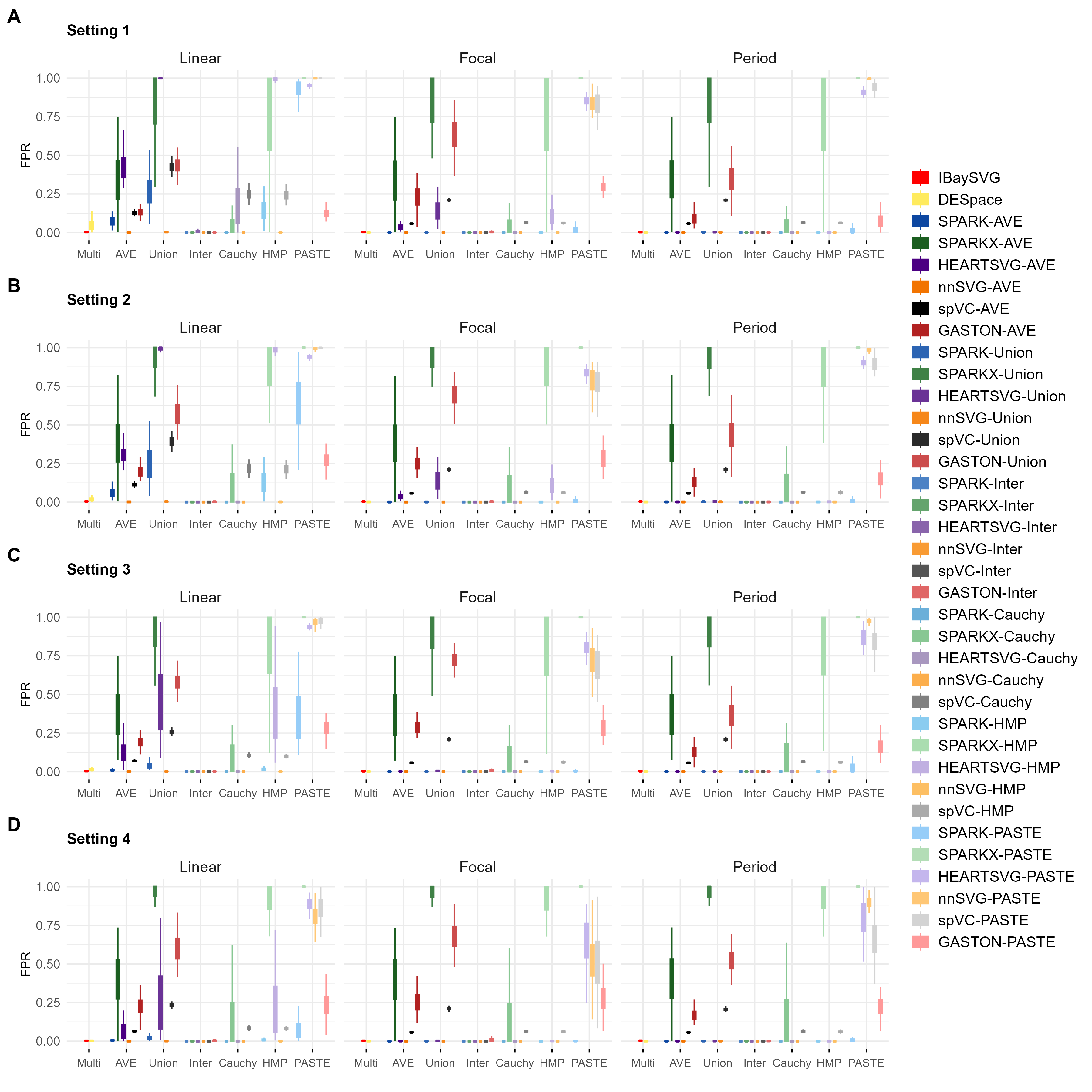}
    \caption{Boxplots of the FPR values over 50 replicates with different methods under the scenarios with the high dropout rate. (A) Setting 1, (B) Setting 2, (C) Setting 3, and (D) Setting 4. In each subfigure, three types of spatial patterns (linear, focal, and periodic) are examined.}
    \label{simu_FPR_inf5}
\end{figure}

We employ three principal metrics to quantify SV gene detection accuracy: $TPR= \frac{TP}{TP+FN}$, $FPR= \frac{FP}{FP+TN}$, and $F1=\frac{2 \cdot TP}{2 \cdot TP+FP+FN}$, where $TP$, $FP$, $TN$, and $FN$ represent true positive, false positive, true negative, and false negative identifications, respectively. Beyond the proposed IBaySVG, six single-sample alternative methods with four types of late integration and two integrated methods are also examined in the simulation. Here, as the deep learning-based approach StarTrail is highly computationally time-consuming, we exclude it from our large-scale simulation study. For the six single-sample methods, in addition to the integration strategies described in Section 3.3 (main text), we also compute the average of each metric across the four samples in each scenario (denoted as AVE). The corresponding results are summarized using boxplots in Figures \ref{simu_F1_inf3}-\ref{simu_FPR_inf5}.

IBaySVG achieves consistently higher F1 scores and demonstrates greater robustness across simulations. Compared with nonparametric approaches (SPARKX, HEARTSVG), IBaySVG provides better false-positive control and greater stability. Although parametric methods (spVC, nnSVG, SPARK) exhibit stable performance, they are consistently outperformed by our framework, demonstrating the effectiveness of our nonparametric integrative approach.

Specifically, under linear patterns with simple structures, single-sample methods (except nnSVG) maintain adequate TPR but suffer from poor FPR control, reducing F1 scores. nnSVG's TPR declines with higher dropout rates or weak signals (Setting 4), while SPARK better controls false positives. Union integration increases false positives across SPARK, SPARKX, HEARTSVG, GASTON, and spVC, while intersection strategies improve specificity at the cost of sensitivity in low-signal samples. Notably, nnSVG responds oppositely: union boosts TPR and F1 but intersection degrades performance. For focal patterns, all conventional methods show limited effectiveness regardless of strategy, while our approach maintains robust performance. Periodic patterns reveal moderate success for SPARK, HEARTSVG, GASTON, and nnSVG under ideal conditions (low dropout, strong signals), with intersection occasionally helping. However, all alternative methods deteriorate markedly from Settings 1 to 4, and union strategies offer minimal gains, especially with high dropout.

The HMP post integration method is generally more effective than the Cauchy combination test in improving the TPR. In contrast, the Cauchy method demonstrates better control over false positives compared to both the union strategy and HMP. Regarding the F1-score performance, the Cauchy method enhances the performance of SPARK, HEARTSVG, and SPARKX under strong signal settings, but may reduce the detection efficiency of nnSVG and spVC. The HMP method, on the other hand, improves the overall detection performance of nnSVG and SPARK, and provides additional gains for HEARTSVG and spVC in low-signal scenarios under the periodic spatial pattern. However, both strategies exhibit very limited improvements in the focal pattern, where the overall detection power is inherently low. Moreover, as both methods are compromise strategies between union and intersect strategies, their performance gains remain marginal.

The PASTE method increases true positives but concurrently elevates false positives by amplifying both SV gene signals and non-SV noise. For SPARKX, HEARTSVG, nnSVG, and spVC, this integration reduces F1 scores below individual analysis baselines in linear/periodic patterns, underscoring concerns about PASTE's compatibility with SV gene detection pipelines. SPARK partially benefits from PASTE under high dropout or weak signal conditions, while GASTON benefits primarily under strong-signal conditions. In focal patterns, PASTE marginally improves detection for methods with minimal baseline efficacy, yet remains substantially inferior to the proposed framework. DESpace performs adequately in linear patterns with high dropout rates but fails in complex nonlinear (focal/periodic) patterns due to its inherent linear spatial effect assumption.

\clearpage
\subsection{Examination on additional spatial structures}
\begin{figure}[!ht]
    \centering
    \includegraphics[width=0.8\linewidth]{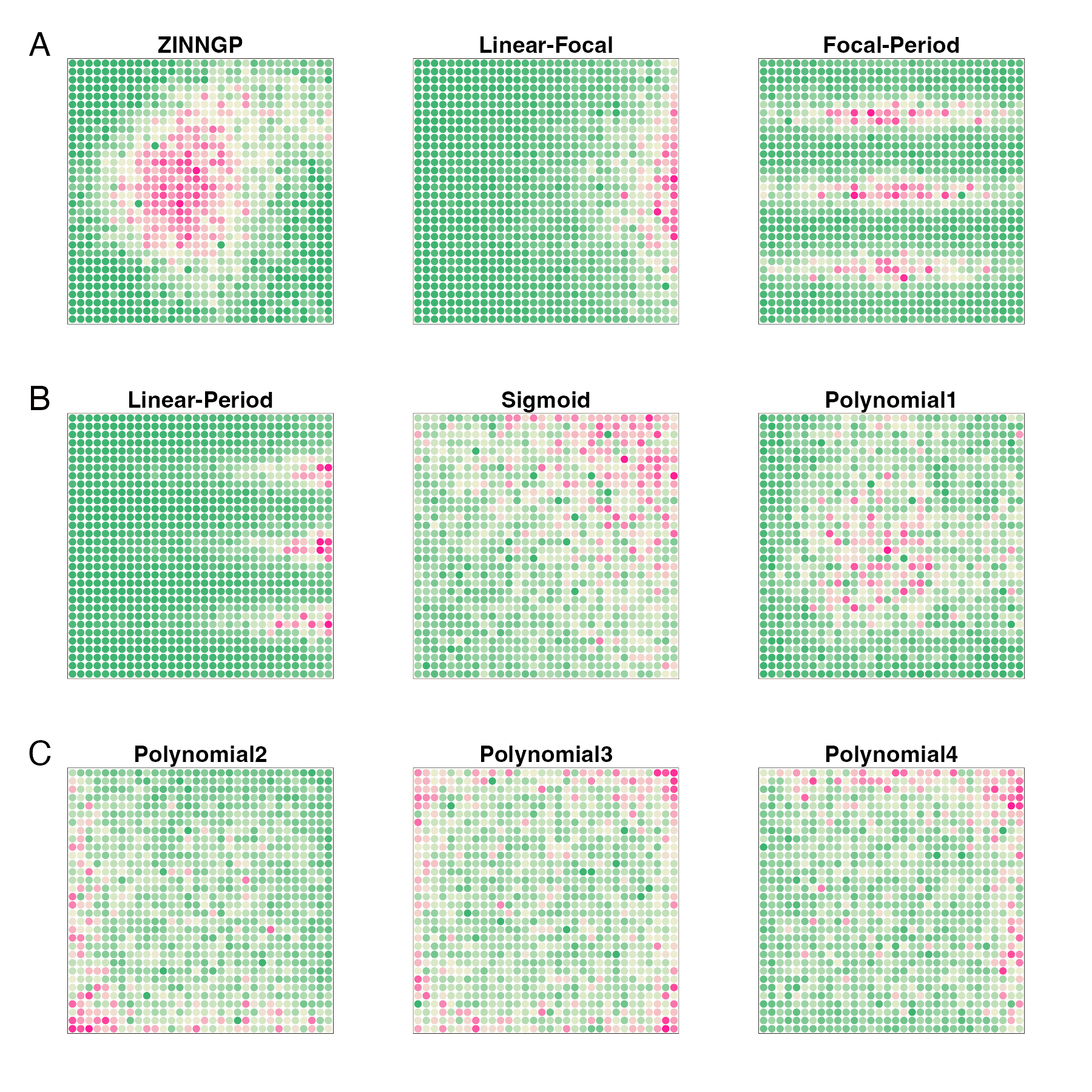} 
    \caption{Examples of spatial expressions under additional spatial structures. To improve representation, the dropout rate and effects of covariates are set to be zero.}
    \label{extend-spatial-pattern}
\end{figure}

\begin{figure}[!ht]
    \centering
    \includegraphics[width=\linewidth]{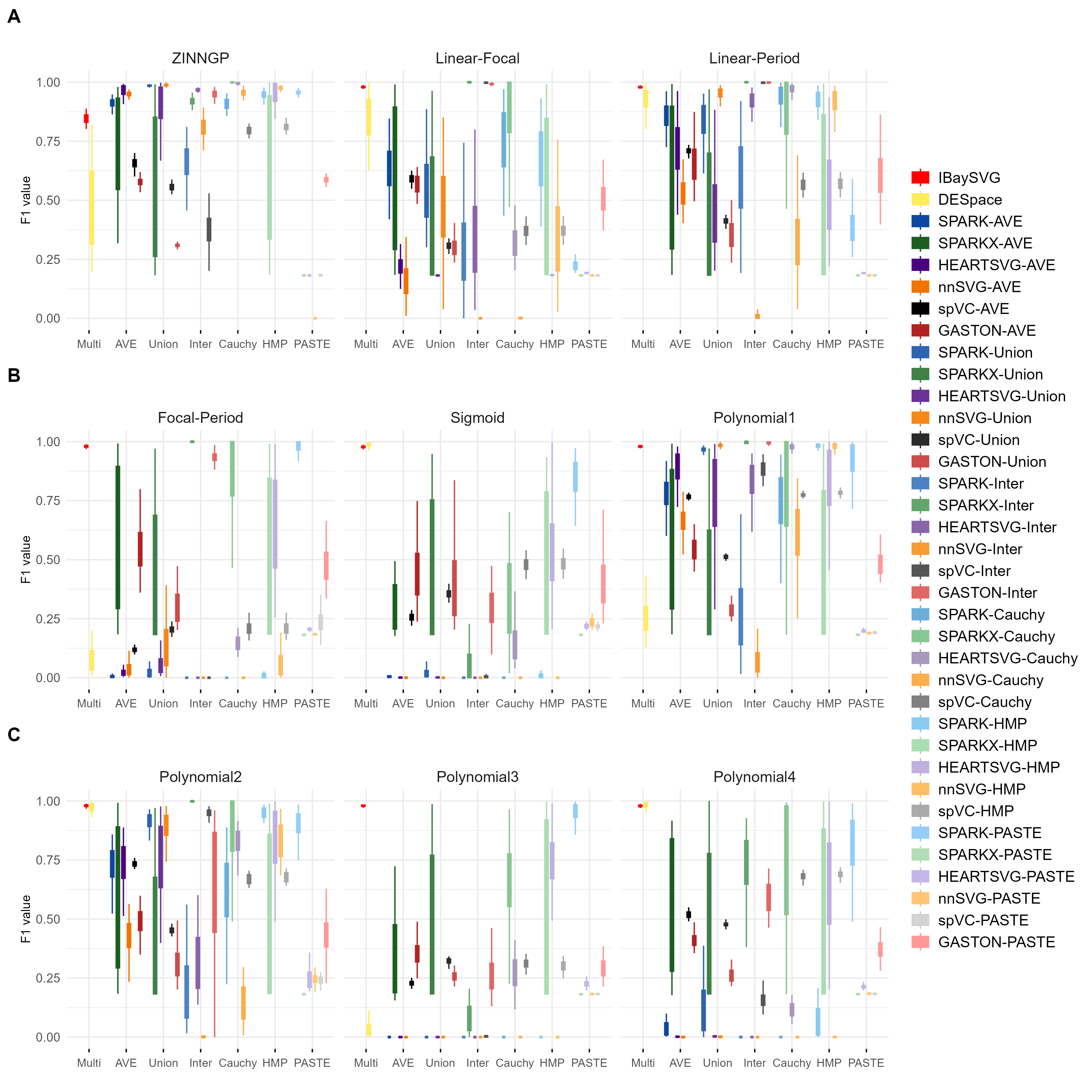}
    \caption{Comparative boxplots of F1-score distributions derived from 50 replicates across competing methodologies under enhanced spatial complexity. This evaluation employs Setting 1 of the cross-slide signal configuration with a medium dropout probability of 0.5, incorporating three distinct spatial architecture paradigms.}
    \label{supple_simu_F1}
\end{figure}

\begin{figure}[!ht]
    \centering
    \includegraphics[width=\linewidth]{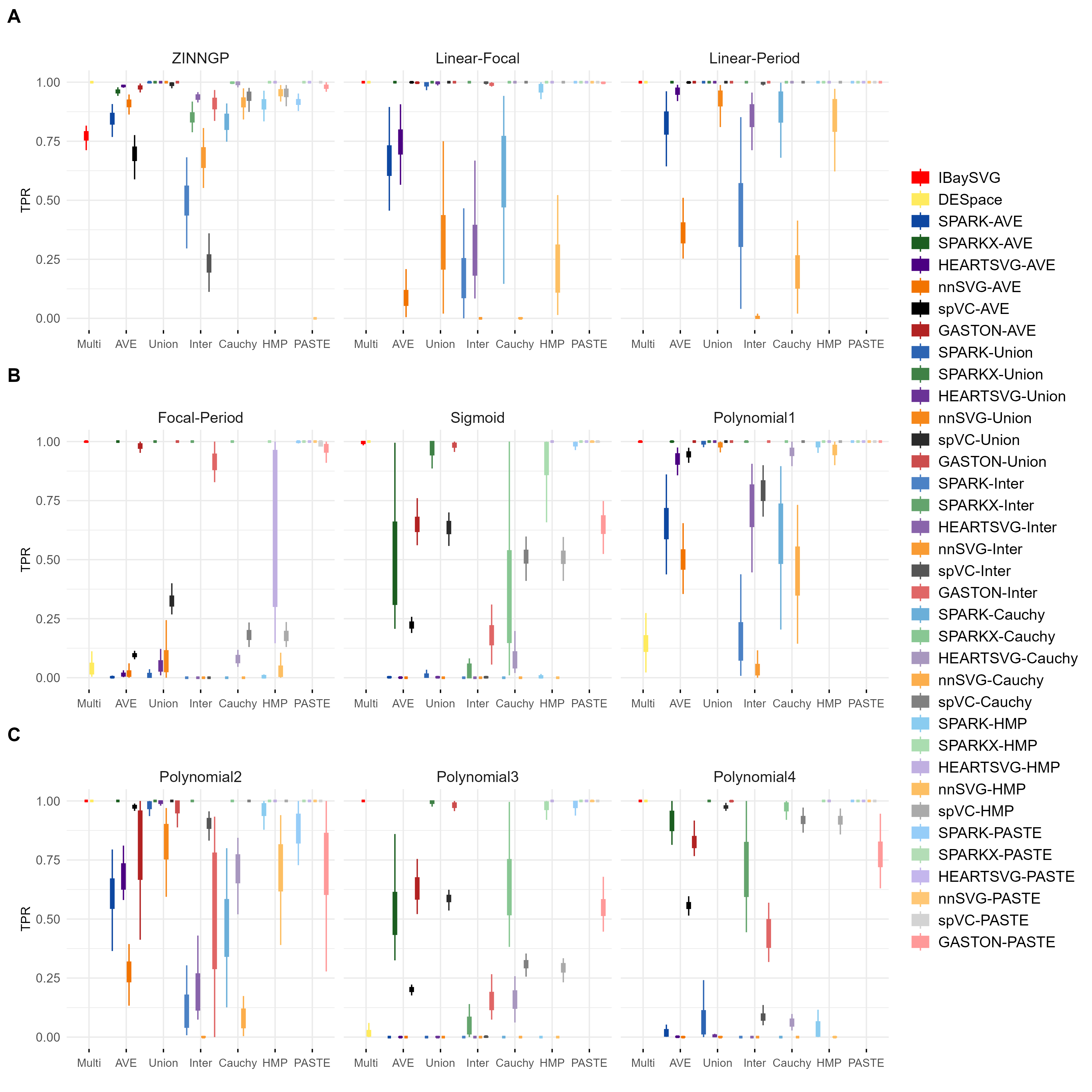}
    \caption{Comparative boxplots of TPR distributions derived from 50 replicates across competing methodologies under enhanced spatial complexity. This evaluation employs Setting 1 of the cross-slide signal configuration with a medium dropout probability of 0.5, incorporating three distinct spatial architecture paradigms.}
    \label{supple_simu_recall}
\end{figure}

\begin{figure}[!ht]
    \centering
    \includegraphics[width=\linewidth]{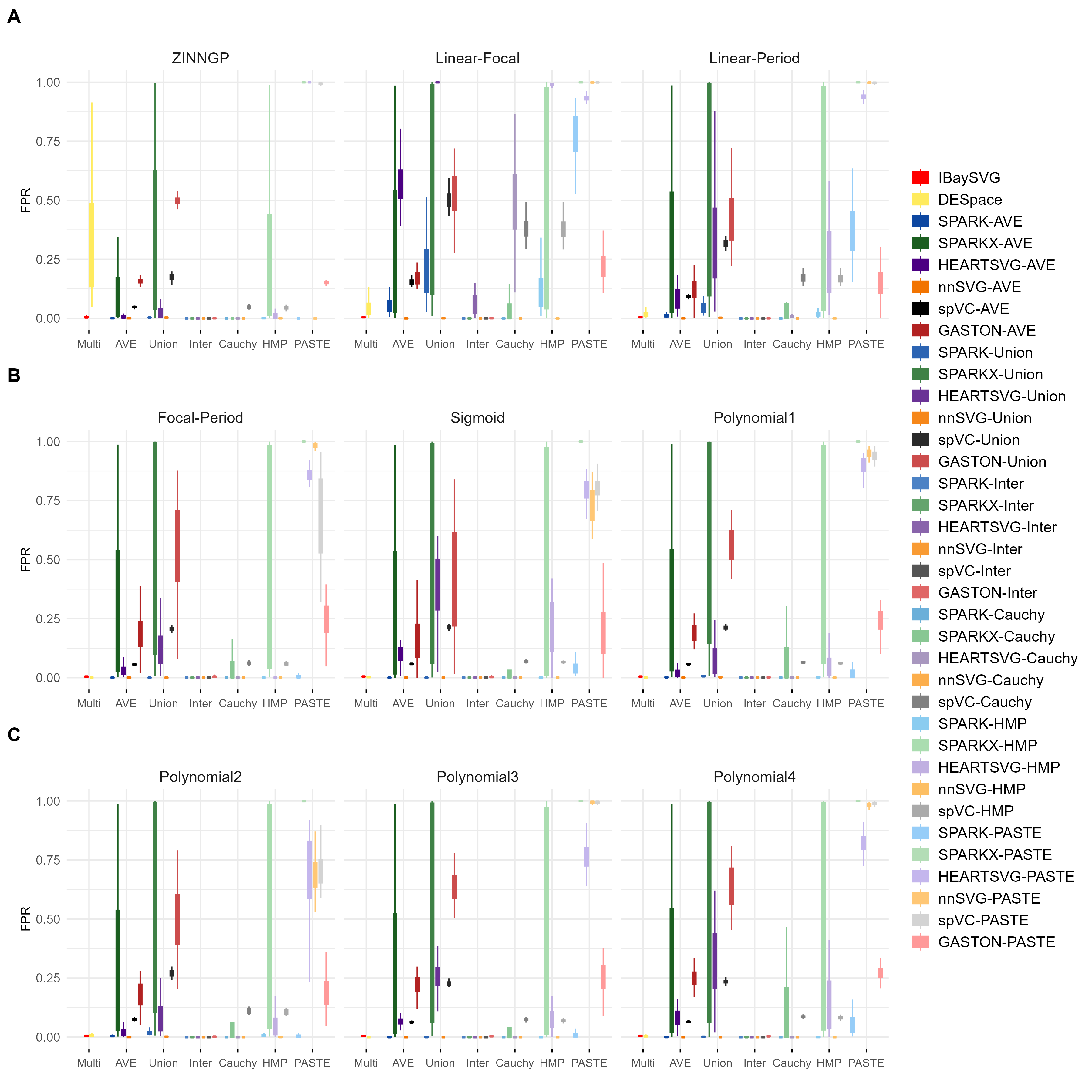}
    \caption{Comparative boxplots of FPR distributions derived from 50 replicates across competing methodologies under enhanced spatial complexity. This evaluation employs Setting 1 of the cross-slide signal configuration with a medium dropout probability of 0.5, incorporating three distinct spatial architecture paradigms.}
    \label{supple_simu_FPR}
\end{figure}

In this section, we extend our evaluation of the effectiveness of the proposed methodology to diverse spatial configurations. Building upon Setting 1 of the cross-slide signal configurations with a medium dropout probability of 0.5, we systematically examine three distinct classes of spatial architectures:\\
(a) The zero-inflated nearest-neighbor Gaussian process (ZINNGP) framework proposed by \cite{weber2023nnsvg}, which employs a covariance matrix structure to account for spatial dependencies. Specifically, it formulates the model as $Y \mid \beta,r,\theta,X\sim (\delta_0)^{r}\left(N(X \beta,C(\theta)+\tau^2 I)\right)^{1-r}$, where the exponential covariance function is defined as
    $$C_{ij}(\theta)=\sigma^2 \exp \left(\frac{-\Vert s_i - s_j \Vert}{l}\right).$$
    The covariance parameters $\theta=(\sigma^2,l)$ determine the length scale and variance of the spatial component and the parameter $\tau^2$ represents the variance of additional non-spatial component.\\
(b) Model (1) considered in the main text incorporating hybrid spatial effects for 2D coordinates, specifically evaluating three combinatorial formulations: 
\begin{itemize}
        \item Linear-focal fusion: $b_{1}(s_{i1})=\beta_0 (s_{i1})$ and $b_{2}(s_{i2})=\beta_0\exp(-s_{ik}^2)/2$,
        \item Linear-periodic fusion:  $b_{1}(s_{i1})=\beta_0(s_{i1})$ and $b_{2}(s_{i2})=\beta_0cos(2 \pi s_{i2})$,
        \item Focal-periodic fusion:  $b_{1}(s_{i1})=\beta_0\exp(-s_{i1}^2)/2$  and $b_{2}(s_{i2}) = \beta_0cos(2 \pi s_{i2})$.
\end{itemize}
(c) Enhanced configurations of Model (1) with sophisticated spatial patterns, including sigmoidal activation patterns and four polynomial functional forms (designated as Polynomial1-4):
\begin{itemize}
        \item  Sigmoid function. The sigmoid function is a mathematical function characterized by an S-shaped curve, frequently utilized in machine learning, statistics, and neuroscience. The formula employed in this context is $b_{k}(s_{ik})=\beta_0/(1+exp(-s_{ik}))$. 
        \item Four polynomial functions:
        \begin{equation*}
            \begin{aligned}
                \text{Polynomial1}:\ &b_{k}(s_{ik})=0.5\beta_0(s_{ik} + 1)(s_{ik} - 0.8)(s_{ik}- 1.6),\\
                \text{Polynomial2}:\ &b_{k}(s_{ik})=\beta_0(-0.5s_{ik}^3+0.3s_{ik}),\\
                \text{Polynomial3}:\ &b_{k}(s_{ik})=\beta_0(0.15s_{ik}^4 - 0.1s_{ik}^2 + 0.7),\\
               \text{Polynomial4}:\  &b_{k}(s_{ik})=\beta_0(0.25s_{ik}^3 + 0.1s_{ik}^2 - 0.15s_{ik} + 0.3).
            \end{aligned}
        \end{equation*}
    \end{itemize}
Visual illustrations of these spatial architectures are available in Figure \ref{extend-spatial-pattern}, which express various spatial patterns clearly. Figures \ref{supple_simu_F1}-\ref{supple_simu_FPR} summarize the aggregated F1 scores, TPR values, and FPR values from 100 experimental replicates.

Our analysis demonstrates that IBaySVG consistently preserves robust identification accuracy and stability across heterogeneous spatial configurations. In particular, despite variations in model architectures, IBaySVG exhibits exceptional proficiency in controlling the FPR, a critical determinant of its superior performance in SV gene detection. Specifically, under the ZINNGP framework, IBaySVG demonstrates marginally inferior performance relative to SPARK, HEARTSVG, and nnSVG, but it generally outperforms other methods. This is as expected since SPARK also employs a Gaussian process for spatial effect modeling but assumes a Poisson distribution, while HEARTSVG, although not directly using a Gaussian process, leverages spatial coordinates to compute marginal expressions and their autocorrelations for SV gene identification, achieving a similar effect. Meanwhile, nnSVG directly adopts the NNGP model. The comparative advantage of IBaySVG becomes particularly evident under the other two classes of spatial architectures, particularly for scenarios with sophisticated spatial patterns. The improvement of the union, intersection, Cauchy, and HMP integration varies across different methods and scenarios. The PASTE augmentation framework amplifies single-sample analysis performance in challenging scenarios, such as the patterns with polynomial and sigmoid functions. Notably, SPARK achieves a more pronounced performance boost due to its superior FPR control in PASTE-generated composite samples, especially for the patterns with sigmoid and polynomial functions. DESpace achieves reliable accuracy primarily in linear or near-linear spatial configurations (e.g., sigmoid, Polynomial2, Polynomial4 patterns), aligning with its algorithmic design principles. These empirical findings corroborate previously established simulation results, reinforcing the methodological validity of IBaySVG for SV gene identification under diverse spatial complexities.

\clearpage

\clearpage

\subsection{Simulation without covariates}

\begin{figure}[!ht]
    \centering
    \includegraphics[width=1\linewidth]{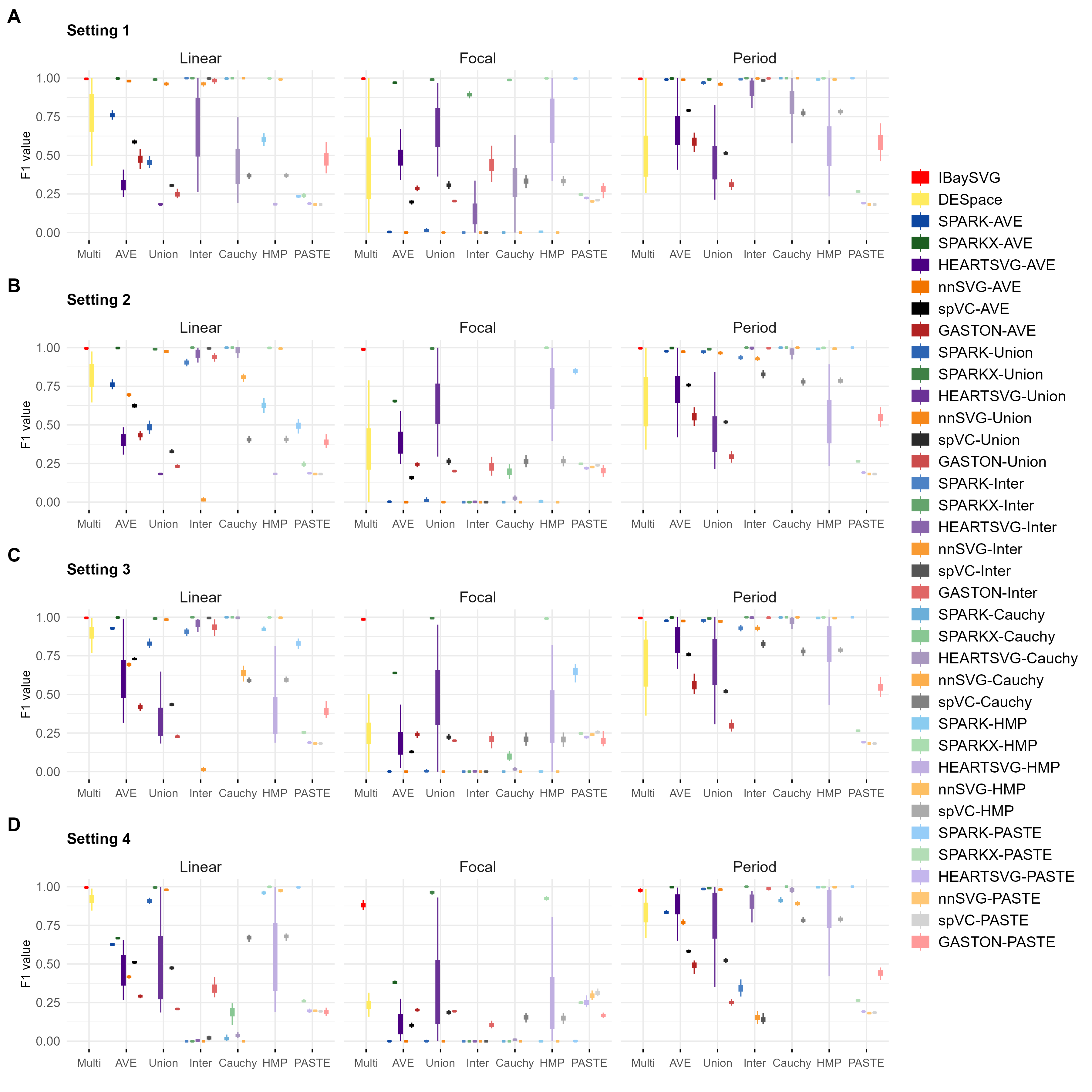}
    \caption{Boxplots of the F1 values over 50 replicates with different methods under scenarios without covariates. (A) Setting 1, (B) Setting 2, (C) Setting 3, and (D) Setting 4. In each subfigure, three types of spatial patterns (linear, focal, and periodic) with the middle dropout rate are examined.}
    \label{simu_F1_nocov_inf3}
\end{figure}

\begin{figure}[!ht]
    \centering
    \includegraphics[width=1\linewidth]{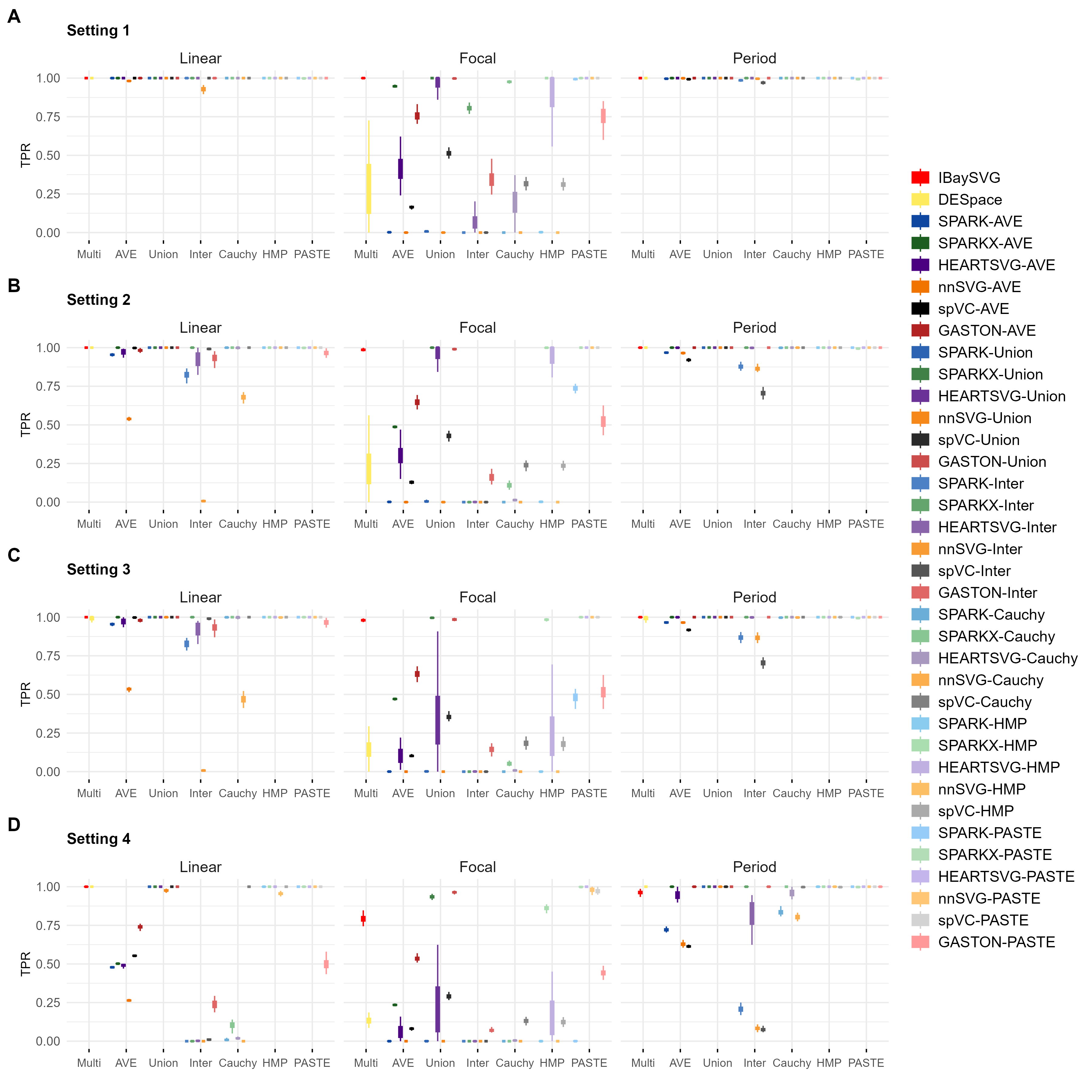}
    \caption{Boxplots of the TPR values over 50 replicates with different methods under scenarios without covariates. (A) Setting 1, (B) Setting 2, (C) Setting 3, and (D) Setting 4. In each subfigure, three types of spatial patterns (linear, focal, and periodic) with the middle dropout rate are examined.}
    \label{simu_tpr_nocov_inf3}
\end{figure}

\begin{figure}[!ht]
    \centering
    \includegraphics[width=1\linewidth]{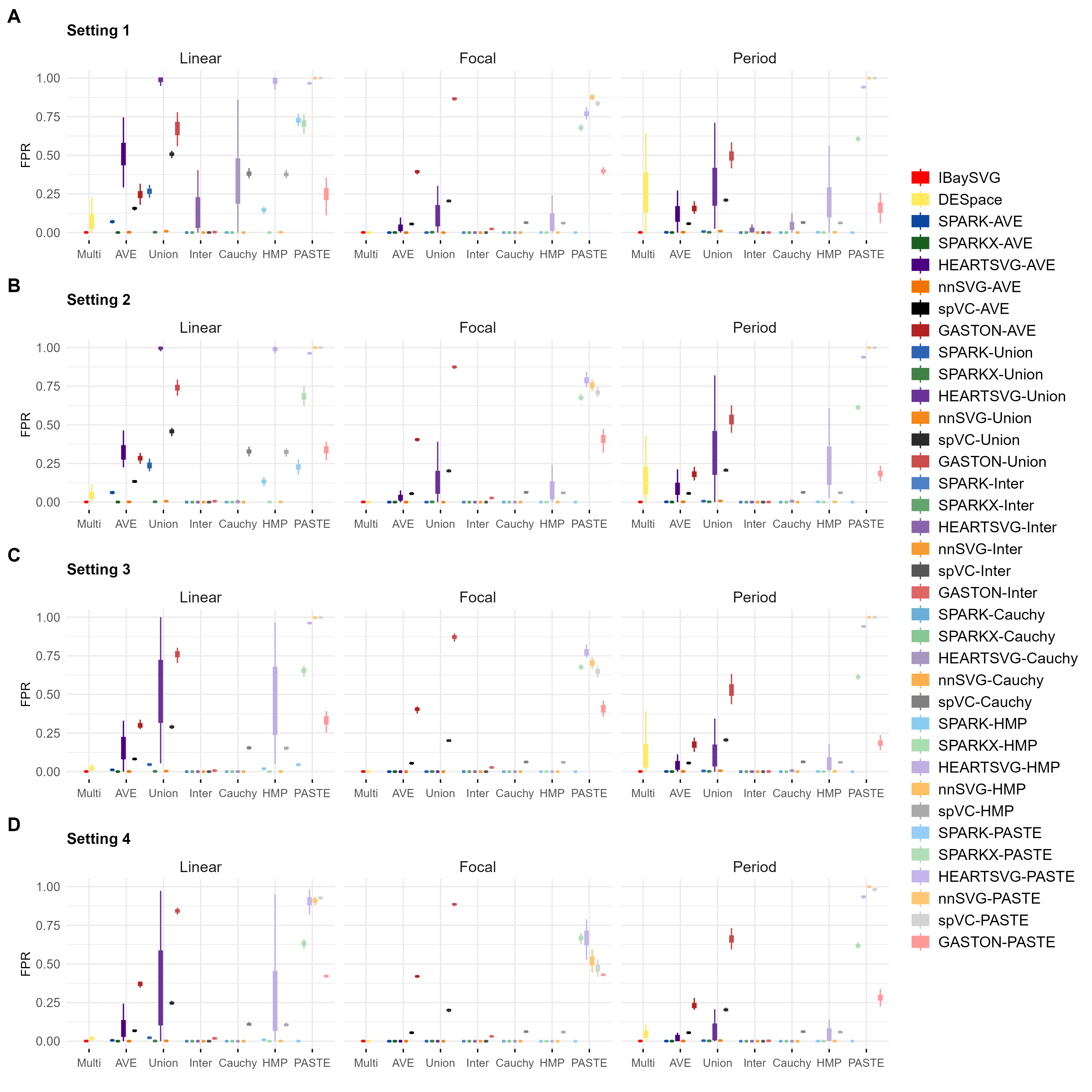}
    \caption{Boxplots of the FPR values over 50 replicates with different methods under scenarios without covariates. (A) Setting 1, (B) Setting 2, (C) Setting 3, and (D) Setting 4. In each subfigure, three types of spatial patterns (linear, focal, and periodic) with the middle dropout rate are examined.}
    \label{simu_fpr_nocov_inf3}
\end{figure}

In practice, covariate information in real datasets sometimes is unavailable or difficult to obtain. To ensure both the broad applicability and usability of our method, we allow for input data without covariate information. Following the setup of our simulation studies, we evaluate the performance of IBaySVG relative to competing approaches in detecting SV genes under moderate dropout conditions. Specifically, for each signal pattern—linear, focal, and periodic—signal strength is parameterized at four levels (high, medium, low, and extremely low) with $\beta_0$ set to $(0.6,0.4,0.2,0.05)$ for linear, $(0.5,0.4,0.2,0.1)$ for focal, and $(1,0.8,0.6,0.4)$ for periodic. To simulate scenarios without covariate information, we set $\psi$=0, while keeping all other settings identical to those in the original simulation framework. Summarized results are provided in Figures \ref{simu_F1_nocov_inf3}-\ref{simu_fpr_nocov_inf3}. Under these simplified scenarios, SPARKX-union and SPARKX-HMP effectively reduce the occurrence of false positives while maintaining a high TPR. However, consistent with the results observed under simulation settings that account for covariance, IBaySVG again exhibits substantially higher detection power than existing approaches—even in the absence of covariate information.

\clearpage

\subsection{Comparison and Evaluation of ZINB and NB Models}
To assess the necessity of the zero-inflation component, we perform an ablation study on simulated datasets by removing the zero-inflation structure from the proposed model. Specifically, the latent zero-inflation indicator $r_{i}^{(m)}$=0 is fixed to 0 throughout all inference iterations, while all other model components and hyperparameters remain unchanged to ensure a fair comparison. Both moderate and high dropout settings are examined. As shown in Tables \ref{tab:inf1} and \ref{tab:inf2}, IBaySVG consistently outperforms its negative binomial (NB) counterpart across all scenarios. Furthermore, as the dropout level increases from moderate to high, the performance of the NB-based model declines sharply, whereas IBaySVG maintains stable and robust results. These findings underscore the importance of explicitly modeling zero inflation for reliable spatially variable gene detection, especially in spatial transcriptomics data with high dropout rates.

\begin{table}[htbp]
\footnotesize
\centering
\caption{Performance of IBaySVG and its NB counterpart without zero‑inflation under moderate dropout scenarios. Values are reported as mean (SD).}
\label{tab:inf1}
\begin{tabular}{lccc ccc}
\toprule
Scenario  & TPR & FPR & F1 &
TPR &FPR & F1 \\
\midrule
&&IBaySVG &&&  NB counterpart  &\\
\midrule
Linear\\
1 & 1.000 (0.000) &  0.003 (0.001) & 0.985 (0.004) & 1.000 (0.000) & 0.011 (0.002) & 0.951 (0.006) \\
2 & 1.000 (0.000) & 0.004 (0.001) & 0.984 (0.005) & 1.000 (0.000) & 0.011 (0.001) & 0.952 (0.005) \\
3 & 1.000 (0.000) & 0.004 (0.001)& 0.983 (0.004) & 1.000 (0.000) & 0.012 (0.003) & 0.947 (0.011) \\
4 & 0.985 (0.032) & 0.003 (0.001) & 0.979 (0.015) & 1.000 (0.000) &  0.013 (0.002)  & 0.945 (0.008) \\
 \midrule
 Focal\\
1 & 1.000 (0.000) & 0.004 (0.001) & 0.984 (0.004) & 0.999 (0.001) & 0.011 (0.002) & 0.950 (0.010) \\
2 & 0.998 (0.004) & 0.003 (0.001) & 0.984 (0.004) & 0.953 (0.014) & 0.011 (0.002) & 0.927 (0.010) \\
3 & 0.995 (0.008) & 0.003 (0.001) & 0.982 (0.005) & 0.869 (0.029) & 0.012 (0.003) & 0.878 (0.019) \\
4 & 0.630 (0.141) & 0.002 (0.001) & 0.756 (0.106) & 0.252 (0.030) & 0.012 (0.002) & 0.363 (0.034) \\
 \midrule
 Period\\
1 & 0.999 (0.002) & 0.004 (0.002) & 0.983 (0.007) & 1.000 (0.000) & 0.012 (0.002) & 0.950 (0.009) \\
2 & 0.993 (0.019) & 0.004 (0.001) & 0.981 (0.011) & 1.000 (0.000) & 0.012 (0.002) & 0.949 (0.009) \\
3 & 0.987 (0.035) & 0.004 (0.002) & 0.976 (0.020) & 1.000 (0.001) & 0.012 (0.002) & 0.950 (0.008) \\
4 & 0.912 (0.137) & 0.003 (0.002) & 0.934 (0.086) & 0.562 (0.067) & 0.010 (0.002) & 0.680 (0.060) \\
\bottomrule
\end{tabular}
\label{noinf1}
\end{table}

\begin{table}[htbp]
\footnotesize
\centering
\caption{ Performance of IBaySVG and its NB counterpart without zero‑inflation under high dropout scenarios. Values are reported as mean (SD).}
\label{tab:inf2}
\begin{tabular}{lccc ccc}
\toprule
Scenario  & TPR & FPR & F1 &
TPR &FPR & F1 \\
\midrule
&&IBaySVG &&& NB counterpart  &\\
\midrule
Linear\\
1 & 1.000 (0.000) & 0.004 (0.001) & 0.982 (0.004) & 1.000 (0.000) & 0.120 (0.011)  & 0.650 (0.020) \\
2 & 1.000 (0.000) & 0.004 (0.001) & 0.982 (0.004) & 1.000 (0.000) & 0.120 (0.006)  & 0.651 (0.013) \\
3 & 1.000 (0.000) & 0.004 (0.001) & 0.982 (0.004) & 1.000 (0.000) &0.120 (0.008)   & 0.648 (0.014) \\
4 & 0.859 (0.093) & 0.002 (0.001) & 0.910 (0.054) & 1.000 (0.000) & 0.126 (0.014)   & 0.640 (0.026) \\
 \midrule
 Focal\\
1 & 0.997 (0.009) & 0.004 (0.001) & 0.980 (0.006) & 1.000 (0.001) &  0.123 (0.010)  & 0.644 (0.019) \\
2 & 0.935 (0.057) & 0.003 (0.001) & 0.951 (0.030) & 0.958 (0.010) & 0.120 (0.009) & 0.632 (0.016) \\
3 & 0.833 (0.127) & 0.003 (0.001) & 0.891 (0.081) & 0.894 (0.023) & 0.124 (0.011)  & 0.595 (0.022) \\
4 & 0.306 (0.113) & 0.002 (0.001) & 0.451 (0.120) & 0.478 (0.062) & 0.122 (0.016)    & 0.371 (0.043) \\
 \midrule
 Period\\
1 & 0.974 (0.054) & 0.004 (0.001) & 0.970 (0.029) & 1.000 (0.000) & 0.118 (0.009)  & 0.651 (0.017) \\
2 & 0.931 (0.102) & 0.003 (0.001) & 0.946 (0.062) & 1.000 (0.000) &  0.118 (0.007)   & 0.654 (0.014) \\
3 & 0.885 (0.158) & 0.003 (0.001) & 0.917 (0.104) & 1.000 (0.001) & 0.119 (0.012)& 0.651 (0.023) \\
4 & 0.672 (0.215) & 0.003 (0.001) & 0.771 (0.179) & 0.914 (0.029) & 0.125 (0.014)   & 0.603 (0.026) \\
\bottomrule
\end{tabular}
\label{noinf2}
\end{table}

\clearpage

\subsection{Sensitivity analysis}\label{Sensitive-simulation}

We perform a sensitivity analysis on simulated spatial transcriptomics datasets under three distinct spatial patterns (linear, focal, and periodic). Specifically, we examine scenarios with a high dropout rate under setting 1. 

The following hyperparameters are examined over the specified ranges, with default values (suggested in Section \ref{hyperpara-set}) highlighted: 
\begin{itemize}
\item Prior for $\pi^{(m)}$: $a_{\pi}$ and $b_{\pi}$ each vary in $\{0.6, 0.8, \mathbf{1}, 1.2, 1.4\}$;
\item Prior for $\phi^{(m)}$: $a_{\phi}$ and $b_{\phi}$ each vary in $\{0.0005, 0.0008, \mathbf{0.001}, 0.003, 0.005\}$;
\item Priors for $\eta^{(m)}$ and $\psi^{(m)}$: $\sigma_{\eta}^{(m)}$ and $\sigma_{\psi}^{(m)}$ each vary in $\{0.6, 0.8, \mathbf{1}, 1.2, 1.4\}$;
\item Bi-level structure parameters: $\Gamma_{1} \in \{\sqrt{0.0006}, \sqrt{0.0008}, \mathbf{\sqrt{0.001}}, \sqrt{0.0012}, \sqrt{0.0014}\}$. For general multi-sample data, $\Gamma_{2} \in \{0.015, 0.0125, \mathbf{0.01}, 0.0075, 0.005\}$;
\item Prior for $q_k$: $c_q$ and $d_q$ each vary in $\{0.6, 0.8, \mathbf{1}, 1.2, 1.4\}$;
\item Prior for $p_k$: $c_p \in \{0.1, 0.15, \mathbf{0.2}, 0.25, 0.3\}$ and $d_p \in \{0.9, 1.35, \mathbf{1.8}, 2.25, 2.7\}$.
\item Prior for $a_k^{(m)}$ (degree‑specific): $A_k$ is selected from $\{0.06, 0.07, \mathbf{0.08}, 0.09, 0.10\}$ (degree=1), $\{0.04, 0.045, \mathbf{0.05}, 0.055, 0.06\}$ (degree=2), $\{0.03, 0.035, \mathbf{0.04}, 0.045, 0.05\}$ (degree=3), and $\{0.02, 0.025, \mathbf{0.03}, 0.035, 0.04\}$ (degree=4).
\end{itemize}

\begin{figure}[!ht]
    \centering
    \includegraphics[width=\linewidth]{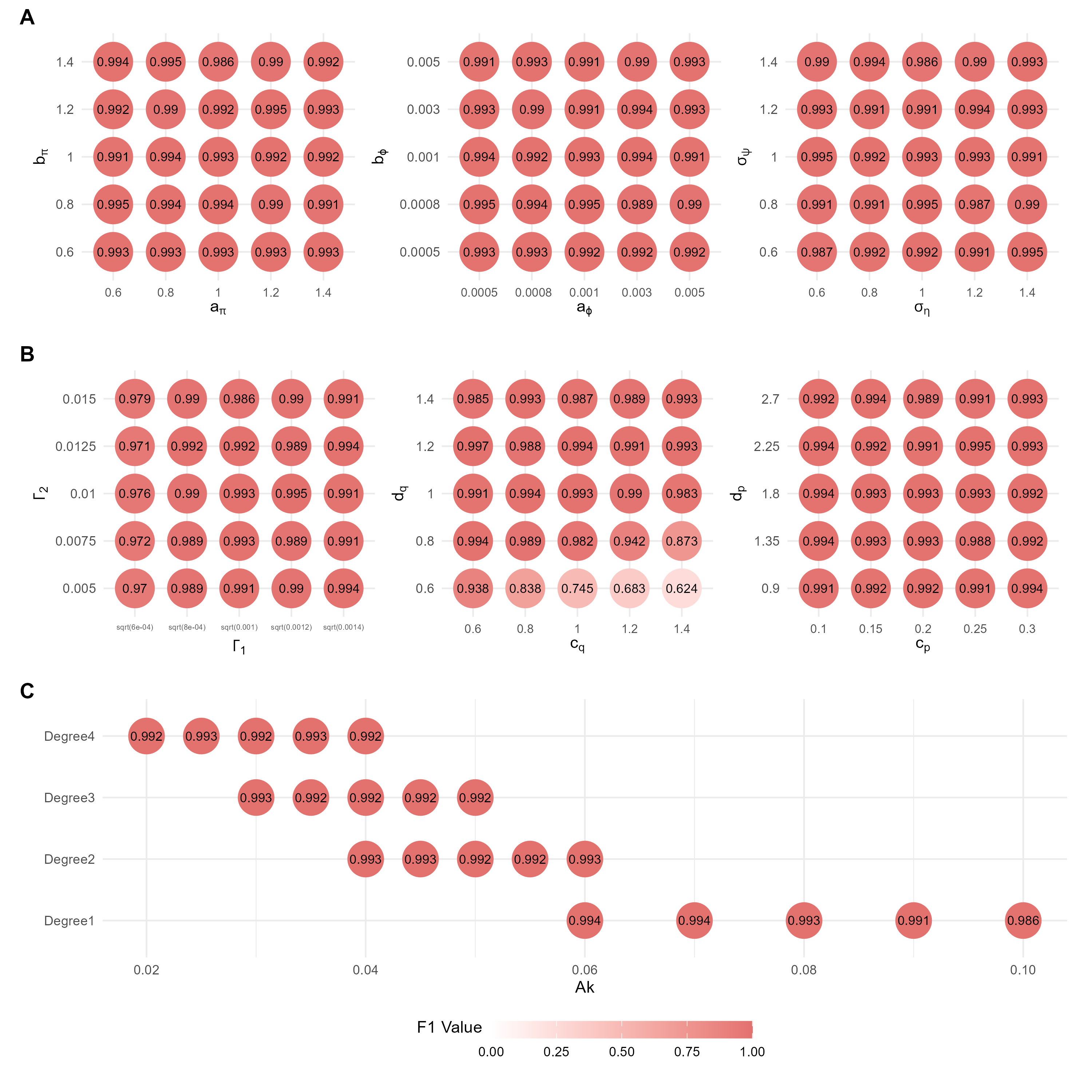}
    \caption{Sensitivity analysis of hyperparameters for simulated linear‑pattern data under high dropout and Setting 1. The results report the F1 score based on the identified spatially variable genes under the range of parameters. (a) $a_{\pi} \sim b_{\pi}$; (b) $a_{\phi} \sim b_{\phi}$; (c) $\sigma_{\eta} \sim \sigma_{\psi}$; (d) $\Gamma_{1}\sim \Gamma_{2}$; (e) $c_{q} \sim d_{q}$; (f) $c_{p} \sim d_{p}$; (g) $A_k$ for different orders of the basis functions (Degrees 1–4).}
    \label{fig:sensitive_linear}
\end{figure}

\begin{figure}[!ht]
    \centering
    \includegraphics[width=\linewidth]{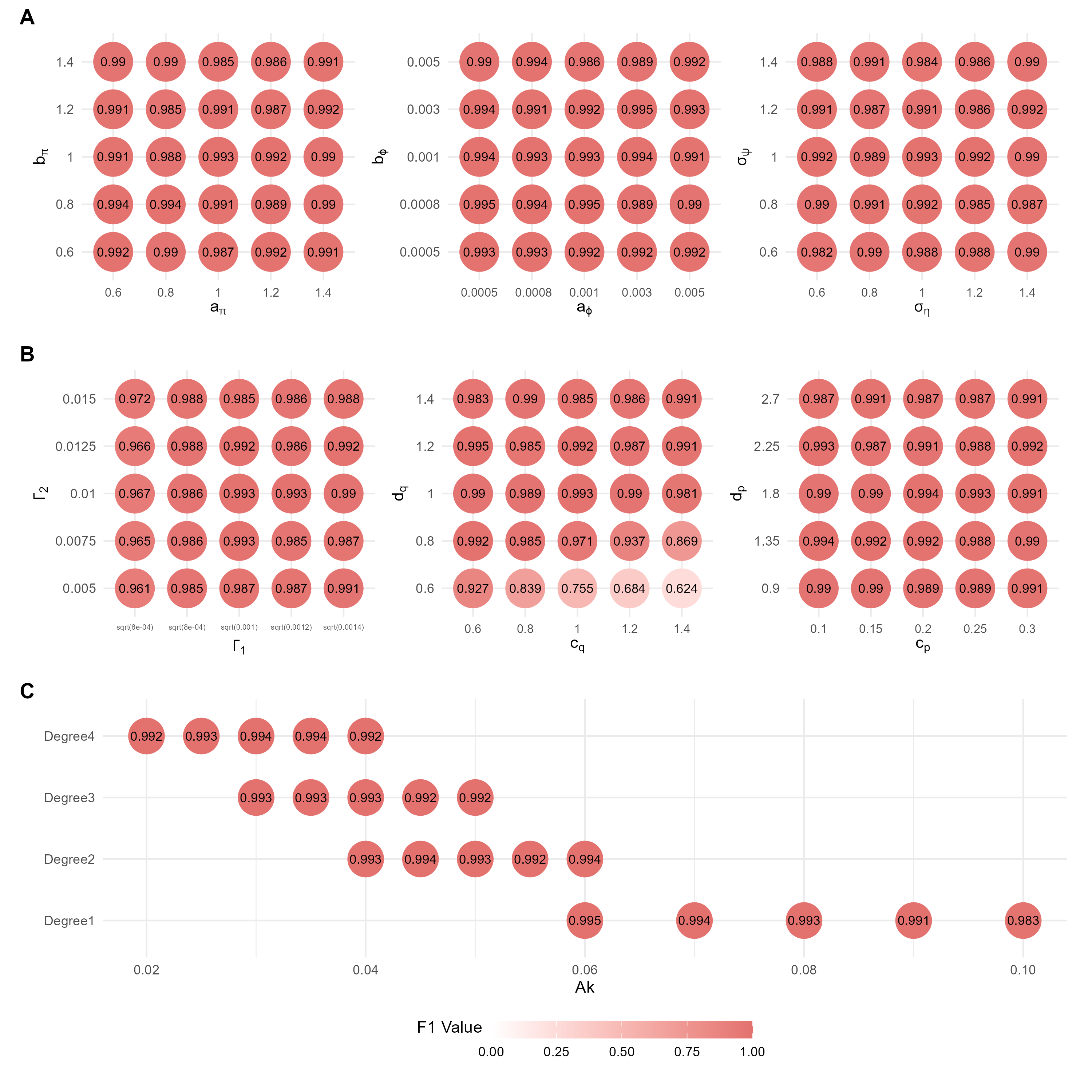}
    \caption{Sensitivity analysis of hyperparameters for simulated focal‑pattern data under high dropout and Setting 1. The results report the F1 score based on the identified spatially variable genes under the range of parameters. (a) $a_{\pi} \sim b_{\pi}$; (b) $a_{\phi} \sim b_{\phi}$; (c) $\sigma_{\eta} \sim \sigma_{\psi}$; (d) $\Gamma_{1}\sim \Gamma_{2}$; (e) $c_{q} \sim d_{q}$; (f) $c_{p} \sim d_{p}$; (g) $A_k$ for different orders of the basis functions (Degrees 1–4).}
    \label{fig:sensitive_focal}
\end{figure}

\begin{figure}[!ht]
    \centering
    \includegraphics[width=\linewidth]{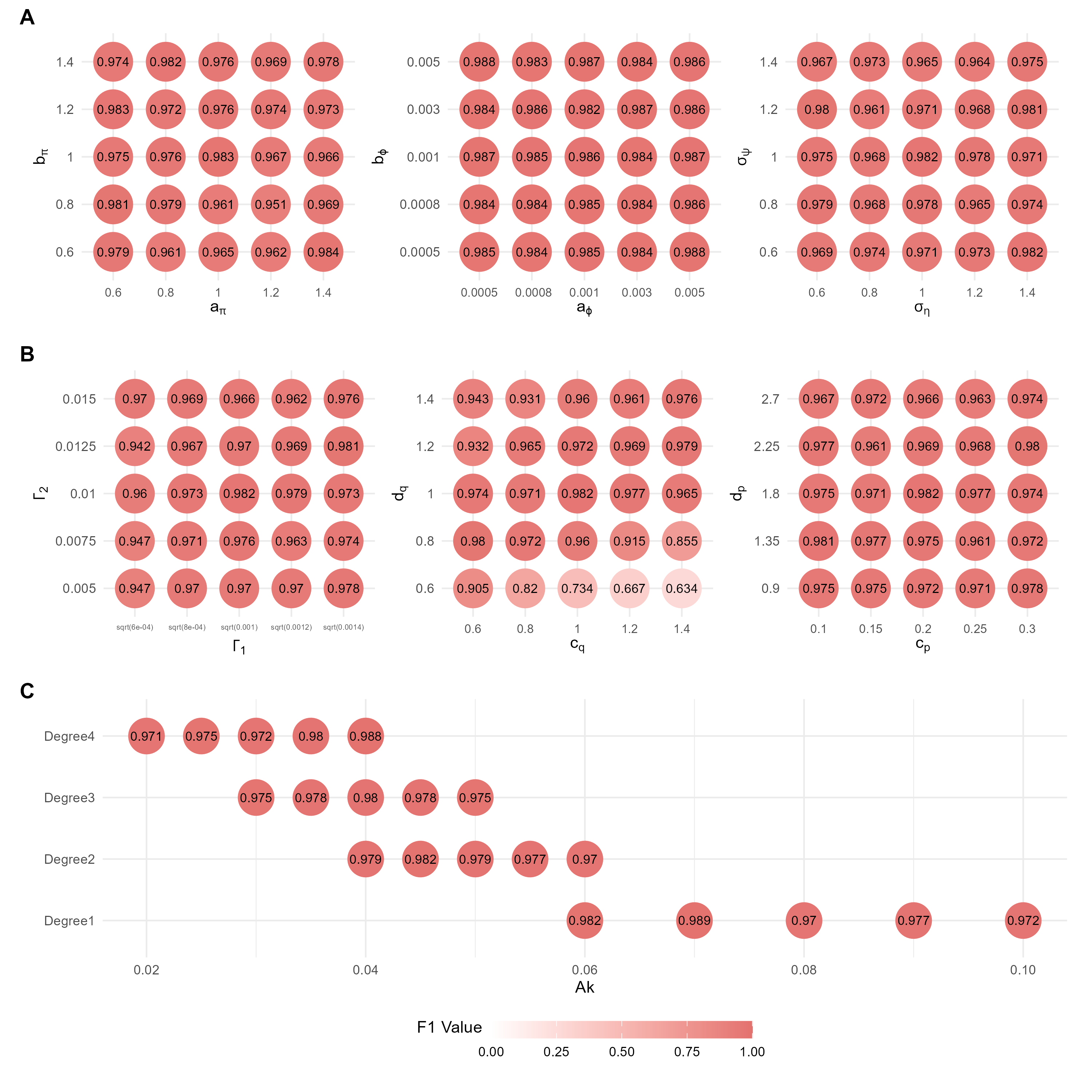}
    \caption{Sensitivity analysis of hyperparameters for simulated period‑pattern data under high dropout and Setting 1. The results report the F1 score based on the identified spatially variable genes under the range of parameters. (a) $a_{\pi} \sim b_{\pi}$; (b) $a_{\phi} \sim b_{\phi}$; (c) $\sigma_{\eta} \sim \sigma_{\psi}$; (d) $\Gamma_{1}\sim \Gamma_{2}$; (e) $c_{q} \sim d_{q}$; (f) $c_{p} \sim d_{p}$; (g) $A_k$ for different orders of the basis functions (Degrees 1–4).}
    \label{fig:sensitive_period}
\end{figure}

The ranges are designed in accordance with their roles. For example, $\Gamma_{1}$ (spike variance) and $\Gamma_{2}$ (probability of being an SV gene when $u_k=0$) are both set to relatively small values. Since ground‑truth labels for the SV genes are available in the simulation, we directly adopt the F1‑score as the stability metric. We report the averaged results over 50 simulation replicates in Figures \ref{fig:sensitive_linear}–\ref{fig:sensitive_period}. As can be observed, the proposed method consistently maintains high F1‑scores across different spatial patterns and hyperparameter choices, and the performance remains stable when the parameters lie within a reasonable range.

\clearpage
\section{Additional data analysis results}

\subsection{Cell-type proportion estimations}

\begin{figure}[!ht]
    \centering
    \includegraphics[width=1\linewidth]{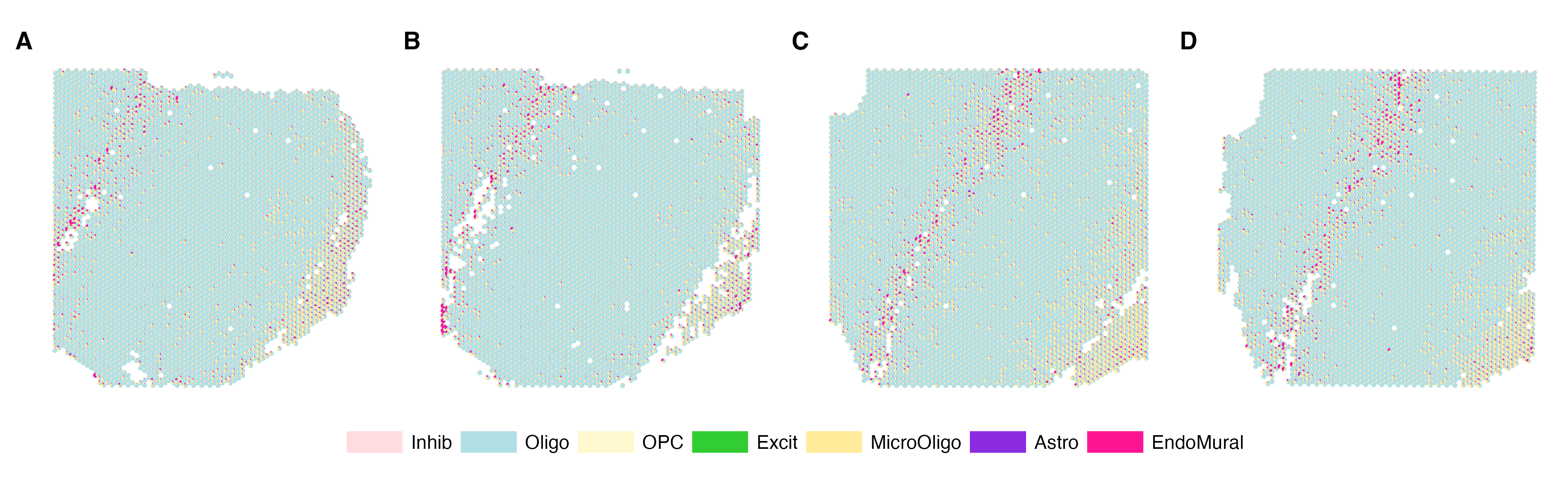}
    \caption{Distribution of cellular composition across spatial domains for the same-donor DLPFC dataset: (A)-(D) pie chart representations for tissue slices A-D with enhanced contrast visualization.}
    \label{fig:celltype}
\end{figure}

\begin{figure}[!ht]
    \centering
    \includegraphics[width=1\linewidth]{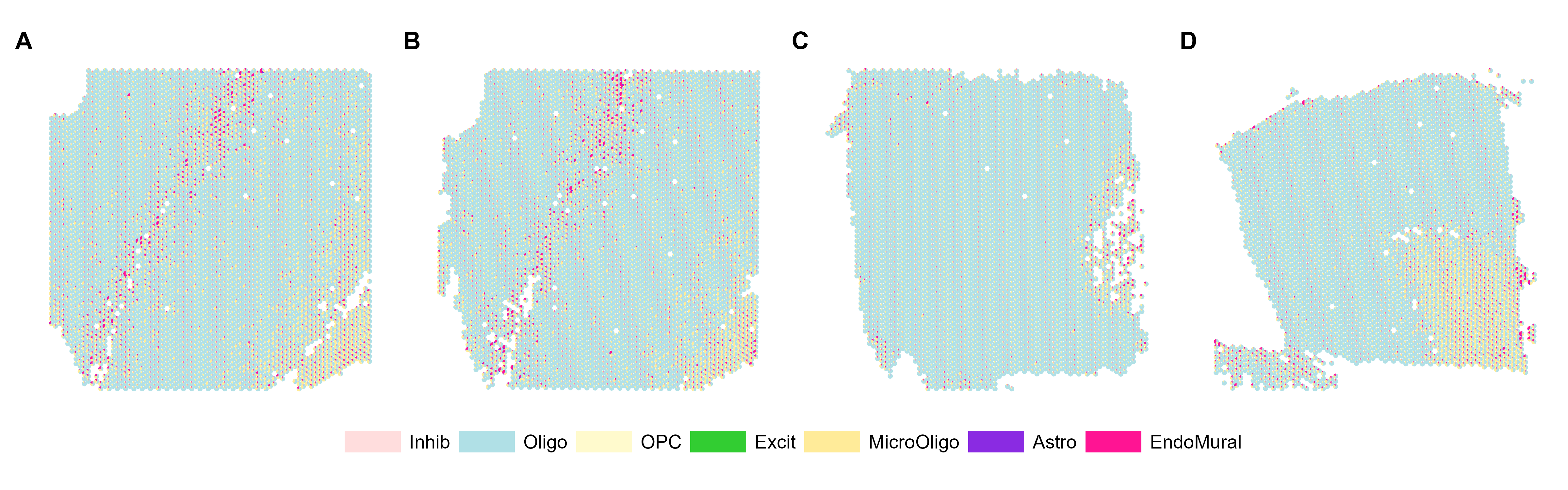}
    \caption{Distribution of cellular composition across spatial domains for DLPFC dataset across donors: (A)-(D) pie chart representations for tissue slices A-D with enhanced contrast visualization.}
    \label{fig:celltype3477}
\end{figure}

\begin{figure}[!ht]
    \centering
    \includegraphics[width=1\linewidth]{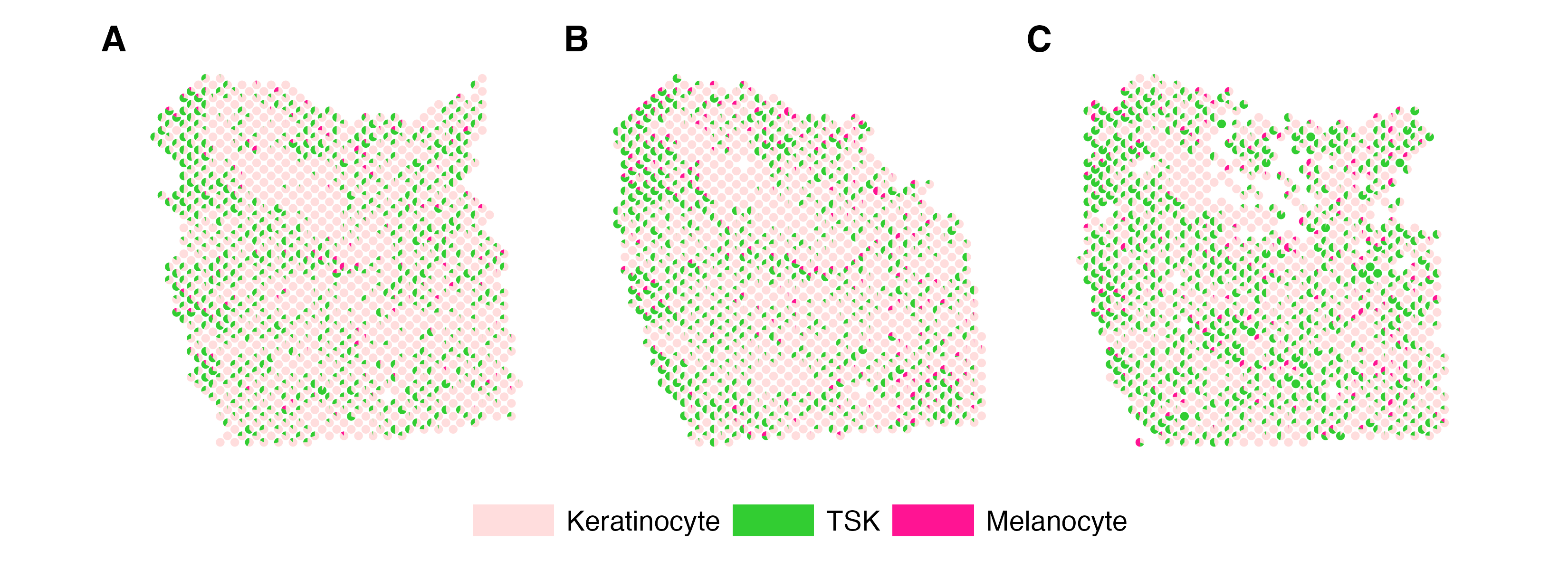}
    \caption{Distribution of cellular composition across spatial domains in SCC dataset: (A)-(C) pie chart representations for tissue slices A-C.}
    \label{scc_cell_type}
\end{figure}

\subsection{Outlier detection}\label{Outlier}
For the three real datasets, as described in Section~\ref{real-details}, we perform quality control procedures that exclude certain spots and genes, potentially rendering the data largely free of obvious outliers. In this section, we further validate this conjecture. Given that the data are count data, conventional outlier detection methods, such as box‑plot‑based (e.g., Tukey’s IQR fences), Grubbs' test, and Z‑score‑based criteria, are no longer applicable. To this end, we adopt randomized quantile residuals (RQRs). In contrast to the model‑free methods mentioned above, RQRs can adequately handle the non‑Gaussian nature of the data. Specifically, RQRs provide standardized values that, under a reasonably specified discrete distribution, follow a standard normal reference, making them particularly suitable for highly sparse and zero‑inflated count data \citep{Bai2021RQR}. Compared with Pearson residuals, RQRs effectively remove the systematic skewness induced by low counts, thus offering a more trustworthy basis for identifying aberrant spatial spots. 

\begin{table}[htbp]
\centering
\caption{Empirical calibration of RQRs
under the fitted ZINB model. Values are reported as medians across all
gene--sample fits. The last column gives the corresponding theoretical
values under a standard normal distribution.}
\label{tab:rqr_calib}
\begin{tabular}{lcccc}
\toprule
Statistic & DLPFC (same donor) & DLPFC (across donor) & SCC & $N(0,1)$ \\
\midrule
Mean($RQR^2$)          & 0.9993 & 0.9994 & 0.9999 & 1.0000 \\
$P(|RQR|>2)$           & 0.0454 & 0.0454 & 0.0455 & 0.0455 \\
$P(|RQR|>3)$           & 0.00273 & 0.00275 & 0.00289 & 0.00270 \\
\bottomrule
\end{tabular}
\end{table}

\begin{figure}[htbp]
    \centering
    \includegraphics[width=\textwidth]{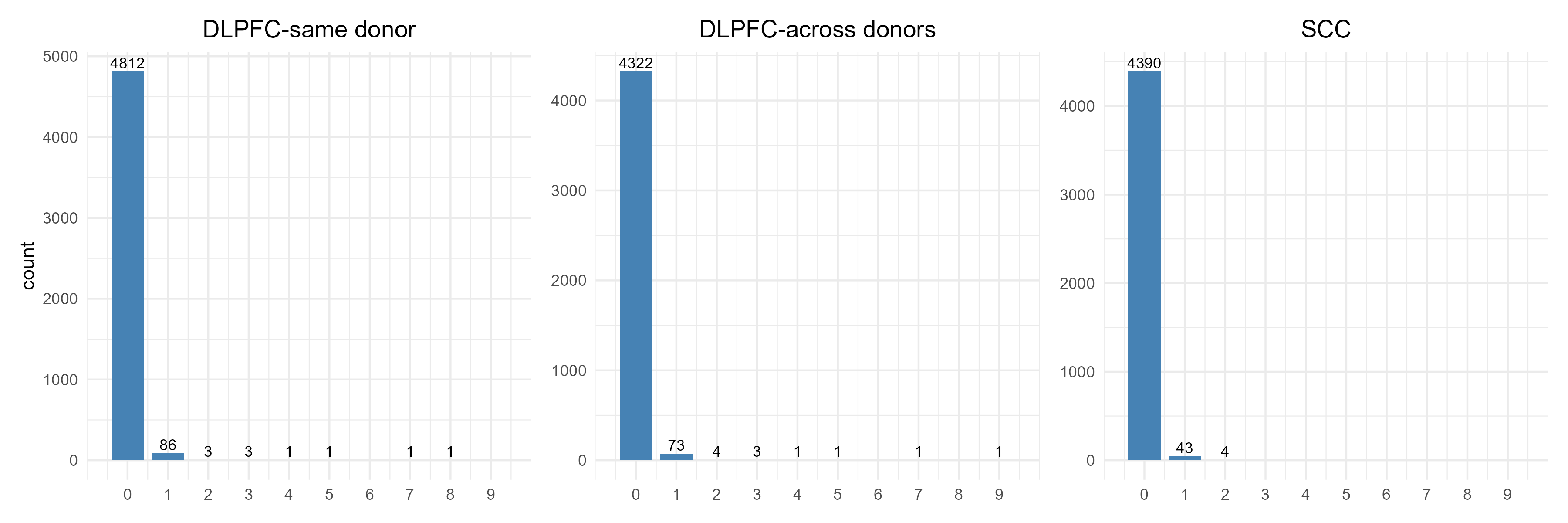}
    \caption{Distribution of the number of outlier spots per gene across the three datasets. The x-axis denotes the number of outlier spots per gene, and the y-axis denotes the number of genes attaining that count.}
    \label{fig:outlier_freq}
\end{figure}

For each gene, we first derive RQRs for all spots on each slice based on the fitted ZINB model, and summarize three statistics from these spot‑wise results. The medians of these three statistics across all genes and slices are then recorded in Table~\ref{tab:rqr_calib}. Across all three real datasets, these empirical quantities are remarkably close to their theoretical counterparts under the standard normal distribution, indicating an appropriate residual calibration. We then identify spot-level outliers as observations with $|RQR|>5$. Figure~\ref{fig:outlier_freq} presents the frequency distribution of genes by the number of detected outliers for the three datasets. Across all three real datasets, the vast majority of genes have zero outliers, indicating that extreme residuals are rare under the fitted model. Further examination reveals that these outliers involve only 103, 93, and 11 spots, and affect 96, 84, and 47 genes, for the DLPFC same-donor, DLPFC across-donor, and SCC dataset, respectively. Moreover, these outliers are highly localized rather than being driven by systematic artifacts; only one, zero, and one spot are shared by five or more genes across the three datasets. This confirms that the preceding quality control procedures have been effective in largely preventing the occurrence of such outliers. 

In addition, a comparison analysis is performed before and after removing the detected outlier spots; the changes in the parameters of the fitted model are reported in Table~\ref{refit_delete_outlier}. The fitted quantities are observed to remain highly stable after outlier removal. Therefore, to maintain the completeness of each slice, we keep all spots and genes in the dataset for subsequent analyses.

\begin{table}[htbp]
\centering
\caption{Comparison analysis before and after removing the detected outlier spots:
median (IQR) change in fitted quantities between the original and outlier-removed
fits (across affected genes).}
\begin{tabular}{lccc}
\toprule
 & DLPFC (same donor) & DLPFC (across donors) & SCC \\
\midrule
%\multicolumn{4}{l}{\emph{Fitted-quantity change}}\\
\quad $\lambda$ relative change  & $0.004$ ($0.001$--$0.016$) & $0.008$ ($0.002$--$0.016$) & $0.032$ ($0.018$--$0.047$) \\
\quad $\pi$ relative change      & $0.000$ ($0.000$--$0.000$) & $0.000$ ($0.000$--$0.000$) & $0.000$ ($0.000$--$0.001$) \\
\quad $|u|$ change               & $0.000$ ($0.000$--$0.000$) & $0.000$ ($0.000$--$0.000$) & $0.002$ ($0.001$--$0.003$) \\
\quad $\phi$ relative change     & $0.000$ ($0.000$--$0.000$) & $0.000$ ($0.000$--$0.000$) & $0.000$ ($0.000$--$0.001$) \\
\bottomrule
\end{tabular}
\label{refit_delete_outlier}
\end{table}

\clearpage

\subsection{Model diagnostics analysis}
\label{model-check}

To evaluate the compatibility and adequacy of the proposed IBaySVG model with real data, we perform a series of model checking analyses on several spatial transcriptomics datasets. For clarity of interpretation, we focus on a simplified setting in which the integrative structure of the full model is omitted and each tissue section is fitted independently. Specifically, we consider the ZINB model in Equation (1) of the main text, expanded via basis functions and incorporating covariate information.

We first examine the necessity of the zero-inflation component. For each gene in each tissue section, we fit both a negative binomial (NB) model using the \textit{glm.nb} function in the MASS package and a zero-inflated negative binomial (ZINB) model using the \textit{zeroinfl} function in the pscl package. For both models, the Akaike Information Criterion (AIC) is used to select the optimal degree of the basis functions. Since the NB model is nested within the ZINB model, likelihood ratio tests are conducted to assess whether adding the zero-inflation component leads to a statistically significant improvement in model fit. We further investigate the generalization ability of the proposed nonparametric spatial structure. Specifically, we replace the nonparametric component in the proposed model with spatial forms commonly adopted in previous studies, including linear, focal, and periodic functions. Analogously, for each parametric specification, we fit gene-wise models using the \textit{zeroinfl} function. As these parametric models are not nested within the proposed ZINB-based model, we employ the Vuong test to compare their relative model adequacy. Statistical significance is assessed based on Vuong test statistics. For both the likelihood ratio and Vuong tests, resulting p-values are adjusted using the Benjamini–Hochberg procedure to control the false discovery rate, accounting for multiple hypothesis testing across genes.

\begin{figure}[htbp]
    \centering
    \includegraphics[width=\linewidth]{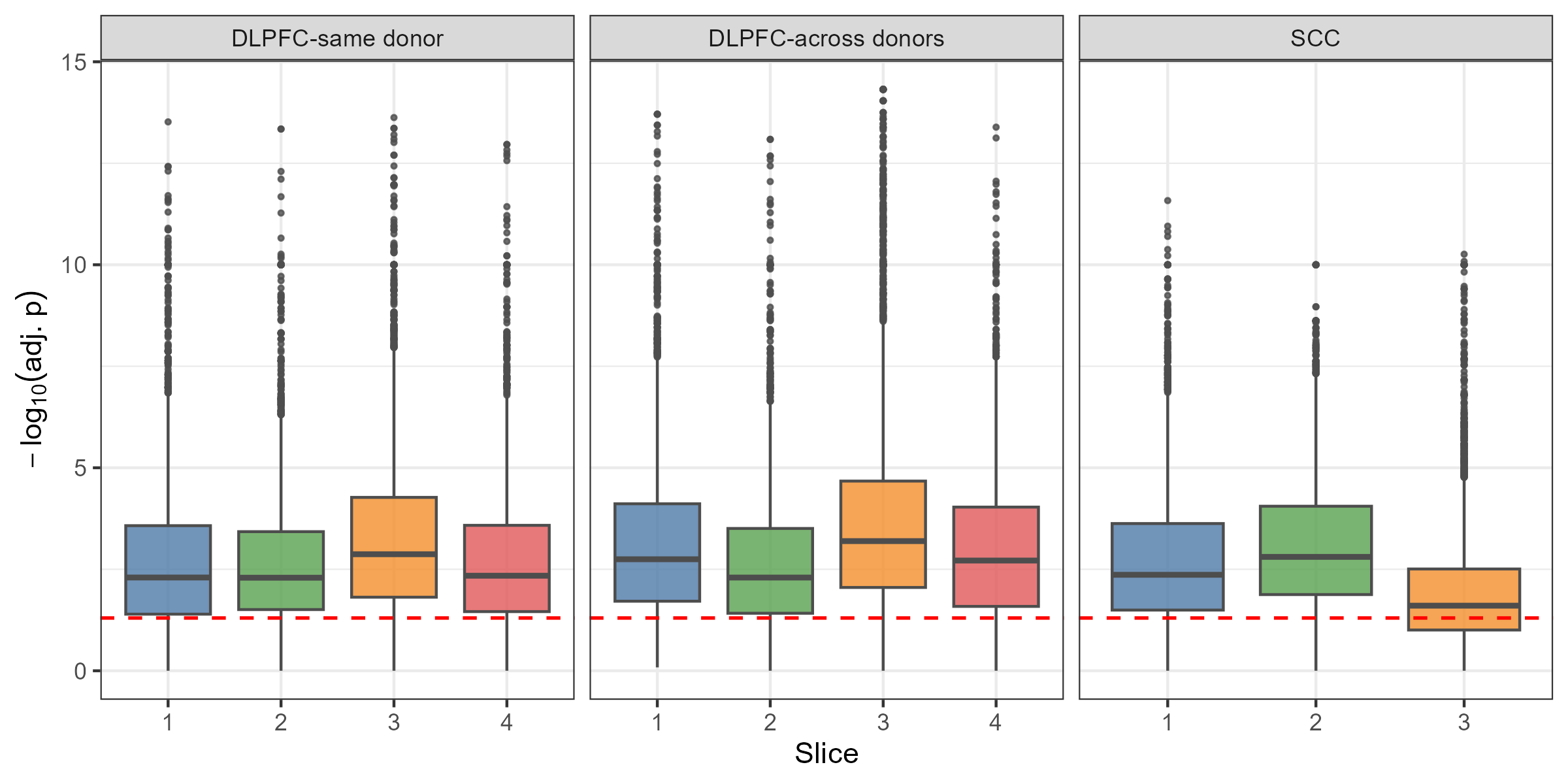}
    \caption{Likelihood ratio test comparison of ZINB and NB models across tissue slices of DLPFC dataset and SCC dataset. Boxplots show the distribution of $-\log_{10}(\mathrm{adjusted}\ p\text{-values})$ for each slice, stratified by dataset.
    The red dashed line denotes the significance threshold where $-\log_{10}(0.05)\approx1.3$.}
    \label{fig:model_checking}
\end{figure}

Results from the likelihood ratio tests are summarized in Figure \ref{fig:model_checking}. Across all three datasets, a majority of genes show p‑values below 0.05 ($-\log_{10}(0.05)> 1.3$), indicating that the ZINB model provides a significantly better fit than the standard NB model for these data. Furthermore, results of the Vuong tests are presented in Tables \ref{tab:significance_comparison}-\ref{tab:significance_comparison3}. Compared to the parametric spatial forms (linear, focal, and periodic), the proposed nonparametric spatial structure demonstrates superior or comparable fit in most cases, while offering greater modeling flexibility.

We further assess how the B-spline representation characterizes spatial transitions near tissue boundaries. Specifically, we investigate whether the smoothness imposed by the B-spline basis leads to overly smoothed expression changes across neighboring spatial regions. To quantify these boundary transitions, we first identify the boundaries between adjacent spatial regions in the real datasets. For the DLPFC data, regions are defined using the annotated cortical layers, whereas for the SCC data, where no ground-truth spatial annotation is available, regions are defined according to the spatial domains identified by IBaySVG. Boundary spots are then identified as locations adjacent to neighboring regions. For each boundary, let $m_a$, $m_b$, and $m_{ab}$ denote the per-gene mean expression in region $a$, region $b$, and the boundary, respectively. We quantify the relative expression changes between the boundary and its two neighboring regions using the normalized measures $2|m_a-m_{ab}|/(m_a+m_{ab})$ and $2|m_b-m_{ab}|/(m_b+m_{ab})$. These measures range from 0 to 2, with values close to 0 indicating smooth transitions between the boundary and neighboring regions, whereas larger values indicating stronger expression shifts and sharper spatial transitions.

\begin{figure}[htbp]
    \centering
    \includegraphics[width=\linewidth]{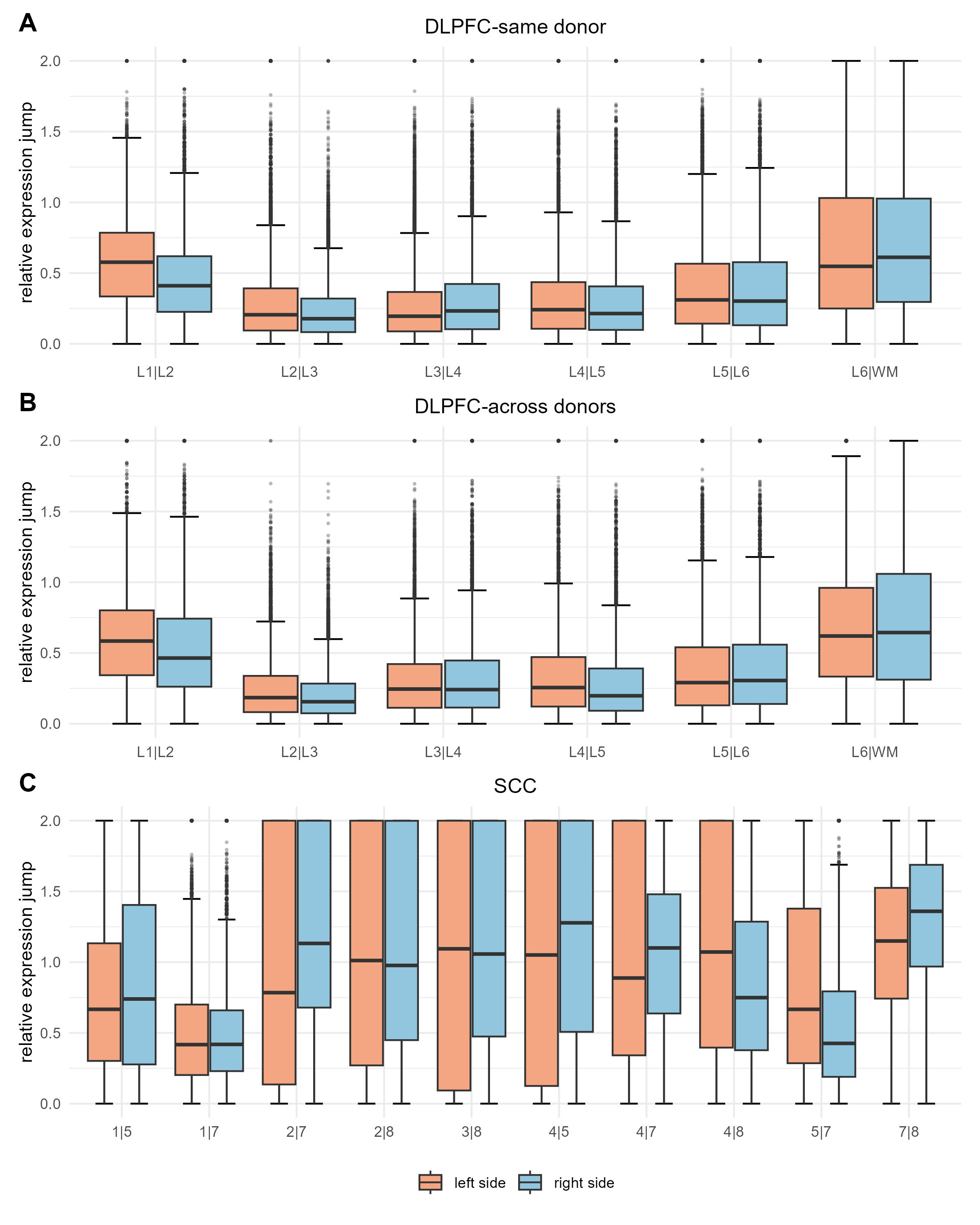}
    \caption{Boxplots of the per-gene relative expression change on each side of the
    region boundaries across three datasets: (A) DLPFC same-donor, (B) DLPFC across-donor,
    and (C) SCC. For each boundary, the orange and blue boxes represent the relative
    change of the left-side and right-side regions relative to the boundary, respectively.}
    \label{real_expression_jump}
\end{figure}

As shown in Figure \ref{real_expression_jump}, all three datasets exhibit some degree of discontinuity in gene expression across spatial domains. Overall, the expression changes at domain boundaries in the two DLPFC datasets are relatively smaller than those observed in the SCC dataset. At the L1$|$L2 and L6$|$WM interfaces in the DLPFC data, as well as at several domain boundaries in the SCC data, a certain fraction of genes exhibit substantial relative expression changes between the boundary and its neighboring regions. The especially large jumps in the SCC data partly reflect its high sparsity: for the many genes that are near-zero in one region but expressed in the adjacent one, the relative-change measure $2|m_a - m_{ab}| / (m_a + m_{ab})$ approaches its upper bound of 2, producing the many boxes concentrated near this value.

\begin{figure}[htbp]
    \centering
    \includegraphics[width=\linewidth]{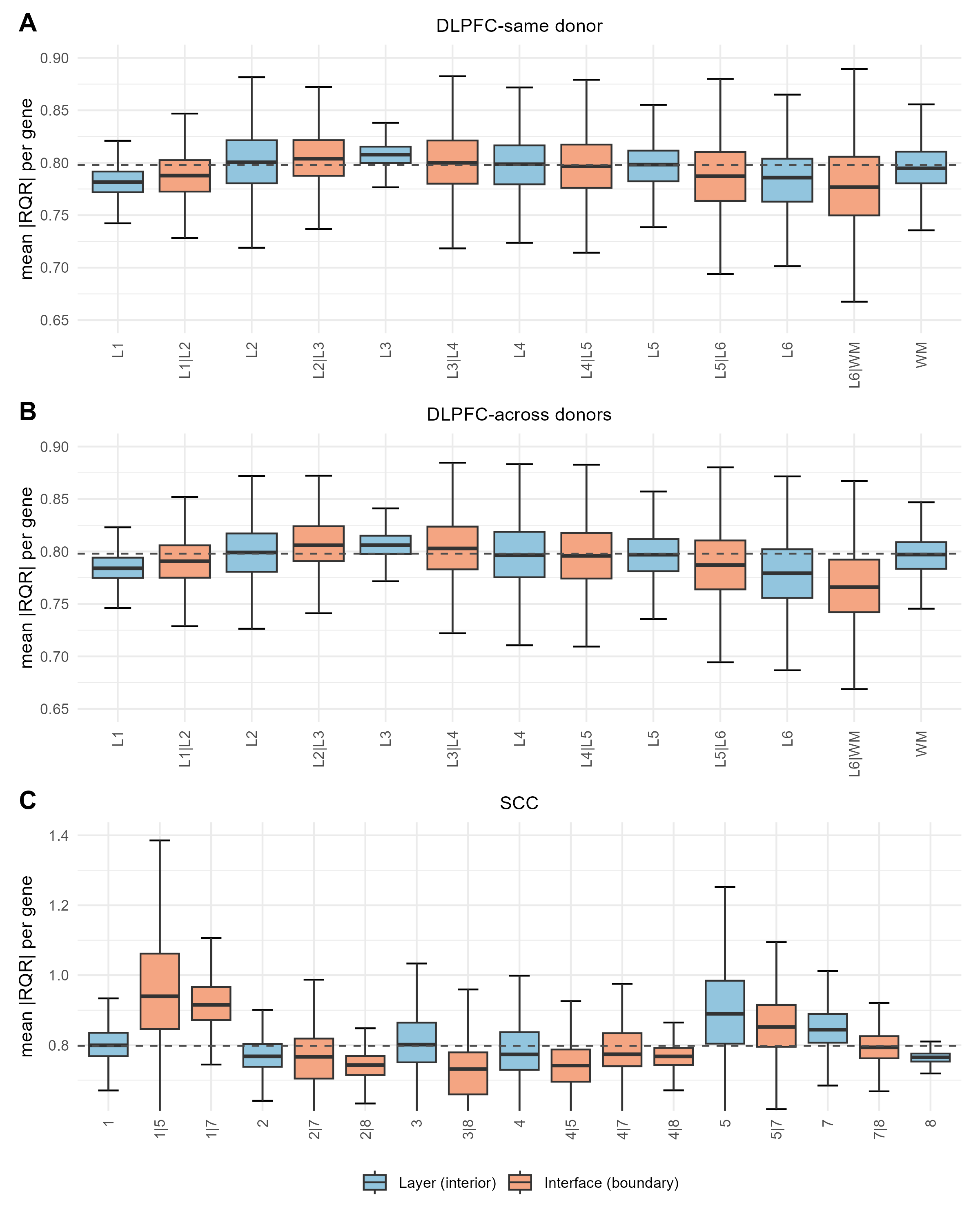}
    \caption{Boxplots of the per-gene mean absolute randomized quantile residuals
    ($|\text{RQR}|$) across interior regions and boundaries in three datasets: (A) DLPFC
    same-donor, (B) DLPFC across-donor, and (C) SCC. Blue and orange boxes denote interior
    layers and boundary interfaces, respectively. The dashed line marks the theoretical
    mean $\sqrt{2/\pi}$ expected under a well-calibrated $\mathcal{N}(0,1)$ residual. }
    \label{real_rqr_estimated}
\end{figure}

To assess whether these spatial transitions affect model fit, we focus on the SV genes identified by our method which are mostly influenced by spatial structure and examine their RQRs from the ZINB model with the nonparametric B-spline component. Figure \ref{real_rqr_estimated} shows the distribution of the per-gene mean $|\text{RQR}|$ across interior regions and boundaries. Since the RQRs are calibrated to $\mathcal{N}(0,1)$ \citep{Bai2021RQR}, a well-fitting model yields values centered near $\sqrt{2/\pi}$. Across all three datasets, the boundary residuals are not systematically elevated relative to the interior, indicating that the model provides comparable fits across regions with different levels of spatial variation.

\begin{figure}[htbp]
    \centering
    \includegraphics[width=\linewidth]{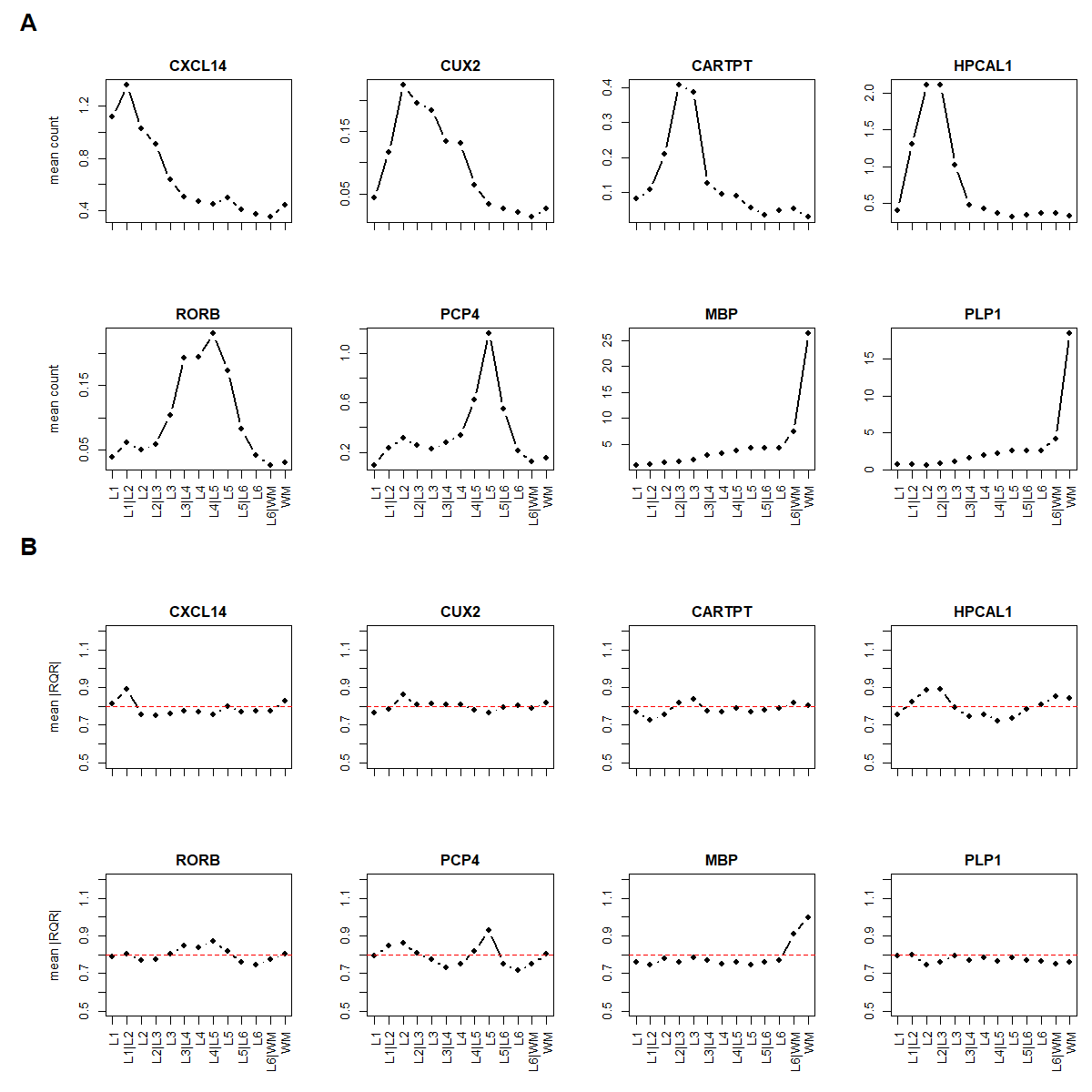}
    \caption{Expression profiles and residuals of eight canonical layer-marker genes along
    the ordered layer/interface sequence in the DLPFC same-donor dataset. (A) Mean
    expression of each gene across layers and interfaces. (B) Corresponding mean
    $|\text{RQR}|$ of each gene, with the dashed line marking $\sqrt{2/\pi}$.}
    \label{profile_rqr_samedonor}
\end{figure}

\begin{figure}[htbp]
    \centering
    \includegraphics[width=\linewidth]{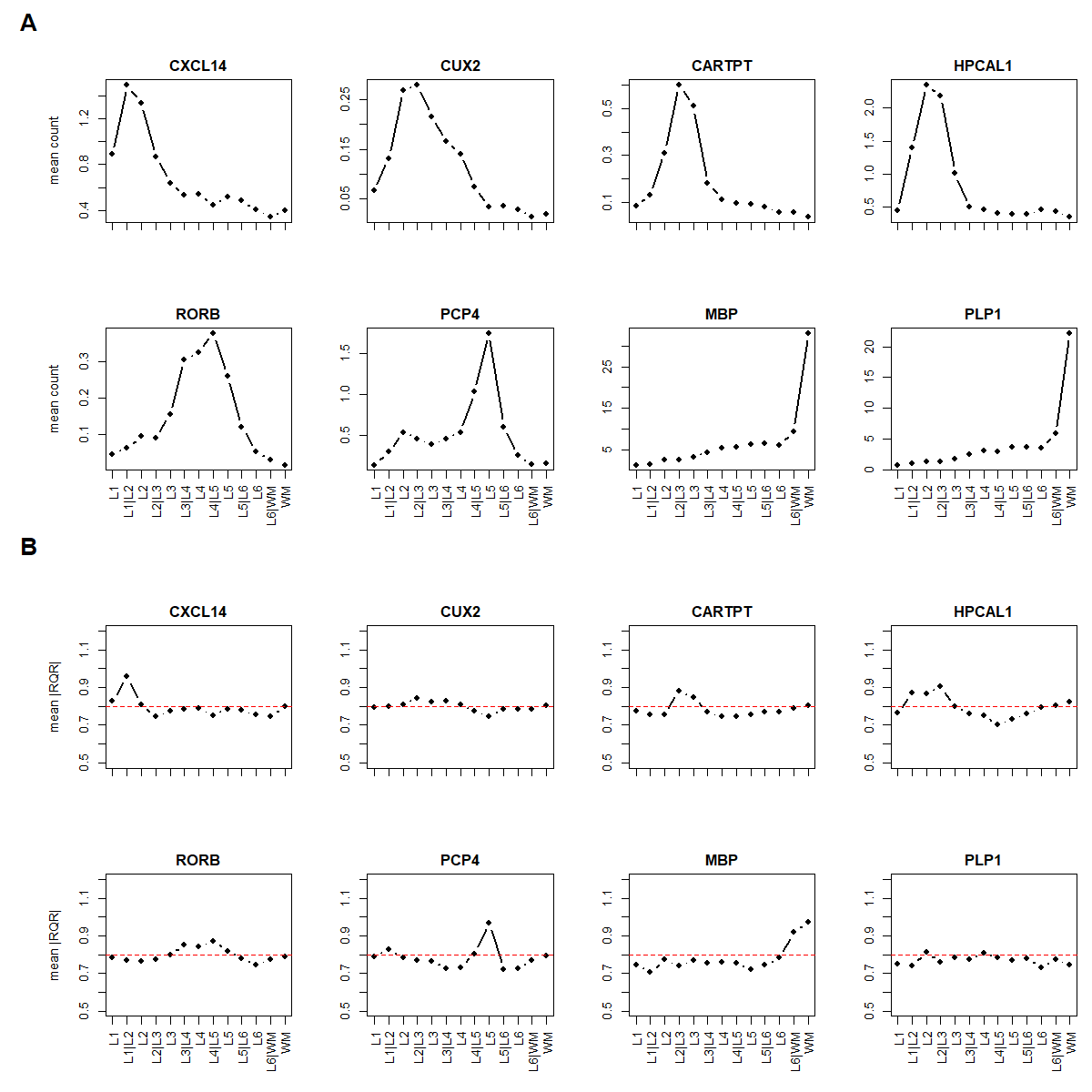}
    \caption{Expression profiles and residuals of eight canonical layer-marker genes along
    the ordered layer/interface sequence in the DLPFC across-donor dataset. (A) Mean
    expression of each gene across layers and interfaces. (B) Corresponding mean
    $|\text{RQR}|$ of each gene, with the dashed line marking $\sqrt{2/\pi}$.}
    \label{profile_rqr_acrossdonor}
\end{figure}

\begin{figure}[htbp]
    \centering
    \includegraphics[width=\linewidth]{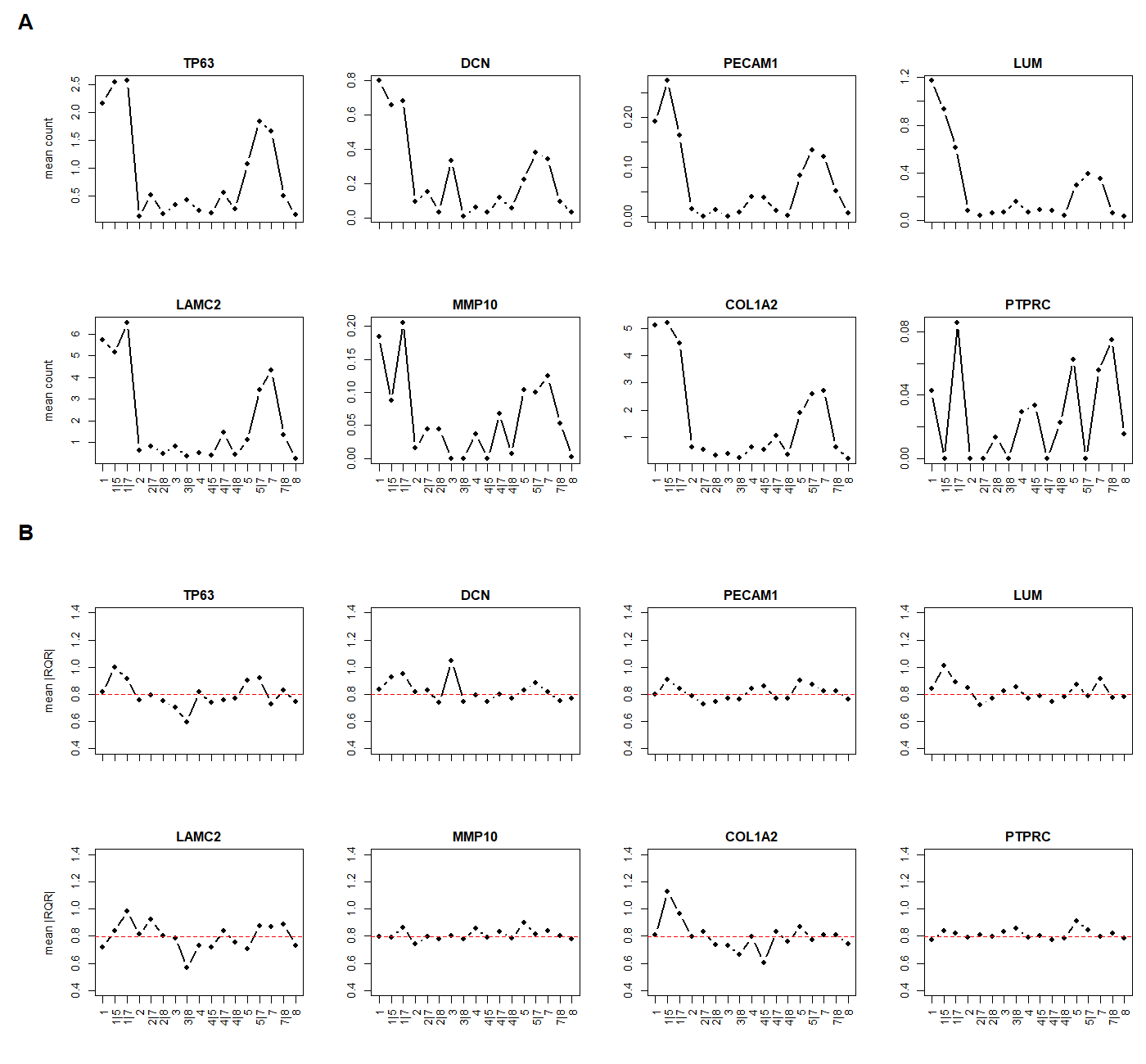}
    \caption{Mean expression profiles of representative marker genes along the
    cluster/boundary sequence in the SCC dataset. (A) Mean
    expression of each gene across layers and interfaces. (B) Corresponding mean
    $|\text{RQR}|$ of each gene, with the dashed line marking $\sqrt{2/\pi}$.}
    \label{profile_rqr_scc}
\end{figure}

We further investigate the specific genes showing the largest boundary-associated expression changes in Figure \ref{real_expression_jump}, which are also well-known marker genes, and examine them individually (Figures \ref{profile_rqr_samedonor}, \ref{profile_rqr_acrossdonor}, and \ref{profile_rqr_scc}). For the two DLPFC datasets, we examine eight canonical layer markers (CXCL14, CUX2, CARTPT, HPCAL1, RORB, PCP4, MBP, and PLP1 \citep{Zhang2024CXCL14,Suzuki2022CUX2,Kanat2026CART,Zhang2019HPCAL1,Lehrer2023RORB,Harashima2011PCP4,Foran1992MBP,Santarelli2019}), and for the SCC dataset we examine eight representative markers (TP63, DCN, PECAM1, LUM, LAMC2, MMP10, COL1A2 and PTPRC \citep{Jiang2024TP63,Rao2025DCN,Tang1993PECAM1,Matsuda2008LUM,Shan2023LAMC2,Mitsui2014IL24,Lin2020COL1A,AlBarashdi2021PTPRC}) . Although these markers display substantial changes in mean expression along the layer or cluster sequence, as illustrated in panel~(A) of the three figures, their mean $|\text{RQR}|$ values in panel~(B) remain stable and close to the theoretical level, confirming that they are well fitted. A small number of genes with the most extreme transitions, such as MBP in the DLPFC data and COL1A2 in SCC, show a modest increase in $|\text{RQR}|$ at the most pronounced boundary transition, yet the residuals stay within a reasonable range. These results suggest that the B-spline basis provides an adequate representation of spatial expression patterns, including regions with substantial boundary-associated variation.

\begin{table}[htbp]
\centering
\caption{Voung test comparisons are conducted in the four sections of same-donor DLPFC dataset between the proposed method and alternative ZINB-linear, ZINB-focal, and ZINB-period where the spatial terms $b_{1}^{(m)}$ and $b_{2}^{(m)}$ are replaced by linear, focal, and periodic functions, respectively. We also compare the best competing model which chooses the best performance across the three spatial modes for comparison. The table summarizes, for each pairwise comparison, the number of genes falling into three categories based on p‑value significance tests: ``Significantly Better" ($p_{\text{proposed}>\text{alternative}} < 0.05$), ``No Significant Difference", and ``Significantly Worse" ($p_{\text{alternative}>\text{proposed}} < 0.05$). All p-values are adjusted using the Benjamini-Hochberg (BH) procedure. }
\label{tab:significance_comparison}
\begin{tabular}{lccc}
\hline
Section & Significantly Better & No Significant Difference & Significantly Worse \\
\midrule
 & \multicolumn{3}{c}{ZINB-linear model}  \\
1 & 1362 & 3335 & 211 \\
2 & 1463 & 3208 & 237 \\
3 & 1651 & 3126 & 131 \\
4 & 1494 & 3237 & 177 \\
\midrule
 & \multicolumn{3}{c}{ZINB-focal model}\\
1 & 1549 & 3324 & 35 \\
2 & 1317 & 3537 & 54 \\
3 & 1667 & 3186 & 55 \\
4 & 1477 & 3392 & 39 \\
\midrule
 &\multicolumn{3}{c}{ZINB-period model}  \\
1 & 1577 & 3296 & 35 \\
2 & 1434 & 3438 & 36 \\
3 & 1889 & 3000 & 19 \\
4 & 1596 & 3276 & 36 \\
\midrule
 &\multicolumn{3}{c}{Best competing model}  \\
1 & 1015 & 3635 & 258 \\
2 & 981  & 3636 & 291 \\
3 & 1258 & 3459 & 191 \\
4 & 1081 & 3589 & 238 \\
\hline
\end{tabular}
\end{table}

\begin{table}[htbp]
\centering
\caption{Voung test comparisons are conducted in the four sections of DLPFC dataset across donors between the proposed method and alternative ZINB-linear, ZINB-focal, and ZINB-period where the spatial terms $b_{1}^{(m)}$ and $b_{2}^{(m)}$ are replaced by linear, focal, and periodic functions, respectively. We also compare the best competing model which chooses the best performance across the three spatial modes for comparison. The table summarizes, for each pairwise comparison, the number of genes falling into three categories based on p‑value significance tests: ``Significantly Better" ($p_{\text{proposed}>\text{alternative}} < 0.05$), ``No Significant Difference", and ``Significantly Worse" ($p_{\text{alternative}>\text{proposed}} < 0.05$). All p-values are adjusted using the Benjamini-Hochberg (BH) procedure.}
\label{tab:significance_comparison2}
\begin{tabular}{lccc}
\hline
Section & Significantly Better & No Significant Difference & Significantly Worse \\
\midrule
 & \multicolumn{3}{c}{ZINB-linear model}  \\
1 & 1458 & 2824 & 124 \\
2 & 1374 & 2850 & 182 \\
3 & 2035 & 2277 & 94  \\
4 & 2114 & 2106 & 186 \\
\midrule
 & \multicolumn{3}{c}{ZINB-focal model}\\
1 & 1515 & 2827 & 64  \\
2 & 1373 & 2974 & 59  \\
3 & 1905 & 2447 & 54  \\
4 & 3267 & 1125 & 14  \\
\midrule
 &\multicolumn{3}{c}{ZINB-period model}  \\
1 & 1665 & 2720 & 21  \\
2 & 1450 & 2903 & 53  \\
3 & 2218 & 2172 & 16  \\
4 & 3239 & 1162 & 5   \\
\midrule
 &\multicolumn{3}{c}{Best competing model}  \\
1 & 1159 & 3050 & 197 \\
2 & 1035 & 3106 & 265 \\
3 & 1661 & 2594 & 151 \\
4 & 1997 & 2210 & 199 \\
\hline
\end{tabular}
\end{table}

\begin{table}[htbp]
\centering
\caption{Voung test comparisons are conducted in three sections of SCC dataset between the proposed method and alternative ZINB-linear, ZINB-focal, and ZINB-period  where the spatial terms $b_{1}^{(m)}$ and $b_{2}^{(m)}$ are replaced by linear, focal, and periodic functions, respectively. We also compare the best competing model which chooses the best performance across the three spatial modes for comparison. The table summarizes, for each pairwise comparison, the number of genes falling into three categories based on p‑value significance tests: ``Significantly Better" ($p_{\text{proposed}>\text{alternative}} < 0.05$), ``No Significant Difference", and ``Significantly Worse" ($p_{\text{alternative}>\text{proposed}} < 0.05$). All p-values are adjusted using the Benjamini-Hochberg (BH) procedure.}
\label{tab:significance_comparison3}
\begin{tabular}{lccc}
\hline
Section & Significantly Better & No Significant Difference & Significantly Worse \\
\midrule
 & \multicolumn{3}{c}{ZINB-linear model}  \\
1 & 1273 & 3099 & 65 \\
2 & 1321 & 2976 & 140 \\
3 & 521  & 3786 & 130 \\
\midrule
 & \multicolumn{3}{c}{ZINB-focal model}\\
1 & 1636 & 2787 & 14 \\
2 & 1457 & 2951 & 29 \\
3 & 1439 & 2991 & 7 \\
\midrule
 &\multicolumn{3}{c}{ZINB-period model}  \\
1 & 2235 & 2195 & 7 \\
2 & 2171 & 2256 & 10 \\
3 & 1567 & 2852 & 18 \\
\midrule
 &\multicolumn{3}{c}{Best competing model}  \\
1 & 1243 & 3116 & 78 \\
2 & 1278 & 2989 & 170 \\
3 & 478  & 3806 & 153 \\
\hline
\end{tabular}
\end{table}

\newpage
\subsection{Stability analysis}
\label{Stability}

We assess the algorithm's stability across all three datasets using two complementary approaches. First, to evaluate stability with respect to sample composition, we perform integrated analyses on subsets of the available tissue slices (e.g., two out of three or four) and compare the similarity of the identified SV genes both across different subset combinations and against the full-set result. Second, to assess robustness to spatial sampling, we randomly subsample 90\% of the spots within each slice while integrating all samples and compare the output to the result from the complete spatial data. Similarity is quantified using the Jaccard index between sets of SV genes. The results, detailed in Tables \ref{inter_robust_dlpfc}-\ref{scc_robust_resample}, consistently show that our method achieves satisfactory stability. Its performance is comparable to established parametric methods like nnSVG and often surpasses several nonparametric or semiparametric alternatives (e.g., spVC, SPARKX). While some methods (e.g., DESpace, HEARTSVG-PASTE) exhibit high Jaccard indices by designating nearly all genes as SV, this yields limited biological interpretability. Overall, the stability of an integrated method largely depends on the underlying single-sample analysis, with our method demonstrating robust and reliable performance across diverse datasets.

\begin{table}[htbp]
\setlength{\belowcaptionskip}{-2pt} 
\centering
\caption{Stability analysis of SV gene detection in same-donor DLPFC dataset through subsampling validation: Jaccard indices ($\text{Jaccard}(X,Y) = \frac{|X \cap Y|}{|X \cup Y|}$) are shown for pairwise comparisons between (A,B) vs (C,D) (Comparison 1), (A,C) vs (B,D) (Comparison 2), and (A,D) vs (B,C) (Comparison 3), as well as between integration of sample pairs (AB/AC/AD/BC/BD/CD) and full integration of four samples (ABCD), with \#SV indicating the total number of spatially variable genes identified by full integration.}
\footnotesize
\begin{tabular}{lcccccccccc}
\toprule
Method& \# SV  & 1 & 2 & 3 &AB&AC&AD&BC&BD&CD\\
\midrule
IBaySVG & 1092 & 0.398 & 0.443 & 0.396 & 0.517 & 0.654 & 0.588 & 0.544 & 0.553 & 0.617 \\
DESpace & 4727 & 0.881 & 0.884 & 0.893 & 0.902 & 0.928 & 0.924 & 0.920 & 0.908 & 0.944 \\
nnSVG-Union & 357 & 0.496 & 0.482 & 0.501 & 0.686 & 0.874 & 0.675 & 0.826 & 0.608 & 0.810 \\
nnSVG-Inter & 96 & 0.615 & 0.596 & 0.623 & 0.807 & 0.658 & 0.756 & 0.780 & 0.865 & 0.722 \\
nnSVG-Cauchy & 167 & 0.527 & 0.547 & 0.562 & 0.740 & 0.693 & 0.744 & 0.738 & 0.780 & 0.699 \\
nnSVG-HMP & 276 & 0.515 & 0.520 & 0.540 & 0.711 & 0.876 & 0.696 & 0.826 & 0.622 & 0.785 \\
nnSVG-PASTE & 86 & 0.839 & 0.940 & 0.744 & 0.711 & 0.761 & 0.688 & 0.904 & 0.768 & 0.606 \\
SPARKX-Union & 2405 & 0.416 & 0.570 & 0.551 & 0.480 & 0.753 & 0.854 & 0.696 & 0.817 & 0.936 \\
SPARKX-Inter & 509 & 0.400 & 0.564 & 0.536 & 0.902 & 0.649 & 0.607 & 0.822 & 0.812 & 0.418 \\
SPARKX-Cauchy & 947 & 0.399 & 0.566 & 0.508 & 0.764 & 0.695 & 0.631 & 0.776 & 0.789 & 0.515 \\
SPARKX-HMP & 1930 & 0.407 & 0.578 & 0.556 & 0.479 & 0.742 & 0.837 & 0.702 & 0.820 & 0.893 \\
SPARKX-PASTE & 4829 & 0.972 & 0.964 & 0.937 & 0.983 & 0.982 & 0.984 & 0.934 & 0.958 & 0.962 \\
SPARK-Union & 1083 & 0.669 & 0.688 & 0.696 & 0.820 & 0.907 & 0.834 & 0.862 & 0.781 & 0.849 \\
SPARK-Inter & 454 & 0.649 & 0.669 & 0.678 & 0.845 & 0.718 & 0.768 & 0.852 & 0.906 & 0.737 \\
SPARK-Cauchy & 661 & 0.649 & 0.669 & 0.685 & 0.808 & 0.801 & 0.830 & 0.818 & 0.826 & 0.792 \\
SPARK-HMP & 969 & 0.672 & 0.697 & 0.703 & 0.827 & 0.906 & 0.848 & 0.849 & 0.784 & 0.840 \\
SPARK-PASTE & 504 & 0.399 & 0.392 & 0.360 & 0.381 & 0.489 & 0.353 & 0.488 & 0.365 & 0.291 \\
spVC-Union & 1984 & 0.285 & 0.289 & 0.291 & 0.588 & 0.675 & 0.642 & 0.649 & 0.614 & 0.697 \\
spVC-Inter & 82 & 0.187 & 0.191 & 0.192 & 0.364 & 0.276 & 0.301 & 0.347 & 0.381 & 0.278 \\
spVC-Cauchy & 1141 & 0.289 & 0.290 & 0.293 & 0.545 & 0.646 & 0.602 & 0.605 & 0.550 & 0.652 \\
spVC-HMP & 1120 & 0.291 & 0.288 & 0.289 & 0.555 & 0.659 & 0.606 & 0.617 & 0.562 & 0.668 \\
spVC-PASTE & 3862 & 0.076 & 0.066 & 0.638 & 0.624 & 0.709 & 0.610 & 0.714 & 0.069 & 0.077 \\
HEARTSVG-Union & 1574 & 0.550 & 0.577 & 0.591 & 0.750 & 0.812 & 0.738 & 0.853 & 0.765 & 0.800 \\
HEARTSVG-Inter & 541 & 0.625 & 0.657 & 0.675 & 0.875 & 0.730 & 0.776 & 0.837 & 0.867 & 0.686 \\
HEARTSVG-Cauchy & 752 & 0.639 & 0.676 & 0.674 & 0.789 & 0.829 & 0.843 & 0.801 & 0.817 & 0.807 \\
HEARTSVG-HMP & 1439 & 0.561 & 0.585 & 0.604 & 0.747 & 0.816 & 0.746 & 0.852 & 0.763 & 0.808 \\
HEARTSVG-PASTE & 4877 & 0.999 & 0.993 & 0.998 & 0.994 & 0.994 & 0.993 & 0.994 & 0.990 & 0.993 \\
GASTON-Union & 4171 & 0.703 & 0.699 & 0.714 & 0.861 & 0.838 & 0.853 & 0.861 & 0.860 & 0.842 \\
GASTON-Inter & 1003 & 0.389 & 0.386 & 0.396 & 0.571 & 0.555 & 0.576 & 0.558 & 0.559 & 0.549 \\
GASTON-PASTE & 1053 & 0.665 & 0.690 & 0.653 & 0.662 & 0.705 & 0.680 & 0.577 & 0.685 & 0.636 \\
StarTrail-Union & 136 & 0.559 & 0.529 & 0.544 & 0.838 & 0.838 & 0.779 & 0.765 & 0.691 & 0.721 \\
StarTrail-Inter & 41 & 0.631 & 0.594 & 0.612 & 0.820 & 0.641 & 0.707 & 0.820 & 0.891 & 0.732 \\
StarTrail-PASTE & 286 & 0.773 & 0.319 & 0.607 & 0.434 & 0.476 & 0.203 & 0.265 & 0.161 & 0.360 \\
\bottomrule
\end{tabular}
\label{inter_robust_dlpfc}
\end{table}

\begin{table}[htbp]
\setlength{\belowcaptionskip}{-2pt} 
\renewcommand{\arraystretch}{1.25}
\centering
\caption{Stability analysis of SV gene detection in same-donor DLPFC dataset through subsampling validation: Jaccard indices ($\text{Jaccard}(X,Y) = \frac{|X \cap Y|}{|X \cup Y|}$) compare integration results using 90\% randomly subsampled spots versus full original spots across four samples, with each cell displaying mean (standard deviation) from 50 sampling iterations. \#SV is the total number of spatially variable genes identified by full integration.}
\footnotesize
\begin{tabular}{lcc|lcc}
\toprule
Method & \#SV & Jaccard & Method & \#SV & Jaccard \\
\midrule
IBaySVG       &1092 &0.747 (0.000) & DESpace        &4727 &0.977 (0.000) \\
nnSVG-Union   &357  &0.732 (0.001) & nnSVG-Inter   &96   &0.842 (0.000) \\
nnSVG-Cauchy  &167  &0.779 (0.001) & nnSVG-HMP     &276  &0.756 (0.000) \\
nnSVG-PASTE   &86   &0.712 (0.001) & SPARKX-Union  &2405 &0.672 (0.000) \\
SPARKX-Inter  &509  &0.509 (0.000) & SPARKX-Cauchy &947  &0.592 (0.000) \\
SPARKX-HMP    &1930 &0.641 (0.000) & SPARKX-PASTE  &4829 &0.983 (0.000) \\
SPARK-Union   &1083 &0.892 (0.000) & SPARK-Inter   &454  &0.894 (0.000) \\
SPARK-Cauchy  &661  &0.906 (0.000) & SPARK-HMP     &969  &0.900 (0.000) \\
SPARK-PASTE   &504  &0.438 (0.006) & spVC-Union    &1984 &0.421 (0.013) \\
spVC-Inter    &82   &0.246 (0.027) & spVC-Cauchy   &1141 &0.382 (0.017) \\
spVC-HMP      &1120 &0.384 (0.011) & spVC-PASTE    &3862 &0.639 (0.003) \\
HEARTSVG-Union &1574 &0.771 (0.001) & HEARTSVG-Inter &541  &0.864 (0.000) \\
HEARTSVG-Cauchy&752  &0.882 (0.000) & HEARTSVG-HMP   &1439 &0.794 (0.001) \\
HEARTSVG-PASTE &4877 &0.994 (0.000) & GASTON-Union   &4171 &0.896 (0.002) \\
GASTON-Inter   &1003 &0.374 (0.001) & GASTON-PASTE   &1053 &0.661 (0.006) \\
StarTrail-Union&136  &0.276 (0.000) & StarTrail-Inter&41   &0.189 (0.003) \\
StarTrail-PASTE&286  &0.244 (0.004) &                &     &              \\
\bottomrule
\end{tabular}
\end{table}

\begin{table}[htbp]
\setlength{\belowcaptionskip}{-2pt} 
\centering
\caption{Stability analysis of SV gene detection in across-donor DLPFC dataset through subsampling validation: Jaccard indices ($\text{Jaccard}(X,Y) = \frac{|X \cap Y|}{|X \cup Y|}$) are shown for pairwise comparisons between (A,B) vs (C,D) (Comparison 1), (A,C) vs (B,D) (Comparison 2), and (A,D) vs (B,C) (Comparison 3), as well as between integration of sample pairs (AB/AC/AD/BC/BD/CD) and full integration of four samples (ABCD), with \#SV indicating the total number of spatially variable genes identified by full integration.}
\footnotesize
\begin{tabular}{lcccccccccc}
\toprule
Method& \# SV  & 1 & 2 & 3 &AB&AC&AD&BC&BD&CD\\
\midrule
IBaySVG & 1590 & 0.352 & 0.768 & 0.750 & 0.568 & 0.580 & 0.542 & 0.576 & 0.547 & 0.540 \\
DESpace & 4094 & 0.862 & 0.873 & 0.879 & 0.937 & 0.932 & 0.914 & 0.933 & 0.911 & 0.880 \\
nnSVG-Union & 457 & 0.389 & 0.497 & 0.488 & 0.534 & 0.764 & 0.832 & 0.656 & 0.733 & 0.856 \\
nnSVG-Inter & 83 & 0.401 & 0.525 & 0.512 & 0.716 & 0.669 & 0.585 & 0.806 & 0.709 & 0.477 \\
nnSVG-Cauchy & 195 & 0.396 & 0.491 & 0.466 & 0.633 & 0.696 & 0.642 & 0.721 & 0.693 & 0.614 \\
nnSVG-HMP & 350 & 0.400 & 0.488 & 0.482 & 0.554 & 0.749 & 0.787 & 0.667 & 0.714 & 0.807 \\
nnSVG-PASTE & 111 & 0.658 & 0.736 & 0.661 & 0.874 & 0.917 & 0.638 & 0.948 & 0.685 & 0.575 \\
SPARKX-Union & 4030 & 0.387 & 0.631 & 0.612 & 0.393 & 0.665 & 0.972 & 0.640 & 0.966 & 0.994 \\
SPARKX-Inter & 650 & 0.271 & 0.459 & 0.436 & 0.942 & 0.554 & 0.487 & 0.807 & 0.729 & 0.276 \\
SPARKX-Cauchy & 1945 & 0.324 & 0.533 & 0.486 & 0.578 & 0.736 & 0.688 & 0.704 & 0.725 & 0.561 \\
SPARKX-HMP & 3773 & 0.337 & 0.577 & 0.554 & 0.349 & 0.613 & 0.961 & 0.586 & 0.957 & 0.974 \\
SPARKX-PASTE & 4400 & 0.998 & 0.988 & 0.995 & 0.997 & 0.987 & 0.999 & 0.994 & 0.999 & 0.999 \\
SPARK-Union & 1374 & 0.507 & 0.582 & 0.560 & 0.557 & 0.667 & 0.926 & 0.634 & 0.915 & 0.951 \\
SPARK-Inter & 381 & 0.517 & 0.601 & 0.574 & 0.728 & 0.828 & 0.620 & 0.886 & 0.686 & 0.640 \\
SPARK-Cauchy & 665 & 0.562 & 0.656 & 0.621 & 0.835 & 0.797 & 0.790 & 0.782 & 0.819 & 0.670 \\
SPARK-HMP & 1240 & 0.511 & 0.582 & 0.557 & 0.565 & 0.666 & 0.917 & 0.630 & 0.908 & 0.935 \\
SPARK-PASTE & 770 & 0.346 & 0.212 & 0.172 & 0.226 & 0.330 & 0.418 & 0.271 & 0.457 & 0.114 \\
spVC-Union & 1827 & 0.401 & 0.435 & 0.432 & 0.655 & 0.756 & 0.681 & 0.752 & 0.679 & 0.746 \\
spVC-Inter & 333 & 0.484 & 0.533 & 0.529 & 0.707 & 0.669 & 0.733 & 0.654 & 0.724 & 0.605 \\
spVC-Cauchy & 1165 & 0.425 & 0.459 & 0.469 & 0.634 & 0.731 & 0.689 & 0.724 & 0.671 & 0.732 \\
spVC-HMP & 1108 & 0.433 & 0.483 & 0.483 & 0.631 & 0.763 & 0.690 & 0.753 & 0.681 & 0.759 \\
spVC-PASTE & 4386 & 0.969 & 0.992 & 0.995 & 0.977 & 0.991 & 0.995 & 0.993 & 0.995 & 0.982 \\
HEARTSVG-Union & 2074 & 0.399 & 0.456 & 0.466 & 0.546 & 0.799 & 0.719 & 0.747 & 0.657 & 0.853 \\
HEARTSVG-Inter & 413 & 0.472 & 0.546 & 0.561 & 0.603 & 0.750 & 0.639 & 0.821 & 0.667 & 0.685 \\
HEARTSVG-Cauchy & 707 & 0.469 & 0.536 & 0.553 & 0.787 & 0.684 & 0.797 & 0.711 & 0.801 & 0.599 \\
HEARTSVG-HMP & 1909 & 0.400 & 0.464 & 0.472 & 0.553 & 0.803 & 0.711 & 0.752 & 0.652 & 0.835 \\
HEARTSVG-PASTE & 4400 & 0.995 & 0.999 & 0.997 & 0.996 & 0.998 & 0.999 & 0.996 & 0.999 & 0.997 \\
GASTON-Union & 3819 & 0.702 & 0.698 & 0.711 & 0.822 & 0.854 & 0.860 & 0.850 & 0.843 & 0.880 \\
GASTON-Inter & 873 & 0.373 & 0.370 & 0.378 & 0.544 & 0.546 & 0.528 & 0.572 & 0.535 & 0.543 \\
GASTON-PASTE & 1047 & 0.548 & 0.621 & 0.562 & 0.654 & 0.718 & 0.675 & 0.705 & 0.732 & 0.655 \\
StarTrail-Union & 129 & 0.519 & 0.457 & 0.550 & 0.868 & 0.853 & 0.837 & 0.713 & 0.605 & 0.651 \\
StarTrail-Inter & 19 & 0.345 & 0.302 & 0.373 & 0.373 & 0.306 & 0.826 & 0.404 & 0.950 & 0.826 \\
StarTrail-PASTE & 216 & 0.571 & 0.505 & 0.778 & 0.452 & 0.231 & 0.407 & 0.481 & 0.390 & 0.350 \\
\bottomrule
\end{tabular}
\label{inter_robust_dlpfc_across}
\end{table}

\begin{table}[htbp]
\setlength{\belowcaptionskip}{-2pt} 
\renewcommand{\arraystretch}{1.25}
\centering
\caption{Stability analysis of SV gene detection in across-donor DLPFC dataset through subsampling validation: Jaccard indices ($\text{Jaccard}(X,Y) = \frac{|X \cap Y|}{|X \cup Y|}$) compare integration results using 90\% randomly subsampled spots versus full original spots across four samples, with each cell displaying mean (standard deviation) from 50 sampling iterations. \#SV is the total number of spatially variable genes identified by full integration.}
\footnotesize
\begin{tabular}{lcc|lcc}
\toprule
Method & \#SV & Jaccard & Method & \#SV & Jaccard \\
\midrule
proposed        &1590 &0.770 (0.002) & DEspace        &4094 &0.977 (0.000) \\
nnSVG-Union     &457  &0.748 (0.000) & nnSVG-Inter    &83   &0.836 (0.001) \\
nnSVG-Cauchy    &195  &0.752 (0.001) & nnSVG-HMP      &350  &0.776 (0.000) \\
nnSVG-PASTE     &111  &0.967 (0.000) & SPARKX-Union   &4030 &0.953 (0.000) \\
SPARKX-Inter    &650  &0.847 (0.000) & SPARKX-Cauchy  &1945 &0.840 (0.000) \\
SPARKX-HMP      &3773 &0.950 (0.000) & SPARKX-PASTE   &4400 &0.998 (0.000) \\
SPARK-Union     &1374 &0.901 (0.000) & SPARK-Inter    &381  &0.894 (0.000) \\
SPARK-Cauchy    &665  &0.910 (0.000) & SPARK-HMP      &1240 &0.916 (0.000) \\
SPARK-PASTE     &770  &0.538 (0.002) & spVC-Union     &1827 &0.668 (0.000) \\
spVC-Inter      &333  &0.840 (0.000) & spVC-Cauchy    &1165 &0.732 (0.000) \\
spVC-HMP        &1108 &0.745 (0.000) & spVC-PASTE     &4386 &0.678 (0.000) \\
HEARTSVG-Union  &2074 &0.774 (0.000) & HEARTSVG-Inter &413  &0.789 (0.000) \\
HEARTSVG-Cauchy &707  &0.861 (0.000) & HEARTSVG-HMP   &1909 &0.780 (0.000) \\
HEARTSVG-PASTE  &4400 &0.999 (0.000) & GASTON-Union   &3819 &0.895 (0.003) \\
GASTON-Inter    &873  &0.309 (0.001) & GASTON-PASTE   &1047 &0.755 (0.004) \\
StarTrail-Union &129  &0.336 (0.006) & StarTrail-Inter&19   &0.247 (0.001) \\
StarTrail-PASTE &216  &0.231 (0.004) &                &     &              \\
\bottomrule
\end{tabular}
\end{table}

\begin{table}[htbp]
\setlength{\belowcaptionskip}{-2pt} 
\centering
\caption{Stability analysis of SV gene detection in SCC dataset through subsampling validation: Jaccard indices ($\text{Jaccard}(X,Y) = \frac{|X \cap Y|}{|X \cup Y|}$) are shown for pairwise comparisons between (A,B) vs (A,C) (Comparison 1), (A,B) vs (B,C) (Comparison 2), and (A,C) vs (B,C) (Comparison 3), as well as between integration of sample pairs (AB/AC/BC) and full integration of three samples (ABC), with \#SV indicating the total number of spatially variable genes identified by full integration.}
\footnotesize
\begin{tabular}{lccccccc}
\toprule
Method& \# SV  & 1 & 2 & 3 &AB&AC&BC\\
\midrule
IBaySVG        & 1051 & 0.624 & 0.588 & 0.519 & 0.562 & 0.552 & 0.590 \\
DESpace         & 3269 & 0.893 & 0.862 & 0.833 & 0.957 & 0.909 & 0.885 \\
nnSVG-Union     & 391  & 0.658 & 0.686 & 0.426 & 0.796 & 0.610 & 0.645 \\
nnSVG-Inter     & 56   & 0.362 & 0.354 & 0.750 & 0.387 & 0.810 & 0.746 \\
nnSVG-Cauchy    & 198  & 0.558 & 0.566 & 0.587 & 0.656 & 0.718 & 0.720 \\
nnSVG-HMP       & 286  & 0.603 & 0.739 & 0.480 & 0.800 & 0.616 & 0.731 \\
nnSVG-PASTE     & 3938 & 0.764 & 0.770 & 0.684 & 0.879 & 0.769 & 0.763 \\
SPARKX-Union    & 3399 & 0.877 & 0.853 & 0.775 & 0.978 & 0.899 & 0.876 \\
SPARKX-Inter    & 1692 & 0.694 & 0.685 & 0.891 & 0.720 & 0.950 & 0.934 \\
SPARKX-Cauchy   & 2565 & 0.768 & 0.759 & 0.848 & 0.819 & 0.925 & 0.915 \\
SPARKX-HMP      & 3015 & 0.852 & 0.843 & 0.764 & 0.951 & 0.873 & 0.874 \\
SPARKX-PASTE    & 4427 & 0.993 & 0.992 & 0.996 & 0.994 & 0.996 & 0.995 \\
SPARK-Union     & 256  & 0.684 & 0.816 & 0.562 & 0.969 & 0.715 & 0.848 \\
SPARK-Inter     & 71   & 0.563 & 0.538 & 0.835 & 0.582 & 0.947 & 0.877 \\
SPARK-Cauchy    & 129  & 0.618 & 0.597 & 0.761 & 0.738 & 0.822 & 0.796 \\
SPARK-HMP       & 208  & 0.699 & 0.788 & 0.586 & 0.930 & 0.736 & 0.835 \\
SPARK-PASTE     & 4429 & 0.995 & 0.995 & 0.995 & 0.998 & 0.996 & 0.996 \\
spVC-Union      & 809  & 0.426 & 0.604 & 0.308 & 0.671 & 0.403 & 0.561 \\
spVC-Inter      & 39   & 0.348 & 0.322 & 0.488 & 0.406 & 0.709 & 0.609 \\
spVC-Cauchy     & 362  & 0.360 & 0.483 & 0.323 & 0.525 & 0.428 & 0.547 \\
spVC-HMP        & 390  & 0.431 & 0.600 & 0.302 & 0.653 & 0.416 & 0.581 \\
spVC-PASTE      & 104  & 0.678 & 0.638 & 0.673 & 0.746 & 0.769 & 0.750 \\
HEARTSVG-Union  & 173  & 0.237 & 0.301 & 0.647 & 0.445 & 0.792 & 0.855 \\
HEARTSVG-Inter  & 4    & 0.444 & 0.267 & 0.333 & 0.500 & 0.800 & 0.364 \\
HEARTSVG-Cauchy & 12   & 0.360 & 0.269 & 0.265 & 0.600 & 0.478 & 0.375 \\
HEARTSVG-HMP    & 145  & 0.245 & 0.293 & 0.656 & 0.416 & 0.792 & 0.845 \\
HEARTSVG-PASTE  & 256  & 0.816 & 0.784 & 0.866 & 0.778 & 0.893 & 0.902 \\
GASTON-Union    & 3736 & 0.787 & 0.748 & 0.778 & 0.879 & 0.908 & 0.870 \\
GASTON-Inter    & 996  & 0.498 & 0.536 & 0.486 & 0.711 & 0.625 & 0.686 \\
GASTON-PASTE    & 1162 & 0.397 & 0.411 & 0.491 & 0.405 & 0.472 & 0.510 \\
StarTrail-Union & 180  & 0.978 & 0.689 & 0.667 & 1.000 & 0.978 & 0.689 \\
StarTrail-Inter & 55   & 0.458 & 0.495 & 0.859 & 0.495 & 0.859 & 1.000 \\
StarTrail-PASTE & 249  & 0.856 & 0.767 & 0.768 & 0.002 & 0.004 & 0.002 \\
\bottomrule
\end{tabular}
\label{tab:scc_robust_Inter}
\end{table}

\begin{table}[ht]
\centering
\renewcommand{\arraystretch}{1.25}
\caption{Stability analysis of SV gene detection in SCC dataset through subsampling validation: Jaccard indices ($\text{Jaccard}(X,Y) = \frac{|X \cap Y|}{|X \cup Y|}$) compare integration results using 90\% randomly subsampled spots versus full original spots across three samples, with each cell displaying mean (standard deviation) from 50 sampling iterations. \#SV is the total number of spatially variable genes identified by full integration.}
\label{scc_robust_resample}
\footnotesize
\begin{tabular}{lcc|lcc}
\toprule
Method & \#SV & Jaccard & Method & \#SV & Jaccard \\
\midrule
IBaySVG & 1051 & 0.851(0.000) & DESpace & 3269 & 0.986(0.000) \\
nnSVG-Union & 391 & 0.802(0.000) & nnSVG-Inter & 56 & 0.762(0.000) \\
nnSVG-Cauchy & 198 & 0.817(0.000) & nnSVG-HMP & 286 & 0.825(0.000) \\
nnSVG-PASTE & 3938 & 0.824(0.000) & SPARKX-Union & 3399 & 0.983(0.000) \\
SPARKX-Inter & 1692 & 0.970(0.000) & SPARKX-Cauchy & 2565 & 0.971(0.000) \\
SPARKX-HMP & 3015 & 0.978(0.000) & SPARKX-PASTE & 4427 & 0.997(0.000) \\
SPARK-Union & 256 & 0.956(0.000) & SPARK-Inter & 71 & 0.964(0.000) \\
SPARK-Cauchy & 129 & 0.950(0.000) & SPARK-HMP & 208 & 0.959(0.000) \\
SPARK-PASTE & 4429 & 0.996(0.000) & spVC-Union & 809 & 0.643(0.000) \\
spVC-Inter & 39 & 0.838(0.000) & spVC-Cauchy & 362 & 0.655(0.000) \\
spVC-HMP & 390 & 0.715(0.000) & spVC-PASTE & 104 & 0.866(0.000) \\
HEARTSVG-Union & 173 & 0.879(0.000) & HEARTSVG-Inter & 4 & 0.858(0.004) \\
HEARTSVG-Cauchy & 12 & 0.834(0.000) & HEARTSVG-HMP & 145 & 0.880(0.000) \\
HEARTSVG-PASTE & 256 & 0.969(0.000) & GASTON-Union & 3736 & 0.852(0.002) \\
GASTON-Inter & 996 & 0.360(0.000) & GASTON-PASTE & 1162 & 0.083(0.004) \\
StarTrail-Union & 180 & 0.333(0.008) & StarTrail-Inter & 55 & 0.239(0.000) \\
StarTrail-PASTE & 249 & 0.285(0.003) & & & \\
\bottomrule
\end{tabular}
\end{table}

\newpage
\subsection{Sensitivity analysis}\label{Sensitive-real}

\begin{figure}[!ht]
    \centering
    \includegraphics[width=\linewidth]{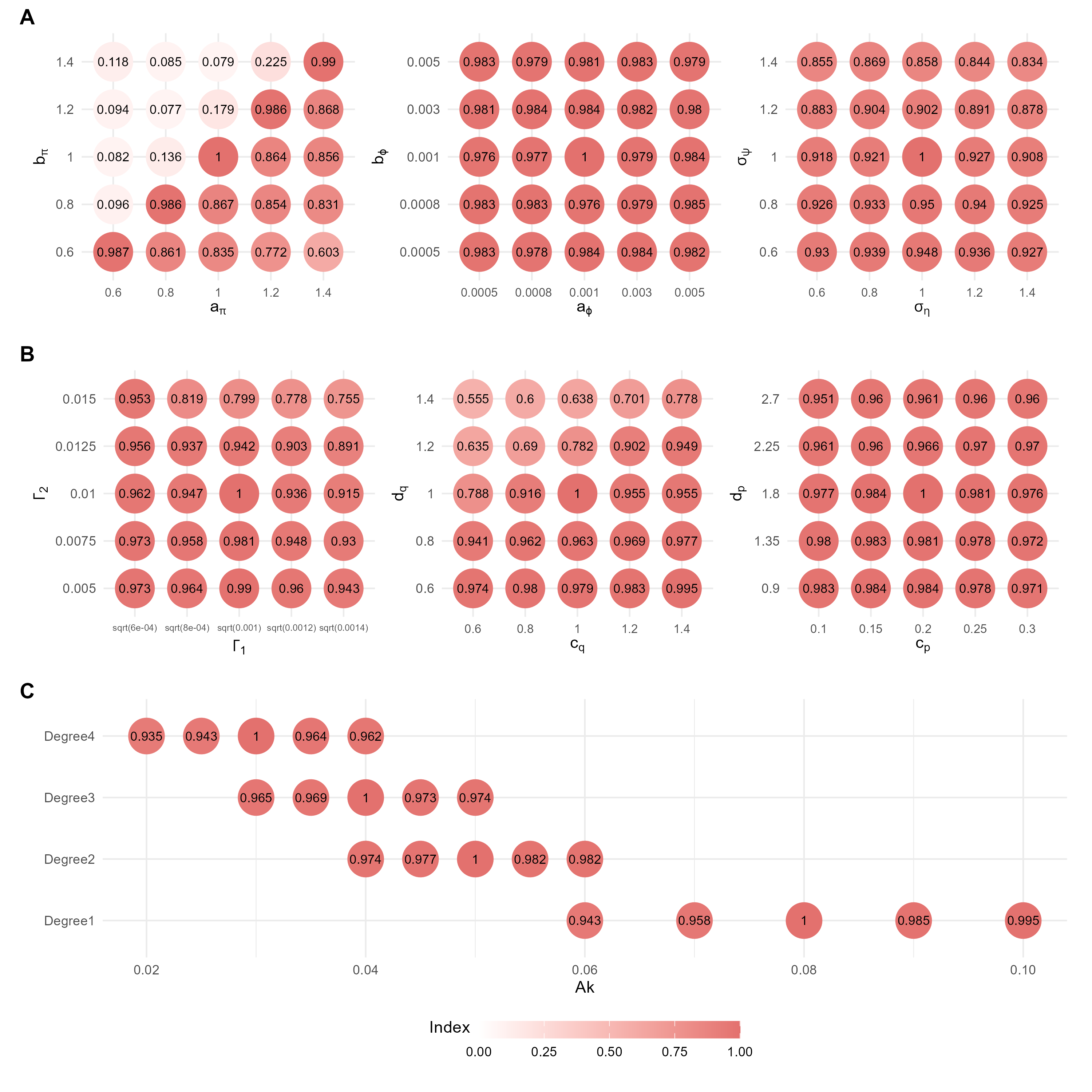}
    \caption{Sensitivity analysis of hyperparameters in the DLPFC dataset (same donor). The results report the overlap proportion based on the identified spatially variable genes under the range of parameters. (a) $a_{\pi} \sim b_{\pi}$; (b) $a_{\phi} \sim b_{\phi}$; (c) $\sigma_{\eta} \sim \sigma_{\psi}$; (d) $\Gamma_{1}\sim \Gamma_{2}$; (e) $c_{q} \sim d_{q}$; (f) $c_{p} \sim d_{p}$; (g) $A_k$ for different orders of the basis functions (Degrees 1–4).}
    \label{fig:sensitive_combined_dlpfc}
\end{figure}

\begin{figure}[!ht]
    \centering
    \includegraphics[width=\linewidth]{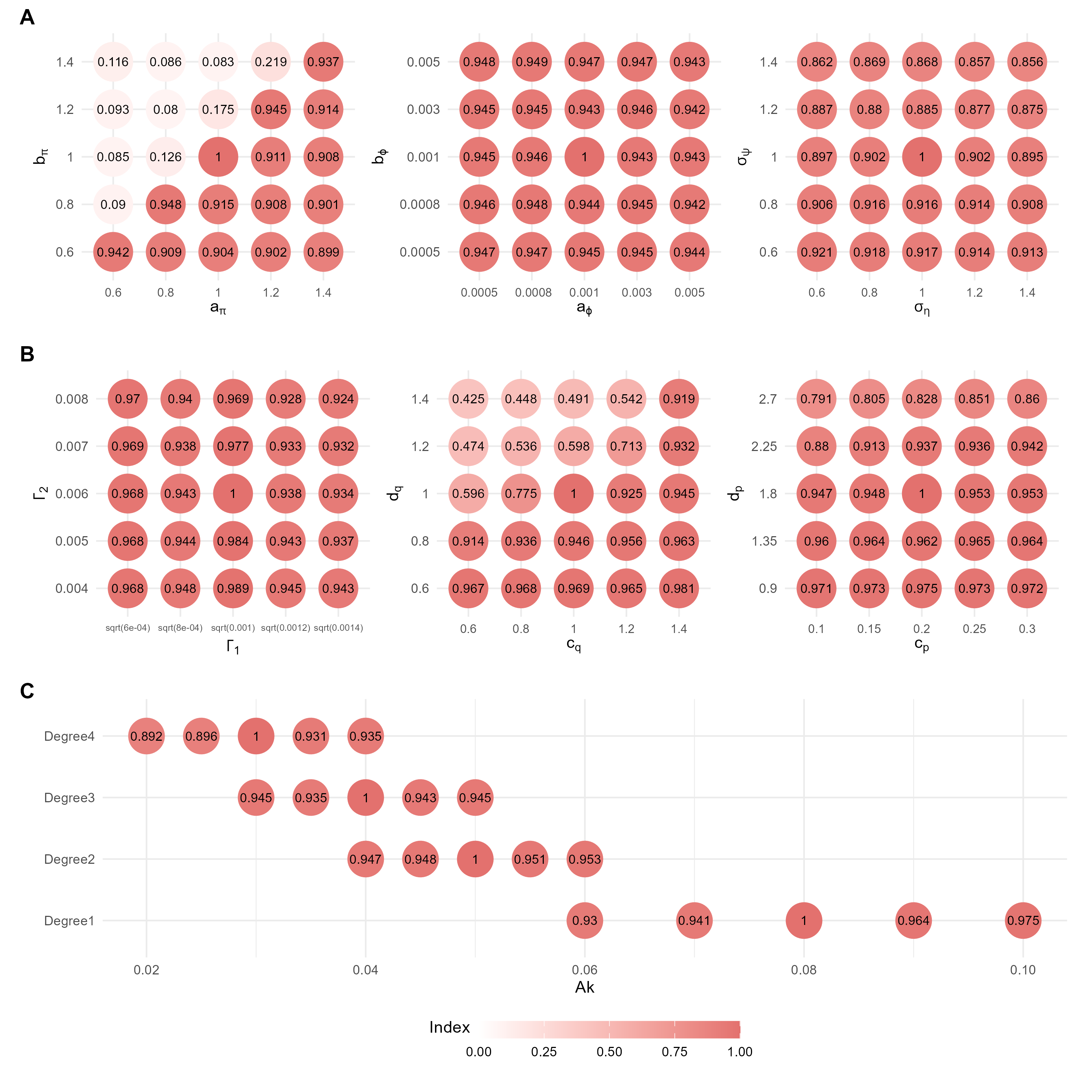}
    \caption{Sensitivity analysis of hyperparameters in the DLPFC dataset (across donors). The results report the overlap proportion based on the identified spatially variable genes under the range of parameters. (a) $a_{\pi} \sim b_{\pi}$; (b) $a_{\phi} \sim b_{\phi}$; (c) $\sigma_{\eta} \sim \sigma_{\psi}$; (d) $\Gamma_{1}\sim \Gamma_{2}$; (e) $c_{q} \sim d_{q}$; (f) $c_{p} \sim d_{p}$; (g) $A_k$ for different orders of the basis functions (Degrees 1–4).}
    \label{fig:sensitive_combined_dlpfc_cross}
\end{figure}

\begin{figure}[!ht]
    \centering
    \includegraphics[width=\linewidth]{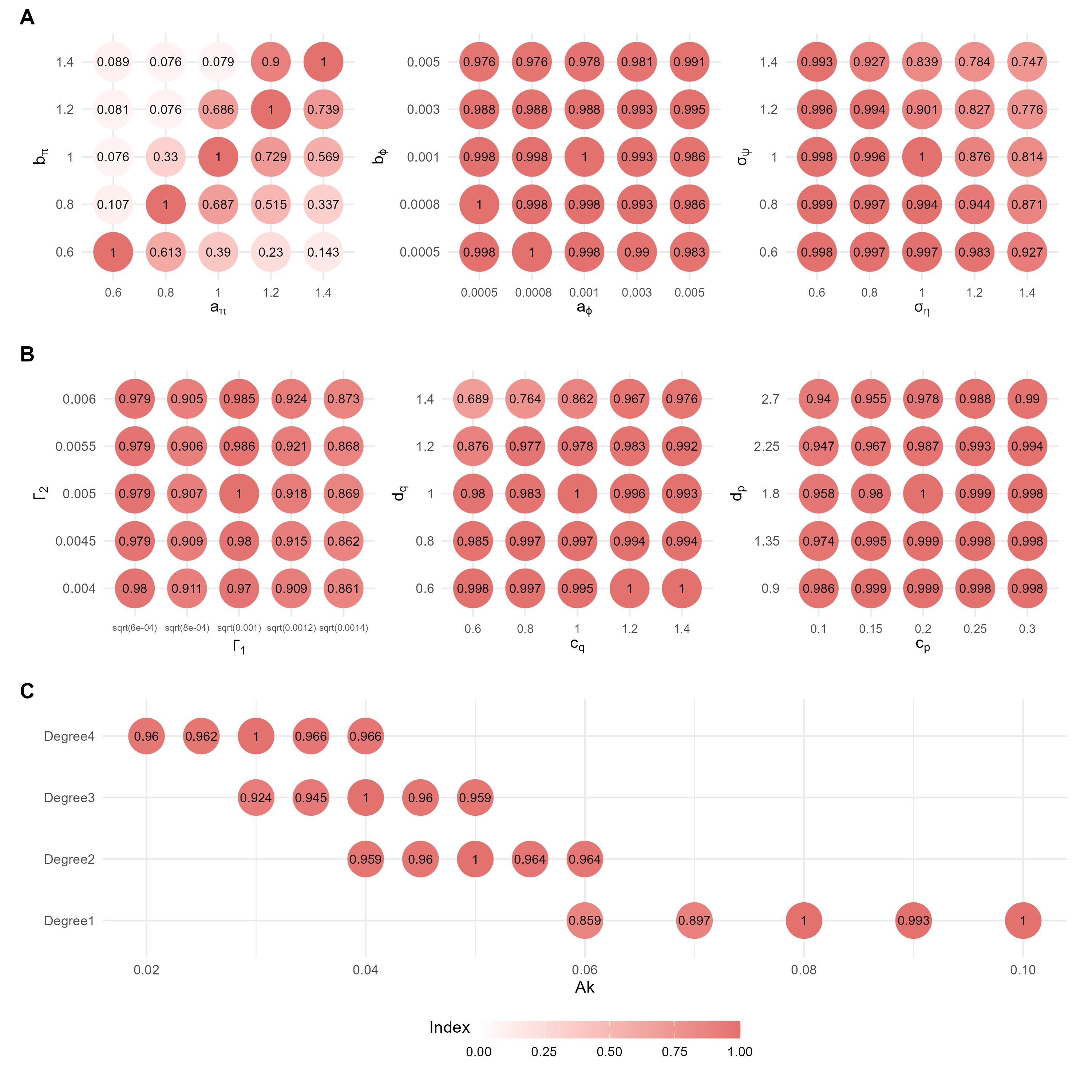}
    \caption{Sensitivity analysis of hyperparameters in the SCC dataset. The results report the overlap proportion based on the identified spatially variable genes under the range of parameters. (a) $a_{\pi} \sim b_{\pi}$; (b) $a_{\phi} \sim b_{\phi}$; (c) $\sigma_{\eta} \sim \sigma_{\psi}$; (d) $\Gamma_{1}\sim \Gamma_{2}$; (e) $c_{q} \sim d_{q}$; (f) $c_{p} \sim d_{p}$; (g) $A_k$ for different orders of the basis functions (Degrees 1–4).}
    \label{fig:sensitive_combined_SCC}
\end{figure}

Similar to the simulation study, we also perform a sensitivity analysis on three spatial transcriptomics datasets. For each sensitivity test, all other hyperparameters are held fixed and the examined ranges remain consistent with those in simulation (Section \ref{Sensitive-simulation}), except that for data involving multiple donors $\Gamma_{2} \in \{0.004, 0.005, \mathbf{0.006}, 0.007, 0.008\}$ to better capture inter‑donor heterogeneity; for integrating three sections such as SCC dataset, $\Gamma_{2} \in \{0.004, 0.0045, \mathbf{0.005}, 0.0055, 0.006\}$ to better balance the contribution of individual slices. Since the true SV gene labels are not available, taking the default combination as the reference, we compute the overlap proportion of identified SV genes between each alternative setting and the reference. Results are shown in Figures \ref{fig:sensitive_combined_dlpfc}-\ref{fig:sensitive_combined_SCC}. Across all three datasets and the considered ranges, the results are robust, with most overlap proportions exceeding 90\%. For the zero‑inflation Beta priors, we recommend $a_{\pi}=b_{\pi}$ (prior expectation 0.5), under which stability is also maintained.

\clearpage

\subsection{Upset plots of SV gene identification results}

\begin{figure}[htbp]
    \centering
    \includegraphics[width=0.65\linewidth]{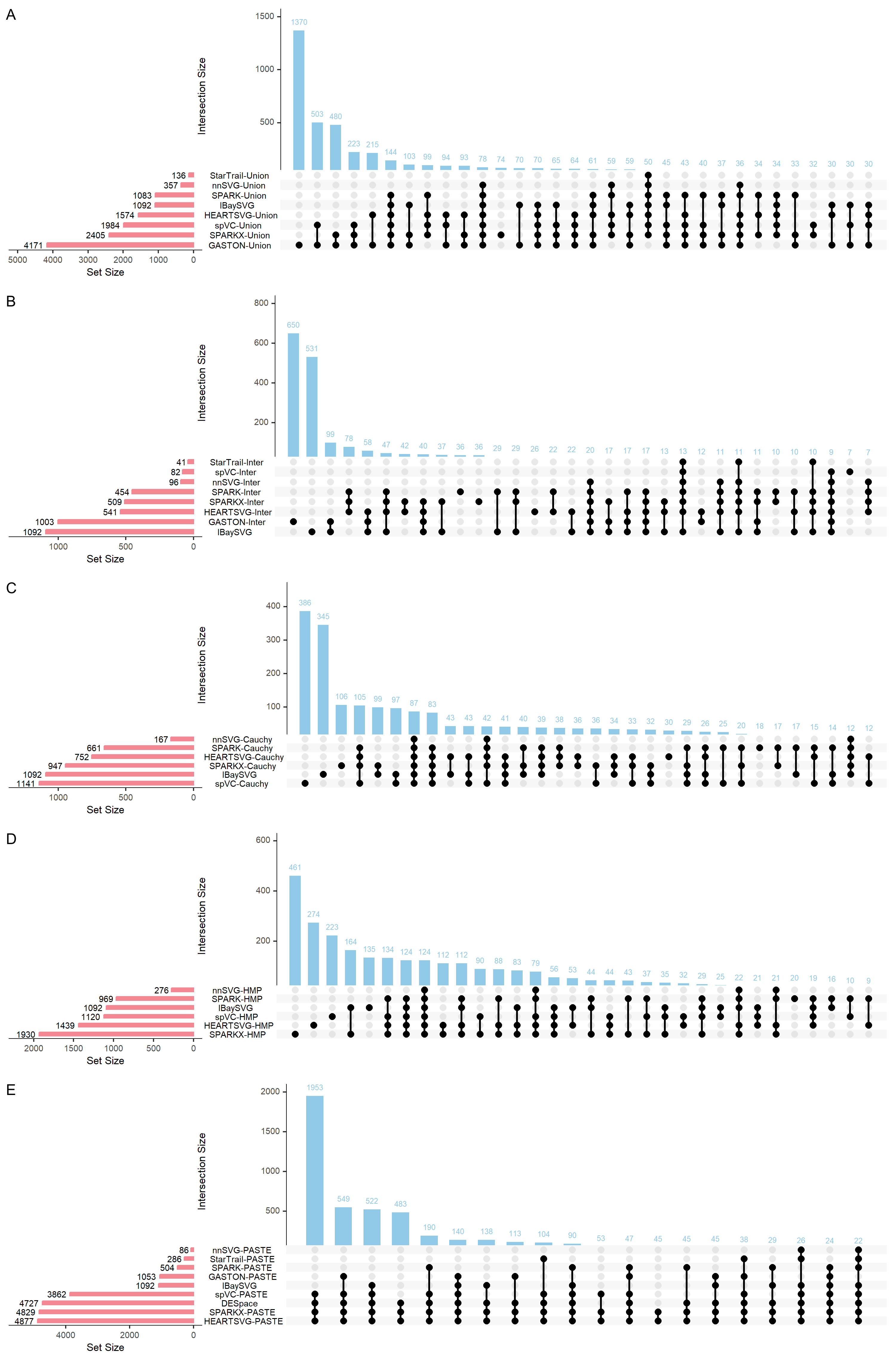}
    \caption{Comparative analysis of SV gene identification through multiple integration strategies for the same-donor DLPFC dataset: (A) proposed method and union approaches, (B) proposed method and intersection approaches, (C) proposed method and Cauchy combination approaches, (D) proposed method and harmonic mean p-value approaches, and (E) proposed method, DESpace, and PASTE frameworks.}
    \label{fig:upset}
\end{figure}

\begin{figure}[!ht]
    \centering
    \includegraphics[width=0.7\linewidth]{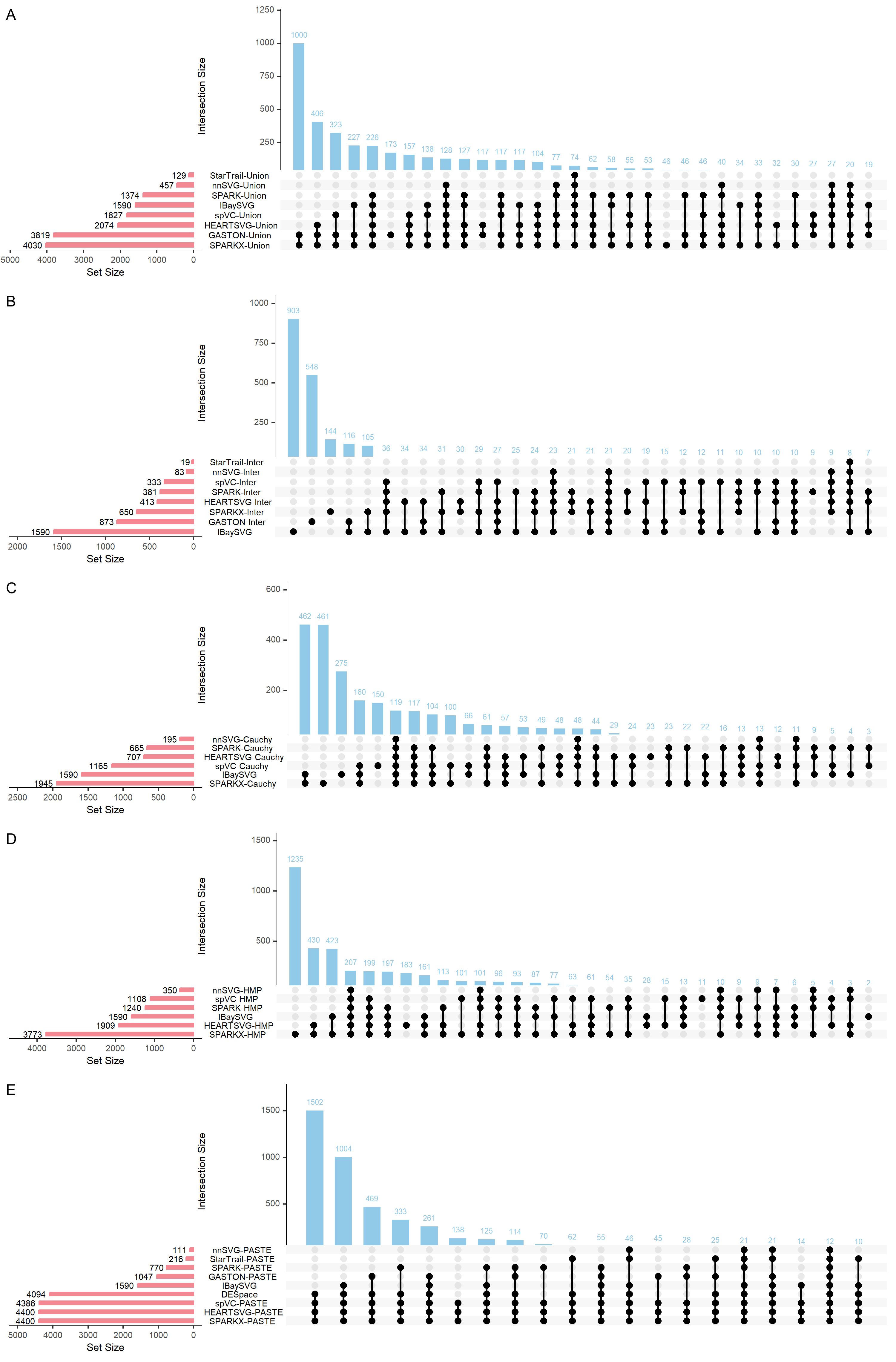}
    \caption{Comparative analysis of SV gene identification through multiple integration strategies for the across-donor DLPFC dataset: (A) proposed method and union approaches, (B) proposed method and intersection approaches, (C) proposed method and Cauchy combination approaches, (D) proposed method and harmonic mean p-value approaches, and (E) proposed method, DESpace, and PASTE frameworks.}
    \label{fig:upset_combined_34711}
\end{figure}

\begin{figure}[!ht]
    \centering
    \includegraphics[width=0.7\linewidth]{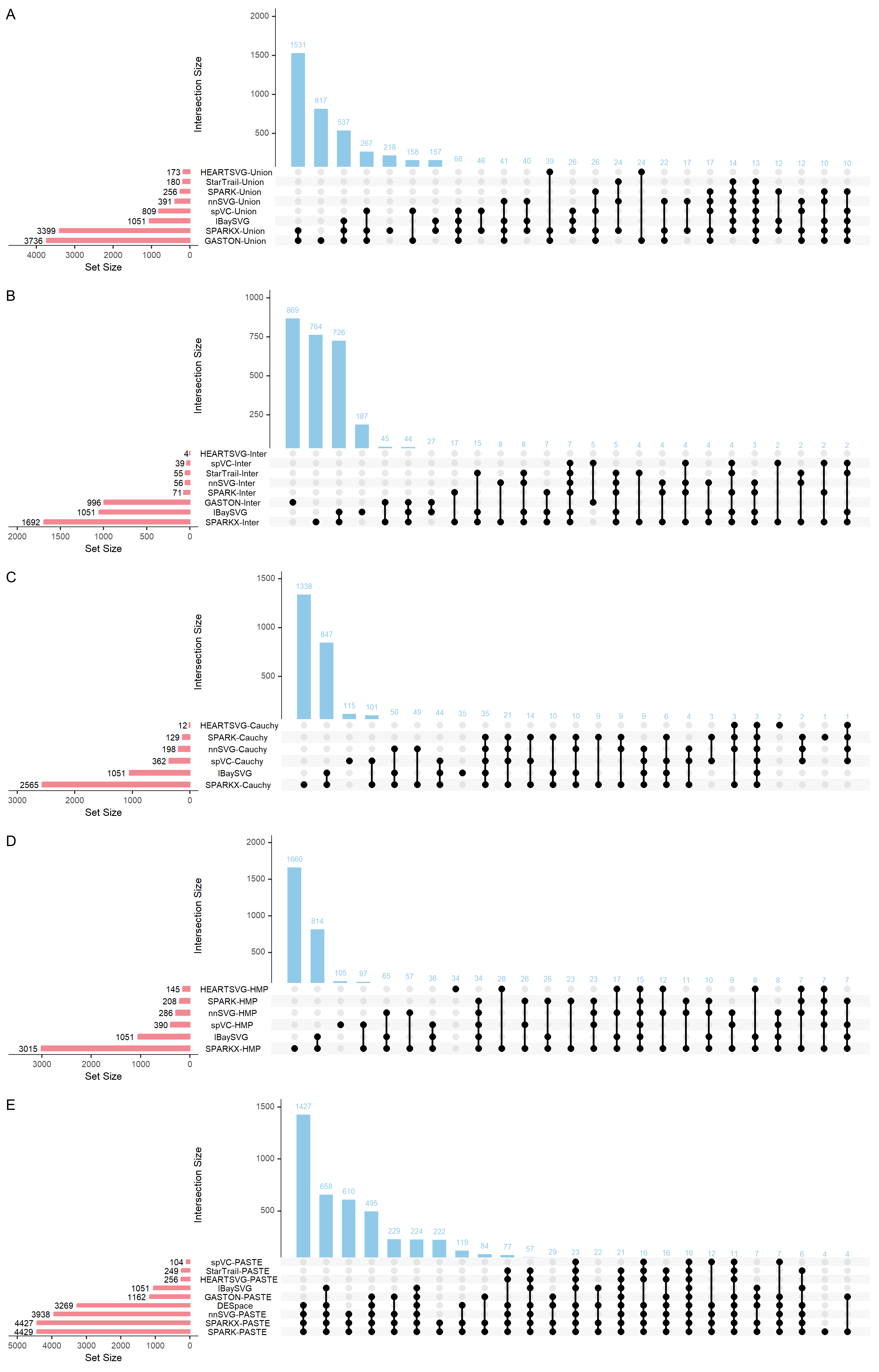}
    \caption{Comparative analysis of SV gene identification through multiple integration strategies in SCC dataset: (A) proposed method and union approaches, (B) proposed method and intersection approaches, (C) proposed method and Cauchy combination approaches, (D) proposed method and harmonic mean p-value approaches, (E) proposed method, DESpace, and PASTE frameworks. }
    \label{fig:upset_scc}
\end{figure}

\clearpage

\subsection{Benchmarking performance in identifying layer-specific marker genes}
% The complete list of marker genes is provided in GitHub repository.
\begin{table}[htbp]
\footnotesize
\renewcommand{\arraystretch}{1}
\centering
\caption{List of selected layer-specific marker genes from two reference datasets.}
\label{tab:performance_combined_list}
\begin{tabular}{cccccccc}
\toprule
 \multicolumn{4}{c}{\textbf{Genes from Maynard et al.~(2021)}} 
& \multicolumn{4}{c}{\textbf{Genes from Zeng et al.~(2012)}}  \\
\midrule
NDRG1 & PTP4A2 & AQP1 & PAQR6 & CTGF & LMO4 & ETV1 & PCP4 \\
ANP32B & JAM3 & PHLDB1 & PMP22 & CXCL14 & NTNG2 & NR4A2 & CUX2 \\
MTUS1 & PLEKHG3 & MAL & LAMP2 & RELN & CALB1 & B3GALT2 & CBLN2 \\
PRKCSH & TMEM206 & CERS2 & NENF & RXFP1 & CDH24 & GABRA5 & CARTPT \\
SLAIN1 & RNASE1 & AMER2 & SLC44A1 & OPRK1 & RPRM & CACNA1E & FEZF2 \\
MT1G & VIM & FABP7 & LINC00052 & HTR2C & PDE1A & VAT1L & GSG1L \\
MT1F & MYL9 & F3 & CDC42EP4 & C1QL2 & MEF2C & BHLHE22 & SEMA3E \\
ATP1A2 & SOX9 & PTN & CLU & TRIB2 & KCNIP2 & KITLG & SLITRK1 \\
MSX1 & CALD1 & MYH11 & MLC1 & SCN3B & KCNK2 & SYNPR & CCK \\
C11orf96 & MT1M & RELN & NCAN & FAM3C & ANXA1 & FXYD6 & RASGRF2 \\
DACT1 & MPL & STXBP6 & SIPA1L1 & DKK3 & RORB & TLE1 & GPR6 \\
MAN1A1 & DDX54 & CAMK2N1 & SERPINE2 & TOX & PCDH17 & POU3F1 & BEND5 \\
PAX2 & GNAL & HPCAL1 & IPO5 & UNC5D & ATP2B4 & CNR1 & TLE4 \\
C1QL2 & CPNE6 & VXN & C12orf29 & CRIM1 & CPNE7 & PENK & SYT6 \\
PSD3 & PDE8B & LAMP5 & KCNIP2 & TH & BCL11B & SATB2 & FOXP2 \\
CARTPT & BAIAP3 & ADCYAP1 & LINC01007 & TBR1 & CRYM & RAC3 & PDYN \\
CNGB1 & SPON2 & FREM3 & CA10 & KCNA1 & SNCG & SYT17 & KCNH4 \\
SNX3 & FAM84A & AC109779.1 & CALB1 & CHRNA7 & MFGE8 & LXN & ADRA2A \\
GRB14 & ADGRF5 & HAPLN4 & OXGR1 & AKR1C2 & CHRNA3 & MARCKSL1 & CYR61 \\
HOPX & GRIA4 & LINC01331 & HS6ST3 & AKR1C3 & NEFH & LMO3 & TMEM163 \\
VAMP1 & GUCA1C & NEFH & LINC01827 & LGALS1 & PRSS12 & SCN4B & PLXND1 \\
SCN1B & NGB & PVALB & TPBG & LHX2 & COL6A1 & NR4A3 & GRIK4 \\
SLC5A12 & NEFM & PLCH1 & TNNT2 & SYT2 & WFS1 & INPP4B & OPN3 \\
SLC25A5 & RORB & PTPRT & SYT2 & ID2 & OMA1 & IGFBP4 & COL24A1 \\
PCP4 & CAMK2D & SMYD2 & TRABD2A & CUX1 & FOXO1 & S100A10 & KCNIP1 \\
TMSB10 & HTR2C & VAT1L & RHO & TLE3 & SEMA3C & DTX4 & MDGA1 \\
PID1 & KIF17 & RAB27B & ETV1 & SOX5 & OTX1 & SV2C & LIX1 \\
IPCEF1 & RPRM & DPY19L1 & FAM3C & PCDH20 & TPBG & RGS8 & IGSF11 \\
SMARCD3 & SSBP3 & BHMT & PKD2L1 & LDB2 & SYT10 & NPY2R & CYP39A1 \\
ISLR & NR4A2 & MCTP2 & DACH1 & CNTN6 & DIAPH3 & PPP1R1B & SYT9 \\
NTNG2 & SMIM32 & OPRK1 & MMD &  &  &  &  \\
KRT17 & THEMIS & NPTX1 & PTPRU &  &  &  &  \\
OLFML2B & TUBA3D & THEM5 & B3GALT2 &  &  &  &  \\
SMYD1 & CPNE5 & MREG & MCTP1 &  &  &  &  \\
\bottomrule
\end{tabular}
\label{tab:marker_genes_combined} \\
\end{table}

\begin{table}[htbp]
\footnotesize
\renewcommand{\arraystretch}{1.25}
\centering
\caption{Performance comparison of different methods in identifying layer-specific marker genes. Results are based on the SV genes identified in DLPFC dataset with same donor.}
\label{tab:performance_combined_dlpfc_same}
\begin{tabular}{lccc|lccc}
\toprule
Method & TPR & FPR & F1 & Method & TPR & FPR & F1 \\
\midrule
IBaySVG  & 0.283 & 0.281 & 0.282 & DESpace  & 0.520 & 0.120 & 0.194 \\
nnSVG-AVE       & 0.076 & 0.026 & 0.129 & nnSVG-Union     & 0.130 & 0.050 & 0.195 \\
nnSVG-Inter     & 0.040 & 0.012 & 0.073 & nnSVG-PASTE     & 0.018 & 0.015 & 0.034 \\
nnSVG-Cauchy    & 0.065 & 0.022 & 0.113 & nnSVG-HMP       & 0.110 & 0.036 & 0.176 \\
SPARKX-AVE      & 0.270 & 0.231 & 0.245 & SPARKX-Union    & 0.407 & 0.452 & 0.253 \\
SPARKX-Inter    & 0.140 & 0.082 & 0.191 & SPARKX-PASTE    & 0.521 & 0.982 & 0.191 \\
SPARKX-Cauchy   & 0.243 & 0.157 & 0.260 & SPARKX-HMP      & 0.364 & 0.354 & 0.262 \\
SPARK-AVE       & 0.240 & 0.110 & 0.286 & SPARK-Union     & 0.292 & 0.176 & 0.292 \\
SPARK-Inter     & 0.183 & 0.059 & 0.258 & SPARK-PASTE     & 0.127 & 0.084 & 0.174 \\
SPARK-Cauchy    & 0.234 & 0.094 & 0.291 & SPARK-HMP       & 0.280 & 0.153 & 0.296 \\
spVC-AVE        & 0.158 & 0.137 & 0.185 & spVC-Union      & 0.333 & 0.374 & 0.236 \\
spVC-Inter      & 0.028 & 0.012 & 0.051 & spVC-PASTE      & 0.419 & 0.785 & 0.184 \\
spVC-Cauchy     & 0.247 & 0.201 & 0.241 & spVC-HMP        & 0.240 & 0.198 & 0.237 \\
HEARTSVG-AVE    & 0.250 & 0.159 & 0.265 & HEARTSVG-Union  & 0.333 & 0.279 & 0.272 \\
HEARTSVG-Inter  & 0.171 & 0.082 & 0.229 & HEARTSVG-PASTE  & 0.521 & 0.993 & 0.190 \\
HEARTSVG-Cauchy & 0.225 & 0.117 & 0.266 & HEARTSVG-HMP    & 0.319 & 0.252 & 0.274 \\
GASTON-AVE      & 0.337 & 0.532 & 0.195 & GASTON-Union    & 0.442 & 0.850 & 0.183 \\
GASTON-Inter    & 0.208 & 0.179 & 0.216 & GASTON-PASTE    & 0.109 & 0.215 & 0.110 \\
StarTrail-AVE   & 0.016 & 0.014 & 0.029 & StarTrail-Union & 0.025 & 0.025 & 0.044 \\
StarTrail-Inter & 0.007 & 0.008 & 0.014 & StarTrail-PASTE & 0.052 & 0.053 & 0.083 \\
\bottomrule
\end{tabular}
\end{table}

\begin{table}[htbp]
\centering
\renewcommand{\arraystretch}{1.25}
\caption{ Performance comparison of different methods in identifying layer-specific marker genes. Results are based on the SV genes identified in DLPFC dataset across donors.}
\label{tab:performance_combined_dlpfc_across}
\footnotesize
\begin{tabular}{lccc|lccc}
\toprule
Method & TPR & FPR & F1 & Method & TPR & FPR & F1 \\
\midrule
IBaySVG       & 0.320 & 0.218 & 0.259 & DESpace       & 0.485 & 0.129 & 0.203 \\
nnSVG-AVE      & 0.093 & 0.034 & 0.152 & nnSVG-Union   & 0.169 & 0.071 & 0.237 \\
nnSVG-Inter    & 0.036 & 0.011 & 0.067 & nnSVG-PASTE   & 0.082 & 0.027 & 0.139 \\
nnSVG-Cauchy   & 0.148 & 0.049 & 0.224 & nnSVG-HMP     & 0.025 & 0.022 & 0.045 \\
SPARK-AVE      & 0.253 & 0.135 & 0.291 & SPARK-Union   & 0.343 & 0.259 & 0.303 \\
SPARK-Inter    & 0.161 & 0.053 & 0.239 & SPARK-PASTE   & 0.233 & 0.106 & 0.289 \\
SPARK-Cauchy   & 0.330 & 0.228 & 0.308 & SPARK-HMP     & 0.127 & 0.163 & 0.149 \\
SPARKX-AVE     & 0.321 & 0.468 & 0.216 & SPARKX-Union  & 0.474 & 0.908 & 0.201 \\
SPARKX-Inter   & 0.150 & 0.126 & 0.188 & SPARKX-PASTE  & 0.338 & 0.408 & 0.242 \\
SPARKX-Cauchy  & 0.463 & 0.845 & 0.207 & SPARKX-HMP    & 0.490 & 0.999 & 0.194 \\
spVC-AVE       & 0.263 & 0.158 & 0.287 & spVC-Union    & 0.393 & 0.362 & 0.293 \\
spVC-Inter     & 0.148 & 0.044 & 0.227 & spVC-PASTE    & 0.334 & 0.207 & 0.323 \\
spVC-Cauchy    & 0.329 & 0.194 & 0.325 & spVC-HMP      & 0.491 & 0.995 & 0.195 \\
HEARTSVG-AVE   & 0.233 & 0.205 & 0.238 & HEARTSVG-Union& 0.330 & 0.443 & 0.227 \\
HEARTSVG-Inter & 0.144 & 0.066 & 0.208 & HEARTSVG-PASTE& 0.206 & 0.125 & 0.250 \\
HEARTSVG-Cauchy& 0.318 & 0.404 & 0.230 & HEARTSVG-HMP  & 0.492 & 0.998 & 0.195 \\
GASTON-AVE     & 0.292 & 0.545 & 0.180 & GASTON-Union  & 0.411 & 0.871 & 0.182 \\
GASTON-Inter   & 0.143 & 0.185 & 0.158 & GASTON-PASTE  & 0.120 & 0.237 & 0.122 \\
StarTrail-AVE  & 0.016 & 0.013 & 0.031 & StarTrail-Union& 0.034 & 0.024 & 0.061 \\
StarTrail-Inter& 0.006 & 0.003 & 0.011 & StarTrail-PASTE& 0.041 & 0.044 & 0.068 \\
\bottomrule
\end{tabular}
\end{table}

\clearpage

\subsection{Additional biological evidence of the identified SV genes for the same-donor DLPFC dataset}\label{supp-DLPFC-same}

\begingroup
\renewcommand{\arraystretch}{0.8}
\singlespacing                        
\footnotesize
\begin{longtable}{p{0.36\linewidth}p{0.65\linewidth}}
\caption{DLPFC same-donor dataset: curated biological evidence summary.}\\
\toprule
Gene & Brief functional summary \\
\midrule
\endfirsthead
\toprule
Gene & Brief functional summary \\
\midrule
\endhead
\midrule
\multicolumn{2}{r}{Continued on next page}\\
\midrule
\endfoot
\bottomrule
\endlastfoot
MAPK4 \citep{Yan2025a} & Regulates neuronal signaling and alpha-synuclein accumulation in models of neurodegeneration. \\
ADGRA3 \citep{Sun2021} & Adhesion GPCR expressed in cochlear sensory cells and spiral-ganglion neurons. \\
POLR2B \citep{Li2023b} & RNA polymerase II subunit involved in glioblastoma growth and altered brain-region transcription. \\
PSMA4 \citep{Ma2018} & Proteasome subunit linked to synaptic protein regulation and schizophrenia-associated cortical dysfunction. \\
UBE2W \citep{Wang2018} & Ubiquitin-conjugating enzyme regulating mutant huntingtin aggregation and neuronal toxicity. \\
UBE2O \citep{Cheng2023} & Ubiquitin-conjugating enzyme supporting neuronal survival in aging and Alzheimer-related neurodegeneration. \\
REXO2 \citep{Wang2021b} & RNA exonuclease associated with glioma heterogeneity and central-nervous-system tumor biology. \\
NUDT16 \citep{Peng2021} & Maintains nucleotide homeostasis and suppresses CAG RNA-induced neuronal DNA damage. \\
MKRN1 \citep{Miroci2012} & RNA-binding ubiquitin ligase regulating dendritic mRNA translation and synaptic plasticity. \\
KPNA6 \citep{Roy2017} & Nuclear transport factor associated with piRNA dysregulation in Alzheimer disease brain. \\
PRPSAP2 \citep{Gui2025} & Purine-biosynthesis regulator linked to generalized epilepsy through brain proteomic and transcriptomic analyses. \\
EMC9 \citep{Marquez2023} & Endoplasmic-reticulum membrane-complex component required for neural-crest and craniofacial development. \\
MBTPS1 \citep{Miao2025} & Protease promoting microglial pyroptosis during neonatal hypoxic-ischemic brain injury. \\
AK4 \citep{Zhong2024} & Regulates neuronal energy metabolism and mitophagy during cerebral ischemic injury. \\
PDXDC1 \citep{Feldcamp2017} & Neurodevelopmental gene affecting sensorimotor gating and ADHD-related prenatal brain risk. \\
RNF175 \citep{Liu2021} & E3 ubiquitin-ligase candidate altered across Alzheimer disease brain regions. \\
FAM120B \citep{Jia2023} & DLPFC protein linked to multiple-sclerosis risk and validated in demyelinated brain tissue. \\
LINC01007 \citep{Duche2025} & Brain-expressed lncRNA included in Alzheimer disease postmortem expression signatures. \\
PRKAG2-AS1 \citep{Zhang2023b} & Angiogenesis-related lncRNA associated with glioblastoma prognosis and immune-treatment signatures. \\
FAM216A \citep{McGuigan2023} & Cortical-thickness-associated gene implicated in first-episode psychosis brain-structure variation. \\
FBLL1 \citep{Zhang2024a} & Neuron-enriched RNA methyltransferase regulating GAP43 expression and neuronal differentiation. \\
SERTM1 \citep{Fullard2018} & Cortical-neuron-enriched gene marked by adult brain chromatin accessibility. \\
TMEM268 \citep{Frye2024} & Cortical transcript associated with diet-linked anti-inflammatory neural states in primates. \\
GARNL3 \citep{SantosCortez2018} & Candidate neurodevelopmental gene linked to autosomal-recessive intellectual disability. \\
FAM81A \citep{Kaizuka2024} & Postsynaptic protein organizing PSD-95 complexes, NMDA receptor signaling, and neuronal firing. \\
PRAG1 \citep{Zolboot2025} & Regulates Purkinje-cell dendritogenesis and climbing-fiber synapse formation. \\
SPAG16 \citep{de2014} & Astrocyte-associated antigen linked to multiple-sclerosis lesions and disease progression. \\
TMEM176B \citep{Melchior2010} & Microglial immune regulator enriched around amyloid plaques and neuroinflammatory lesions. \\
C1QC \citep{Datta2020} & Complement component enriched in aged DLPFC neurons vulnerable to spine loss. \\
CPLX3 \citep{Viswanathan2017} & Marks subplate/L6b neurons supporting cortico-thalamic circuit organization. \\
DLGAP2 \citep{Hsieh2023} & Postsynaptic scaffold regulating spine morphology, synaptic proteins, and spatial memory. \\
AKAP5 \citep{Robertson2009} & Scaffolds postsynaptic signaling complexes regulating dendritic-spine structure and synaptic maturation. \\
DOC2B \citep{Groffen2010} & High-affinity calcium sensor mediating spontaneous synaptic vesicle release. \\
SRGAP3 \citep{Endris2002} & Rho-GTPase regulator enriched in cortical neurons and linked to intellectual disability. \\
DPYSL4 \citep{Xie2025} & Microtubule-associated protein regulating neurite outgrowth and cytoskeletal dynamics. \\
RIMKLB \citep{Becker2021} & NAAG synthetase supporting neural peptide metabolism and recognition memory. \\
ZIC2 \citep{Murillo2015} & Forebrain transcription factor controlling neuronal migration during cortical development. \\
CSNK2A1 \citep{Wu2021} & Neurodevelopmental syndrome gene associated with congenital neuropsychiatric phenotypes. \\
CNTFR \citep{Dutta2007} & CNTF receptor mediating cortical neuroprotective signaling in multiple sclerosis. \\
KDM1A \citep{Zibetti2010} & Chromatin regulator controlling cortical development and neurite morphogenesis. \\
KATNAL1 \citep{Banks2018} & Microtubule-severing enzyme required for neurological and ciliary function. \\
RHOT1 \citep{LopezDomenech2021} & Mitochondrial trafficking protein maintaining neuronal mitophagy and stress responses. \\
TSNAX \citep{Cannon2005} & DISC1-associated factor linked to prefrontal structure and memory phenotypes. \\
BCKDK \citep{Jishi2024} & Mitochondrial kinase supporting complex I activity and limiting alpha-synuclein aggregation. \\
CCDC85C \citep{Mori2012} & Radial-glia junction protein required for cortical development and ventricular-zone integrity. \\
SLC24A4 \citep{Hassan2025} & Neuronal sodium-calcium exchanger supporting calcium homeostasis in brain circuits. \\
HEY1 \citep{Harada2021} & Notch-regulated transcription factor maintaining slowly dividing neural stem cells. \\
FGF9 \citep{Lin2009} & Neurotrophic factor supporting Bergmann glia scaffolds and granule-neuron migration. \\
MYO1B \citep{Iuliano2018} & Regulates actin-wave propagation and growth-cone dynamics during axon formation. \\
PRKCI \citep{Roberts2009} & Apical polarity kinase required for maintaining spinal-cord neural precursors. \\
NAA20 \citep{Morrison2021} & NatB acetyltransferase subunit linked to developmental delay, intellectual disability, and microcephaly. \\
DLD \citep{Aloulou2025} & Mitochondrial energy-metabolism enzyme linked to pediatric neurological metabolic disease. \\
RUSC2 \citep{Alwadei2016} & Vesicle-transport regulator required for neurodevelopment and normal head growth. \\
FEM1C \citep{Dubey2023} & Protein-homeostasis factor associated with severe neurodevelopmental motor and speech phenotypes. \\
TMEM120A \citep{Del2022} & Negative regulator of PIEZO2 mechanosensory channels in peripheral sensory neurons. \\
TCOF1 \citep{Dixon2006} & Ribosome-biogenesis factor required for neural-crest formation and proliferation. \\
SOX15 \citep{Choi2023a} & Transcription factor promoting neural fate during stem-cell differentiation. \\
\end{longtable}
\label{unique_bio_samedonor}
\endgroup

\subsection{Deeper insights into the identified SV genes for the across-donor DLPFC dataset}\label{supp-DLPFC-across}

\begin{figure}[htbp]
    \centering
    \includegraphics[width=0.7\linewidth]{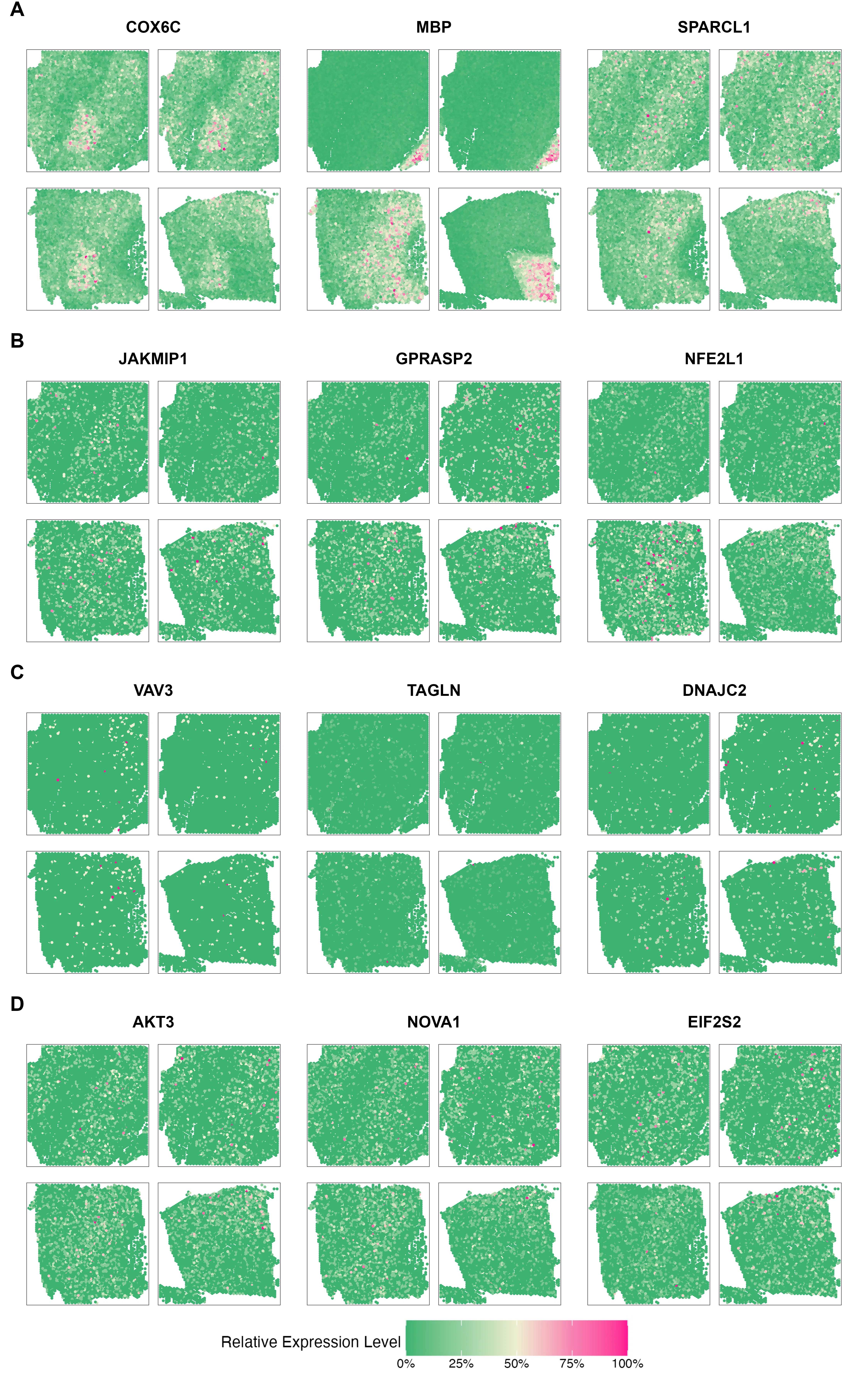}
    \caption{Spatial expression patterns of representative genes identified through comparative analysis for DLPFC dataset across donors.
(A) Three genes identified by the proposed method and all single-sample approaches using intersection-based integration, as well as the Cauchy and HMP integration. (B) Three genes detected by our method but overlooked by all single-sample approaches with intersection-based integration. (C) Three genes recognized by three or more single-sample methods using union-based integration but undetected by the proposed method. (D) Three genes uniquely identified by the proposed method but absent in all single-sample analyses (SPARKX and GASTON, which select almost all genes, are excluded from this comparison) across individual slice examinations.}
    \label{fig:spatial-expression_across donors}
\end{figure}

For the across-donor DLPFC dataset, and in contrast to its same-donor counterpart, a larger set of 55 genes is jointly identified by the proposed method and the five conventional single-sample methods (SPARK, SPARKX, HEARTSVG, nnSVG, spVC) with ``Inter'', ``Cauchy'', and ``HMP'' integration. Of these, GASTON-Inter and StarTrail-Inter manage to recover only 24 and 11 genes, respectively. The spatial expression patterns of three representative genes (COX6C, MBP, and SPARCL1) are visualized in Figure \ref{fig:spatial-expression_across donors} (A). As expected, these genes exhibit strong spatial differential expression, underpinning their consistent detection across slices. Here, COX6C, a key subunit of mitochondrial complex IV, maintains oxidative phosphorylation and energy metabolism in neurons, which is essential for brain function and neuronal survival \citep{Wang2022COX6c}. Myelin basic protein (MBP) exerts profound effects on the morphology, function, and growth of axons in the mammalian central nervous system \citep{Foran1992MBP}. In addition, SPARCL1 is expressed in neurons of the human brainstem and sensory ganglia, where it contributes to the transmission and modulation of motor and sensory signals \citep{Hashimoto2016}. It is noteworthy that although these genes exhibit spatially variable expression across slices, the expression patterns of the same gene can be either conserved across donors (e.g., COX6C) or show notable differences (e.g., MBP and SPARCL1), posing challenges for the integration of multi-donor datasets.

Furthermore, among the 1,549 genes detected by at least one conventional single-sample method under the intersection-based integration strategy, 687 are also identified by the proposed method. In addition, the proposed method discovers 903 genes that are missed by all single-sample methods. When evaluated using the Cauchy and HMP combination approaches, 398 and 399 of these 903 genes are detected by at least one single-sample method, respectively, leaving nearly 500 genes undetected. A closer inspection shows that these genes often display pronounced spatial variation in only part of the four slices, while showing little to no spatial signal in the others. The spatial expression patterns of three representative genes, JAKMIP1, Gprasp2, and NFE2L1, are illustrated in Figure \ref{fig:spatial-expression_across donors} (B). Specifically, JAKMIP1 and Gprasp2 demonstrate spatial differential expression in two slices, but their effects diminish in the remaining slice, while NFE2L1 indicates strong spatial differential expression in the third slice. Hence, they cannot be identified by the single-sample method with intersection integration. However, JAKMIP1 regulates synaptic translation and neuronal development, and its dysregulation is associated with autism-related phenotypes \citep{Berg2015JAKMIP1}. Gprasp2 is a gene associated with neurodevelopmental disorders that regulates post-endocytic sorting of G protein-coupled receptors. Loss of Gprasp2 leads to autism-like behaviors and alters synaptic communication \citep{Edfawy2019}. Furthermore, NFE2L1 plays a multifaceted role in maintaining protein homeostasis, alleviating oxidative stress, and supporting neuronal health, making it a highly attractive therapeutic target for neurodegenerative diseases \citep{Khodadadi2025}. 

Using the union-based integration strategy, 2,672 genes are detected by at least three conventional single-sample methods. Of these, 1,404 genes are missed by the proposed approach. When evaluated under the HMP framework, 1399 of the 1,404 genes are still supported by three or more single-sample methods, and even with the more conservative Cauchy approach, 428 genes continue to be detected. Closer inspection shows that these genes are highly sparse and exhibit outlier-like expression in one or two of the four slices. Representative spatial maps are shown in Figure \ref{fig:spatial-expression_across donors} (C).

We further observe that the proposed method identifies 261 genes that are not captured by any of the other single-sample approaches or 1-D structure-based methods (with the exception of SPARKX and GASTON, which classify nearly all genes as SV genes). The spatial expression patterns of three illustrative genes are shown in Figure \ref{fig:spatial-expression_across donors} (D). These genes display only subtle spatial structure across all four slices, leading single-sample methods to overlook them. By integrating information across slices, our framework enables effective borrowing of strength and enhances the underlying biological signals through empirical Bayes estimation. Notably, the majority of these uniquely detected genes are involved in biologically meaningful processes, with 153 of them independently corroborated by previous studies. As shown in Table S22, these genes are enriched in biological processes related to neuronal development, synaptic function, axon guidance, myelination, cellular homeostasis, and neuroimmune regulation. They have also been previously associated with a broad spectrum of neurological and neuropsychiatric disorders, including Alzheimer's disease, schizophrenia, epilepsy, stroke, and neurodevelopmental disorders. Specifically, AKT3 is the predominant AKT isoform during brain development, regulating neural progenitor proliferation and cortical formation. Brain-restricted activating somatic mutations in AKT3 can lead to focal brain overgrowth and are associated with epilepsy \citep{Poduri2012AKT3}. Additionally, NOVA1 is a brain-enriched RNA-binding protein that orchestrates alternative splicing of key neurodevelopmental genes, thereby regulating neuronal differentiation, synapse formation, and the establishment of functional neural circuits \citep{Piton2025NOVA}. Furthermore, EIF2S2 regulates the proliferation, migration, and apoptosis of glial cells through its downstream target gene WNT5A, and is closely associated with the progression and poor prognosis of primary brain tumors, making it a potential prognostic biomarker and therapeutic target \citep{Fan2025EIF2S2}.

Finally, as observed in the same-donor analysis, methods such as DESpace, HEARTSVG-PASTE, spVC-PASTE, and SPARKX-PASTE again select genes extensively, covering most genes identified by our approach.  In contrast, nnSVG-PASTE, consistently identifying a relatively small gene set across both datasets, detects 111 genes here, all of which are contained within the 885 genes detected by at least one single-sample method with intersection integration. Although SPARK and GASTON identify a similar number of genes via the ``PASTE'' and ``Inter'' integration strategies, their overlap is relatively low ($<$30\%). Furthermore, while most genes identified by StarTrail-Union are also detected by StarTrail-PASTE, the additional genes uniquely identified by StarTrail-Union exhibit a high level of noise.

\begingroup
\footnotesize
\renewcommand{\arraystretch}{0.8}
\singlespacing 
\begin{longtable}{p{0.35\linewidth}p{0.65\linewidth}}
\caption{DLPFC across-donor dataset: curated biological evidence summary.}\\
\toprule
Gene & Brief functional summary \\
\midrule
\endfirsthead
\toprule
Gene & Brief functional summary \\
\midrule
\endhead
\midrule
\multicolumn{2}{r}{Continued on next page}\\
\midrule
\endfoot
\bottomrule
\endlastfoot
KDM1A \citep{Wei2024a} & Chromatin demethylase modulating mPFC stress-related pain and emotional behavior. \\
AZIN2 \citep{MartinezdelaTorre2018} & Marks hindbrain neuronal populations with region- and cell-type-specific expression. \\
NRDC \citep{Ohno2009} & Regulates axonal maturation and myelination in central and peripheral nerves. \\
PIGK \citep{Chen2021a} & GPI-anchor biosynthesis factor required for neuronal survival and infant brain development. \\
USP33 \citep{Xu2025} & Stabilizes CNR1 to promote oligodendrocyte precursor differentiation and remyelination. \\
RTCA \citep{Song2015} & Regulates RNA repair pathways controlling axon regeneration after injury. \\
S1PR1 \citep{Chand2022} & Sphingosine-1-phosphate receptor altered in schizophrenia DLPFC molecular subtypes. \\
STRIP1 \citep{Ahmed2022} & Suppresses Jun-mediated retinal ganglion-cell apoptosis during neural circuit formation. \\
SRGAP2 \citep{LibPhilippot2024} & Balances cortical neuron synaptic maturation through SRGAP2-SYNGAP1 signaling. \\
SIPA1L2 \citep{AndresAlonso2019} & Controls TrkB-containing amphisome trafficking and presynaptic neurotrophin signaling. \\
AKT3 \citep{Rivire2012} & PI3K-AKT kinase controlling brain growth in megalencephaly syndromes. \\
SMYD3 \citep{Williams2023} & Histone methyltransferase modulating NMDAR function and cognition in tauopathy. \\
TRAPPC12 \citep{Milev2017} & Membrane-trafficking factor whose mutations cause progressive childhood encephalopathy. \\
ADCY3 \citep{Yang2021} & Adenylyl cyclase regulating interneuron subtype function and affective behavior. \\
EFEMP1 \citep{McGrath2022} & Extracellular matrix protein associated with brain aging and dementia biomarkers. \\
RETSAT \citep{Hu2026} & Lipid-metabolism enzyme modulating microglial activation under hypoxic stress. \\
DUSP2 \citep{Shao2023} & Phosphatase restricting early axon regeneration after neuronal injury. \\
NPAS2 \citep{Guo2026} & Circadian transcription factor regulating mPFC dopamine synthesis and nap behavior. \\
DPP10 \citep{Wang2015} & Kv4 channel auxiliary subunit localized to cortical and hippocampal neurons. \\
KCNJ3 \citep{Anderson2021} & GIRK channel subunit regulating infralimbic cortical excitability and cognitive flexibility. \\
CERS6 \citep{Hammerschmidt2023} & Ceramide synthase driving hypothalamic neuronal ER and mitochondrial stress. \\
C2orf69 \citep{Oh2025} & Biallelic variants cause early-onset neurodegeneration, leukoencephalopathy, and autoinflammation. \\
ICA1L \citep{Traylor2021} & Locus linked to lacunar stroke susceptibility and cerebral small-vessel disease. \\
NR1D2 \citep{Nadarajah2026} & Circadian nuclear receptor regulating astrocyte reactivity and proteostatic function. \\
EPM2AIP1 \citep{Ianzano2003} & Laforin-interacting protein linked to progressive myoclonus epilepsy pathways. \\
SACM1L \citep{Rapoport2015} & Phosphoinositide-metabolism gene coordinated across human DLPFC development and aging. \\
LZTFL1 \citep{Gana2024} & Ciliary transport gene linked to Bardet-Biedl syndrome and intellectual disability. \\
PPP4R2 \citep{Bosio2012} & Protein phosphatase regulator promoting neuronal differentiation and survival with SMN. \\
SNX4 \citep{Poppinga2024} & Endosomal sorting protein limiting synaptic vesicle docking and release. \\
PLXND1 \citep{TomasRoca2015} & Semaphorin receptor required for hindbrain motor-neuron migration and cranial nerve development. \\
PIK3CB \citep{Zhou2022} & PI3K subunit downregulated in Alzheimer pathways involving apoptosis and axon guidance. \\
MRPS22 \citep{Smits2011} & Mitochondrial ribosomal protein required for brain development and mitochondrial translation. \\
KPNA4 \citep{Sakurai2024} & Nuclear import factor whose deficiency induces neuroinflammation and psychiatric-like behaviors. \\
DNAJC19 \citep{Ucar2017} & Mitochondrial chaperone linked to ataxia, basal-ganglia lesions, and hearing loss. \\
OPA1 \citep{Lei2024} & Mitochondrial fusion GTPase required for retinal ganglion-cell differentiation and function. \\
ADGRL3 \citep{Wang2024b} & Adhesion GPCR regulating cortical neurodevelopment, alternative splicing, and synapse formation. \\
BBS7 \citep{Zhang2013} & BBSome component required for ciliary membrane-protein trafficking and neurodevelopmental phenotypes. \\
KIAA1109 \citep{Gueneau2018} & Neurodevelopmental gene required for brain development and associated with arthrogryposis. \\
NR3C2 \citep{Wang2024a} & Mineralocorticoid receptor promoting ER stress and apoptosis after cerebral ischemia. \\
LMBRD2 \citep{Malhotra2021} & Developmental gene linked to motor delay, dysmorphism, and brain-structure abnormalities. \\
PART1 \citep{Ning2022} & LncRNA regulating blood-brain-barrier permeability in brain endothelial cells. \\
MRPS36 \citep{Jiang2025b} & Mitochondrial enzyme-complex component linked to Leigh syndrome. \\
EFNA5 \citep{Iguchi2021} & Ephrin ligand organizing region-specific corticopontine projections during cortical circuit formation. \\
FBXL17 \citep{Kang2022} & F-box protein regulating spastin and axonal stability in hereditary spastic paraplegia. \\
ATG12 \citep{Sarkar2020} & Autophagy factor involved in lysosomal damage and neurodegeneration after brain trauma. \\
NR3C1 \citep{McGowan2009} & Glucocorticoid receptor epigenetically altered in human hippocampus after childhood adversity. \\
RAB24 \citep{Schwarz2025} & Autophagy-related GTPase linked to cerebellar ataxia. \\
RANBP9 \citep{Palavicini2013} & Scaffolding protein required for neonatal cortical and hippocampal development. \\
SRF \citep{Lu2011} & Activity-dependent transcription factor required for cortical axon growth. \\
TMEM63B \citep{Vetro2023} & Stretch-activated ion channel linked to developmental epileptic encephalopathy and neurodegeneration. \\
ELOVL4 \citep{He2021} & Very-long-chain fatty-acid elongase linked to frontotemporal dementia lipid dysregulation. \\
SNX14 \citep{Zhang2021b} & Sorting nexin maintaining axonal mitochondrial transport in Purkinje neurons. \\
PREP \citep{Etelinen2023} & Prolyl oligopeptidase modulating tau clearance and tauopathy-related cognitive decline. \\
AHI1 \citep{Wang2022} & Ciliary scaffold protein regulating serotonin production and depression-related behaviors. \\
UTRN \citep{Adams2010} & Postsynaptic scaffold component organizing acetylcholine receptors at neuromuscular junctions. \\
PPP1R14C \citep{Drgonova2010} & Protein phosphatase inhibitor modulating morphine analgesia and tolerance. \\
TIAM2 \citep{Rao2024} & RhoGEF regulating glutamatergic synaptic transmission in hippocampal pyramidal neurons. \\
TRA2A \citep{Liu2020a} & Splicing factor controlling blood-brain-barrier permeability in Alzheimer-like microenvironment. \\
MPP6 \citep{Khoury2021} & Sleep-quality GWAS gene functioning in sleep center neurons. \\
RALA \citep{Brown2024} & Small GTPase regulating brain development and organ growth downstream of DYRK1A. \\
BLVRA \citep{Choi2023b} & Biliverdin reductase protecting dopaminergic neurons from oxidative stress. \\
VOPP1 \citep{Braune2025} & Fusion partner activating NF-kappaB signaling in glioneural tumors. \\
ORAI2 \citep{Stegner2019} & Calcium-entry channel whose loss protects against acute ischemic stroke injury. \\
AKR1B1 \citep{Rao2023} & Aldose reductase driving retinal microglial activation and optic-nerve degeneration. \\
NLGN4X \citep{Lehr2025} & Postsynaptic adhesion molecule regulating spinogenesis and excitatory synaptic function. \\
FRMPD4 \citep{Piard2018} & Postsynaptic scaffold required for dendritic spine morphogenesis and cognition. \\
SMS \citep{Akinyele2024} & Polyamine-synthesis enzyme required for normal neuroanatomy and behavior. \\
EIF2S3 \citep{Skopkova2017} & Translation initiation factor required for neurodevelopment and MEHMO syndrome biology. \\
UBQLN2 \citep{Wu2020} & Proteostasis factor maintaining autophagosome acidification in ALS and FTD pathways. \\
ZNF711 \citep{van2017} & Transcriptional regulator implicated in X-linked intellectual disability. \\
XIAP \citep{Katz2001} & Apoptosis inhibitor involved in caspase regulation after hypoxic brain injury. \\
FBXO25 \citep{Harich2020} & F-box protein linked to ADHD and comorbid psychiatric phenotypes. \\
CSMD1 \citep{Baum2024} & Complement regulator controlling synapse pruning and circuit development. \\
MSRA \citep{Oien2008} & Oxidative-repair enzyme influencing behavior and brain dopamine levels. \\
YTHDF3 \citep{Zhang2025b} & M6A reader mediating microglial pyroptosis and cognitive impairment. \\
ARFGEF1 \citep{Teoh2020} & Vesicle-trafficking regulator controlling postsynaptic GABAA receptor surface levels. \\
SDC2 \citep{Hu2016} & Postsynaptic proteoglycan coordinating FGF22 signaling and bidirectional synapse formation. \\
EBAG9 \citep{Rder2005} & Vesicle-exocytosis regulator controlling dense-core neurosecretory granule release. \\
FAM49B \citep{Martin2025} & Parkinson-risk factor regulating age-related microglial inflammatory priming. \\
FOCAD \citep{Brand2020} & Cytoskeletal regulator affecting microtubule assembly in astrocytic gliomas. \\
RFK \citep{Zou2025} & Riboflavin kinase with neuroprotective effects in cerebral ischemia. \\
RAD23B \citep{Chen2026} & DNA-repair and proteasome factor exacerbating aggregates in spinocerebellar ataxia. \\
ELP1 \citep{Jackson2014} & Familial dysautonomia gene required for sympathetic and sensory innervation. \\
ENDOG \citep{Li2018} & Mitochondrial nuclease involved in ischemia-reperfusion neuronal injury pathways. \\
CCKBR \citep{Shi2025} & Cholecystokinin receptor modulating long-term potentiation signaling. \\
SNX32 \citep{Sugatha2023} & Sorting nexin involved in cargo sorting and neurite outgrowth. \\
MTMR2 \citep{Chua2022} & Neuronal phosphatase controlling autophagy and membrane homeostasis. \\
PTS \citep{Wang2006} & Tetrahydrobiopterin enzyme linked to neurodevelopmental and neuroradiological abnormalities. \\
SC5D \citep{Jiang2010} & Cholesterol-biosynthesis enzyme linked to altered metabolic pathways in neurodevelopmental disease. \\
ZNF32 \citep{Wei2016} & Transcription factor limiting lateral-line neural regeneration through SOX2 regulation. \\
ANXA7 \citep{Li2023c} & Autophagy regulator reducing neuronal apoptosis after spinal-cord injury. \\
IFIT1 \citep{Szretter2012} & Antiviral effector affecting neurotropic viral restriction in vivo. \\
BLOC1S2 \citep{Zhou2016} & BLOC-1 component modulating Notch signaling during neural progenitor development. \\
CACNA1C \citep{Datta2024} & Calcium-channel subunit with key roles in prefrontal neurons altered in cognitive disorders. \\
NDUFA9 \citep{Magrinelli2025} & Mitochondrial complex I subunit causing dystonia and progressive neurodevelopmental disease. \\
CREBL2 \citep{Li2023e} & Schwann-cell regulator involved in peripheral nerve repair after injury. \\
CMAS \citep{Qu2020} & Sialic-acid biosynthesis enzyme linked to autosomal-recessive intellectual disability. \\
ERGIC2 \citep{ak2012} & Otoferlin-interacting brain protein implicated in neural membrane trafficking. \\
CAPRIN2 \citep{Konopacka2015} & RNA-binding protein regulating central osmotic defense responses. \\
PRICKLE1 \citep{Ban2021} & Planar-cell-polarity protein maintaining glutamatergic synapses. \\
ACVR1B \citep{Zheng2021} & Activin receptor promoting oligodendroglial remyelination after ischemic stroke. \\
MFSD5 \citep{Perland2016} & Brain-enriched solute transporter potentially involved in neural energy homeostasis. \\
PIP4K2C \citep{Noch2021} & Phosphoinositide kinase localized in brain and implicated in neuronal lipid signaling. \\
PPP1R12A \citep{Hughes2020} & Myosin phosphatase subunit linked to holoprosencephaly-spectrum phenotypes. \\
SLC6A15 \citep{Kohli2011} & Neuronal amino-acid transporter conferring major-depression risk. \\
PPTC7 \citep{Yang2026} & Mitochondrial phosphatase affecting brain endothelial function and diabetic cognitive dysfunction. \\
FBXO21 \citep{Long2025} & Ubiquitin ligase regulating NMNAT2-dependent axon survival after nerve injury. \\
VPS33A \citep{Chintala2009} & Vesicle-trafficking gene regulating cerebellar Purkinje cell number and behavior. \\
AACS \citep{Hasegawa2012} & Ketone-body metabolism enzyme required for normal neuronal development. \\
ATP8A2 \citep{Liu2025a} & Phospholipid flippase altered in prefrontal cortex after immune activation. \\
BCL2L2 \citep{Shi2012} & Anti-apoptotic factor protecting neurons after ischemic injury. \\
NOVA1 \citep{Tajima2025} & Neuron-specific splicing factor influencing vocal communication and neural transcript processing. \\
TRAPPC6B \citep{Almousa2024} & TRAPP-II trafficking component causing neurodevelopmental disease when biallelically disrupted. \\
ERO1A \citep{Yeewa2024} & ER oxidoreductase contributing to neuronal ER stress in UBQLN2-related models. \\
CNIH1 \citep{Drummond2012} & AMPA-receptor auxiliary protein upregulated in schizophrenia DLPFC. \\
KCNH5 \citep{Yu2024} & Potassium channel regulating neuronal growth through AKT-mTOR signaling. \\
ZC3H14 \citep{Lancaster2024} & RNA-binding protein controlling axon and dendrite morphology through Trio regulation. \\
TTBK2 \citep{Bowie2020} & Ciliary kinase required for Purkinje neuron connectivity and survival. \\
TUBGCP4 \citep{Scheidecker2015} & Microtubule-organizing component causing microcephaly with chorioretinopathy. \\
BLOC1S6 \citep{Ito2018} & Endosomal pathway protein regulating dendritic formation in neurons. \\
AAGAB \citep{Dai2021} & PTEN-trafficking regulator promoting recovery after hypoxic-ischemic brain damage. \\
NTAN1 \citep{Kwon2000} & N-end rule pathway amidase influencing activity, social behavior, and spatial memory. \\
ATXN2L \citep{Key2025} & Frontal-cortex neuronal splicing regulator affecting spontaneous mobility. \\
ARL2BP \citep{Davidson2013} & ADP-ribosylation-factor binding protein causing retinitis pigmentosa. \\
GOT2 \citep{German2025} & Mitochondrial transaminase causing progressive neurodevelopmental disorder with epilepsy. \\
DOC2B \citep{Guiberson2024} & Synaptic calcium sensor depleted by disease-linked Munc18-1 mutations. \\
VPS53 \citep{Feinstein2014} & Vesicle-tethering gene causing progressive cerebello-cerebral atrophy. \\
SLC43A2 \citep{Taslimifar2022} & Amino-acid transporter active in human blood-brain-barrier models. \\
PSMD11 \citep{Deb2024} & Proteasome subunit linked to neurobehavioral phenotypes and interferon response. \\
NMT1 \citep{Lacrampe2025} & Myristoyltransferase regulating TMEM106B trafficking and turnover. \\
APPBP2 \citep{Li2023d} & APP-binding protein targeted in Alzheimer pathways affecting tau and amyloid-beta. \\
PSMD12 \citep{Li2024a} & Proteasome subunit modulating NF-kappaB immune responses after brain ischemia. \\
RAC3 \citep{Sugawara2025} & Small GTPase required for cortical neuron migration and axon elongation. \\
SEH1L \citep{Weckhuysen2016} & GATOR-complex component linked to familial focal epilepsy and cortical dysplasia. \\
CSNK2A1 \citep{Okur2016} & Kinase mutated in Okur-Chung neurodevelopmental syndrome. \\
ATRN \citep{Liu2023a} & Attractin required for working-memory performance. \\
PANK2 \citep{Zhou2001} & Pantothenate kinase defective in neurodegeneration with brain iron accumulation. \\
TOP1 \citep{King2013} & Topoisomerase facilitating transcription of long autism-linked genes. \\
DPM1 \citep{Kim2000} & Glycosylation enzyme whose deficiency causes neurological congenital glycosylation disorder. \\
CHRNA4 \citep{Steinlein1995} & Nicotinic acetylcholine receptor subunit mutated in nocturnal frontal lobe epilepsy. \\
SLC39A3 \citep{Qian2011} & Zinc transporter whose loss attenuates seizure-induced hippocampal neurodegeneration. \\
ANKRD24 \citep{Hu2023} & Autophagy-associated factor mediating ketone-body-induced tau degradation. \\
TNFAIP8L1 \citep{Ha2014} & Autophagy regulator interacting with FBXW5 in Parkinsonian stress models. \\
CARM1 \citep{Ishino2022} & Arginine methyltransferase controlling corpus-callosum oligodendrocyte differentiation. \\
AKAP8L \citep{Zhang2024c} & Microglial scaffold promoting autophagy inhibition and neuroinflammation in diabetic cognitive impairment. \\
KCTD15 \citep{Raymundo2023} & Ectodermal and neural-crest regulator forming KCTD1-KCTD15 complexes. \\
HNRNPL \citep{Jiang2025a} & Splicing factor targeted during oligodendrocyte differentiation via nonsense-mediated decay. \\
NUMBL \citep{Petersen2004} & Maintains cortical neural progenitor pools during neurogenesis. \\
GEMIN7 \citep{Baccon2002} & SMN-complex component linked to motor-neuron RNA-processing machinery. \\
MEIS3 \citep{Elkouby2012} & Hindbrain transcriptional regulator promoting neuronal differentiation and tissue maturation. \\
SPHK2 \citep{Hait2014} & Sphingosine kinase target facilitating fear extinction memory. \\
CRKL \citep{Park2008} & Reelin-pathway adaptor required for neuronal positioning during brain development. \\
USP25 \citep{Li2023a} & Deubiquitinase suppressing neuroinflammation after ischemic stroke. \\
\end{longtable}
\label{tab:unique_bio_acrossdonor}
\endgroup

\subsection{Deeper insights into the identified SV genes for the SCC dataset}\label{supp-SCC}

\begin{figure}[htbp]
    \centering
    \includegraphics[width=0.8\linewidth]{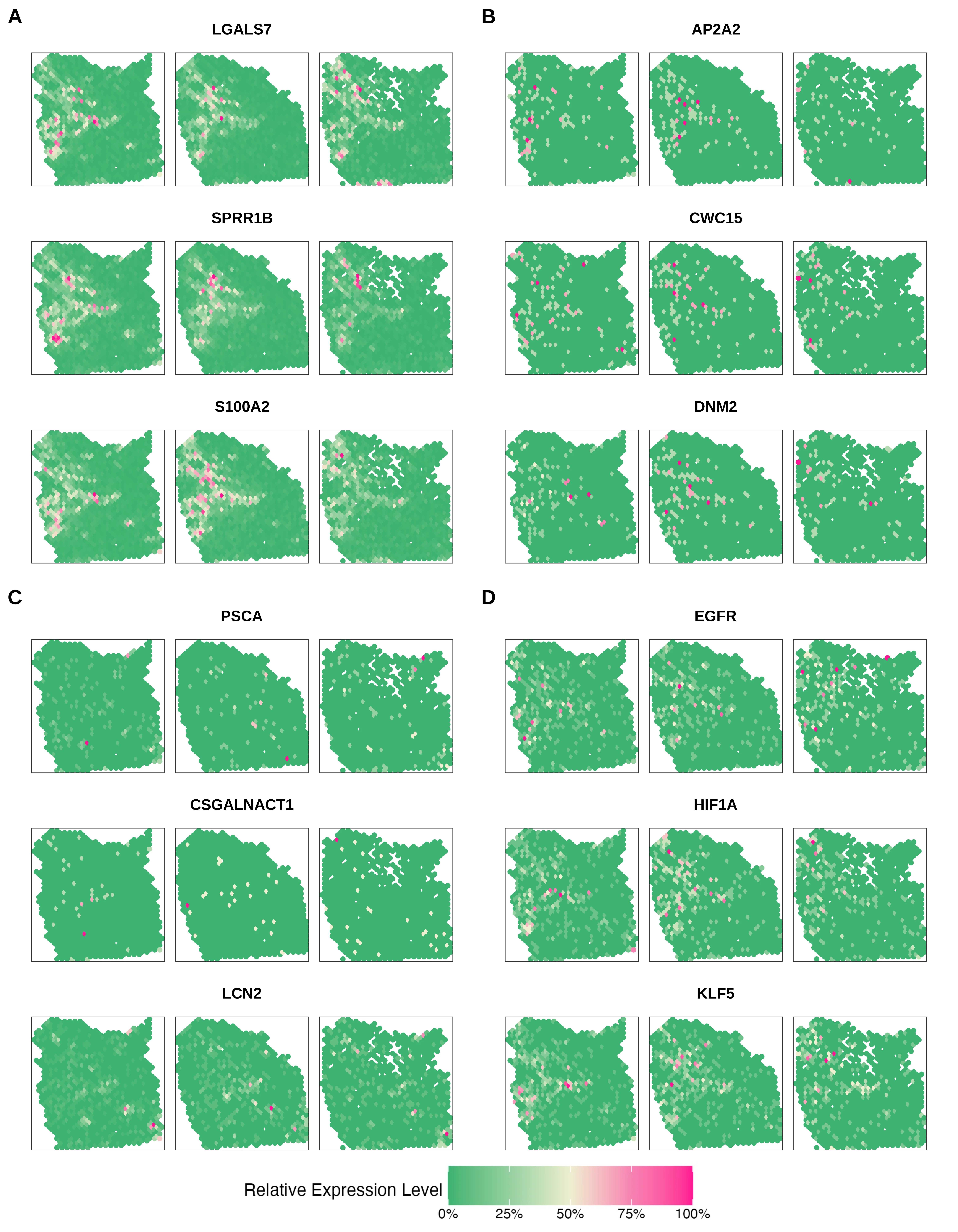}
    \caption{Spatial expression patterns of representative genes identified through comparative analysis for SCC dataset. (A) Three genes identified by the proposed method and most single-sample approaches using intersection-based integration. (B) Three genes detected by our method but overlooked by all single-sample approaches with intersection-based integration. (C) Three genes recognized by three or more single-sample methods using union-based integration but undetected by the proposed method. (D) Three genes uniquely identified by the proposed method but absent in all single-sample analyses (SPARKX, which selects almost all genes, is excluded from this comparison) across individual slice examinations.}
    \label{problem_scc}
\end{figure}

For the SCC dataset, only one gene S100A2 is identified by the proposed method and the five conventional single-sample methods across ``Inter'', ``Cauchy'', and ``HMP'' integrations, which has been shown to act as a tumor suppressor in certain squamous cell carcinoma tissues by inhibiting cellular migratory properties \citep{nagy2001s100a2}. Additionally, 10 additional genes are detected by the proposed method and the five conventional single-sample methods except HEARTSVG under these integrations. Although StarTrail-Inter identifies a relatively small number of genes, it successfully recovers 8 out of the 10 key genes. In contrast, GASTON-Inter detects nearly 1,000 genes, yet only one overlapped with this set. The spatial expression patterns of three representative genes LGALS7, SPRR1B, and S100A2 of these genes are provided in Figure \ref{problem_scc} (A). These genes demonstrate pronounced spatial differential expression across all three slices, which contributes to their high detectability. LGALS7, which encodes galectin-7, plays multiple regulatory roles in the differentiation and development of epithelial tissues, and is also critically involved in various aspects of cancer progression \citep{saussez2006galectin7}. SPRR1B has been found to be highly expressed in well-differentiated squamous cell carcinoma tissues and promotes cancer cell proliferation by activating the p38 MAPK signaling pathway \citep{sasahira2021identification}. S100A2 can exert tumor suppressor effects in certain epithelial tissues by inhibiting cell migration \citep{nagy2001s100a2}.

%problem2
Furthermore, among the 2,600 genes detected by at least one single-sample method utilizing intersection integration, 864 of them (82.2\% of 1051) are detected by our proposed method, indicating strong consistency with single-sample approaches. The proposed method also identifies additional 187 genes that are overlooked by all single-sample methods with intersection-based integration. A detailed analysis of these genes indicates that they often display pronounced spatial differential expression in one or two slices among the three, while showing negligible spatial patterns in the others. Under both the Cauchy and HMP strategies, nearly no single-sample method except SPARKX is able to detect these 187 genes. It is noted that SPARKX identifies a large number of genes, which almost entirely encompass those identified in our analysis. The spatial expression patterns of three representative genes, AP2A2, CWC15, and DNM2, are illustrated in Figure \ref{problem_scc} (B). It is observed that AP2A2 and CWC15 exhibit prominent spatial patterns in both slices A and B, whereas DNM2 shows significant spatially variable expression only in slice B. AP2A2 plays a critical role in cancer treatment and progression (including migration and invasion) by regulating the transcription of specific target genes \citep{orso2007ap}. DNM2 (Dynamin-2) plays a crucial role in the formation and trafficking of intracellular vesicles, cytokinesis, and receptor-mediated endocytosis, and is closely associated with cancer metastasis \citep{Hemant2014dynamin}. CWC15 plays a critical role in mRNA splicing and miRNA regulation in tumor cells. Its expression is associated with prognosis across multiple cancer types, suggesting its potential as a prognostic biomarker for cancer \citep{uhlen2015tissue}. 

%problem3
Using the union-based integration approach, 815 genes are detected by at least three of the single-sample methods. Among these, 540 genes are not identified by the proposed method. These genes exhibit high sparsity and display potential outlier expression patterns in some of the three slices. Under the HMP strategy, 107 out of the 540 genes are still identified by at least three single-sample methods. Even with the relatively conservative Cauchy strategy, 56 genes are still consistently detected by at least three single-sample methods. Figure \ref{problem_scc} (C) presents the spatial expression patterns of three representative genes. This suggests that union-based and p-value integration approaches are prone to introducing noise when applying single-sample methods to multi-sample data.

%problem4
We additionally find that the proposed method detects 157 genes that are missed by other general single-sample and two structure-informed SV gene identification methods (except for SPARKX, which identifies nearly all genes as SV genes) in individual slice analyses. Figure \ref{problem_scc} (D) illustrates the spatial expression patterns of three representative genes: EGFR, HIF1a, and KLF5. These genes exhibit consistently weak spatial signals across the three slices, resulting in their frequent omission by single-sample analysis methods. As illustrated in Table S23 (84 of 157 genes), we further find that these uniquely identified genes are enriched in biological processes underlying tumor growth, invasion and metastasis, epithelial-mesenchymal transition, metabolic remodeling, autophagy, and treatment response, and have been extensively associated with esophageal squamous cell carcinoma and other squamous cell carcinomas, including oral, head-and-neck, lung, laryngeal, and cutaneous subtypes. Specifically, aberrant expression of epidermal growth factor receptor (EGFR) and its ligands has been observed, making EGFR a potential biomarker for targeted therapy \citep{Uribe2011EGFR}. HIF1A inhibition can enhance the efficacy of anti-PD-1 antibody treatment in immune-cold HNSCC, highlighting its role in modulating the tumor immune microenvironment \citep{ONeill2025HIF1a}. Additionally, KLF5 has emerged as a novel diagnostic biomarker and therapeutic target in SCC, contributing to tumor progression and malignancy \citep{Yang2021KLF5}.

Finally, comparative analysis reveals our method's gene discoveries are largely encompassed by DESpace, nnSVG-PASTE, SPARK-PASTE, and SPARKX-PASTE, which detect over 3,900 genes, approaching the total number analyzed. On the other hand, 86 of the 104 genes identified by spVC-PASTE, as well as all genes detected by HEARTSVG-PASTE and StarTrail-PASTE, are included in the set of 1,704 genes identified by at least one single-sample method using intersection integration. Thus, these methods with the PASTE integration remains limited in detecting SV genes that are expressed in only a small subset of tissue slices. 

% Although GASTON-PASTE and GASTON-Inter identified a similar number of genes, the overlap between their gene sets is less than 30\%. These observations indicate that, as an integration strategy for single-slice methods, PASTE still exhibits considerable incompatibility in SCC dataset.

\begingroup
\footnotesize
\renewcommand{\arraystretch}{0.9}
\singlespacing 
\begin{longtable}{p{0.35\linewidth}p{0.65\linewidth}}
\caption{SCC dataset: curated biological evidence summary.}\\
\toprule
Gene & Brief functional summary \\
\midrule
\endfirsthead
\toprule
Gene & Brief functional summary \\
\midrule
\endhead
\midrule
\multicolumn{2}{r}{Continued on next page}\\
\midrule
\endfoot
\bottomrule
\endlastfoot
SYTL1 \citep{Lyu2025} & Candidate transcriptomic marker distinguishing esophageal adenocarcinoma from esophageal squamous cell carcinoma. \\
UQCRH \citep{Liu2020b} & Mitochondrial respiratory-chain protein altered during growth inhibition of well-differentiated laryngeal squamous carcinoma cells. \\
WLS \citep{Jung2020} & Mediates Wnt ligand secretion regulated by miR-31 in oral cancer signaling networks. \\
RAP1A \citep{Li2019} & Activates AKT signaling to promote migration and metastasis in esophageal squamous cell carcinoma. \\
TPM3 \citep{Chen2021b} & Promotes esophageal cancer cell migration and invasion through MMP2/MMP9 regulation and epithelial-mesenchymal transition. \\
VDAC2 \citep{Zhang2026} & Supports esophageal squamous carcinoma proliferation and glycolysis downstream of SEMA5B. \\
WEE1 \citep{Li2024b} & Cell-cycle checkpoint kinase whose inhibition sensitizes TP53-mutant lung squamous carcinoma to cisplatin. \\
PSMA1 \citep{Liu2023b} & Proteasome subunit promoting lung squamous carcinoma growth and associated with poor prognosis. \\
DDB1 \citep{McHugh2020} & DNA-damage-binding factor identified as a functional therapeutic-target candidate in cutaneous squamous carcinoma. \\
POLR2G \citep{Zhang2025a} & Patient-stratification marker in TP53-mutant hypopharyngeal squamous carcinoma multi-omics prognostic models. \\
MAP3K11 \citep{Liu2022MAP3K11} & Enhances autophagy and malignant phenotypes in oral squamous carcinoma cells. \\
RELA \citep{Yang2023a} & NF-kappaB subunit driving TFAP2A-Wnt/beta-catenin signaling during oral squamous carcinoma progression. \\
SDHD \citep{Lintell2007} & Mitochondrial complex II component analyzed in actinic keratosis, a precursor lesion of cutaneous squamous carcinoma. \\
CAPRIN2 \citep{Ai2020} & Wnt/beta-catenin effector activated by LINC00941 to promote oral squamous carcinoma growth and migration. \\
SHMT2 \citep{Liao2021} & Mitochondrial one-carbon enzyme supporting cell-cycle progression and growth in tongue squamous carcinoma. \\
CSRP2 \citep{Zhang2023a} & Promotes stem-like properties and tumor maintenance in head-and-neck squamous carcinoma cells. \\
OAS3 \citep{Caglar2024} & Interferon-pathway gene associated with paclitaxel resistance in head-and-neck squamous carcinoma. \\
KLF5 \citep{Wang2023b} & Transcriptionally activates FGFBP1 to promote esophageal squamous carcinoma growth. \\
MBNL2 \citep{Zhang2025c} & RNA-binding splicing regulator linked to expression, upstream regulation, and prognosis in esophageal squamous carcinoma. \\
HNRNPC \citep{Zhou2023} & Stabilizes GLI2 mRNA through circ-FIRRE interaction to promote esophageal squamous carcinoma progression. \\
MMP14 \citep{Huang2025} & Mediates chemotherapy-induced epithelial-mesenchymal transition and metastasis in oral squamous carcinoma. \\
HIF1A \citep{Wang2023a} & Hypoxia regulator disrupting lysosomal homeostasis and enhancing extracellular-vesicle release in head-and-neck squamous carcinoma. \\
NDUFB1 \citep{Wang2021a} & Mitochondrial complex I subunit included in prognostic expression signatures for early lung squamous carcinoma. \\
SERF2 \citep{Dogra2025} & Computationally prioritized prognostic marker for oral squamous carcinoma from transcriptomic and explainable-AI analyses. \\
EIF3C \citep{Zhao2022} & Translation-initiation factor supporting nasopharyngeal carcinoma cell growth within the head-and-neck squamous carcinoma spectrum. \\
MT1E \citep{BrazoSilva2015} & Metallothionein family member whose altered oral-cancer expression associates with metastasis and patient outcome. \\
VPS4A \citep{Peng2025} & Endosomal-sorting ATPase mediating EPHB2-driven autophagy and progression in oral squamous carcinoma. \\
KARS \citep{Hsu2019} & Aminoacyl-tRNA synthetase detected in oral squamous carcinoma interstitial-fluid proteomic marker panels. \\
SLC7A5 \citep{Chen2024} & TP63-activated amino-acid transporter suppressing ferroptosis and supporting head-and-neck squamous carcinoma survival. \\
APRT \citep{Wintergerst2018} & Purine-metabolism enzyme included in chemoradiotherapy outcome signatures for head-and-neck squamous carcinoma. \\
SPAG7 \citep{Singh2020} & Transcript isoform marker associated with well-differentiated keratinizing oral squamous carcinoma. \\
FLOT2 \citep{Li2022} & Membrane-scaffold protein suppressed by oncogenic miR-486-3p in cutaneous squamous carcinoma proliferation and metastasis models. \\
TAF15 \citep{Ren2020} & RNA-binding protein recruited by PITPNA-AS1 to stabilize HMGB3 and promote lung squamous carcinoma growth and migration. \\
ACLY \citep{Zhang2024b} & Citrate lyase activated by SIRT2 deacetylation to drive lipid-metabolic progression in esophageal squamous carcinoma. \\
DDX5 \citep{Shi2024} & RNA helicase stabilizing VAV3 mRNA to promote esophageal squamous carcinoma growth. \\
SRSF2 \citep{Yu2019} & Splicing factor interacting with AC091729.7 to promote sinonasal squamous carcinoma proliferation and invasion. \\
ANAPC11 \citep{Xu2024} & Cell-cycle regulator associated with expression-based prognosis in lung squamous carcinoma. \\
NDUFA11 \citep{Wu2016} & Mitochondrial respiratory-chain subunit embedded in lung squamous carcinoma lncRNA-mRNA regulatory networks. \\
HNRNPL \citep{Wu2022} & M6A-regulated RNA-binding protein promoting esophageal squamous carcinoma proliferation and chemoresistance. \\
UBE2S \citep{Yoshimura2017} & Ubiquitin-conjugating enzyme promoting p21 degradation and proliferation in oral squamous carcinoma. \\
SPTBN1 \citep{Grnhj2018} & Genomic candidate mutated in HPV-positive locally advanced oropharyngeal squamous carcinoma. \\
EFEMP1 \citep{Guo2025HIPK4} & Extracellular-matrix protein suppressed by HIPK4-TAp63 signaling during cutaneous squamous carcinoma progression. \\
TGOLN2 \citep{Yang2020} & MiRNA-target network component associated with expression and prognosis in esophageal squamous carcinoma. \\
AGPS \citep{Fang2026a} & Susceptibility locus regulated by allele-specific MEIS3 binding and NF-kappaB activation in esophageal squamous carcinoma. \\
SSFA2 \citep{Zhu2021} & Pro-tumor target of miR-363-3p influencing oral squamous carcinoma cell proliferation and invasion. \\
HSPE1 \citep{Feng2017} & Mitochondrial chaperonin overexpressed in oral squamous carcinoma and associated with poor prognosis. \\
COPS9 \citep{Zhou2021} & COP9 signalosome component whose expression associates with prognosis in head-and-neck squamous carcinoma. \\
JAG1 \citep{Osathanon2016} & Notch ligand influencing oral squamous carcinoma cell behavior and squamous epithelial differentiation signaling. \\
EIF6 \citep{Zhao2021eIF6} & Translation factor activating AKT signaling to promote malignant behavior in oral squamous carcinoma. \\
TPD52L2 \citep{Lu2024} & Overexpressed tumor-progression marker associated with prognosis and immune infiltration in head-and-neck squamous carcinoma. \\
ADAMTS1 \citep{Chien2024} & Matrix protease driving L1CAM-EGFR signaling, epithelial-mesenchymal transition, and lymph-node metastasis in oral squamous carcinoma. \\
TIAM1 \citep{Yao2025} & CAF-derived Rac activator promoting oral squamous carcinoma growth and metastasis through ZEB2 regulation. \\
DYRK1A \citep{Martin2021} & Kinase required for cancer stemness and tumor-initiating capacity in oral and oropharyngeal squamous carcinoma. \\
MCM5 \citep{Jiang2022} & DNA-replication licensing factor interacting with lnc-POP1-1 to promote cisplatin resistance in head-and-neck squamous carcinoma. \\
TMA7 \citep{Yang2023b} & Autophagy-related translational factor regulated by IGF2BP3-m6A signaling during laryngeal carcinoma cisplatin resistance. \\
MST1R \citep{Song2022} & EMT-related receptor included in chemoradiotherapy resistance and prognosis signatures for esophageal squamous carcinoma. \\
PCNP \citep{Zhang2022} & Differentiation-associated nuclear protein modulated by tissue mechanical states in oral squamous carcinoma. \\
ITGB5 \citep{Chan2024} & Integrin subunit mediating ENAH-driven PI3K/AKT/beta-catenin and Src signaling in oral carcinoma migration and growth. \\
RPN1 \citep{Chen2025} & Ribosome-associated glycoprotein stabilized by CERS6 to promote esophageal squamous carcinoma proliferation. \\
RBP1 \citep{Gao2020} & Retinoid-binding protein activating CKAP4-dependent oncogenic autophagy in oral squamous carcinoma. \\
KPNA4 \citep{Zhang2019} & Nuclear-import factor supporting cutaneous squamous carcinoma proliferation and cisplatin resistance downstream of miR-3619-5p. \\
POLR2H \citep{Su2023} & Ferroptosis-related prognostic and immune-stratification marker in lung squamous carcinoma. \\
G3BP2 \citep{Zhang2021a} & Stress-granule RNA-binding protein forming a BAALC-AS1-c-Myc loop that promotes esophageal squamous carcinoma proliferation. \\
HNRNPH1 \citep{Liang2023} & RNA-processing factor contributing mutation-transcript features for oral tongue squamous carcinoma prognosis. \\
DEK \citep{Zhao2025} & Fusion partner defining papillary sinonasal squamous carcinoma with distinct morphology and molecular features. \\
DDR1 \citep{Hu2025} & Collagen receptor whose inhibition enhances ferroptosis, immunogenic cell death, and radiotherapy response in head-and-neck squamous carcinoma. \\
RPS10 \citep{Gao2024} & Ribosomal protein altered in metastatic laryngeal squamous carcinoma proteomic profiles. \\
EIF3B \citep{Tan2025} & Translation-initiation subunit stabilizing MAP2K2 and activating ERK signaling in laryngeal squamous carcinoma. \\
GNA12 \citep{Fang2026b} & G12-family G protein regulating oral squamous carcinoma cell proliferation, migration, and invasion. \\
TBRG4 \citep{Ramasubramanian2023} & Overexpressed biomarker promoting oral squamous carcinoma progression and invasion. \\
EGFR \citep{Zhou2025} & Receptor tyrosine kinase driving local invasion and cetuximab response in head-and-neck squamous carcinoma. \\
YWHAG \citep{Ramesh2025} & 14-3-3 protein dysregulated in oral squamous carcinoma and associated with metastasis and poor prognosis. \\
SYPL1 \citep{Ding2022} & Effector of the YY1-KCNQ1OT1-miR-506-3p axis promoting oral squamous carcinoma progression. \\
SND1 \citep{Yan2025b} & PIM-enhanced chromatin/RNA-associated factor promoting esophageal squamous carcinoma proliferation, metastasis, and chemoresistance. \\
SQLE \citep{Liu2025b} & Cholesterol-biosynthesis enzyme amplified to accumulate 2,3-oxidosqualene and suppress Hippo signaling in esophageal squamous carcinoma. \\
BAG1 \citep{Zhao2024} & Co-chaperone mediating cisplatin resistance in oral squamous carcinoma. \\
STOML2 \citep{Wang2020coexist} & Stomatin-family protein co-overexpressed with STOML1 and associated with oral squamous carcinoma pathology. \\
TLE1 \citep{Liu2020c} & Effector of the RASSF8-AS1-miR-664b-3p axis suppressing laryngeal squamous carcinoma progression. \\
CIZ1 \citep{Zhou2018} & Candidate tissue biomarker identified for lung squamous carcinoma. \\
TUBB4B \citep{Dharmapal2021} & Microtubule isoform regulating cancer stem-cell maintenance in squamous carcinoma models. \\
CASK \citep{Liu2019} & CeRNA-network component identified in laryngeal squamous carcinoma transcriptomic analyses. \\
EFNB1 \citep{Chen2023} & Ephrin signaling component in abnormal epithelial-cell interaction networks driving esophageal squamous carcinoma progression. \\
MORF4L2 \citep{Cheng2025} & Chromatin regulator identified in aberrant regulatory networks of head-and-neck squamous carcinoma. \\
PGRMC1 \citep{Huang2020} & Progesterone-receptor membrane component promoting oral-cancer migration and invasion in models relevant to oral squamous carcinoma. \\
\end{longtable}
\label{unique_bio_scc}
\endgroup

\clearpage
\subsection{Detailed gene clustering results}

\begin{figure}[!ht]
    \centering
    \includegraphics[width=0.8\linewidth]{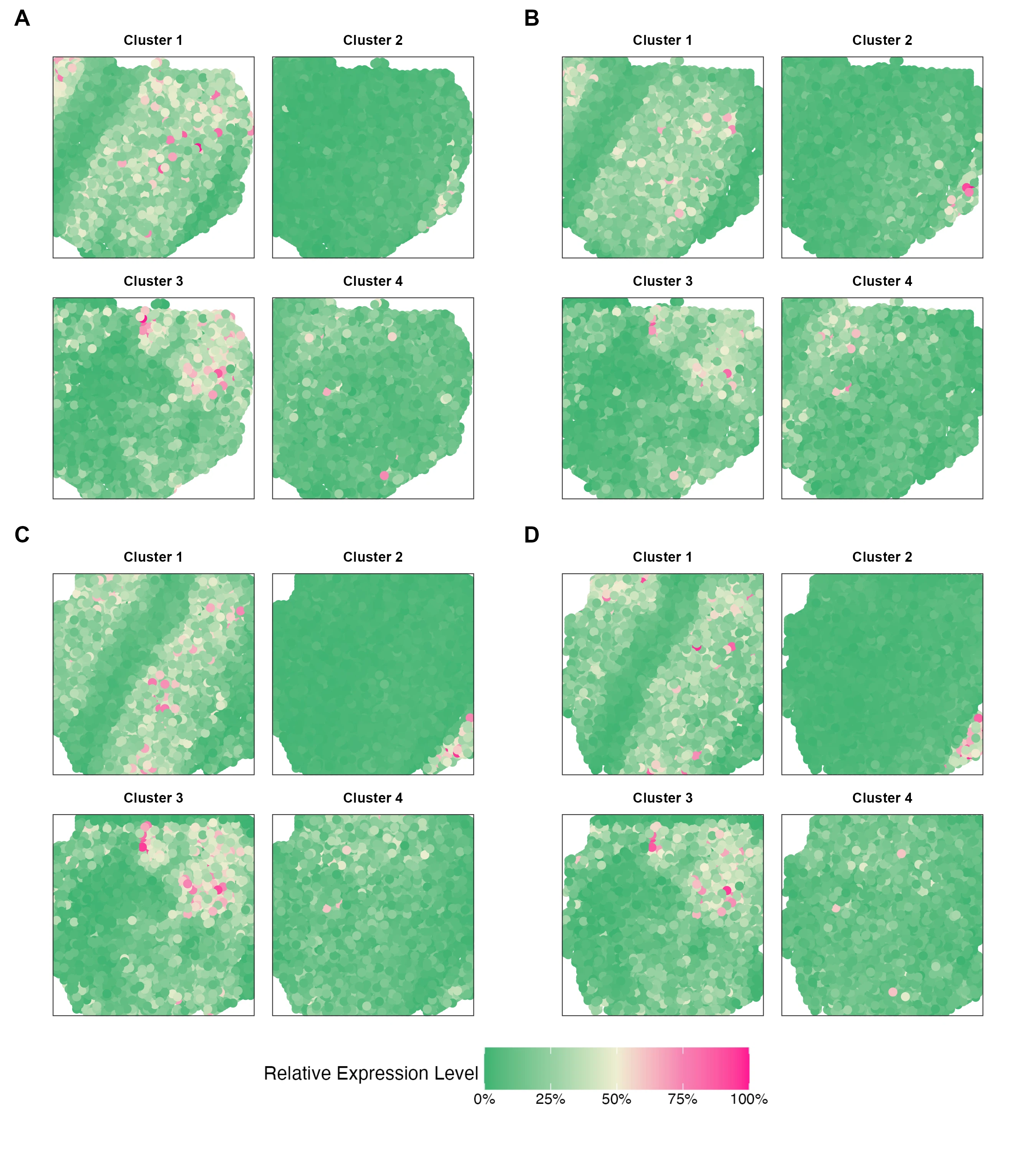}
    \caption{Four dominant gene clusters of SV genes identified by IBaySVG in the same-donor DLPFC dataset. (A–D) correspond to tissue slices A to D.}
    \label{genecluster}
\end{figure}

\begin{figure}[htbp]
    \centering
    \includegraphics[width=\linewidth]{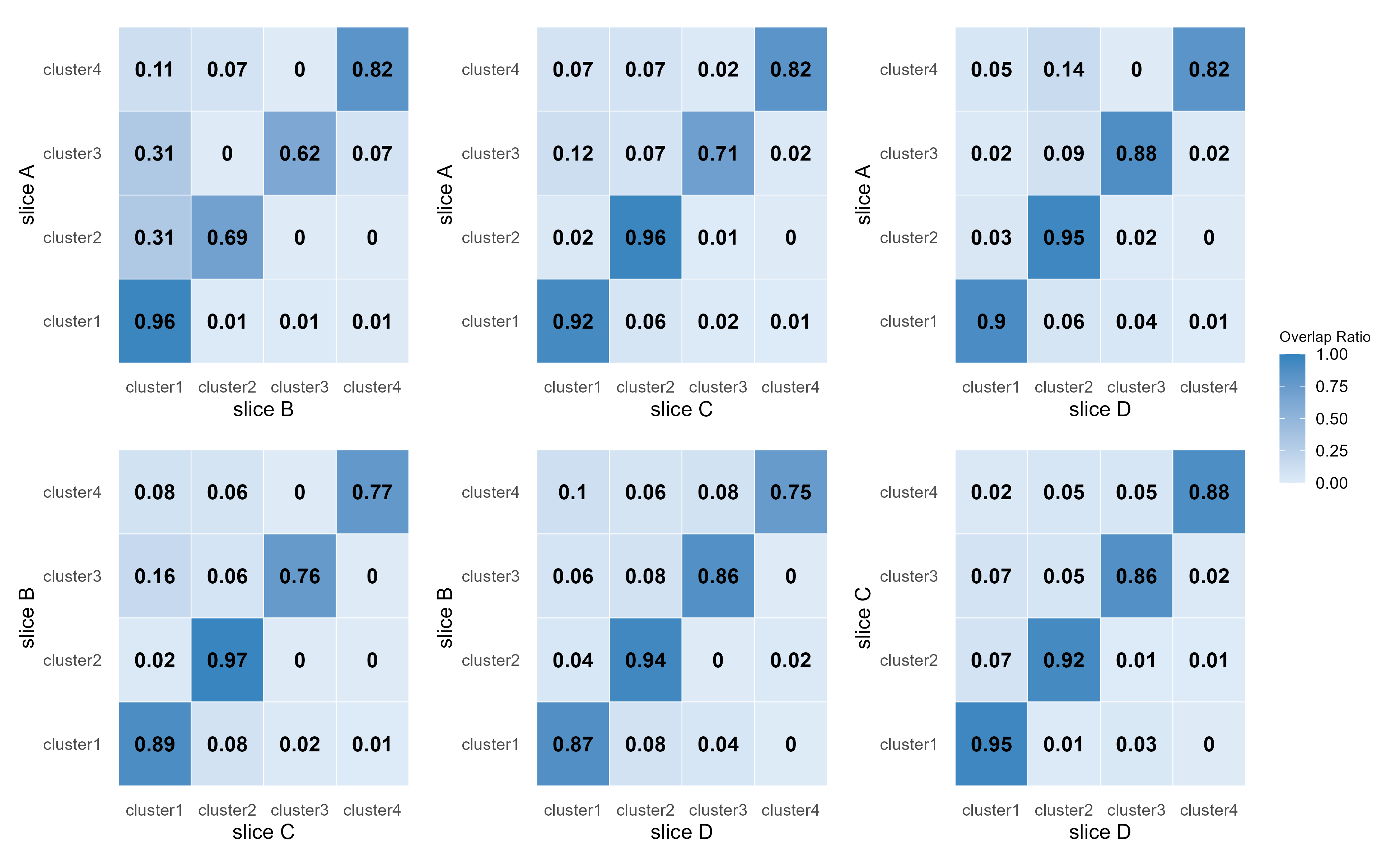}
    \caption{Heatmaps showing the overlap ratio of four dominant gene clusters identified by IBaySVG across different tissue slices in the same-donor DLPFC dataset. Each tile represents the proportion of genes in a cluster from one slice that are also assigned to a cluster in another slice. Higher overlap ratios (darker blue) indicate stronger correspondence between gene clusterings across slices.}
    \label{DLPFC_cluster_heatmap}
\end{figure}

\begin{figure}[!ht]
    \centering
    \includegraphics[width=0.8\linewidth]{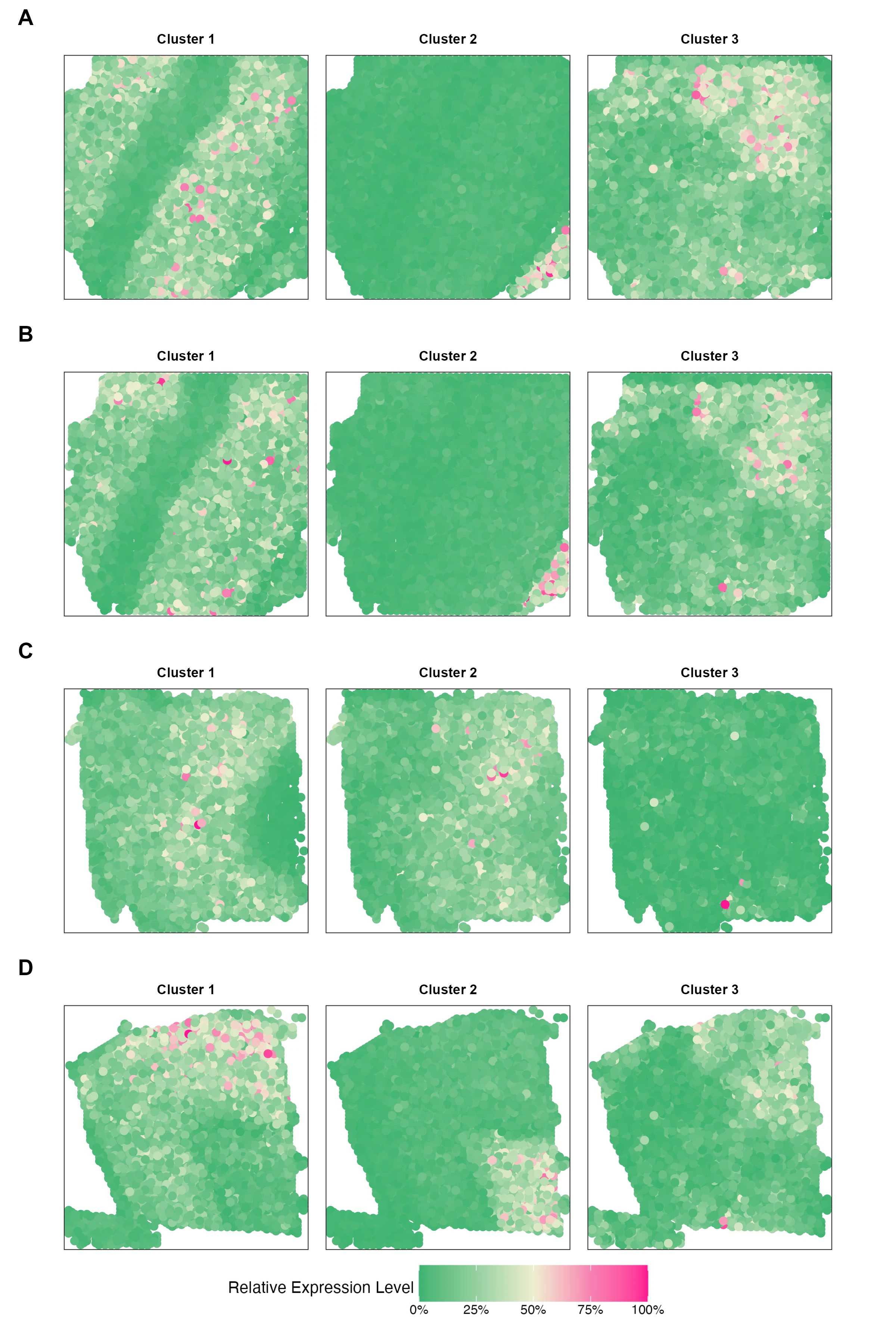}
    \caption{Three gene clusters of SV genes identified by IBaySVG in the across-donor DLPFC dataset. (A–D) correspond to tissue slices A to D.}
    \label{genecluster_34711}
\end{figure}

\begin{figure}[!ht]
    \centering
    \includegraphics[width=\linewidth]{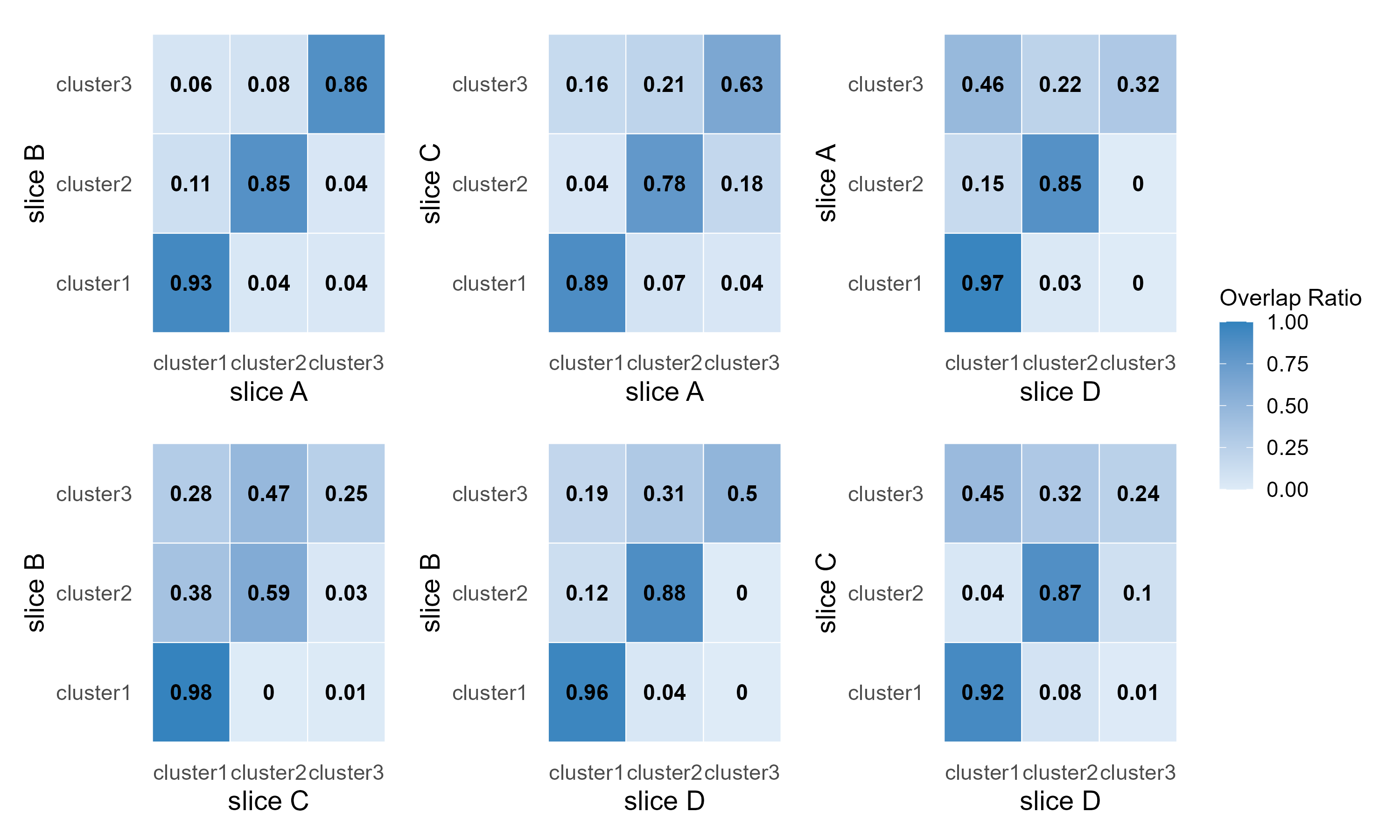}
    \caption{Heatmaps showing the overlap ratio of three gene clusters identified by IBaySVG across different tissue slices in the across-donor DLPFC dataset. Each tile represents the proportion of genes in a cluster from one slice that are also assigned to a cluster in another slice. Higher overlap ratios (darker blue) indicate stronger correspondence between gene clusterings across slices.}
    \label{dlpfc34711_cluster_heatmap}
\end{figure}

\begin{figure}[!ht]
    \centering
        \includegraphics[width=0.8\linewidth]{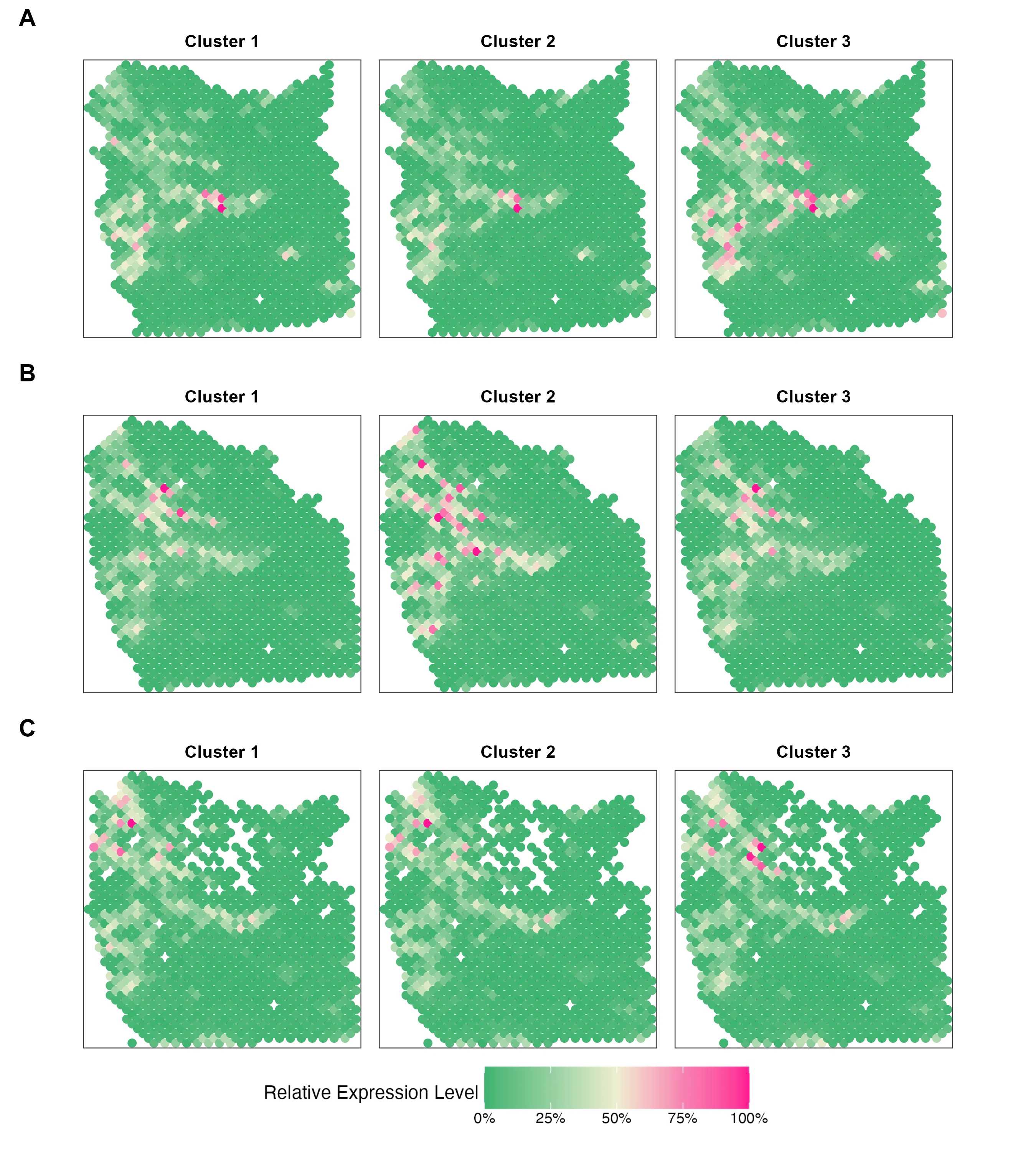}
    \caption{Three gene clusters of SV genes identified by IBaySVG in the SCC dataset. (A–C) correspond to tissue slices A to C.}
    \label{fig:genecluster_scc}
\end{figure}

\begin{figure}[!ht]
    \centering
    \includegraphics[width=\linewidth]{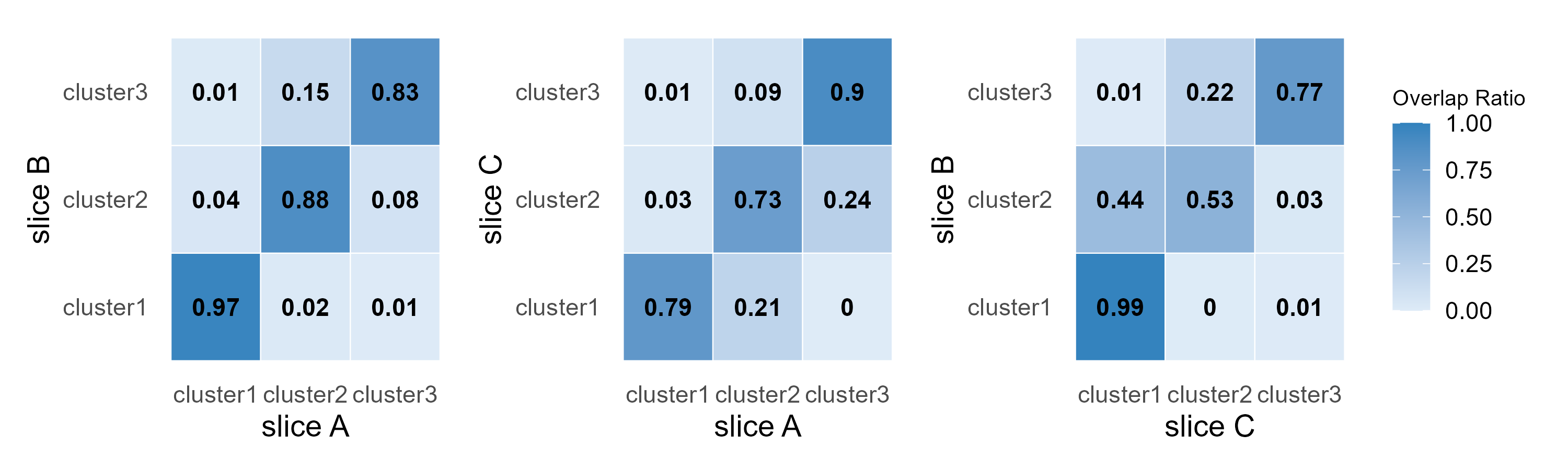}
    \caption{Heatmaps showing the overlap ratio of three gene clusters identified by IBaySVG across different tissue slices in the SCC dataset. Each tile represents the proportion of genes in a cluster from one slice that are also assigned to a cluster in another slice. Higher overlap ratios (darker blue) indicate stronger correspondence between gene clusterings across slices.}
    \label{scc_cluster_heatmap}
\end{figure}

\begin{table}[!ht]
\centering
\caption{Top five enriched GO terms and their adjusted p-values (the maximum values observed across the four slices) for the four dominant gene clusters of the four slices in the same-donor DLPFC dataset.}
\begin{tabular}{ccc}
\toprule
\textbf{GO Term} & \textbf{Description} & \textbf{p.adjust} \\ 
\midrule
 \multicolumn{3}{c}{Cluster 1}\\
GO:0050803 &regulation of synapse structure or activity &$1.31 \times 10^{-9}$   \\ 
GO:0050807 &         regulation of synapse organization& $5.82 \times 10^{-9}$   \\
GO:0042391  &          regulation of membrane potential &$1.52 \times 10^{-8}$   \\
GO:0010975 & regulation of neuron projection development& $3.98 \times 10^{-8}$   \\ 
GO:0007409  &                              axonogenesis &$5.55 \times 10^{-8} $   \\
\midrule
\multicolumn{3}{c}{Cluster 2}\\
GO:0008366 & Axon ensheathment & $7.31 \times 10^{-19}$ \\
GO:0007272 & Ensheathment of neurons & $7.31 \times 10^{-19}$ \\ 
GO:0042552 & Myelination & $1.36 \times 10^{-17}$ \\ 
GO:0048709 & Oligodendrocyte differentiation & $4.49 \times 10^{-14}$ \\ 
GO:0042063 & Gliogenesis & $1.52 \times 10^{-11}$ \\ 
\midrule
\multicolumn{3}{c}{Cluster 3}\\
GO:0015670  &                                          carbon dioxide transport &$4.24 \times 10^{-3}$ \\     
GO:0015671 &                                                   oxygen transport &$4.24 \times 10^{-3}$ \\ 
GO:0098869  &                                   cellular oxidant detoxification &$7.35 \times 10^{-3}$ \\     
GO:0010273  &                                      detoxification of copper ion &$8.36 \times 10^{-3}$ \\   
GO:0010744 &positive regulation of macrophage derived  &$8.36 \times 10^{-3}$ \\ 
& foam cell differentiation\\
\midrule
\multicolumn{3}{c}{Cluster 4}\\
GO:0002429 &immune response-activating cell surface  &$4.93 \times 10^{-3}$ \\ 
&receptor signaling pathway\\
GO:0002768 &immune response-regulating cell surface  &$4.93 \times 10^{-3}$ \\ 
&receptor signaling pathway\\
GO:0098743 &                                                  cell aggregation &$6.10 \times 10^{-4}$ \\ 
GO:0030199   &                                    collagen fibril organization &$6.10 \times 10^{-4}$ \\ 
GO:0001101  &                                        response to acid chemical &$6.10 \times 10^{-4}$ \\  
\bottomrule
\end{tabular}
\label{tab:go_enrichment}
\end{table}

\begin{table}[!ht]
\centering
\caption{Top five enriched GO terms and their adjusted p-values (the maximum values observed across the four slices) for the three gene clusters of the four slices in the across-donor DLPFC dataset.}
\begin{tabular}{ccc}
\toprule
\textbf{GO Term} & \textbf{Description} & \textbf{p.adjust} \\ 
\midrule
 \multicolumn{3}{c}{Cluster 1}\\
GO:0050807 & regulation of synapse organization & $2.40 \times 10^{-14}$ \\
GO:0050803 & regulation of synapse structure or activity & $2.40 \times 10^{-14}$ \\
GO:0099003 & vesicle-mediated transport in synapse & $9.84 \times 10^{-14}$ \\
GO:0007416 & synapse assembly & $3.00 \times 10^{-13}$ \\
GO:0042391 & regulation of membrane potential & $2.87 \times 10^{-11}$ \\
\midrule
\multicolumn{3}{c}{Cluster 2}\\
GO:0008366 & axon ensheathment & $1.84 \times 10^{-18}$ \\
GO:0007272 & ensheathment of neurons & $1.84 \times 10^{-18}$ \\
GO:0042552 & myelination & $1.51 \times 10^{-17}$ \\
GO:0048709 & oligodendrocyte differentiation & $9.56 \times 10^{-14}$ \\
GO:0042063 & gliogenesis & $5.45 \times 10^{-12}$ \\
\midrule
\multicolumn{3}{c}{Cluster 3}\\
GO:0090183 & regulation of kidney development & $1.37 \times 10^{-3}$ \\
GO:0070542 & response to fatty acid & $1.37 \times 10^{-3}$ \\
GO:0097006 & regulation of plasma lipoprotein particle levels & $2.84 \times 10^{-3}$ \\
GO:0002526 & acute inflammatory response & $3.89 \times 10^{-3}$ \\
GO:0042476 & odontogenesis & $4.46 \times 10^{-3}$ \\
\bottomrule
\end{tabular}
\label{tab:go_enrichment_34711}
\end{table}

\begin{table}[!ht]
\centering
\caption{Top five enriched GO terms and their adjusted p-values (the maximum values observed across the three slices) for the three clusters of the three slices in the SCC dataset.}
\begin{tabular}{ccc}
\toprule
\textbf{GO Term} & \textbf{Description} & \textbf{p.adjust} \\ 
\midrule
\multicolumn{3}{c}{Cluster 1}\\
GO:0045116 & protein neddylation &$3.72 \times 10^{-4}$ \\
GO:0046034 & ATP metabolic process &$2.28 \times 10^{-3}$ \\
GO:0009060 & aerobic respiration &$2.78 \times 10^{-3}$ \\
GO:0009141 & nucleoside triphosphate metabolic process &$2.78 \times 10^{-3}$ \\
GO:0006091 & generation of precursor metabolites and energy &$3.02 \times 10^{-3}$ \\
\midrule
\multicolumn{3}{c}{Cluster 2}\\
GO:0006119 & oxidative phosphorylation &$4.72 \times 10^{-4}$ \\
GO:0045229 & external encapsulating structure organization &$5.96 \times 10^{-4}$ \\
GO:0030198 & extracellular matrix organization &$5.96 \times 10^{-4}$ \\
GO:0043062 & extracellular structure organization &$5.96 \times 10^{-4}$ \\
GO:0002011 & morphogenesis of an epithelial sheet &$7.65 \times 10^{-4}$ \\
\midrule
\multicolumn{3}{c}{Cluster 3}\\
GO:0002181 & cytoplasmic translation &$1.38 \times 10^{-23}$ \\
GO:0042773 & ATP synthesis coupled electron transport &$1.36 \times 10^{-4}$ \\
GO:0042775 & mitochondrial ATP synthesis coupled electron transport &$1.36 \times 10^{-4}$ \\
GO:0022904 & respiratory electron transport chain &$1.36 \times 10^{-4}$ \\
GO:0022900 & electron transport chain &$1.42 \times 10^{-4}$ \\
\bottomrule
\end{tabular}
\label{tab:go_enrichment_scc}
\end{table}

\clearpage
\subsection{Details of the spot clustering results}

Denote $Y \in \mathbb{R}^{n \times p}$ with $n$ samples and $p$ features. Let  $D \in \mathbb{R}^{n \times n}$  be the pairwise distance matrix where each entry $D_{ij}$ denotes the distance between samples $i$ and $j$ based on their feature vectors $y_i$ and $y_j$,  $C = \{C_1, \dots, C_K\}$  be the clustering result with $K$ clusters and a reference partition $G = \{G_1, \dots, G_L\}$. Let $n_{kl} = |C_k \cap G_l|$ denote the number of samples that are assigned to cluster $C_k$ and reference group $G_l$. Define
$a_k = \sum_{l=1}^{L} n_{kl}$ and $b_l = \sum_{k=1}^{K} n_{kl}$ as the marginal
sums of the contingency table. For each cluster $C_k$ , define the medoid center as:
\[
u_k = \arg\min_{i \in C_k} \sum_{j \in C_k} D_{ij}.
\]
\begin{itemize}
\item { ARI. It measures the agreement between two partitions by comparing all pairs of samples and correcting for chance:
\begin{equation}
\mathrm{ARI}=\frac{\sum_{k=1}^{K}\sum_{l=1}^{L} \binom{n_{kl}}{2}-
\frac{\sum_{k=1}^{K} \binom{a_k}{2}\sum_{l=1}^{L} \binom{b_l}{2}}{
\binom{n}{2}}}{\frac{1}{2}\left(\sum_{k=1}^{K} \binom{a_k}{2}+\sum_{l=1}^{L} \binom{b_l}{2}
\right)-\frac{\sum_{k=1}^{K} \binom{a_k}{2}\sum_{l=1}^{L} \binom{b_l}{2}}{\binom{n}{2}}}.
\end{equation}
where $\binom{m}{2} = m(m-1)/2$ denotes the number of unordered pairs that
can be formed from $m$ elements. 
}

\item { NMI. It quantifies the amount of shared information between two partitions, normalized to account for differences in cluster sizes. It is defined based on the contingency table $\{n_{kl}\}$ as
\begin{equation}
\mathrm{NMI}=\frac{\sum_{k=1}^{K}\sum_{l=1}^{L}n_{kl}\log\left(\frac{n_{kl} \, n}{a_k b_l}\right)}{\sqrt{\left(\sum_{k=1}^{K} a_k \log \frac{a_k}{n}\right)
\left(\sum_{l=1}^{L} b_l \log \frac{b_l}{n}\right)}}.
\end{equation}
}

\item DBI. It evaluates clustering quality based on the trade-off between compactness and separation, focusing on the regularity of cluster shapes and the clarity of cluster boundaries:
\[
\text{DBI} = \frac{1}{K} \sum_{i=1}^K \max_{j \ne i} \left( \frac{S_i + S_j}{D_{u_i u_j}} \right),
\]
where $S_k = \frac{1}{n_k} \sum_{i \in C_k} D_{iu_k}$ is average distance from samples in $C_k$ to the medoid $u_k$.
\item CH. It utilizes a variance ratio approach, making it particularly suitable for comparing different numbers of clusters:
\[
\text{CH} = \frac{B / (K - 1)}{W / (n - K)},
\]
where $W = \sum_{k=1}^K \frac{1}{2 n_k} \sum_{i,j \in C_k} D_{ij}^2$ and $B = \frac{1}{2n} \sum_{k=1}^K \sum_{l \ne k} \frac{1}{n_k n_l} \sum_{i \in C_k} \sum_{j \in C_l} D_{ij}^2$ represent the within-cluster and between-cluster dispersion, respectively.
\item Silhouette Coefficient. It provides a more intuitive measure of clustering effectiveness at the individual sample level, facilitating the identification of potential misclassified samples. Here, we computed the mean Silhouette Coefficient across all samples:
\[
\bar{s} = \frac{1}{n} \sum_{i=1}^n \frac{b(i) - a(i)}{\max\{a(i), b(i)\}},
\]
where $a(i) = \frac{1}{n_k - 1} \sum_{\substack{j \in C_k \\ j \ne i}} D_{ij}$ is average distance from sample $i$ to other samples in its own cluster and
$b(i) = \min_{l \ne k} \left( \frac{1}{n_l} \sum_{j \in C_l} D_{ij} \right)$ is minimum average distance from sample $i$ to all other clusters.

\item ANOVA-derived F statistic. It is a testing statistic to assess the statistical significance of inter-cluster differences:
\[
        F = \frac{\sum_{k=1}^K n_k (\bar{M}_k - \bar{M})^2 / (K - 1)}{\sum_{k=1}^K \sum_{i \in C_k} (M_i - \bar{M}_k)^2 / (n - K)},
\]
where $M_i = \frac{1}{p} \sum_{j=1}^p Y_{ij}$, $\bar{M}_k = \frac{1}{n_k} \sum_{i \in C_k} M_i$, and $\bar{M} = \frac{1}{n} \sum_{i=1}^n M_i$.

We used the F-value in place of the p-value as an evaluation metric to enhance interpretability and facilitate cross-method comparison.
\end{itemize}

Here, superior clustering performance is indicated by higher ARI, NMI, CH Index, Silhouette Coefficient, and F statistic values, coupled with lower DBI scores. In our study, the reference partition for ARI and NMI is based on the manually annotated spot domains from \cite{Maynard2021}, which divide the human dorsolateral prefrontal cortex into eight regions: layers ``L1'' through ``L6'', white matter (``WM''), and ``others''.

\begin{table}[htbp]
\setlength{\belowcaptionskip}{-2pt} 
\centering
\renewcommand{\arraystretch}{1} 
\caption{Comparative evaluation of spot clustering performance via ARI on the same-donor DLPFC dataset. In each part, the single-layer method yields six ARI results based on SV genes 
from each, multiple-slice union set, multiple-slice intersection set, Cauchy combination set, harmonic mean p-value set, and PASTE integration slice.}
\footnotesize
\begin{tabular}{l c cccccc} 
\toprule
\textbf{Sample/Method} & \textbf{Multi-layer} & \textbf{Single-layer} & \textbf{Union} & \textbf{Intersect} &\textbf{Cauchy} &\textbf{HMP} & \textbf{PASTE} \\ 
\midrule
\textbf{Section 1} & & & & & &&\\
IBaySVG & 0.506  &       &       &       &       &       &       \\
DESpace      & 0.443  &       &       &       &       &       &       \\
Full transcriptome & 0.427  &       &       &       &       &       &       \\
nnSVG        &        & 0.413 & 0.428 & 0.432 & 0.385 & 0.371 & 0.369 \\
SPARKX       &        & 0.488 & 0.424 & 0.538 & 0.473 & 0.452 & 0.412 \\
SPARK        &        & 0.459 & 0.479 & 0.440 & 0.472 & 0.446 & 0.423 \\
spVC         &        & 0.376 & 0.455 & 0.428 & 0.395 & 0.480 & 0.409 \\
HEARTSVG     &        & 0.524 & 0.439 & 0.392 & 0.387 & 0.488 & 0.441 \\
GASTON && 0.380 &0.450 &0.2780 &&&0.461\\
StarTrail && 0.450 &0.383 &0.293&&&0.390\\
\midrule
\textbf{Section 2} & & & & & &&\\
IBaySVG & 0.499  &       &       &       &       &       &       \\
DESpace      & 0.405  &       &       &       &       &       &       \\
Full transcriptome & 0.409  &       &       &       &       &       &       \\
nnSVG        &        & 0.414 & 0.433 & 0.425 & 0.382 & 0.386 & 0.381 \\
SPARKX       &        & 0.448 & 0.410 & 0.539 & 0.490 & 0.433 & 0.398 \\
SPARK        &        & 0.441 & 0.494 & 0.439 & 0.444 & 0.442 & 0.394 \\
spVC         &        & 0.503 & 0.443 & 0.412 & 0.404 & 0.464 & 0.403 \\
HEARTSVG     &        & 0.521 & 0.438 & 0.383 & 0.401 & 0.453 & 0.420 \\
GASTON &&0.352 &0.448 &0.247 &&&0.470\\
StarTrail &&0.439 &0.393 &0.309  &&&0.404  \\
\midrule
\textbf{Section 3} & & & & & &&\\
IBaySVG & 0.567  &       &       &       &       &       &       \\
DESpace      & 0.414  &       &       &       &       &       &       \\
Full transcriptome & 0.443  &       &       &       &       &       &       \\
nnSVG        &        & 0.469 & 0.473 & 0.494 & 0.454 & 0.411 & 0.364 \\
SPARKX       &        & 0.503 & 0.453 & 0.535 & 0.497 & 0.494 & 0.447 \\
SPARK        &        & 0.493 & 0.512 & 0.488 & 0.427 & 0.492 & 0.434 \\
spVC         &        & 0.438 & 0.486 & 0.415 & 0.414 & 0.523 & 0.432 \\
HEARTSVG     &        & 0.476 & 0.483 & 0.445 & 0.424 & 0.542 & 0.390 \\
GASTON&& 0.378 &0.502 &0.231 &&&0.411\\
StarTrail&&0.473 &0.478 &0.404  &&&0.359  \\
\midrule
\textbf{Section 4} & & & & & &&\\
IBaySVG & 0.567  &       &       &       &       &       &       \\
DESpace      & 0.427  &       &       &       &       &       &       \\
Full transcriptome & 0.449  &       &       &       &       &       &       \\
nnSVG        &        & 0.484 & 0.490 & 0.474 & 0.491 & 0.426 & 0.369 \\
SPARKX       &        & 0.438 & 0.484 & 0.581 & 0.515 & 0.525 & 0.465 \\
SPARK        &        & 0.482 & 0.551 & 0.522 & 0.422 & 0.507 & 0.427 \\
spVC         &        & 0.429 & 0.511 & 0.380 & 0.422 & 0.496 & 0.457 \\
HEARTSVG     &        & 0.554 & 0.501 & 0.451 & 0.450 & 0.561 & 0.399 \\
GASTON &&0.322 &0.496 &0.242 &&&0.410\\
StarTrail&& 0.523 &0.484 &0.322 &&&0.375\\   
\hline
\end{tabular}
\end{table}

\begin{table}[htbp]
\setlength{\belowcaptionskip}{-2pt} 
\centering
\renewcommand{\arraystretch}{1} 
\caption{Comparative evaluation of spot clustering performance via NMI on the same-donor DLPFC dataset. In each part, the single-layer method yields six NMI results based on SV genes 
from each, multiple-slice union set, multiple-slice intersection set, Cauchy combination set, harmonic mean p-value set, and PASTE integration slice.}
\footnotesize
\begin{tabular}{l c cccccc} 
\toprule
\textbf{Sample/Method} & \textbf{Multi-layer} & \textbf{Single-layer} & \textbf{Union} & \textbf{Intersect} &\textbf{Cauchy} &\textbf{HMP} & \textbf{PASTE} \\ 
\midrule
\textbf{Section 1} & & & & & &&\\
IBaySVG  & 0.619 &  &  &  &  &  &  \\
DESpace   & 0.630 &  &  &  &  &  &  \\
Full transcriptome & 0.650  &       &       &       &       &       &       \\
nnSVG     &  & 0.578 & 0.601 & 0.572 & 0.569 & 0.560 & 0.582 \\
SPARKX    &  & 0.647 & 0.639 & 0.645 & 0.635 & 0.651 & 0.609 \\
SPARK     &  & 0.639 & 0.645 & 0.607 & 0.578 & 0.655 & 0.619 \\
spVC      &  & 0.579 & 0.622 & 0.565 & 0.635 & 0.618 & 0.632 \\
HEARTSVG  &  & 0.648 & 0.599 & 0.621 & 0.639 & 0.602 & 0.650 \\
GASTON    &  & 0.578 & 0.545 & 0.432 &  &  & 0.581 \\
StarTrail &  & 0.427 & 0.561 & 0.406 &  &  &  0.582\\

\midrule
\textbf{Section 2} & & & & & &&\\
IBaySVG  & 0.609 &  &  &  &  &  &  \\
DESpace   & 0.604 &  &  &  &  &  &  \\
Full transcriptome & 0.613  &       &       &       &       &       &       \\
nnSVG     &  & 0.565 & 0.595 & 0.563 & 0.544 & 0.562 & 0.581 \\
SPARKX    &  & 0.607 & 0.608 & 0.650 & 0.607 & 0.637 & 0.580 \\
SPARK     &  & 0.597 & 0.628 & 0.597 & 0.543 & 0.619 & 0.595 \\
spVC      &  & 0.632 & 0.593 & 0.510 & 0.607 & 0.594 & 0.595 \\
HEARTSVG  &  & 0.645 & 0.589 & 0.601 & 0.607 & 0.604 & 0.602 \\
GASTON    &  & 0.543 & 0.537 & 0.440 &  &  &  0.572\\
StarTrail &  & 0.538 & 0.551 & 0.381 &  &  &  0.563\\

\midrule
\textbf{Section 2} & & & & & &&\\
IBaySVG  & 0.650 &  &  &  &  &  &  \\
DESpace   & 0.590 &  &  &  &  &  &  \\
Full transcriptome & 0.596 &       &       &       &       &       &       \\
nnSVG     &  & 0.589 & 0.610 & 0.576 & 0.551 & 0.582 & 0.596 \\
SPARKX    &  & 0.613 & 0.582 & 0.643 & 0.593 & 0.635 & 0.604 \\
SPARK     &  & 0.627 & 0.633 & 0.609 & 0.583 & 0.597 & 0.618 \\
spVC      &  & 0.592 & 0.603 & 0.544 & 0.591 & 0.585 & 0.589 \\
HEARTSVG  &  & 0.598 & 0.603 & 0.618 & 0.589 & 0.604 & 0.621 \\
GASTON    &  & 0.492 & 0.571 & 0.512 &  &  &  0.579\\
StarTrail &  & 0.446 & 0.577 & 0.349 &  &  & 0.508 \\

\midrule
\textbf{Section 2} & & & & & &&\\
IBaySVG  & 0.627 &  &  &  &  &  &  \\
DESpace   & 0.559 &  &  &  &  &  &  \\
Full transcriptome & 0.591  &       &       &       &       &       &       \\
nnSVG     &  & 0.599 & 0.620 & 0.557 & 0.545 & 0.599 & 0.606 \\
SPARKX    &  & 0.582 & 0.598 & 0.649 & 0.592 & 0.635 & 0.628 \\
SPARK     &  & 0.618 & 0.645 & 0.627 & 0.579 & 0.587 & 0.620 \\
spVC      &  & 0.555 & 0.626 & 0.517 & 0.598 & 0.570 & 0.577 \\
HEARTSVG  &  & 0.619 & 0.629 & 0.601 & 0.584 & 0.610 & 0.626 \\
GASTON    &  & 0.569 & 0.571 & 0.437 &  &  & 0.564 \\
StarTrail &  & 0.438 & 0.584 & 0.358 &  &  & 0.519 \\
\hline
\end{tabular}
\end{table}

\begin{table}[htbp]
\setlength{\belowcaptionskip}{-2pt} 
\centering
\renewcommand{\arraystretch}{1} 
\caption{Comparative evaluation of spot clustering performance in slice A of same-donor DLPFC dataset. In each part, the single-layer method yields six results based on SV genes 
from slice A, multiple-slice union set, multiple-slice intersection set, Cauchy combination set, harmonic mean p-value set, and PASTE integration slice. 
}
\footnotesize
\begin{tabular}{l c ccccccc} 
\toprule
\textbf{Index/Test} &\textbf{Multi-layer} & \textbf{Single-layer} & \textbf{Union} & \textbf{Intersect}  & \textbf{Cauchy} &\textbf{HMP} & \textbf{PASTE} \\ % 表头
\midrule
DBI        & &  &  &  &  & &\\  
IBaySVG   & 1.858  & &    &     &    &     &    \\
 DESpace    & 1.867 &  &    &     &     &     &     \\
nnSVG    &  & 1.872 & 1.877 & 1.870 & 1.863 & 1.874 & 1.893 \\
SPARKX   &  & 1.858  & 1.879 & 1.883 & 1.863 & 1.875 & 1.875 \\
SPARK    &  & 1.876  & 1.886 & 1.863 & 1.909 & 1.871 & 1.877 \\
 spVC    &   & 1.887  & 1.879 & 1.860 & 1.864 & 1.889 & 1.893 \\
 HEARTSVG &  & 1.880  & 1.878 & 1.868 & 1.904 & 1.900 & 1.888 \\
 GASTON  &  & 1.872 &1.86& 1.865& &&1.874  \\
  StarTrail  &  &1.924&1.90& 1.935 &&&1.863  \\

\midrule
CH index & &  &  &  &  & &\\ 
 IBaySVG   &  327.965 & & & & & \\
 DESpace    & 312.984 & & & & & \\
nnSVG    &  & 281.514 & 320.745 & 289.977 & 328.428 & 304.014 & 306.755 \\
SPARKX   &  & 309.151 & 306.345 & 316.791 & 308.460 & 330.214 & 333.501 \\
 SPARK    &  & 334.301 & 318.878 & 305.882 & 275.070 & 309.632 & 329.100 \\
spVC     &  & 320.350 & 339.793 & 322.250 & 305.153 & 314.880 & 322.758 \\
HEARTSVG &  & 318.101 & 327.937 & 314.544 & 307.256 & 313.411 & 323.045 \\
GASTON  &  & 309.374  &309.935 &263.664 &&&342.098  \\
StarTrail&  & 206.316 &332.219 &176.251 &&&323.933  \\

\midrule
Silhouette & &  &  &  &  & &\\ 
IBaySVG   & -0.034  & & & & & & \\
DESpace    & -0.088  & & & & & & & \\
 nnSVG   &   & -0.052  & -0.078 & -0.041 & -0.078 & -0.081 & -0.088 \\
SPARKX   &  & -0.057  & -0.093 & -0.033 & -0.092 & -0.072 & -0.078 \\
 SPARK   &   & -0.070  & -0.073 & -0.080 & -0.063 & -0.082 & -0.074 \\
spVC     &  & -0.063  & -0.068 & -0.047 & -0.091 & -0.068 & -0.045 \\
HEARTSVG &  & -0.033  & -0.076 & -0.082 & -0.072 & -0.086 & -0.048 \\
GASTON   &&-0.040 &5 -0.037 &-0.093 &&&-0.072 \\
StarTrail &&-0.077  &-0.038 &-0.087 &&&-0.037  \\

\midrule
ANOVA F value & &  &  &  &  & &\\
 IBaySVG   & 412.684 & & & & & \\
 DESpace    & 371.460 & & & & & \\
nnSVG     & & 334.798 & 402.662 & 335.045 & 414.574 & 370.885 & 368.558 \\
SPARKX   &  & 374.396 & 380.565 & 389.256 & 379.115 & 400.237 & 414.203 \\
SPARK    &  & 411.090 & 387.489 & 382.019 & 328.162 & 378.387 & 412.641 \\
spVC     &  & 411.118 & 420.755 & 410.013 & 373.119 & 396.754 & 405.716 \\
 HEARTSVG  & & 381.108 & 411.718 & 394.580 & 378.044 & 380.468 & 403.709 \\
 GASTON  & &   366.657 & 371.859& 375.284&&& 452.101  \\
 StarTrail  & & 241.455 &416.000 &204.387 &&&428.505  \\

\bottomrule
\end{tabular}
\label{cluster_table2}
\end{table}

\begin{table}[htbp]
\setlength{\belowcaptionskip}{-2pt} 
\centering
\renewcommand{\arraystretch}{1} 
\caption{Comparative evaluation of spot clustering performance in slice B of same-donor DLPFC dataset. In each part, the single-layer method yields six results based on SV genes 
from slice B, multiple-slice union set, multiple-slice intersection set, Cauchy combination set, harmonic mean p-value set, and PASTE integration slice. }
\footnotesize
\begin{tabular}{l c cccccc} 
\toprule
\textbf{Index/Test} & \textbf{Multi-layer} & \textbf{Single-layer} & \textbf{Union} & \textbf{Intersect} &\textbf{Cauchy} &\textbf{HMP} & \textbf{PASTE} \\ 
\midrule
DBI & & & & &&& \\  
IBaySVG & 1.853 &      &      &      &      &      \\
DESpace  & 1.869 &      &      &      &      &      \\
nnSVG   & & 1.867 & 1.869 & 1.861 & 1.864 & 1.880 & 1.890 \\
SPARKX  & & 1.870 & 1.870 & 1.887 & 1.857 & 1.873 & 1.889 \\
SPARK   & & 1.877 & 1.891 & 1.869 & 1.884 & 1.861 & 1.880 \\
spVC    & & 1.874 & 1.872 & 1.840 & 1.858 & 1.879 & 1.900 \\
HEARTSVG& & 1.881 & 1.876 & 1.868 & 1.893 & 1.887 & 1.912 \\
GASTON  & &  1.874 & 1.844 &1.883 &&& 1.877  \\
StarTrail & & 1.904 & 1.890 &1.946 &&& 1.855  \\

\midrule
CH index & & & & & &&\\ 
IBaySVG &410.122&      &      &      &      &      \\
DESpace  & 334.901 &      &      &      &      &      \\
nnSVG   & & 394.686 & 344.182 & 362.118 & 283.128 & 298.463 & 290.845 \\
SPARKX  & & 335.786 & 331.043 & 381.052 & 339.933 & 391.467 & 429.955 \\
SPARK   & & 433.383 & 373.277 & 330.548 & 347.537 & 319.568 & 432.711 \\
spVC    & & 363.158 & 439.870 & 286.728 & 340.927 & 308.232 & 404.288 \\
HEARTSVG& & 373.361 & 427.034 & 313.705 & 326.920 & 320.084 & 408.343 \\
GASTON    & & 309.024 & 287.64 &333.303&&&  332.210  \\
StarTrail & & 269.037 & 351.75 &162.935 &&& 317.718  \\

\midrule
Silhouette & & & & & &&\\ 
IBaySVG & -0.038 &      &      &      &      &      \\
DESpace  & -0.094 &      &      &      &      &      \\
nnSVG   & & -0.067 & -0.081 & -0.057 & -0.100 & -0.091 & -0.101 \\
SPARKX  & & -0.083 & -0.105 & -0.041 & -0.102 & -0.072 & -0.059 \\
SPARK   & & -0.060 & -0.072 & -0.083 & -0.068 & -0.093 & -0.059 \\
spVC    & & -0.080 & -0.058 & -0.086 & -0.101 & -0.085 & -0.052 \\
HEARTSVG && -0.077 & -0.064 & -0.105 & -0.090 & -0.099 & -0.053 \\
GASTON   & & -0.055 & -0.072 &-0.086 &&& -0.092  \\
StarTrail & & -0.087 & -0.046 &-0.111 &&& -0.046   \\

\midrule
ANOVA F value & & & & & &&\\
IBaySVG & 476.987 &      &      &      &      &      \\
DESpace  & 368.490 &      &      &      &      &      \\
nnSVG   & & 439.944 & 387.762 & 379.365 & 343.345 & 346.386 & 335.152 \\
SPARKX &  & 361.768 & 358.173 & 440.767 & 367.820 & 441.553 & 475.574 \\
SPARK   & & 477.283 & 422.323 & 363.353 & 366.350 & 360.214 & 474.046 \\
spVC    & & 427.732 & 502.048 & 365.378 & 366.159 & 362.341 & 481.059 \\
HEARTSVG& & 440.261 & 483.774 & 360.626 & 362.712 & 347.315 & 477.996 \\
GASTON  & & 379.810 & 324.989 &451.694&&&  413.733  \\
StarTrail & & 329.419 & 403.129& 175.448&&&  371.260  \\

\bottomrule
\end{tabular}
\label{cluster_table3}
\end{table}

\begin{table}[htbp]
\setlength{\belowcaptionskip}{-2pt}
\centering
\renewcommand{\arraystretch}{1}
\caption{Comparative evaluation of spot clustering performance in slice C of same-donor DLPFC dataset. In each part, the single-layer method yields six results based on SV genes 
from slice C, multiple-slice union set, multiple-slice intersection set, Cauchy combination set, harmonic mean p-value set, and PASTE integration slice.}
\footnotesize
\begin{tabular}{l c cccccc}
\toprule
\textbf{Index/Test} & \textbf{Multi-layer} & \textbf{Single-layer} & \textbf{Union} & \textbf{Intersect}  & \textbf{Cauchy} &\textbf{HMP} & \textbf{PASTE}\\ 

\midrule
DBI & & & & & &&\\  
IBaySVG & 1.827     &      &      &      &      &      \\
DESpace  &  1.891     &      &      &      &      &      \\
nnSVG  &  & 1.906 & 1.897 & 1.915 & 1.878 & 1.882 & 1.895 \\
SPARKX  & & 1.903 & 1.888 & 1.849 & 1.882 & 1.835 & 1.875 \\
SPARK  &  & 1.828 & 1.824 & 1.897 & 1.911 & 1.885 & 1.837 \\
spVC    & & 1.870 & 1.863 & 1.869 & 1.886 & 1.894 & 1.885 \\
HEARTSVG& & 1.889 & 1.871 & 1.893 & 1.895 & 1.900 & 1.839 \\
GASTON    & & 1.873 & 1.890 &1.874 &&& 1.873  \\
StarTrail & & 1.909 & 1.919 &1.929 &&& 1.871 \\

\midrule
CH index & & & & & &&\\ 
IBaySVG &  518.117    &      &      &      &      &      \\
DESpace  &  374.721    &      &      &      &      &      \\
nnSVG  &  & 341.052 & 375.526 & 351.595 & 321.536 & 339.914 & 366.488 \\
SPARKX &  & 369.947 & 377.564 & 391.441 & 381.579 & 400.143 & 406.246 \\
SPARK  &  & 405.268 & 420.365 & 371.505 & 366.768 & 380.610 & 406.081 \\
spVC    & & 349.690 & 409.951 & 328.883 & 391.741 & 357.211 & 405.048 \\
HEARTSVG& & 364.707 & 400.842 & 363.577 & 378.867 & 368.740 & 403.322 \\
GASTON  & &  236.524 &323.743 &274.552 &&& 334.982   \\
StarTrail  & & 234.040 & 365.442 &194.206&&&  292.177 \\

\midrule
Silhouette & & & & & &&\\ 
IBaySVG &  -0.034    &      &      &      &      &      \\
DESpace  &  -0.068    &      &      &      &      &      \\
nnSVG   & & -0.055 & -0.063 & -0.012 & -0.075 & -0.069 & -0.078 \\
SPARKX &  & -0.020 & -0.075 & -0.035 & -0.069 & -0.072 & -0.068 \\
SPARK   & & -0.068 & -0.070 & -0.066 & -0.054 & -0.066 & -0.064 \\
spVC    & & -0.066 & -0.066 & -0.065 & -0.065 & -0.058 & -0.039 \\
HEARTSVG& & -0.063 & -0.069 & -0.069 & -0.062 & -0.081 & -0.032 \\
GASTON  & &  -0.108 &-0.022 &-0.095 &&& -0.075   \\
StarTrail & & -0.071 & -0.011& -0.080 &&& -0.039   \\

\midrule
ANOVA F value && & & & &&\\
IBaySVG & 574.263     &      &      &      &      &      \\
DESpace  & 521.455     &      &      &      &      &      \\
nnSVG   & & 481.149 & 530.815 & 466.148 & 467.962 & 487.500 & 516.324 \\
SPARKX &  & 502.817 & 532.267 & 546.803 & 541.916 & 564.660 & 572.431 \\
SPARK  &  & 574.268 & 593.000 & 522.993 & 496.346 & 526.214 & 579.934 \\
spVC    & & 494.612 & 596.218 & 482.998 & 552.063 & 499.891 & 563.275 \\
HEARTSVG& & 528.146 & 583.114 & 520.115 & 518.159 & 519.134 & 587.595 \\
GASTON   & &  371.049 & 426.529 &475.502 &&& 507.007  \\
StarTrail & & 308.922 & 545.865 &262.134 &&& 441.569  \\

\bottomrule
\end{tabular}
\label{cluster_table4}
\end{table}

\begin{table}[htbp]
\setlength{\belowcaptionskip}{-2pt} 
\centering
\renewcommand{\arraystretch}{1} 
\caption{Comparative evaluation of spot clustering performance in slice D of same-donor DLPFC dataset. In each part, the single-layer method yields six results based on SV genes 
from slice D, multiple-slice union set, multiple-slice intersection set, Cauchy combination set, harmonic mean p-value set, and PASTE integration slice.}
\footnotesize
\begin{tabular}{l c cccccc} 
\toprule
\textbf{Index/Test} & \textbf{Multi-layer} & \textbf{Single-layer} & \textbf{Union} & \textbf{Intersect}  & \textbf{Cauchy} &\textbf{HMP} & \textbf{PASTE}\\ 

\midrule
DBI & & & & & &&\\  
IBaySVG & 1.872 &  &  &  &  &  \\
DESpace  & 1.868 &  &  &  &  &  \\
nnSVG  &  & 1.864 & 1.864 & 1.860 & 1.868 & 1.878 & 1.863 \\
SPARKX &  & 1.872 & 1.886 & 1.870 & 1.878 & 1.877 & 1.889 \\
SPARK   & & 1.871 & 1.889 & 1.880 & 1.882 & 1.871 & 1.880 \\
spVC   &  & 1.867 & 1.872 & 1.869 & 1.872 & 1.867 & 1.860 \\
HEARTSVG& & 1.903 & 1.884 & 1.879 & 1.873 & 1.879 & 1.905 \\
GASTON    & & 1.857 &1.852 &1.886&&&  1.875  \\
StarTrail & & 1.920& 1.873 &1.927 &&& 1.863 \\

\midrule
CH index & & & & & &&\\ 
IBaySVG & 421.654 &  &  &  &  &  \\
DESpace  & 334.925 &  &  &  &  &  \\
nnSVG   & & 320.998 & 336.636 & 321.479 & 315.937 & 329.614 & 331.922 \\
SPARKX &  & 339.799 & 341.407 & 334.718 & 342.924 & 339.783 & 336.141 \\
SPARK  &  & 333.335 & 339.383 & 329.685 & 316.254 & 326.615 & 335.936 \\
spVC    & & 337.005 & 335.055 & 327.552 & 343.100 & 316.193 & 335.839 \\
HEARTSVG& & 331.639 & 330.893 & 339.251 & 325.221 & 333.233 & 336.925 \\
GASTON   & & 321.510 &329.458 &276.258&&&  313.337\\   
StarTrail & &264.873 &333.104& 202.579 &&& 331.050  \\

\midrule
Silhouette & & & & & &&\\ 
IBaySVG & -0.017 &  &  &  &  &  \\
DESpace  & -0.064 &  &  &  &  &  \\
nnSVG   & & -0.043 & -0.058 & -0.014 & -0.064 & -0.057 & -0.071 \\
SPARKX &  & -0.068 & -0.069 & -0.018 & -0.067 & -0.056 & -0.071 \\
SPARK   & & -0.059 & -0.054 & -0.058 & -0.054 & -0.062 & -0.055 \\
spVC    & & -0.063 & -0.064 & -0.043 & -0.068 & -0.055 & -0.031 \\
HEARTSVG& & -0.034 & -0.069 & -0.075 & -0.057 & -0.070 & -0.033 \\
GASTON   & &  0.000 &-0.005 &-0.082 &&& -0.064  \\
StarTrail & &-0.027 &-0.008 &-0.062 &&& -0.010  \\

\midrule
ANOVA F value & & & & & &&\\
IBaySVG & 457.843 &  &  &  &  &  \\
DESpace  & 444.328 &  &  &  &  &  \\
nnSVG  &  & 411.487 & 441.014 & 399.701 & 437.287 & 426.975 & 435.679 \\
SPARKX  & & 461.218 & 452.340 & 435.330 & 460.278 & 444.189 & 452.111 \\
SPARK  &  & 437.994 & 443.503 & 430.438 & 416.783 & 425.232 & 448.138 \\
spVC   &  & 446.049 & 443.350 & 447.047 & 457.597 & 423.190 & 460.638 \\
HEARTSVG && 440.124 & 443.484 & 452.274 & 424.818 & 437.792 & 461.003 \\
GASTON    & & 409.401& 411.240 &434.479 &&& 448.143 \\
StarTrail & & 388.462 &439.376 &257.362&&& 473.320  \\

\bottomrule
\end{tabular}
\label{cluster_table}
\end{table}

\begin{table}[htbp]
\setlength{\belowcaptionskip}{-2pt} 
\centering
\renewcommand{\arraystretch}{1} 
\caption{Comparative evaluation of spot clustering performance via ARI on the across-donor DLPFC dataset. In each part, the single-layer method yields six ARI results based on SV genes 
from each, multiple-slice union set, multiple-slice intersection set, Cauchy combination set, harmonic mean p-value set, and PASTE integration slice.}
\footnotesize
\begin{tabular}{l c cccccc} 
\toprule
\textbf{Sample/Method} & \textbf{Multi-layer} & \textbf{Single-layer} & \textbf{Union} & \textbf{Intersect} &\textbf{Cauchy} &\textbf{HMP} & \textbf{PASTE} \\ 
\midrule
\textbf{Section 1} & & & & & &&\\
IBaySVG  & 0.499 &        &       &       &       &       &       \\
DESpace   & 0.468 &        &       &       &       &       &       \\
Full transcriptome&0.441&       &       &       &       &       &       \\
nnSVG     &      & 0.362 & 0.506 & 0.502 & 0.367 & 0.439 & 0.475 \\
SPARKX    &      & 0.432 & 0.447 & 0.437 & 0.363 & 0.346 & 0.373 \\
SPARK     &      & 0.452 & 0.506 & 0.486 & 0.505 & 0.467 & 0.316 \\
spVC      &      & 0.480 & 0.558 & 0.421 & 0.474 & 0.468 & 0.491 \\
HEARTSVG  &      & 0.458 & 0.457 & 0.383 & 0.423 & 0.457 & 0.449 \\
Gaston &&0.319 &0.527 &0.167 &&&0.352\\
StarTrail  && 0.407 &0.482 &0.428  &&&0.362  \\
\midrule
\textbf{Section 2} & & & & & &&\\
IBaySVG  & 0.497 &       &       &       &       &       &       \\
DESpace   & 0.465 &       &       &       &       &       &       \\
Full transcriptome&0.436&       &       &       &       &       &       \\
nnSVG     &      & 0.412 & 0.508 & 0.466 & 0.360 & 0.411 & 0.430 \\
SPARKX    &      & 0.402 & 0.443 & 0.425 & 0.378 & 0.335 & 0.342 \\
SPARK     &      & 0.504 & 0.489 & 0.496 & 0.507 & 0.474 & 0.259 \\
spVC      &      & 0.380 & 0.560 & 0.454 & 0.486 & 0.486 & 0.453 \\
HEARTSVG  &      & 0.419 & 0.459 & 0.361 & 0.412 & 0.461 & 0.470 \\
Gaston &&0.298 &0.466 &0.183 &&&0.317\\
StarTrail   &&0.346 &0.445 &0.327 &&&0.321  \\
\midrule
\textbf{Section 3} & & & & & &&\\
IBaySVG  & 0.319 &       &       &       &       &       &       \\
DESpace   & 0.308 &       &       &       &       &       &       \\
Full transcriptome&0.269&       &       &       &       &       &       \\
nnSVG     &      & 0.253 & 0.340 & 0.274 & 0.263 & 0.288 & 0.155 \\
SPARKX    &      & 0.238 & 0.287 & 0.202 & 0.244 & 0.199 & 0.205 \\
SPARK     &      & 0.335 & 0.311 & 0.348 & 0.361 & 0.286 & 0.275 \\
spVC      &      & 0.339 & 0.349 & 0.262 & 0.337 & 0.350 & 0.211 \\
HEARTSVG  &      & 0.324 & 0.276 & 0.277 & 0.270 & 0.330 & 0.294 \\
Gaston &&0.212 &0.309 &0.195 &&&0.074\\
StarTrail  && 0.207 &0.181 &0.114   &&&0.325 \\
\midrule
\textbf{Section 4} & & & & & &&\\
IBaySVG  & 0.521 &       &       &       &       &       &       \\
DESpace   & 0.515 &       &       &       &       &       &       \\
Full transcriptome&0.498&       &       &       &       &       &       \\
nnSVG     &      & 0.430 & 0.517 & 0.495 & 0.388 & 0.455 & 0.308 \\
SPARKX    &      & 0.501 & 0.485 & 0.428 & 0.455 & 0.451 & 0.473 \\
SPARK     &      & 0.597 & 0.497 & 0.510 & 0.528 & 0.462 & 0.439 \\
spVC      &      & 0.498 & 0.524 & 0.470 & 0.518 & 0.516 & 0.562 \\
HEARTSVG  &      & 0.452 & 0.463 & 0.399 & 0.394 & 0.476 & 0.482 \\
Gaston &&0.401 &0.532 &0.334 &&&0.362\\
StarTrail   && 0.256& 0.366 &0.266   &&&0.471\\ 

\hline
\end{tabular}
\label{spotcluster_ari_34711}
\end{table}

\begin{table}[htbp]
\setlength{\belowcaptionskip}{-2pt} 
\centering
\renewcommand{\arraystretch}{1} 
\caption{Comparative evaluation of spot clustering performance via NMI on the across-donor DLPFC dataset. In each part, the single-layer method yields six NMI results based on SV genes 
from each, multiple-slice union set, multiple-slice intersection set, Cauchy combination set, harmonic mean p-value set, and PASTE integration slice.}
\footnotesize
\begin{tabular}{l c cccccc} 
\toprule
\textbf{Sample/Method} & \textbf{Multi-layer} & \textbf{Single-layer} & \textbf{Union} & \textbf{Intersect} &\textbf{Cauchy} &\textbf{HMP} & \textbf{PASTE} \\ 
\midrule
\textbf{Section 1} & & & & & & & \\
IBaySVG  & 0.583 &  &  &  &  &  &  \\
DESpace   & 0.565 &  &  &  &  &  &  \\
Full transcriptome&0.578&       &       &       &       &       &       \\
nnSVG     &  & 0.535 & 0.630 & 0.602 & 0.544 & 0.582 & 0.570 \\
SPARKX    &  & 0.607 & 0.558 & 0.595 & 0.549 & 0.543 & 0.557 \\
SPARK     &  & 0.602 & 0.592 & 0.624 & 0.637 & 0.568 & 0.469 \\
spVC      &  & 0.616 & 0.646 & 0.540 & 0.613 & 0.618 & 0.613 \\
HEARTSVG  &  & 0.593 & 0.584 & 0.549 & 0.570 & 0.588 & 0.563 \\
GASTON    &  & 0.483 & 0.600 & 0.269 &  &  & 0.501 \\
StarTrail &  & 0.531 & 0.623 & 0.483 &  &  & 0.528 \\
\midrule
\textbf{Section 2} & & & & & & & \\
IBaySVG  & 0.581 &  &  &  &  &  &  \\
DESpace   & 0.560 &  &  &  &  &  &  \\
Full transcriptome&0.556&       &       &       &       &       &       \\
nnSVG     &  & 0.553 & 0.617 & 0.567 & 0.523 & 0.554 & 0.534 \\
SPARKX    &  & 0.566 & 0.560 & 0.572 & 0.545 & 0.513 & 0.541 \\
SPARK     &  & 0.611 & 0.586 & 0.602 & 0.626 & 0.564 & 0.440 \\
spVC      &  & 0.549 & 0.615 & 0.551 & 0.607 & 0.608 & 0.586 \\
HEARTSVG  &  & 0.564 & 0.571 & 0.530 & 0.540 & 0.580 & 0.572 \\
GASTON    &  & 0.341 & 0.552 & 0.270 &  &  & 0.474 \\
StarTrail &  & 0.463 & 0.584 & 0.397 &  &  & 0.511 \\

\midrule
\textbf{Section 3} & & & & & & & \\
IBaySVG  & 0.432 &  &  &  &  &  &  \\
DESpace   & 0.458 &  &  &  &  &  &  \\
Full transcriptome&0.416&       &       &       &       &       &       \\
nnSVG     &  & 0.359 & 0.444 & 0.385 & 0.367 & 0.397 & 0.316 \\
SPARKX    &  & 0.418 & 0.443 & 0.363 & 0.427 & 0.343 & 0.411 \\
SPARK     &  & 0.452 & 0.477 & 0.455 & 0.459 & 0.459 & 0.361 \\
spVC      &  & 0.478 & 0.458 & 0.419 & 0.441 & 0.476 & 0.389 \\
HEARTSVG  &  & 0.438 & 0.443 & 0.366 & 0.387 & 0.470 & 0.453 \\
GASTON    &  & 0.313 & 0.398 & 0.238 &  &  & 0.282 \\
StarTrail &  & 0.272 & 0.348 & 0.337 &  &  & 0.451 \\

\midrule
\textbf{Section 4} & & & & & & & \\
IBaySVG  & 0.594 &  &  &  &  &  &  \\
DESpace   & 0.620 &  &  &  &  &  &  \\
Full transcriptome&0.605&       &       &       &       &       &       \\
nnSVG     &  & 0.531 & 0.628 & 0.569 & 0.501 & 0.558 & 0.467 \\
SPARKX    &  & 0.606 & 0.584 & 0.546 & 0.567 & 0.527 & 0.565 \\
SPARK     &  & 0.659 & 0.616 & 0.614 & 0.632 & 0.583 & 0.505 \\
spVC      &  & 0.580 & 0.629 & 0.564 & 0.627 & 0.606 & 0.660 \\
HEARTSVG  &  & 0.568 & 0.600 & 0.514 & 0.481 & 0.597 & 0.600 \\
GASTON    &  & 0.516 & 0.614 & 0.359 & 0.486 &  &  \\
StarTrail &  & 0.391 & 0.493 & 0.408 & 0.575 &  &  \\

\hline
\end{tabular}
\label{spotcluster_nmi_34711}
\end{table}

\begin{table}[htbp]
\setlength{\belowcaptionskip}{-2pt} 
\centering
\renewcommand{\arraystretch}{1} 
\caption{Comparative evaluation of spot clustering performance in slice A of across-donor DLPFC dataset. In each part, the single-layer method yields six results based on SV genes 
from slice A, multiple-slice union set, multiple-slice intersection set, Cauchy combination set, harmonic mean p-value set, and PASTE integration slice. 
}
\footnotesize
\begin{tabular}{l c ccccccc} 
\toprule
\textbf{Index/Test} &\textbf{Multi-layer} & \textbf{Single-layer} & \textbf{Union} & \textbf{Intersect}  & \textbf{Cauchy} &\textbf{HMP} & \textbf{PASTE} \\ % 表头
\midrule
DBI        & &  &  &  &  & &\\ 
IBaySVG  & 1.890  &  & & &  & & \\
DEspace   & 1.877  &  & & &  & &\\
nnSVG    & & 1.867 & 1.836 & 1.841 & 1.870 & 1.872 & 1.857 \\
SPARKX    && 1.863 & 1.892 & 1.866 & 1.862 & 1.880 & 1.918 \\
SPARK   &  & 1.864 & 1.870 & 1.838 & 1.837 & 1.879 & 1.883 \\
spVC     & & 1.845 & 1.844 & 1.893 & 1.844 & 1.827 & 1.852 \\
HEARTSVG & & 1.876 & 1.887 & 1.870 & 1.894 & 1.883 & 1.877 \\
StarTrail && 1.877 & 1.858 & 1.795  &  & & 1.888 \\
GASTON  && 1.879 & 1.858 & 1.921  &  &  & 1.881\\
\midrule
CH index & &  &  &  &  & &\\ 
IBaySVG  & 280.737 &  & & &  & & \\
DEspace   & 273.320  &  & & &  & &\\
nnSVG   &  & 260.961 & 285.382 & 279.059 & 254.932 & 255.861 & 259.336 \\
SPARKX  &  & 274.428 & 265.078 & 263.822 & 275.756 & 186.674 & 251.470 \\
SPARK   &  & 265.350 & 281.537 & 289.547 & 292.255 & 282.028 & 241.684 \\
spVC     & & 278.383 & 274.689 & 255.000 & 283.626 & 272.030 & 282.732 \\
HEARTSVG  && 279.724 & 272.272 & 253.663 & 230.189 & 269.804 & 272.383 \\
StarTrail& & 246.441 & 273.808 & 267.672 & &  & 253.423\\
GASTON  & & 257.344 & 300.813 & 113.509  &  &  & 253.607\\
\midrule
Silhouette & &  &  &  &  & &\\ 
IBaySVG  & -0.042 &  & & &  & & \\
DEspace   & -0.054 &  &  & &  & & \\
nnSVG   &  & -0.050 & -0.046 & -0.030 & -0.050 & -0.048 & -0.043 \\
SPARKX   & & -0.049 & -0.062 & -0.049 & -0.055 & -0.083 & -0.066 \\
SPARK   &  & -0.044 & -0.051 & -0.052 & -0.049 & -0.054 & -0.061 \\
spVC    &  & -0.051 & -0.038 & -0.061 & -0.049 & -0.066 & -0.060 \\
HEARTSVG & & -0.063 & -0.052 & -0.047 & -0.080 & -0.066 & -0.056 \\
StarTrail && -0.058 & -0.047 & -0.065  &  & & -0.058\\
GASTON   & & -0.043 & -0.044 & -0.054  &  & & -0.051\\
\midrule
ANOVA F value & &  &  &  &  & &\\
IBaySVG  & 590.968 &  &  &  & &  &\\
DEspace   & 581.021 &  &  &  &  & & \\
nnSVG     &  & 494.730 & 585.586 & 559.600 & 490.153 & 504.089 & 487.717 \\
SPARKX   &  & 535.282 & 524.568 & 507.998 & 557.686 & 396.359 & 512.535 \\
SPARK    &  & 515.250 & 570.237 & 552.453 & 572.621 & 585.827 & 481.792 \\
spVC     &  & 558.474 & 561.754 & 478.739 & 583.174 & 540.555 & 565.476 \\
HEARTSVG &  & 556.886 & 540.334 & 494.897 & 512.157 & 572.184 & 568.886 \\
StarTrail&  & 565.156 & 549.679 & 583.092 &      &      & 508.832 \\
GASTON   &  & 494.710 & 580.950 & 140.514 &     &       & 565.412 \\
\bottomrule
\end{tabular}
\label{cluster_table2_34711}
\end{table}

\begin{table}[htbp]
\setlength{\belowcaptionskip}{-2pt} 
\centering
\renewcommand{\arraystretch}{1} 
\caption{Comparative evaluation of spot clustering performance in slice B of across-donor DLPFC dataset. In each part, the single-layer method yields six results based on SV genes 
from slice B, multiple-slice union set, multiple-slice intersection set, Cauchy combination set, harmonic mean p-value set, and PASTE integration slice. }
\footnotesize
\begin{tabular}{l c cccccc} 
\toprule
\textbf{Index/Test} & \textbf{Multi-layer} & \textbf{Single-layer} & \textbf{Union} & \textbf{Intersect} &\textbf{Cauchy} &\textbf{HMP} & \textbf{PASTE} \\ 
\midrule
DBI & & & & &&& \\  
IBaySVG   & 1.882 &       &       &       &       &     &  \\
DEspace    & 1.891 &       &       &       &       &    &   \\
nnSVG   &   & 1.877 & 1.875 & 1.873 & 1.876 & 1.878 & 1.871 \\
SPARKX  &   & 1.878 & 1.890 & 1.881 & 1.878 & 1.881 & 1.918 \\
SPARK    &  & 1.879 & 1.892 & 1.874 & 1.873 & 1.893 & 1.893 \\
spVC     &  & 1.883 & 1.893 & 1.900 & 1.880 & 1.875 & 1.879 \\
HEARTSVG &  & 1.871 & 1.889 & 1.872 & 1.895 & 1.896 & 1.887 \\
StarTrail & & 1.890 & 1.875 & 1.861  &       &    & 1.895   \\
GASTON   &  & 1.901 & 1.878 & 1.935 &       &      & 1.888  \\
\midrule
CH index & & & & & &&\\ 
IBaySVG   & 233.283 &        &        &        &        &   &     \\
DEspace    & 222.944 &        &        &        &        &    &    \\
nnSVG  &    & 223.624 & 238.818 & 229.057 & 224.994 & 229.694 & 219.452 \\
SPARKX   &  & 225.855 & 239.775 & 230.011 & 229.607 & 196.347 & 219.486 \\
SPARK   &   & 238.636 & 239.137 & 236.206 & 239.353 & 240.738 & 213.744 \\
spVC     &  & 223.173 & 238.833 & 224.476 & 236.773 & 233.835 & 232.202 \\
HEARTSVG  & & 229.539 & 227.457 & 224.573 & 197.283 & 233.571 & 239.664 \\
StarTrail & & 204.475 & 231.365 & 208.605  &        &     & 219.835   \\
GASTON   &  & 124.567 & 235.116 & 103.097  &        &   & 208.429     \\
\midrule
Silhouette & & & & & &&\\ 
IBaySVG   & -0.043 &        &        &        &        &   &     \\
DEspace    & -0.053 &        &        &        &        &    &    \\
nnSVG   &   & -0.045 & -0.044 & -0.013 & -0.046 & -0.045 & -0.044 \\
SPARKX  &   & -0.049 & -0.057 & -0.047 & -0.056 & -0.061 & -0.064 \\
SPARK   &   & -0.044 & -0.054 & -0.045 & -0.044 & -0.055 & -0.054 \\
spVC     &  & -0.052 & -0.031 & -0.056 & -0.047 & -0.057 & -0.009 \\
HEARTSVG  & & -0.048 & -0.051 & -0.047 & -0.072 & -0.064 & -0.056 \\
StarTrail&  & -0.065 & -0.061 & -0.056  &        &     & -0.059   \\
GASTON  & & -0.028 & -0.009 & -0.057  &        &     & -0.058   \\
\midrule
ANOVA F value & & & & & &&\\
IBaySVG   & 422.359 &         &    &     &         &         &         \\
DEspace    & 431.884 &         &    &     &         &         &         \\
nnSVG      &   & 413.822 & 442.336 & 397.758 & 417.978 & 430.313 & 402.513 \\
SPARKX     &   & 419.362 & 452.815 & 426.088 & 453.511 & 407.033 & 425.044 \\
SPARK      &   & 430.688 & 448.123 & 421.140 & 434.930 & 463.988 & 405.598 \\
spVC       &   & 419.506 & 438.227 & 411.516 & 429.730 & 426.189 & 418.307 \\
HEARTSVG   &   & 426.816 & 420.001 & 421.646 & 417.440 & 442.326 & 448.922 \\
StarTrail  &   & 440.629 & 453.804 & 440.714 &         &         & 421.153 \\
GASTON     &   & 154.595 & 427.379 & 146.322 &         &         & 445.939 \\
\bottomrule
\end{tabular}
\label{cluster_table3_34711}
\end{table}

\begin{table}[htbp]
\setlength{\belowcaptionskip}{-2pt}
\centering
\renewcommand{\arraystretch}{1}
\caption{Comparative evaluation of spot clustering performance in slice C of across-donor DLPFC dataset. In each part, the single-layer method yields six results based on SV genes 
from slice C, multiple-slice union set, multiple-slice intersection set, Cauchy combination set, harmonic mean p-value set, and PASTE integration slice.}
\footnotesize
\begin{tabular}{l c cccccc}
\toprule
\textbf{Index/Test} & \textbf{Multi-layer} & \textbf{Single-layer} & \textbf{Union} & \textbf{Intersect}  & \textbf{Cauchy} &\textbf{HMP} & \textbf{PASTE}\\ 

\midrule
DBI & & & & & &&\\  
IBaySVG   & 1.887 &       &       &       &       &    &   \\
DEspace    & 1.899 &       &       &       &       &    &   \\
nnSVG  &    & 1.883 & 1.878 & 1.940 & 1.883 & 1.885 & 1.891 \\
SPARKX   &  & 1.877 & 1.911 & 1.883 & 1.892 & 1.894 & 1.902 \\
SPARK    &  & 1.914 & 1.894 & 1.885 & 1.872 & 1.873 & 1.879 \\
spVC     &  & 1.861 & 1.903 & 1.913 & 1.895 & 1.912 & 1.938 \\
HEARTSVG &  & 1.901 & 1.871 & 1.899 & 1.894 & 1.867 & 1.877 \\
StarTrail & & 1.927 & 1.902 & 1.896  &       &  & 1.890     \\
GASTON    &  & 1.925 & 1.909 & 1.954  &       &  & 1.916     \\
\midrule
CH index & & & & & &&\\ 
IBaySVG   & 342.485 &         &         &         &         &    &     \\
DEspace    & 359.165 &         &         &         &         &     &    \\
nnSVG  &    & 264.476 & 305.565 & 242.616 & 284.797 & 288.675 & 249.729 \\
SPARKX   &  & 262.540 & 320.463 & 272.192 & 273.988 & 263.974 & 323.210 \\
SPARK    &  & 330.730 & 347.654 & 291.937 & 296.119 & 348.305 & 270.457 \\
spVC     &  & 322.623 & 328.807 & 298.167 & 318.354 & 322.669 & 241.735 \\
HEARTSVG  & & 326.271 & 332.009 & 292.365 & 266.492 & 345.524 & 334.331 \\
StarTrail & & 283.430 & 329.958 & 288.124 &         &      & 308.626    \\
GASTON   & & 219.841 & 308.024 & 101.118 &         &       & 316.486   \\
\midrule
Silhouette & & & & & &&\\ 
IBaySVG   & -0.056 &        &        &        &        &      &  \\
DEspace    & -0.068 &        &        &        &        &     &   \\
nnSVG   &   & -0.054 & -0.081 & -0.087 & -0.081 & -0.076 & -0.105 \\
SPARKX   &  & -0.085 & -0.089 & -0.080 & -0.083 & -0.097 & -0.095 \\
SPARK   &   & -0.084 & -0.070 & -0.085 & -0.091 & -0.069 & -0.065 \\
spVC     &  & -0.071 & -0.062 & -0.091 & -0.092 & -0.086 & -0.092 \\
HEARTSVG  & & -0.084 & -0.070 & -0.072 & -0.084 & -0.069 & -0.073 \\
StarTrail & & -0.082 & -0.061 & -0.041  &        &     & -0.081   \\
GASTON   &  & -0.099 & -0.071 & -0.102  &        &    & -0.081    \\
\midrule
ANOVA F value && & & & &&\\
IBaySVG   & 465.301 &         &         &         &         &         &       \\
DEspace    & 447.211 &         &         &         &         &         &       \\
nnSVG      &         & 328.948 & 402.036 & 328.647 & 360.608 & 375.183 & 362.637 \\
SPARKX     &         & 343.553 & 417.823 & 339.576 & 353.371 & 345.206 & 431.708 \\
SPARK      &         & 450.091 & 447.009 & 403.733 & 412.483 & 443.942 & 378.587 \\
spVC       &         & 423.637 & 459.615 & 404.697 & 423.533 & 426.670 & 320.778 \\
HEARTSVG   &         & 448.918 & 420.159 & 385.102 & 347.630 & 444.804 & 438.596 \\
StarTrail  &         & 396.933 & 452.012 & 373.822 &         &         & 424.881 \\
GASTON     &         & 287.993 & 407.670 & 137.686 &         &         & 364.511 \\
\bottomrule
\end{tabular}
\label{cluster_table4_34711}
\end{table}

\begin{table}[htbp]
\setlength{\belowcaptionskip}{-2pt} 
\centering
\renewcommand{\arraystretch}{1} 
\caption{Comparative evaluation of spot clustering performance in slice D of across-donor DLPFC dataset. In each part, the single-layer method yields six results based on SV genes 
from slice D, multiple-slice union set, multiple-slice intersection set, Cauchy combination set, harmonic mean p-value set, and PASTE integration slice.}
\footnotesize
\begin{tabular}{l c cccccc} 
\toprule
\textbf{Index/Test} & \textbf{Multi-layer} & \textbf{Single-layer} & \textbf{Union} & \textbf{Intersect}  & \textbf{Cauchy} &\textbf{HMP} & \textbf{PASTE}\\ 
\midrule
DBI & & & & & &&\\  
IBaySVG   & 1.829 &       &       &       &       &      & \\
DEspace    & 1.832 &       &       &       &       &    &   \\
nnSVG    &  & 1.882 & 1.895 & 1.879 & 1.878 & 1.873 & 1.851 \\
SPARKX   &  & 1.877 & 1.870 & 1.894 & 1.836 & 1.886 & 1.898 \\
SPARK    &  & 1.894 & 1.821 & 1.889 & 1.896 & 1.800 & 1.849 \\
spVC      & & 1.877 & 1.853 & 1.888 & 1.878 & 1.854 & 1.886 \\
HEARTSVG  & & 1.882 & 1.829 & 1.880 & 1.908 & 1.843 & 1.849 \\
StarTrail&  & 1.831 & 1.849 & 1.795  &       &   & 1.873    \\
GASTON   &  & 1.869 & 1.855 & 1.868  &       &    & 1.860   \\
\midrule
CH index & & & & & &&\\ 
IBaySVG   & 370.039 &         &         &         &         &      &   \\
DEspace    & 341.191 &         &         &         &         &      &   \\
nnSVG   &   & 349.909 & 380.447 & 339.319 & 330.357 & 359.983 & 363.273 \\
SPARKX   &  & 438.767 & 444.389 & 369.385 & 457.982 & 245.580 & 381.430 \\
SPARK    &  & 363.714 & 430.031 & 372.903 & 373.493 & 444.737 & 379.301 \\
spVC     &  & 344.420 & 377.786 & 323.309 & 380.265 & 355.394 & 387.299 \\
HEARTSVG &  & 359.858 & 378.094 & 349.316 & 313.342 & 390.492 & 438.553 \\
StarTrail & & 407.627 & 389.462 & 360.486  &         &     & 355.772    \\
GASTON    & & 378.473 & 372.546 & 259.973  &         &      & 366.641   \\
\midrule
Silhouette & & & & & &&\\ 
IBaySVG   & -0.065 &        &        &        &        &    &    \\
DEspace    & -0.078 &        &        &        &        &   &     \\
nnSVG   &   & -0.052 & -0.054 & -0.044 & -0.063 & -0.061 & -0.073 \\
SPARKX   &  & -0.079 & -0.079 & -0.042 & -0.064 & -0.087 & -0.066 \\
SPARK   &   & -0.029 & -0.078 & -0.050 & -0.051 & -0.080 & -0.055 \\
spVC     &  & -0.063 & -0.055 & -0.080 & -0.053 & -0.073 & -0.037 \\
HEARTSVG  & & -0.055 & -0.078 & -0.063 & -0.079 & -0.073 & -0.072 \\
StarTrail & & -0.039 & -0.069 & -0.067 &        &      & -0.069   \\
GASTON    & & -0.058 & -0.054 & -0.050  &        &   & -0.073     \\
\midrule
ANOVA F value & & & & & &&\\
IBaySVG   & 287.245 &         &         &         &         &     &    \\
DEspace    & 306.824 &         &         &         &         &    &     \\
nnSVG    &  & 225.842 & 231.844 & 200.715 & 208.023 & 237.893 & 272.480 \\
SPARKX   &  & 319.672 & 326.140 & 240.523 & 366.124 & 147.187 & 246.678 \\
SPARK    &  & 216.410 & 300.626 & 221.644 & 225.856 & 332.837 & 266.423 \\
spVC     &  & 206.232 & 227.275 & 187.670 & 232.723 & 216.125 & 230.825 \\
HEARTSVG  & & 230.465 & 262.950 & 226.532 & 227.196 & 246.335 & 321.703 \\
StarTrail & & 320.405 & 305.033 & 278.978  &         &      & 217.487   \\
GASTON   & & 243.473 & 231.413 & 134.362 &         &     & 252.461    \\
\bottomrule
\end{tabular}
\label{cluster_table_34711}
\end{table}

\begin{table}[htbp]
\setlength{\belowcaptionskip}{-2pt} 
\centering
\renewcommand{\arraystretch}{1} 
\caption{Comparative evaluation of spot clustering performance in slice A of SCC dataset. In each part, the single-layer method yields six results based on SV genes 
from slice A, multiple-slice union set, multiple-slice intersection set, Cauchy combination set, harmonic mean p-value set, and PASTE integration slice.}
\footnotesize
\begin{tabular}{l c cccccc} 
\toprule
\textbf{Index/Test} & \textbf{Multi-layer} & \textbf{Single-layer} & \textbf{Union} & \textbf{Intersect} & \textbf{Cauthy} &\textbf{HMP} & \textbf{PASTE} \\ 

\midrule
DBI & & & & & & \\
IBaySVG  & 1.930  & & & & &  \\
DESpace   & 1.771  & & & & &  \\
nnSVG    & & 1.899 & 1.863 & 1.828 & 1.876 & 1.832 & 1.927 \\
SPARKX   & & 1.935 & 1.839 & 1.835 & 1.887 & 1.850 & 1.907 \\
SPARK    & & 1.872 & 1.819 & 1.863 & 1.827 & 1.727 & 1.782 \\
spVC     & & 1.873 & 1.905 & 1.796 & 1.871 & 1.786 & 1.808 \\
HEARTSVG & & 1.939 & 1.916 & 1.864 & 1.946 & 1.890 & 1.858 \\
GASTON   & & 1.778 & 1.742 & 1.920 &&& 1.838 \\
StarTrail && 1.832 & 1.864 & 1.913 &&& 1.855 \\

\midrule
CH index & & & & & & \\
IBaySVG  & 80.581  & & & & & \\
DESpace   & 65.629  & & & & & \\
nnSVG   &  & 62.219 & 72.956 & 53.417 & 82.779 & 67.504 & 67.283 \\
SPARKX  &  & 69.343 & 69.688 & 83.039 & 70.510 & 70.531 & 71.147 \\
SPARK    & & 80.234 & 65.215 & 57.427 & 69.861 & 68.737 & 70.299 \\
spVC    &  & 64.906 & 67.356 & 41.712 & 62.327 & 53.172 & 69.269 \\
HEARTSVG & & 36.560 & 57.531 & 54.655 & 83.452 & 50.091 & 52.732 \\
GASTON  &  & 74.152 & 62.888 & 39.975 &&& 54.762 \\
StarTrail && 69.172 & 66.808 & 42.001 &&& 58.370 \\

\midrule
Silhouette & & & & & & \\
IBaySVG  & -0.120  & & & & & \\
DESpace   & -0.179  & & & & & \\
nnSVG   &  & -0.148 & -0.136 & -0.175 & -0.134 & -0.219 & -0.148 \\
SPARKX   & & -0.159 & -0.104 & -0.180 & -0.090 & -0.108 & -0.120 \\
SPARK   &  & -0.102 & -0.172 & -0.180 & -0.161 & -0.204 & -0.186 \\
spVC     & & -0.165 & -0.160 & -0.197 & -0.169 & -0.221 & -0.178 \\
HEARTSVG & & -0.278 & -0.184 & -0.297 & -0.136 & -0.263 & -0.188 \\
GASTON   & & -0.183 & -0.231 & -0.205 &&& -0.204 \\
StarTrail& & -0.147 & -0.131 & -0.217 &&& -0.134 \\

\midrule
ANOVA F value & & & & & & \\
IBaySVG  & 121.104  & & & & & \\
DESpace   & 87.035  & & & & & \\
nnSVG    & & 83.105 & 105.125 & 75.527 & 110.280 & 89.183 & 91.676 \\
SPARKX   & & 95.423 & 93.492 & 111.022 & 94.277 & 96.151 & 96.783 \\
SPARK   &  & 106.841 & 86.952 & 71.566 & 89.084 & 88.718 & 90.619 \\
spVC    &  & 85.503 & 88.752 & 57.601 & 80.296 & 68.723 & 91.180 \\
HEARTSVG & & 57.064 & 82.896 & 67.344 & 120.024 & 77.479 & 74.858 \\
GASTON   & & 101.076 & 82.255 & 61.865 &&& 70.877 \\
StarTrail& & 92.684 & 84.384 & 57.603 & &&77.558 \\

\bottomrule
\end{tabular}
\label{cluster_table_scc1}
\end{table}

\begin{table}[htbp]
\setlength{\belowcaptionskip}{-2pt} 
\centering
\renewcommand{\arraystretch}{1} 
\caption{Comparative evaluation of spot clustering performance in slice B of SCC dataset. In each part, the single-layer method yields six results based on SV genes 
from slice B, multiple-slice union set, multiple-slice intersection set, Cauchy combination set, harmonic mean p-value set, and PASTE integration slice.}
\footnotesize
\begin{tabular}{l c cccccc} 
\toprule
\textbf{Index/Test} & \textbf{Multi-layer} & \textbf{Single-layer} & \textbf{Union} & \textbf{Intersect} & \textbf{Cauthy} &\textbf{HMP}  & \textbf{PASTE}\\ 

\midrule
DBI & & & & & & \\
IBaySVG  & 1.912 & & & & &  \\
DESpace   & 1.799  & & & & & \\
nnSVG    & & 1.885 & 1.870 & 1.847 & 1.882 & 1.816 & 1.876 \\
SPARKX  & & 1.835 & 1.796 & 1.895 & 1.854 & 1.884 & 1.851 \\
SPARK   &  & 1.864 & 1.843 & 1.914 & 1.883 & 1.910 & 1.906 \\
spVC    &  & 1.886 & 1.880 & 1.891 & 1.866 & 1.865 & 1.808 \\
HEARTSVG & & 1.961 & 1.920 & 1.923 & 1.883 & 1.901 & 1.922 \\
GASTON   & & 1.784 & 1.814 & 1.888 & &&1.832 \\
StarTrail& & 1.865 & 1.895 & 1.794 & &&1.853 \\

\midrule
CH index & & & & & & \\
IBaySVG  & 87.192  & & & & & \\
DESpace   & 65.828  & & & & & \\
nnSVG   &  & 69.481 & 63.910 & 55.480 & 70.101 & 61.398 & 65.013 \\
SPARKX   & & 64.704 & 64.305 & 65.699 & 64.445 & 71.320 & 62.787 \\
SPARK    & & 66.798 & 65.784 & 62.964 & 67.205 & 77.865 & 70.402 \\
spVC     & & 63.946 & 72.113 & 44.785 & 56.595 & 58.768 & 66.904 \\
HEARTSVG & & 28.953 & 59.526 & 36.390 & 65.363 & 22.555 & 57.429 \\
GASTON   & & 51.486 & 58.979 & 38.822 &&& 63.116 \\
StarTrail& & 51.990 & 65.061 & 44.385 & &&53.608 \\

\midrule
Silhouette & & & & & & \\
IBaySVG  & -0.118 & & & & &  \\
DESpace   & -0.159  & & & & & \\
nnSVG   &  & -0.099 & -0.101 & -0.186 & -0.132 & -0.088 & -0.149 \\
SPARKX  &  & -0.175 & -0.110 & -0.151 & -0.111 & -0.095 & -0.182 \\
SPARK   &  & -0.148 & -0.162 & -0.169 & -0.159 & -0.155 & -0.166 \\
spVC     & & -0.150 & -0.164 & -0.219 & -0.183 & -0.195 & -0.136 \\
HEARTSVG & & -0.237 & -0.159 & -0.210 & -0.126 & -0.271 & -0.161 \\
GASTON   & & -0.156 & -0.135 & -0.168 &&& -0.172 \\
StarTrail& & -0.179 & -0.149 & -0.198 &&& -0.131 \\

\midrule
ANOVA F value & & & & & & \\
IBaySVG  & 110.760  & & & & & \\
DESpace   & 87.798  & & & & & \\
nnSVG   &  & 95.615 & 88.384 & 84.225 & 95.063 & 78.664 & 83.089 \\
SPARKX  &  & 87.053 & 87.964 & 86.565 & 87.173 & 98.279 & 83.584 \\
SPARK   &  & 93.077 & 86.460 & 85.404 & 89.663 & 98.993 & 94.639 \\
spVC     & & 84.929 & 88.700 & 63.374 & 75.891 & 76.061 & 85.066 \\
HEARTSVG & & 40.651 & 87.876 & 54.487 & 96.614 & 31.570 & 81.625 \\
GASTON  &  & 70.360 & 79.152 & 66.395 &&& 79.566 \\
StarTrail& & 73.445 & 83.226 & 60.402 & &&70.032 \\
\bottomrule
\end{tabular}
\label{cluster_table_scc2}
\end{table}

\begin{table}[htbp]
\setlength{\belowcaptionskip}{-2pt} 
\centering
\renewcommand{\arraystretch}{1} 
\renewcommand{\tabcolsep}{0.2pc}
\caption{Comparative evaluation of spot clustering performance in slice C of SCC dataset. In each part, the single-layer method yields six results based on SV genes 
from slice C, multiple-slice union set, multiple-slice intersection set, Cauchy combination set, harmonic mean p-value set, and PASTE integration slice.}
\footnotesize
\begin{tabular}{l c cccccc} 
\toprule
\textbf{Index/Test} & \textbf{Multi-layer} & \textbf{Single-layer} & \textbf{Union} & \textbf{Intersect} & \textbf{Cauthy} &\textbf{HMP} & \textbf{PASTE} \\ 

\midrule
DBI & & & & & & &\\  
IBaySVG  & 1.774 &       &       &       &       &       \\
DESpace   & 1.747 &       &       &       &       &       \\
nnSVG  &   & 1.840 & 1.861 & 1.837 & 1.880 & 1.882 & 1.872 \\
SPARKX  &  & 1.814 & 1.796 & 1.907 & 1.878 & 1.865 & 1.886 \\
SPARK  &   & 1.876 & 1.864 & 1.914 & 1.885 & 1.883 & 1.912 \\
spVC   &   & 1.841 & 1.860 & 1.807 & 1.897 & 1.847 & 1.903 \\
HEARTSVG & & 1.932 & 1.889 & 1.919 & 1.924 & 1.981 & 1.877 \\
GASTON   & & 1.602 & 1.845 & 1.793 &  &       &   1.823    \\
StarTrail& & 1.798 & 1.870 & 1.820 &  &       &   1.838    \\

\midrule
CH index & & & & & & &\\ 
IBaySVG  & 91.830 &        &        &        &        &        \\
DESpace   & 69.755 &        &        &        &        &        \\
nnSVG   &  & 60.449 & 76.729 & 50.088 & 83.379 & 80.280 & 69.522 \\
SPARKX  &  & 82.807 & 74.802 & 69.391 & 83.040 & 82.438 & 80.698 \\
SPARK   &  & 53.186 & 63.957 & 57.478 & 73.410 & 65.640 & 68.036 \\
spVC    &  & 59.446 & 66.883 & 51.615 & 55.418 & 59.398 & 69.028 \\
HEARTSVG & & 71.520 & 61.846 &  8.971 & 85.615 & 21.824 & 62.205 \\
GASTON   & & 57.630 & 60.828 & 27.862 & &        &     54.001    \\
StarTrail && 50.639 & 58.347 & 49.627 &  &        &   61.644     \\

\midrule
Silhouette & & & & & & &\\ 
IBaySVG  & -0.009 &        &        &        &        &        \\
DESpace   & -0.071 &        &        &        &        &        \\
nnSVG   &  & -0.073 & -0.040 & -0.077 & -0.049 & -0.047 & -0.080 \\
SPARKX  &  & -0.083 & -0.056 & -0.076 & -0.021 &  0.041 & -0.062 \\
SPARK   &  & -0.074 & -0.117 & -0.075 & -0.072 & -0.072 & -0.070 \\
spVC    &  & -0.117 & -0.089 & -0.121 & -0.060 & -0.158 & -0.064 \\
HEARTSVG & & -0.077 & -0.079 & -0.006 & -0.059 & -0.279 & -0.072 \\
GASTON   & & -0.106 & -0.032 & -0.179 & &        &    -0.147     \\
StarTrail& & -0.104 & -0.083 & -0.105 &  &        &   -0.057     \\

\midrule
ANOVA F value & & & & & & &\\
IBaySVG  & 195.686 &        &        &        &        &        \\
DESpace   & 151.413 &        &        &        &        &        \\
nnSVG   &  & 129.321 & 164.910 & 110.500 & 185.711 & 179.476 & 147.973 \\
SPARKX  &  & 181.518 & 165.018 & 153.889 & 186.137 & 183.262 & 179.376 \\
SPARK   &  & 109.783 & 138.124 & 108.219 & 156.238 & 136.135 & 145.226 \\
spVC    &  & 121.699 & 142.160 & 105.920 & 105.262 & 124.796 & 141.102 \\
HEARTSVG & & 157.855 & 140.419 &  14.567 & 186.032 &  44.712 & 137.826 \\
GASTON   & & 120.250 & 135.790 &  58.663 &  &        &  112.120      \\
StarTrail && 108.722 & 128.950 & 103.269 &  &        &   130.979     \\

\bottomrule
\end{tabular}
\label{cluster_table_scc3}
\end{table}

\newpage

\subsection{Contribution of IBaySVG‑unique genes to spatial domain recognition}\label{IBaySVG‑unique}

As an indirect validation of the biological significance of our uniquely identified SV genes, we evaluate their impact on spatial domain identification. To this end, we conduct spatial clustering with the complete set of SV genes selected by IBaySVG, and subsequently repeat it after removing those unique genes. Here, unique genes are defined as IBaySVG‑selected but not selected by any other method, after first excluding overly permissive methods (GASTON for with‑donor DLPFC; SPARKX and GASTON for across‑donor DLPFC; SPARKX for SCC) to ensure meaningful specificity. The clustering outcomes are compared using the same set of metrics, namely the ARI, NMI, DBI, CH index, silhouette coefficient, and ANOVA‑based F‑statistic.

As summarized in Tables \ref{unique_domain1}, \ref{unique_domain2} and \ref{unique_domain3}, removing the uniquely identified SV genes generally degrades clustering performance across all datasets. In both DLPFC datasets, substantial reductions are consistently observed in ARI and NMI across all tissue sections, accompanied by decreases in the CH index and ANOVA statistic and less favorable silhouette coefficients. Although individual internal clustering metrics exhibit minor fluctuations in several SCC sections, the overall clustering quality likewise deteriorates after removing the uniquely identified SV genes. These results indicate that the SV genes uniquely identified by IBaySVG provide biologically meaningful spatial information that is not fully captured by competing methods and contribute directly to improved spatial domain identification.

\begin{table}[htbp]
\centering
\small
\caption{Effect of removing uniquely identified SV genes on spatial domain identification in the DLPFC same-donor dataset.}
\begin{tabular}{lcccccc}
\toprule
Slice & ARI & NMI & DBI & CH & Silhouette & ANOVA \\
\midrule
Sample 1 (Original) & 0.506 & 0.619 & 1.858 &  327.965 & -0.034 & 412.684  \\
Sample 1 (Removed)  & 0.395 & 0.575 & 1.887 & 322.434 & -0.076 & 392.607 \\

Sample 2 (Original) & 0.499 & 0.609 & 1.853 & 410.122 & -0.038 & 476.987 \\
Sample 2 (Removed)  & 0.423 & 0.584 & 1.892 & 283.861 & -0.095 & 334.791 \\

Sample 3 (Original) & 0.567 & 0.650 & 1.827 &  518.117 &  -0.034  & 574.263  \\
Sample 3 (Removed)  & 0.398 & 0.595 & 1.895 & 362.306 & -0.079 &  509.073\\

Sample 4 (Original) & 0.567 & 0.627 & 1.872 & 421.654 & -0.017 & 457.843  \\
Sample 4 (Removed)  & 0.393 & 0.595 & 1.868 & 327.606 & -0.072 &  432.524\\
\bottomrule
\end{tabular}
\label{unique_domain1}
\end{table}

\begin{table}[htbp]
\centering
\small
\caption{Effect of removing uniquely identified SV genes on spatial domain identification in the DLPFC across-donor dataset.}
\begin{tabular}{lcccccc}
\toprule
Slice & ARI & NMI & DBI & CH & Silhouette & ANOVA \\
\midrule
Sample 1 (Original) & 0.499 & 0.583 & 1.890 & 280.737 & -0.042 & 590.968  \\
Sample 1 (Removed)  &0.340  &0.527  & 1.804 & 242.823 & -0.092 & 509.322 \\

Sample 2 (Original) & 0.497 & 0.581 & 1.882 & 233.283 & -0.043 & 422.359  \\
Sample 2 (Removed)  & 0.330 &0.509  & 1.914 & 248.117 & -0.062 &  520.157 \\

Sample 3 (Original) & 0.319 & 0.432 & 1.887 & 342.485 & -0.056 & 465.301  \\
Sample 3 (Removed)  &0.185  & 0.383 & 1.71 & 286.4 & -0.093 & 380.828 \\

Sample 4 (Original) & 0.521 & 0.594 & 1.829 & 370.039 & -0.065 & 287.245 \\
Sample 4 (Removed)  &0.408  & 0.484 & 1.866 & 368.477 & -0.09 & 254.793 \\
\bottomrule
\end{tabular}
\label{unique_domain2}
\end{table}

\begin{table}[htbp]
\centering
\small
\caption{Effect of removing uniquely identified SV genes on spatial domain identification in the scc dataset.}
\begin{tabular}{lcccccc}
\toprule
Slice & ARI & NMI & DBI & CH & Silhouette & ANOVA \\
\midrule
Sample 1 (Original) &--  &--  & 1.930 & 80.581 & -0.120 & 121.104  \\
Sample 1 (Removed)  & -- & -- & 1.939 & 78.281 & -0.170 & 120.042\\
Sample 2 (Original) & -- & -- & 1.912 & 87.192 & -0.118 & 110.760  \\
Sample 2 (Removed)  & -- & -- & 1.896 & 74.614 & -0.141 & 102.712\\
Sample 3 (Original) & -- & -- & 1.774 & 91.830 & -0.009 & 195.686  \\
Sample 3 (Removed)  & -- & -- & 1.906 & 89.340 & -0.028 & 194.508 \\
\bottomrule
\end{tabular}
\label{unique_domain3}
\end{table}

\clearpage

\bibliographystyle{agsm}

\bibliography{Bibliography-MM-MC}